\documentclass[journal]{IEEEtran}
\usepackage[utf8]{inputenc}
\usepackage[T1]{fontenc}
\usepackage{textcomp}
\usepackage{amsmath,amssymb,amsthm}
\usepackage{graphicx}
\usepackage{booktabs}
\usepackage{array}
\usepackage{longtable}
\usepackage{algorithm}
\usepackage{algorithmic}
\usepackage{xcolor}
\usepackage{cite}
\usepackage[hidelinks]{hyperref}
\newtheorem{theorem}{Theorem}
\newtheorem{corollary}{Corollary}
\newtheorem{proposition}{Proposition}

\newcommand{\algname}{SN-EDA}
\begin{document}
\title{SN-EDA: Skew-Normal Estimation of Distribution Algorithms}
\author{Jun~Zhang,~\IEEEmembership{Fellow,~IEEE}, Tong~Qian, Hua~Ding, Weijie~Yu, and Zhi-Hui~Zhan,~\IEEEmembership{Fellow,~IEEE}%
\thanks{This paper is version~2 of arXiv:2611.26938. Version~1 was titled ``MG-EDA: Maturity-Gated Skew-Normal Estimation of Distribution Algorithms.''}%
\thanks{This work was supported in part by the National Research Foundation of Korea under Grants RS-2025-00555463 and RS-2025-25456394. (Corresponding author: Jun Zhang.)}%
\thanks{Jun Zhang is with Hanyang University ERICA, South Korea; Nankai University, China; and Chaoyang University of Technology, Taiwan (e-mail: junzhang@ieee.org).}%
\thanks{Tong Qian, Hua Ding, and Zhi-Hui Zhan are with Nankai University, Tianjin, China (e-mail: qiantongnku@163.com; dinghua\_ieee@163.com).}%
\thanks{Weijie Yu is with Sun Yat-sen University, Guangzhou, China (e-mail: yuweijie6@mail.sysu.edu.cn).}}
\markboth{Preprint}{%
Skew-Normal Estimation of Distribution Algorithms}
\maketitle
\begin{abstract}
Gaussian estimation of distribution algorithms (EDAs) converge prematurely on slopes, because truncation selection removes most of the variance along the gradient. The usual remedies enlarge the variance or shift the mean. We propose the skew-normal EDA (SN-EDA), which corrects the third moment instead. It samples from a skew-normal distribution whose skewness points along the cumulative path of the mean shifts, while the mean and covariance remain exactly the maximum-likelihood estimates. The estimation step is unchanged, and the skewness needs no control. On a linear function it raises the fraction of the variance that survives selection from 0.29 to 0.53. In a pre-registered test on 24 benchmark functions in 10, 20 and 40 dimensions the skewness, applied through a gate that the test finds dispensable, improves the Gaussian EDA significantly in five of six settings. When the population starts far from the optimum, the error, averaged logarithmically over a run, falls by one to three orders of magnitude, and the gain grows with the magnitude of the skewness as predicted. The skewness adds nothing to the anticipated mean shift, whose offspring distribution we show to be skewed already, and SN-EDA remains behind algorithms that also correct the second moment.
\end{abstract}
\begin{IEEEkeywords}
Anticipated mean shift, estimation of distribution algorithms, matrix-based evolutionary computation, premature convergence, skew-normal distribution, third moment
\end{IEEEkeywords}
\IEEEpeerreviewmaketitle

\section{Introduction}
\label{sec:intro}

Estimation of distribution algorithms (EDAs) replace variation operators by a probability model that is learned from the selected solutions and sampled to produce the next population \cite{muhlenbein1996,larranaga2002,hauschild2011,larranaga2024}. In continuous search spaces the most common model is a Gaussian whose mean and covariance are the maximum-likelihood estimates from the selected solutions; with a full covariance matrix this is the estimation of multivariate normal algorithm (EMNA) \cite{larranaga2002}, which current EDA libraries still provide \cite{soloviev2024b}. Gaussian EDAs are known to converge prematurely when the population has to travel along a slope: under truncation selection the variance along the gradient shrinks by a constant factor per generation, so the total progress is bounded \cite{grahl2005,posik2008}. The standard remedies act on the first two moments of the sampling distribution. Adaptive variance scaling (AVS) enlarges the covariance, with triggers based on correlation or on the standard-deviation ratio \cite{grahl2006,bosman2007}; the anticipated mean shift (AMS) moves part of the offspring along the recent shift of the mean \cite{bosman2008}; together they form AMaLGaM \cite{bosman2009,bosman2013}. The covariance matrix adaptation evolution strategy (CMA-ES) lengthens its covariance along an evolution path and controls its step size with a second one \cite{hansen2001,hansen2016}; recent variants also adapt its learning rates \cite{nomura2025}.

The situation these remedies address is asymmetric. On a slope all improvement lies on one side of the mean, and the solutions that survive truncation are themselves a strongly skewed sample, dense at the truncation threshold with a long tail toward better values. A Gaussian fitted to them discards this shape: its tail in the direction of improvement is as short as its tail in the opposite direction, and the next truncation cuts the variance along the gradient by the same factor again. Sampling from a distribution that is skewed toward the direction of movement is a natural alternative. It has a precedent in the directed mutation of evolution strategies, which draws mutations from a skew-normal distribution \cite{berlik2005,berlik2006}, but its value for EDAs has not been established. What does a skewed sampler do to the variance loss on a slope? How much is it worth on benchmark problems? And what does it add to the corrections of the first two moments that are already in use? This paper proposes such a sampler and answers these three questions.

We propose the skew-normal EDA (\algname{}). Its sampler is a standardized form of the multivariate skew-normal distribution \cite{azzalini1996,azzalini1999,azzalini2014} whose mean and covariance are exactly the maximum-likelihood estimates of EMNA, whatever the direction and magnitude of its skewness (Theorem \ref{thm:mom}). The sampler therefore changes the third moment and nothing below it: the estimation step of EMNA is untouched, and any difference between the two algorithms is due to the asymmetry of the sampling distribution alone. The skewness points along a cumulative path of the mean shifts, because the third moment of the selected solutions is too noisy to be estimated at the population sizes in use, and its magnitude is a constant close to the upper limit of the family. No statistic, threshold or schedule decides when the skewness acts. Since it biases neither estimated moment, it is switched on in the second generation and stays on, so that \algname{} is EMNA with one changed sampling step.

The findings are of three kinds. \emph{In theory}, on a linear function with an infinite population, the skewness raises the fraction of the variance along the gradient that survives truncation from 0.288 to 0.534 and the progress that is possible without amplification from 2.28 to 4.12 standard deviations, and it lowers the amplification that is needed to keep moving from 3.47 to 1.87 (Section \ref{sec:theory}). \emph{In experiments} on the 24 BBOB functions \cite{hansen2009,hansen2021}, with hypotheses formulated on a training set and tested under a protocol fixed in advance on instances not used before (10, 20 and 40 dimensions, two starting conditions), the skewness improves EMNA with variance scaling significantly in five of six settings. The effect is large when the population starts far from the optimum, where the anytime performance, a logarithmic measure of the error, improves by 1.0, 3.3 and 3.3 for $D=10$, 20 and 40; it follows the predicted dose--response relation, and with the standard start it is concentrated on the linear slope of the benchmark. The protocol tested the sampler with a gate that scales the skewness by a calibrated measure of movement; the same test showed the gate to be dispensable, and \algname{} is the ungated form. \emph{The limits} are stated as plainly. Added to AMS or to AMaLGaM the skewness changes nothing that we can detect, and the analysis shows why: the offspring distribution of AMS is a mixture whose skewness on a slope (0.91) already exceeds that of our sampler, and whose variance along the direction of movement is enlarged in addition (Proposition \ref{prop:ams}). \algname{} improves on EMNA with variance scaling but stays behind EMNA with AMS, AMaLGaM and CMA-ES, which also correct the second moment. The contribution is thus a correction of the third moment that is simple, exact in the first two moments and measurably effective, together with a characterization of what it is worth next to the corrections already in use.

We formulate the method in the framework of matrix-based evolutionary computation \cite{zhan2022}, in which a population is a matrix and every step of the algorithm is a matrix operation; the skewness adds one triangular solve, one matrix--vector product and one rank-one update per generation to the matrix-based form of EMNA (Section \ref{sec:bg}). The contributions are:
\begin{enumerate}
\item \algname{}, a Gaussian EDA whose sampler is skewed along the cumulative path of the mean shifts and nevertheless has exactly the estimated mean and covariance, with a closed-form third cumulant (Theorem \ref{thm:mom}); it changes one sampling step of EMNA, has two parameters of its own and needs no gate.
\item An analysis on a linear function: the one-generation constants of the skewed sampler, and of AMS viewed as an asymmetric sampler, both confirmed by runs to within 1\% (Section \ref{sec:theory}).
\item A pre-registered assessment of the skewness---alone, at three magnitudes, with and without a gate, and on top of AMS and AMaLGaM---with paired statistics, an equivalence margin fixed in advance and the negative results reported in full (Section \ref{sec:expts}).
\end{enumerate}

The paper is organized as follows. Section \ref{sec:bg} reviews Gaussian EDAs and the corrections along the direction of movement, Section \ref{sec:method} presents the algorithm, Section \ref{sec:theory} analyses it, Section \ref{sec:expts} reports the experiments, and Section \ref{sec:conc} concludes.

\section{Background and Related Work}
\label{sec:bg}

This section states the slope problem of Gaussian EDAs in matrix form and reviews the corrections that act along the direction of movement, by the moment of the sampling distribution that they address.

\subsection{Gaussian EDAs, Their Matrix Form and Premature Convergence on Slopes}

Let $X\in\mathbb{R}^{N\times D}$ hold a population of $N$ solutions in its rows and let $\mathbf{f}=F(X)\in\mathbb{R}^{N}$ be their objective values (minimization). Truncation selection keeps the $M=\lfloor\tau N\rfloor$ best rows $X_{s}$. EMNA estimates
\begin{equation}
\boldsymbol{\mu}=\frac{1}{M}X_{s}^{T}\mathbf{1},\qquad \Sigma=\frac{1}{M}X_{c}^{T}X_{c}
\label{eq:emna}
\end{equation}
\noindent with $X_{c}=X_{s}-\mathbf{1}\boldsymbol{\mu}^{T}$ and the Cholesky factor $L=\mathrm{chol}(\Sigma)$, and samples the next population from $\mathcal{N}(\boldsymbol{\mu},s\Sigma)$, $X=\mathbf{1}\boldsymbol{\mu}^{T}+\sqrt{s}\,ZL^{T}$ with standard normal $Z\in\mathbb{R}^{N\times D}$ and a variance multiplier $s$. On a linear function $f(\mathbf{x})=\mathbf{v}^{T}\mathbf{x}$ with an infinite population, one generation moves the mean by $\lambda_{\tau}$ standard deviations along $-\Sigma\mathbf{v}$ and multiplies the variance along $\mathbf{v}$ by $c_{\tau}=1+z_{\tau}\lambda_{\tau}-\lambda_{\tau}^{2}$, where $z_{\tau}=\Phi^{-1}(1-\tau)$ and $\lambda_{\tau}=\varphi(z_{\tau})/\tau$; this is the one-dimensional result of \cite{grahl2005} applied along the gradient. For $\tau=0.35$, $c_{\tau}=0.288$; without amplification the total progress is bounded by $\lambda_{\tau}/(1-\sqrt{c_{\tau}})=2.28$ standard deviations, and a constant multiplier must satisfy $s\ge1/c_{\tau}=3.47$ to keep moving \cite{grahl2005,posik2008}. A large constant multiplier, however, slows down the convergence near an optimum, which is why AVS increases $s$ after improvements and decreases it otherwise \cite{grahl2006,bosman2007}. Whenever AVS is used in this paper without qualification, it is the simplified rule: after an improvement $s$ is divided by a decay factor of 0.8 and otherwise multiplied by it, with $s\in[1,10]$; no correlation or standard-deviation-ratio trigger is used. The parameters are listed in Section \ref{sec:alg}.

Matrix-based evolutionary computation, proposed by J. Zhang \cite{zhan2022}, writes a population as a matrix and each step as a matrix operation; matrix-based forms of other evolutionary algorithms have followed \cite{bai2025,duan2025,sun2026}, and the formulation suits GPU-accelerated evolutionary computation \cite{huang2025,liang2026}. For EMNA this gives the matrix-based EDA (MEDA) of (\ref{eq:emna}): selection is an index gather of the rows with the $M$ smallest values, the mean is a matrix--vector product, the covariance is one matrix product, and sampling is one matrix product $ZL^{T}$. Independent runs add a batch dimension, so $R$ runs form an $R\times N\times D$ tensor and every step becomes a batched linear-algebra call. The same holds for the skew-normal model below, whose implementation reuses the selection, estimation and Cholesky routines of MEDA.

\subsection{Corrections Along the Direction of Movement}
\label{sec:corr}

\emph{First moment.} AMaLGaM \cite{bosman2008,bosman2009,bosman2013} adds to a fraction $\alpha_{\mathrm{AMS}}$ of the offspring the vector $c_{\mathrm{mult}}\delta_{\mathrm{AMS}}(\boldsymbol{\mu}_{t}-\boldsymbol{\mu}_{t-1})$, with $\delta_{\mathrm{AMS}}=2$ and $c_{\mathrm{mult}}$ the current variance multiplier; the shift is the raw difference of the selected means (the incremental variant smooths it). Although AMS is designed as a correction of the mean, moving only part of the offspring also changes the second and third moments of their distribution (Section \ref{sec:amsmix}). Its variance scaling (SDR-AVS) enlarges the multiplier only when the improvements lie more than one standard deviation from the mean and lets it fall below one after a number of generations without improvement; aspects of AMaLGaM were later built into a gene-pool optimal mixing algorithm for large problems \cite{bouter2017}, which is still being developed \cite{andreadis2024,scholman2025}. \emph{Second moment.} The rank-one update of CMA-ES adds $c_{1}\mathbf{p}_{c}\mathbf{p}_{c}^{T}$ to the covariance, where $\mathbf{p}_{c}$ is an evolution path of the mean shift \cite{hansen2001,hansen2016}, an update that has a natural-gradient interpretation \cite{hamano2024}; its cumulative step-size adaptation compares the length of a whitened path with its expectation under random selection. Among Gaussian EDAs, anisotropic variance scaling changes the variances separately along the eigendirections of the covariance \cite{ren2018}, and archive-based variants estimate the covariance with the help of solutions from earlier generations \cite{liang2020,zhao2025}, so that the direction of evolution enters the model \cite{liang2020}; others estimate the covariance from more of the promising solutions than the mean \cite{yang2021}, combine several modifications of the distribution \cite{wang2023}, use Kalman filtering and learning from the search history \cite{zhu2024}, or use a mixture sampling model \cite{dang2022}. These variants keep the Gaussian family and change the second moment. \emph{Third moment.} Directed mutation \cite{berlik2005} samples mutations from a skew-normal distribution whose skewness is self-adapted; Berlik's dissertation \cite{berlik2006} also considers estimating the skewness direction from an accumulated path of whitened mean shifts across generations. The sampler of Section \ref{sec:skew} differs in that its mean and covariance are held at the estimates of EMNA (a centered parametrization \cite{azzalini2014}), so that it changes the third moment only, and in that its magnitude is constant rather than self-adapted. Semiparametric EDAs replace the Gaussian density of some variables by kernel density estimates \cite{soloviev2024}, and other recent EDAs use a variational autoencoder as the model \cite{zhang2026} or zero-inflated Gaussians for sparse solutions \cite{faust2026}; these relax the shape of the model as a whole, whereas the correction studied here acts along the direction of movement only.

The multivariate skew-normal distribution \cite{azzalini1996,azzalini1999,azzalini2014} extends the normal distribution by a shape vector and has a stochastic representation as a normal vector plus a half-normal component along one direction \cite{henze1986,azzalini1996}. In \cite{berlik2005} the one-dimensional skew-normal mutation is standardized to unit variance but not to zero mean; the multivariate sampler of the directed CMA-ES \cite{berlik2006} draws from $\mathrm{SN}_{D}(\boldsymbol{\mu},C,\boldsymbol{\alpha})$ with the mean and covariance matrix of the CMA-ES as location and scale parameters, so that the mean and covariance of the samples differ from them, and its shape vector is updated within a generation from the fitness of each offspring relative to the mean.

\section{The Skew-Normal EDA}
\label{sec:method}

\algname{} changes the sampling step of EMNA and nothing else. Section \ref{sec:skew} constructs the skew-normal sampler, whose mean and covariance are exactly the estimates of EMNA; Section \ref{sec:dir} gives the direction of its skewness; Section \ref{sec:alg} states the complete algorithm and explains why it needs no gate; and Section \ref{sec:diff} lists the differences from existing methods.

\subsection{The Skew-Normal Sampler}
\label{sec:skew}

The algorithm needs a skewed distribution whose mean and covariance remain the estimates of EMNA. The $D$-dimensional skew-normal density is \cite{azzalini1996,azzalini1999}
\begin{equation}
p(\mathbf{x})=2\,\varphi_{D}(\mathbf{x}-\boldsymbol{\xi};\Omega)\,\Phi(\boldsymbol{\alpha}^{T}\omega^{-1}(\mathbf{x}-\boldsymbol{\xi}))
\label{eq:sn}
\end{equation}
\noindent where $\varphi_{D}(\cdot;\Omega)$ is the normal density with covariance $\Omega$, $\Phi$ the standard normal distribution function, $\omega$ the diagonal matrix of the standard deviations in $\Omega$, and $\boldsymbol{\alpha}$ the shape vector; $\boldsymbol{\alpha}=\mathbf{0}$ gives the normal distribution. In this parametrization the mean and covariance both depend on $\boldsymbol{\alpha}$, which does not fit the maximum-likelihood estimates of EMNA.

We use the stochastic representation of \cite{henze1986,azzalini1996} in a standardized form. Let $z_{0}\sim\mathcal{N}(0,1)$ and $\mathbf{z}\sim\mathcal{N}(\mathbf{0},I_{D})$ be independent and define
\begin{equation}
\mathbf{y}=(|z_{0}|-b)\,\mathbf{d}+Q\mathbf{z},\qquad \mathbf{x}=\boldsymbol{\mu}+L\mathbf{y}
\label{eq:conv}
\end{equation}
\noindent with $b=\sqrt{2/\pi}$ and $QQ^{T}=I-(1-b^{2})\mathbf{d}\mathbf{d}^{T}$. The half-normal variable $|z_{0}|$ has mean $b$ and variance $1-b^{2}$, so $\mathbf{y}$ has mean zero, and $Q$ makes its covariance the identity. A closed form is
\begin{equation}
Q=I-\beta\,\hat{\mathbf{d}}\hat{\mathbf{d}}^{T},\qquad \beta=1-\sqrt{1-(1-b^{2})\|\mathbf{d}\|^{2}}
\label{eq:Q}
\end{equation}
\noindent with $\hat{\mathbf{d}}=\mathbf{d}/\|\mathbf{d}\|$, since $QQ^{T}=I-(2\beta-\beta^{2})\hat{\mathbf{d}}\hat{\mathbf{d}}^{T}$ and $2\beta-\beta^{2}=1-(1-\beta)^{2}=(1-b^{2})\|\mathbf{d}\|^{2}$. $QQ^{T}$ is positive definite if and only if $\|\mathbf{d}\|<D_{\lim}=1/\sqrt{1-b^{2}}\approx1.659$. We restrict $\|\mathbf{d}\|\le d_{\max}=\rho_{\max}D_{\lim}$ with $\rho_{\max}=0.95$ ($d_{\max}=1.576$), which keeps the smallest eigenvalue of $QQ^{T}$ at least $1-\rho_{\max}^{2}$. Given $|z_{0}|$, $\mathbf{y}+b\mathbf{d}$ is normal, and $\mathbf{y}+b\mathbf{d}$ is skew-normal with scale matrix $I+b^{2}\mathbf{d}\mathbf{d}^{T}$ \cite{azzalini1999}.
\begin{theorem}[Moments]
\label{thm:mom}
Let $\mathbf{x}$ be generated by (\ref{eq:conv}) and (\ref{eq:Q}) with $\|\mathbf{d}\|<D_{\lim}$. Then $\mathbb{E}\mathbf{x}=\boldsymbol{\mu}$, $\mathrm{Cov}\,\mathbf{x}=LL^{T}$, and the third cumulant of $\mathbf{x}$ is $K_{3}(L\mathbf{d})\otimes(L\mathbf{d})\otimes(L\mathbf{d})$ with $K_{3}=b(2b^{2}-1)\approx0.2180$. For $\mathbf{d}=\mathbf{0}$, $\mathbf{x}\sim\mathcal{N}(\boldsymbol{\mu},LL^{T})$.
\end{theorem}
\begin{proof}
The half-normal variable satisfies $\mathbb{E}|z_{0}|=b$, $\mathbb{E}z_{0}^{2}=1$ and $\mathbb{E}|z_{0}|^{3}=2b$. Hence $\mathbb{E}(|z_{0}|-b)=0$, $\mathrm{Var}|z_{0}|=1-b^{2}$ and $\mathbb{E}(|z_{0}|-b)^{3}=2b-3b+3b^{3}-b^{3}=b(2b^{2}-1)$. Because $z_{0}$ and $\mathbf{z}$ are independent, $\mathbb{E}\mathbf{y}=\mathbf{0}$ and $\mathrm{Cov}\,\mathbf{y}=(1-b^{2})\mathbf{d}\mathbf{d}^{T}+QQ^{T}=I$. The term $Q\mathbf{z}$ is symmetric and independent of $z_{0}$, so it contributes nothing to the third cumulant, which is therefore $K_{3}\mathbf{d}\otimes\mathbf{d}\otimes\mathbf{d}$. Cumulants transform multilinearly under the affine map $\mathbf{x}=\boldsymbol{\mu}+L\mathbf{y}$. For $\mathbf{d}=\mathbf{0}$, $\beta=0$, $Q=I$ and $\mathbf{y}=\mathbf{z}$.
\end{proof}

Thus the mean and covariance do not depend on $\mathbf{d}$; the estimation step of EMNA is unchanged and only the sampler changes. ``Preserving'' refers to the first two moments: the third cumulant is the one stated, and the cumulants of order four and higher also change with $\mathbf{d}$, although they are not used. The matrix $Q=I-\beta\hat{\mathbf{d}}\hat{\mathbf{d}}^{T}$ contracts the Gaussian component along $\hat{\mathbf{d}}$ by $1-\beta$ (0.31 at $d_{\max}$) to make room for the variance of the half-normal component.
\begin{figure*}[!t]
\centering
\includegraphics[width=0.76\textwidth]{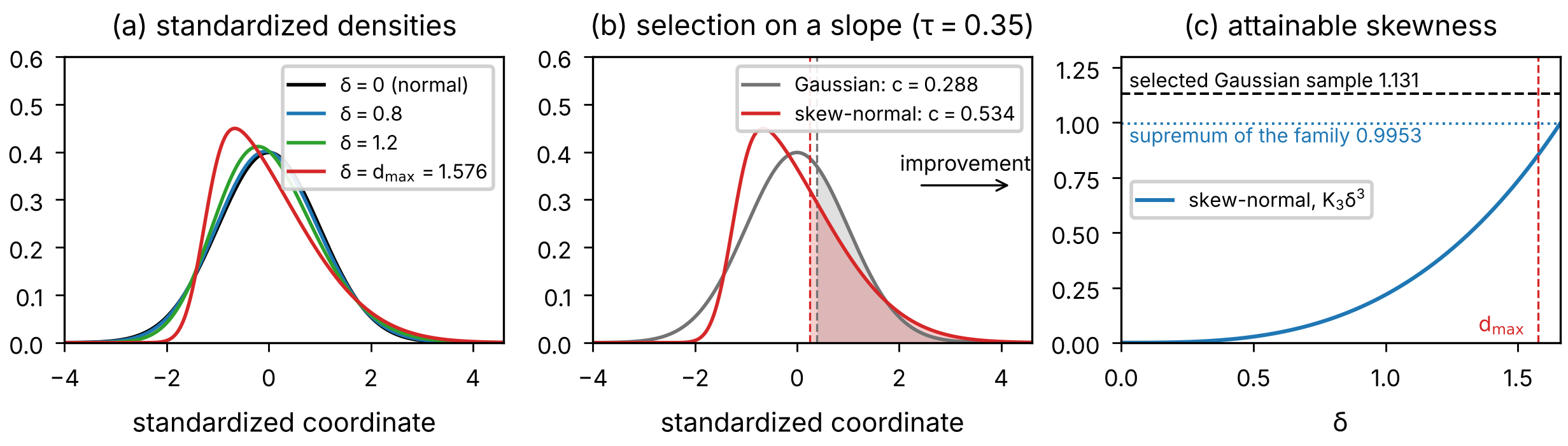}
\caption{(a) Densities of the standardized skew-normal component along its skewness direction (mean 0, variance 1) for four magnitudes $\delta=\|\mathbf{d}\|$; $\delta=0$ is the normal density. (b) Truncation selection with $\tau=0.35$ on a linear function: the shaded part is selected; a sampler skewed toward the improvement raises the ratio of selected to sampled variance from 0.288 to 0.534. (c) Skewness attainable by the family and the skewness of a truncated Gaussian sample.}
\label{fig:sn}
\end{figure*}

In whitened coordinates the skewness along $\hat{\mathbf{d}}$ is $\gamma=K_{3}\|\mathbf{d}\|^{3}$; it equals 0.853 at $d_{\max}$ and tends to the supremum 0.9953 of the skew-normal family as $\|\mathbf{d}\|\to D_{\lim}$ \cite{azzalini2014} (Fig. \ref{fig:sn}(c)). The family contains the normal distribution, so a skew-normal EDA contains EMNA as the special case $\mathbf{d}=\mathbf{0}$, but it is far from all distributions: its third cumulant is a rank-one tensor, so it skews along one direction only; its one-dimensional skewness is below 0.9953; and it is unimodal with Gaussian tails. The solutions selected from a Gaussian on a slope already have skewness $-$1.131 along the gradient (that of the best 35\% of a normal sample), whose magnitude is beyond the range of the family, so a moment-matched skewness would always sit at the boundary. The skewness is therefore set to its cap rather than matched to the selected solutions.

\subsection{The Direction of the Skewness}
\label{sec:dir}

Why not estimate the skewness from the selected solutions, as the mean and covariance are? Their third-moment vector $\mathbf{g}=\frac{1}{M}\sum_{k}\|\mathbf{y}_{k}\|^{2}\mathbf{y}_{k}$ in whitened coordinates has the expectation $K_{3}\|\mathbf{d}\|^{2}\mathbf{d}$, up to a finite-sample bias of order $1/M$, but under normality a noise of $\mathbb{E}\|\mathbf{g}\|^{2}\approx2D(D+2)/M$ (Theorem \ref{S-thm:mardia} and Table \ref{S-tab:mardia} of the supplementary material): at $d_{\max}$ the signal is $\|\mathbb{E}\mathbf{g}\|^{2}=0.728$, whereas the measured noise is 4.05 for $D=10$, $M=38$ and 10.36 for $D=40$, $M=271$, that is 5.6 and 14.2 times the signal. The direction therefore comes from the mean-shift path, whose noise is reduced by the averaging over generations in (\ref{eq:Pt}).

Let $\Delta\boldsymbol{\mu}_{t}=\boldsymbol{\mu}_{t}-\boldsymbol{\mu}_{t-1}$. The path is the exponentially smoothed mean shift, and the direction is the path whitened with the current Cholesky factor:
\begin{equation}
\begin{gathered}
\mathbf{P}_{t}=(1-c_{p})\,\mathbf{P}_{t-1}+c_{p}\Delta\boldsymbol{\mu}_{t}\\
\mathbf{u}_{t}=L_{t}^{-1}\mathbf{P}_{t},\qquad \mathbf{w}_{t}=\frac{\mathbf{u}_{t}}{\|\mathbf{u}_{t}\|}
\end{gathered}
\label{eq:Pt}
\end{equation}
\noindent with $\mathbf{P}_{1}=\mathbf{0}$. In the first generation no shift exists and the sampler is Gaussian ($\mathbf{w}_{1}=\mathbf{0}$); the same holds whenever $\mathbf{u}_{t}=\mathbf{0}$. If the mean only fluctuates, successive shifts point in random directions and largely cancel in $\mathbf{P}_{t}$; if it keeps moving, they add up, so that $\mathbf{w}_{t}$ follows the direction of persistent movement with a memory of about $1/c_{p}$ generations, as the evolution path of the rank-one update of CMA-ES does \cite{hansen2001}. In the original coordinates the skewness points along the path itself, since $L_{t}\mathbf{w}_{t}=\mathbf{P}_{t}/(\mathbf{P}_{t}^{T}\Sigma_{t}^{-1}\mathbf{P}_{t})^{1/2}$, and the sampling distribution of Section \ref{sec:alg} depends on the factor $L_{t}$ only through $\Sigma_{t}=L_{t}L_{t}^{T}$: replacing $L_{t}$ by $L_{t}R$ with an orthogonal $R$ replaces $\mathbf{w}_{t}$ by $R^{T}\mathbf{w}_{t}$ and the Gaussian variates by an equally distributed rotation. The sampler is therefore equivariant under affine transformations of the search space, as that of EMNA is.

\subsection{The Complete Algorithm}
\label{sec:alg}

Let $\mathbf{z}_{0}\in\mathbb{R}^{N}$ and $Z\in\mathbb{R}^{N\times D}$ be standard normal, $s$ the variance multiplier and $\mathbf{w}$ the direction of the current generation. \algname{} samples
\begin{equation}
\begin{gathered}
X=\mathbf{1}\boldsymbol{\mu}^{T}+\sqrt{s}\,\big(Z+\mathbf{h}\mathbf{w}^{T}\big)L^{T}\\
\mathbf{h}=d_{\max}\,(|\mathbf{z}_{0}|-b\mathbf{1})-\beta Z\mathbf{w},\qquad \beta=1-\sqrt{1-\rho_{\max}^{2}}
\end{gathered}
\label{eq:sample}
\end{equation}
\noindent The rows of $Z+\mathbf{h}\mathbf{w}^{T}$ are the standardized skew-normal vectors of (\ref{eq:conv}) with $\mathbf{d}=d_{\max}\mathbf{w}$, for which $(1-b^{2})\|\mathbf{d}\|^{2}=\rho_{\max}^{2}$. By Theorem \ref{thm:mom} and the multilinearity of cumulants, every row of $X$ has mean $\boldsymbol{\mu}$, covariance $s\,LL^{T}$ and third cumulant $s^{3/2}K_{3}d_{\max}^{3}\,(L\mathbf{w})\otimes(L\mathbf{w})\otimes(L\mathbf{w})$, and for $\mathbf{w}=\mathbf{0}$ the sampler is that of EMNA. The skewness is one rank-one term of a matrix equation. A unit test with $2\times10^{6}$ rows, a random non-diagonal $L$ and an arbitrary $\mathbf{w}$ reproduces the three moments within sampling error (supplementary Section \ref{S-sec:checks}).

\emph{No gate.} The skewness is on from the second generation and has the same magnitude throughout the run. Two properties make a control unnecessary. First, by Theorem \ref{thm:mom} the skewness biases neither the mean nor the covariance of the sampling distribution, whatever its direction: when the population has stopped moving, the path is noise and the distribution is skewed along an arbitrary direction, with the two estimated moments unchanged. Second, the skewness does not feed the path that directs it: under neutral selection the expected shift of the mean is zero and the expected covariance is that of EMNA (Proposition \ref{prop:drift}). An earlier form of the algorithm scaled the skewness by an immaturity $\iota_{t}\in[0,1]$, computed from a movement statistic whose null law is exact (supplementary Section \ref{S-sec:gate}). The confirmatory test of Section \ref{sec:expts} was planned with this gated form and found it practically equivalent to the ungated one in all six settings, and we adopt the simpler algorithm.

The scale of the distribution is set by the multiplier $s=c_{\mathrm{mult}}$ of an improvement-triggered AVS: $c_{\mathrm{mult}}\leftarrow\min(c_{\mathrm{mult}}/0.8,10)$ if the best value improved in the generation, else $\max(0.8\,c_{\mathrm{mult}},1)$. Three rules that let a movement statistic drive the scale instead were all worse in an earlier sensitivity study, in which a floor of one on $c_{\mathrm{mult}}$ was important and the decay 0.8 was slightly better than 0.9 (supplementary Section \ref{S-sec:param}).

\begin{algorithm}[!t]
\caption{\algname{} (one run)}
\label{alg:sn}
\begin{algorithmic}[1]
\REQUIRE matrix objective $F$; $N,M,D$; budget $E$; $\rho_{\max}<1$; $c_{p}$
\ENSURE best solution $\mathbf{x}_{bs}$ and $f(\mathbf{x}_{bs})$
\STATE Sample $X$ (uniform in a box, or $\mathcal{N}(\mathbf{c}_{0},\sigma_{0}^{2}I)$); $\mathbf{f}\leftarrow F(X)$; $\mathbf{x}_{bs}\leftarrow$ best row; $\mathbf{P},\mathbf{w}\leftarrow\mathbf{0}$; $c_{\mathrm{mult}}\leftarrow1$
\FOR[while $N-1$ evaluations remain]{$t=1,2,\ldots$}
\STATE $X_{s}\leftarrow$ rows of $X$ with the $M$ smallest $\mathbf{f}$; $\boldsymbol{\mu},\Sigma$ by (\ref{eq:emna}); $\Sigma\leftarrow\Sigma+\epsilon I$; $L\leftarrow\mathrm{chol}(\Sigma)$
\IF{$t>1$}
\STATE $\mathbf{P},\mathbf{w}$ by (\ref{eq:Pt}) \COMMENT{$\mathbf{w}\leftarrow\mathbf{0}$ if $L^{-1}\mathbf{P}=\mathbf{0}$}
\ENDIF
\STATE draw $Z,\mathbf{z}_{0}$; $X\leftarrow$ (\ref{eq:sample}) with $s=c_{\mathrm{mult}}$ \COMMENT{EMNA if $\mathbf{w}=\mathbf{0}$}
\STATE first row $\leftarrow\mathbf{x}_{bs}$ (not re-evaluated); evaluate the other $N-1$ rows; update $\mathbf{x}_{bs}$
\STATE AVS: $c_{\mathrm{mult}}\leftarrow\min(c_{\mathrm{mult}}/0.8,10)$ if improved, else $\max(0.8c_{\mathrm{mult}},1)$
\ENDFOR
\end{algorithmic}
\end{algorithm}

Algorithm \ref{alg:sn} gives one run; Table \ref{tab:params} lists the parameters with the values used. Per generation the work is that of EMNA, $\Theta(NC_{F}+N\log N+(M+N)D^{2}+D^{3})$, where $C_{F}$ is the cost of one function evaluation, plus $O(ND)$ for the rank-one term and one triangular solve of $O(D^{2})$ for the direction; the memory is that of EMNA, $O(ND+D^{2})$. \emph{Budget:} the initial population costs $N$ evaluations and every generation $N-1$, because the elitist row keeps its stored value, so a budget of $E$ gives $\lfloor(E-N)/(N-1)\rfloor$ generations; a last partial generation is not run. \emph{Ties:} solutions are sorted by value and, among equal values, by row index, so the elitist row precedes its ties. \emph{Regularization:} $\epsilon=10^{-14}\mathrm{tr}(\Sigma)/D$; if the Cholesky factorization still fails, $\epsilon$ is multiplied by 10 until it succeeds (this never happened in the experiments reported).
\begin{table}[!t]
\caption{Parameters of \algname{} and the values used}
\label{tab:params}
\centering \footnotesize
\begin{tabular}{@{}>{\raggedright\arraybackslash}p{0.143\columnwidth}>{\raggedright\arraybackslash}p{0.268\columnwidth}>{\raggedright\arraybackslash}p{0.483\columnwidth}@{}}
\toprule
Parameter & Value & Remark \\
\midrule
$N$; $\tau$ & $\lfloor17+3D^{1.5}\rfloor$; 0.35 & population size; truncation ratio \cite{bosman2009} \\
$\rho_{\max}$ & 0.95 & magnitude of the skewness: $d_{\max}=\rho_{\max}D_{\lim}=1.576$, skewness 0.853 \\
$c_{p}$ & 0.3 & learning rate of the path \\
AVS & $\times1/0.8$, $\times0.8$ & multiplier in [1, 10] \\
\bottomrule
\end{tabular}
\end{table}

\subsection{Differences from Existing Methods}
\label{sec:diff}

Directed mutation in evolution strategies samples from a skew-normal distribution \cite{berlik2005,berlik2006}, and AMS \cite{bosman2008,bosman2013} makes the sampling distribution of a Gaussian EDA asymmetric by moving part of the offspring. \algname{} differs from both in three points, the first being the main one.
\begin{enumerate}
\item \emph{Asymmetry without a change of the estimated moments.} The sampler is skewed along the direction of movement and nevertheless has exactly the mean and the covariance that EMNA estimates (Theorem \ref{thm:mom}), so the skewness is a correction of the third moment alone. The samplers of \cite{berlik2005,berlik2006} are not standardized to these two moments, and AMS, as a mixture of the model and a shifted copy, changes them together with the third moment (Proposition \ref{prop:ams}).
\item \emph{No control of the skewness.} Its magnitude is a constant and it is never switched off, which the first point makes possible. Directed mutation self-adapts the skewness. In the earlier version of this work, a mean shift along the path needed a correction of the path for its own shift, and a covariance elongation needed a gate (supplementary Sections \ref{S-sec:gate} and \ref{S-sec:elong}).
\item \emph{A direction from the path.} The direction is the accumulated mean-shift path, not a third-moment estimate, whose noise exceeds the signal (Section \ref{sec:dir}).
\end{enumerate}

\section{Analysis}
\label{sec:theory}

This section analyses the sampler on a linear function, which is the slope problem of Section \ref{sec:bg} and the model on which step-size adaptation has been analysed \cite{chotard2012}, compares it there with AMS, and considers a plateau and global convergence. Two models are kept apart. Theorem \ref{thm:slope}, Corollary \ref{cor:prog} and the fixed point of Section \ref{sec:amsmix} hold in the \emph{infinite-population model}, in which the selected set is replaced by the truncated sampling distribution and the estimates by its moments, and are checked with $N=10^{5}$. Theorem \ref{thm:mom} and Propositions \ref{prop:ams} and \ref{prop:drift} are statements about the sampling distribution and hold for any population size; Theorem \ref{thm:glob} holds for a variant of Algorithm \ref{alg:sn} with a fixed variance floor.

\subsection{What the Skewness Does on a Slope}
\label{sec:slope}
For $0\le\delta<D_{\lim}$ let $Y_{\delta}=\delta(|z_{0}|-b)+\sqrt{1-(1-b^{2})\delta^{2}}\,z_{1}$ with independent standard normal $z_{0},z_{1}$; this is the component of the standardized skew-normal vector along its skewness direction. Let $q$ be its upper $\tau$-quantile and
\begin{equation}
\lambda(\delta)=\mathbb{E}[Y_{\delta}\mid Y_{\delta}\ge q],\qquad c(\delta)=\mathrm{Var}[Y_{\delta}\mid Y_{\delta}\ge q]
\label{eq:lamc}
\end{equation}
\noindent For $\delta=0$ these are $\lambda_{\tau}$ and $c_{\tau}$ of Section \ref{sec:bg}.
\begin{theorem}[One generation on a linear function]
\label{thm:slope}
In the infinite-population model, let $f(\mathbf{x})=\mathbf{v}^{T}\mathbf{x}$ be minimized, and let generation $t$ sample from the skew-normal sampler (\ref{eq:conv}) with $\mathbf{d}=\delta\mathbf{w}_{t}$, which for $\delta=d_{\max}$ is (\ref{eq:sample}), with mean $\boldsymbol{\mu}_{t}$, covariance estimate $\Sigma_{t}=L_{t}L_{t}^{T}$, multiplier $s$ and the downhill direction $\mathbf{w}_{t}=-L_{t}^{T}\mathbf{v}/\sigma_{\mathbf{v},t}$, where $\sigma_{\mathbf{v},t}^{2}=\mathbf{v}^{T}\Sigma_{t}\mathbf{v}$ is the variance along the gradient. Then
\begin{equation}
\begin{gathered}
\boldsymbol{\mu}_{t+1}=\boldsymbol{\mu}_{t}-\sqrt{s}\,\lambda(\delta)\,\frac{\Sigma_{t}\mathbf{v}}{\sigma_{\mathbf{v},t}} \\
\Sigma_{t+1}=s\Big(\Sigma_{t}-\big(1-c(\delta)\big)\frac{\Sigma_{t}\mathbf{v}\mathbf{v}^{T}\Sigma_{t}}{\sigma_{\mathbf{v},t}^{2}}\Big)
\end{gathered}
\label{eq:slopemap}
\end{equation}
\noindent Hence $\sigma_{\mathbf{v},t+1}^{2}=s\,c(\delta)\,\sigma_{\mathbf{v},t}^{2}$: the selected variance along the gradient is multiplied by $s\,c(\delta)$ and the mean progresses by $\sqrt{s}\,\lambda(\delta)$ standard deviations. Moreover $\Sigma_{t+1}\mathbf{v}\parallel\Sigma_{t}\mathbf{v}$, and if the first generation is Gaussian, the direction of (\ref{eq:Pt}) equals the downhill direction in every later generation, for any sequence of multipliers $s_{t}$.
\end{theorem}
\emph{Sketch of proof} (details, for a sampler that may also shift the mean, in supplementary Section \ref{S-sec:proof}). Along $\mathbf{w}$ a whitened sample is $Y_{\delta}$; its components orthogonal to $\mathbf{w}$ are independent of this value and unaffected by selection. The selected whitened samples therefore have mean $\lambda(\delta)\mathbf{w}$ and covariance $I-(1-c(\delta))\mathbf{w}\mathbf{w}^{T}$, and (\ref{eq:slopemap}) follows by transforming back with $L_{t}$. Every mean shift is then parallel to $\Sigma_{1}\mathbf{v}$, and so is the path.

\begin{corollary}[Progress and amplification threshold]
\label{cor:prog}
With a constant multiplier $s=\eta$, let $\rho=\eta\,c(\delta)$. The improvement in generation $t$ is $\sqrt{\eta}\,\lambda(\delta)\,\sigma_{\mathbf{v},t}$ and $\sigma_{\mathbf{v},t+1}=\sqrt{\rho}\,\sigma_{\mathbf{v},t}$, so $f(\boldsymbol{\mu}_{1})-f(\boldsymbol{\mu}_{t})=\sqrt{\eta}\,\lambda(\delta)\,\sigma_{\mathbf{v},1}(1-\rho^{(t-1)/2})/(1-\sqrt{\rho})$ for $\rho\ne1$ and $\sqrt{\eta}\,\lambda(\delta)\,\sigma_{\mathbf{v},1}(t-1)$ for $\rho=1$. The total progress is bounded, by $\sqrt{\eta}\,\lambda(\delta)\,\sigma_{\mathbf{v},1}/(1-\sqrt{\rho})$, if and only if $\rho<1$; it grows linearly for $\rho=1$ and geometrically for $\rho>1$. The distribution keeps moving if and only if $\eta\ge1/c(\delta)$.
\end{corollary}
\begin{table}[!t]
\caption{One-generation constants on a linear function ($\tau=0.35$, $s=1$): theory (Theorem \ref{thm:slope} with numerical integration of (\ref{eq:lamc}); for AMS the fixed point of Section \ref{sec:amsmix}) and a run with $D=10$, $N=10^{5}$, no elitism (mean of generations 5--13)}
\label{tab:slope}
\centering \footnotesize
\begin{tabular}{@{}>{\raggedright\arraybackslash}p{0.33\columnwidth}ccccc@{}}
\toprule
 & \multicolumn{2}{c}{variance ratio} & \multicolumn{2}{c}{step} & \\
\cmidrule(lr){2-3}\cmidrule(lr){4-5}
Sampler & theory & run & theory & run & $1/c$ \\
\midrule
Gaussian (EMNA) & 0.288 & 0.287 & 1.058 & 1.059 & 3.47 \\
skewness, $\rho_{\max}=0.5$ & 0.324 & 0.324 & 1.065 & 1.065 & 3.08 \\
skewness, $\rho_{\max}=0.75$ & 0.413 & 0.412 & 1.080 & 1.080 & 2.42 \\
skewness, $\rho_{\max}=0.95$ (\algname{}) & 0.534 & 0.531 & 1.109 & 1.108 & 1.87 \\
skewness, $\rho_{\max}=0.99$ & 0.562 & 0.558 & 1.118 & 1.117 & 1.78 \\
\midrule
Gaussian + AMS & 1.894 & 1.894 & 2.500 & 2.500 & -- \\
skewness 0.95 + AMS & 1.972 & 1.970 & 2.635 & 2.633 & -- \\
\bottomrule
\end{tabular}
\par\vspace{2pt}\begin{minipage}{\columnwidth}\scriptsize Variance ratio: variance of the selected solutions along the gradient divided by that of the model from which they were sampled; $c(\delta)$ for the first five rows and $\rho_{\mathrm{A}}$ for AMS. Above one the variance grows without amplification. Step: progress of the mean in standard deviations of the model, $\lambda(\delta)$ or $\Lambda_{\mathrm{A}}$. $1/c$: amplification needed to keep moving (none for AMS).\end{minipage}
\end{table}

Table \ref{tab:slope} gives the constants for $\tau=0.35$, computed by numerical integration of (\ref{eq:lamc}) and measured in a run of the algorithm with $N=10^{5}$; the two agree to within 1\%. The skewness raises the fraction of the variance along the gradient that survives selection from 0.288 to 0.534 at $\rho_{\max}=0.95$, the bound on the progress without amplification from 2.28 to 4.12 standard deviations, and lowers the amplification that is needed to keep moving from 3.47 to 1.87. The gain comes from the variance that the selected solutions retain, not from longer steps: $\lambda$ grows only from 1.058 to 1.109. The effect is strongly nonlinear in the magnitude, since the skewness of the sampler is $K_{3}\delta^{3}$: at $\rho_{\max}=0.5$ the retained fraction is still 0.324, at 0.75 it is 0.413. The table also shows the limit of what a third-moment correction can do. It halves the logarithmic rate at which the variance is lost ($\ln c$ rises from $-$1.25 to $-$0.63), but the ratio stays below one even at $\rho_{\max}=0.99$, because the skewness of the family is bounded: the variance along the gradient still has to be restored by the multiplier.
\begin{figure}[!t]
\centering
\includegraphics[width=\columnwidth]{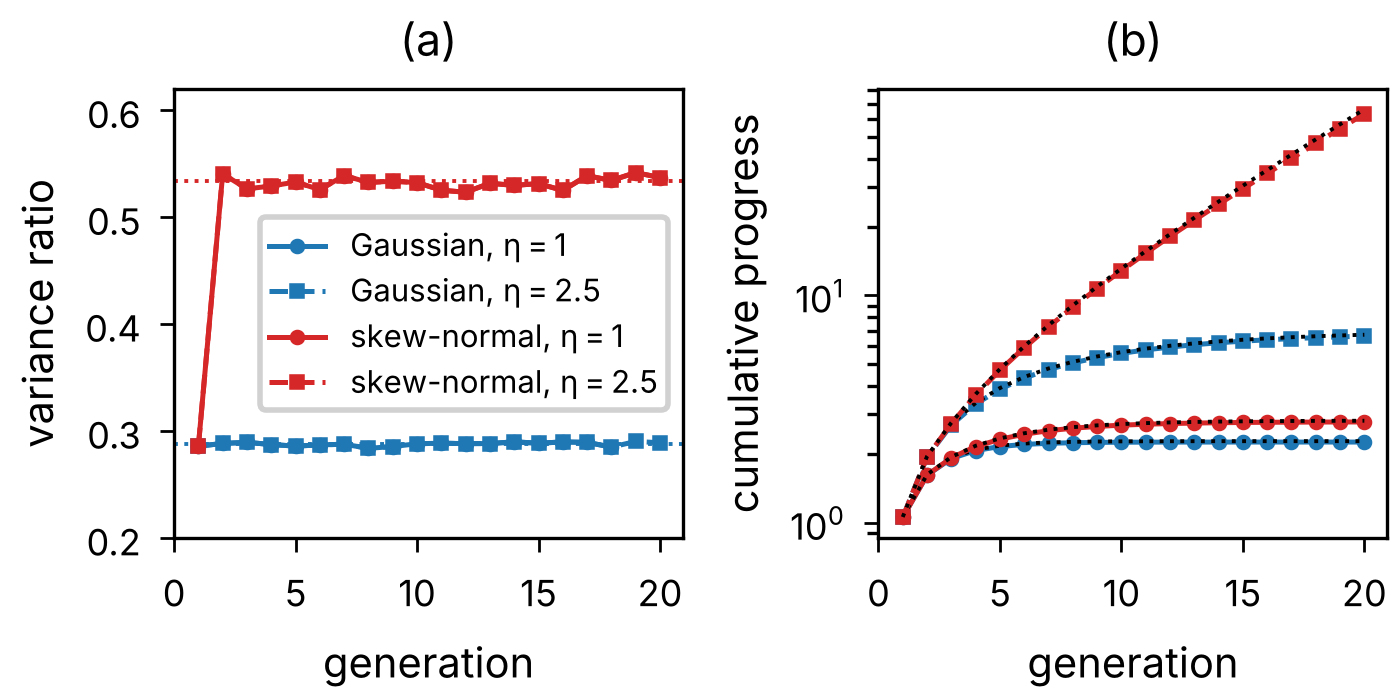}
\caption{Dynamics on a linear function ($D=10$, $N=10^{5}$, $\tau=0.35$, constant multiplier $\eta$). (a) Ratio of selected to sampled variance along the gradient for the Gaussian and the skew-normal sampler ($\rho_{\max}=0.95$); dotted lines are $c(\delta)$ of Table \ref{tab:slope}. (b) Cumulative progress in units of the initial standard deviation; dotted curves are the predictions of Corollary \ref{cor:prog}.}
\label{fig:slope}
\end{figure}

Fig. \ref{fig:slope} checks the dynamics with $D=10$, $N=10^{5}$ and $\tau=0.35$ (no elitism). With $\eta=1$ the Gaussian model stops at 2.26 initial standard deviations (2.28 predicted) and the skew-normal model at 2.77 (2.81 predicted; the initial population and the first sampled generation, in which the direction is not yet defined, are symmetric, which is why this is below the fixed-point bound 4.12). With $\eta=2.5$, $\rho=0.72<1$ for the Gaussian and $\rho=1.34>1$ for the skew-normal model: after 20 generations the progress is 6.6 against 63 (6.7 and 66 predicted), the latter growing geometrically.

The alignment statement of Theorem \ref{thm:slope} is one about the infinite-population model: with a finite population the path direction fluctuates around the downhill direction, and on a function whose gradient turns the path lags behind it; Section \ref{sec:diag} measures the angle.

\subsection{The Anticipated Mean Shift as an Asymmetric Sampler}
\label{sec:amsmix}
AMS is usually described as a correction of the mean. Its offspring distribution, however, is not a shifted Gaussian: a fraction $\alpha$ of the offspring is moved by a vector $\mathbf{a}$ and the rest is not, so the offspring follow the mixture $(1-\alpha)\,\mathcal{N}(\boldsymbol{\mu},s\Sigma)+\alpha\,\mathcal{N}(\boldsymbol{\mu}+\mathbf{a},s\Sigma)$.
\begin{proposition}[Moments of the offspring distribution of AMS]
\label{prop:ams}
The mixture has mean $\boldsymbol{\mu}+\alpha\mathbf{a}$, covariance $s\Sigma+\alpha(1-\alpha)\,\mathbf{a}\mathbf{a}^{T}$ and third cumulant $\alpha(1-\alpha)(1-2\alpha)\,\mathbf{a}\otimes\mathbf{a}\otimes\mathbf{a}$.
\end{proposition}
\begin{proof}
An offspring is $\mathbf{x}=\mathbf{g}+B\mathbf{a}$ with $\mathbf{g}\sim\mathcal{N}(\boldsymbol{\mu},s\Sigma)$ and an independent Bernoulli variable $B$ with $\mathbb{E}B=\alpha$. Cumulants of independent summands add; $\mathbf{g}$ contributes its mean and covariance and no third cumulant, and $B$ has variance $\alpha(1-\alpha)$ and third cumulant $\alpha(1-\alpha)(1-2\alpha)$.
\end{proof}

AMS thus shifts the mean, enlarges the variance along the shift and, for $\alpha<1/2$, skews the distribution toward it, all with one strength. On a linear function its stationary state can be computed as in Theorem \ref{thm:slope}. Without scaling ($s=1$), let the mean progress by $\Lambda_{\mathrm{A}}$ standard deviations of the model per generation and let the variance of the model along the gradient be multiplied by $\rho_{\mathrm{A}}$. The shift $\mathbf{a}=\delta_{\mathrm{AMS}}(\boldsymbol{\mu}_{t}-\boldsymbol{\mu}_{t-1})$ then has the length $d=\delta_{\mathrm{AMS}}\Lambda_{\mathrm{A}}/\sqrt{\rho_{\mathrm{A}}}$ in current standard deviations, and $\Lambda_{\mathrm{A}}$ and $\rho_{\mathrm{A}}$ are the mean and the variance of the upper fraction $\tau$ of the one-dimensional mixture with separation $d$. Solving this fixed point numerically for $\delta_{\mathrm{AMS}}=2$, $\alpha=0.175$ (that is $\lfloor M/2\rfloor$ of $N$ offspring) and $\tau=0.35$ gives $d=3.63$, $\Lambda_{\mathrm{A}}=2.50$ and $\rho_{\mathrm{A}}=1.89$; a run with $N=10^{5}$ gives 2.50 and 1.89 (Table \ref{tab:slope}). In this state the offspring distribution has, along the direction of movement, a mean shifted by 0.64 standard deviations of the model, a variance enlarged by the factor 2.91 and a skewness of 0.91, which exceeds the 0.853 of the skew-normal sampler at $\rho_{\max}=0.95$. AMS therefore already contains a third-moment correction of the size studied in this paper, together with an enlargement of the variance that turns the loss on a slope into a gain. Replacing its Gaussian component by the skew-normal sampler raises $\rho_{\mathrm{A}}$ only from 1.89 to 1.97 (run: 1.97), a change of 4\%, whereas the same skewness raises the ratio of a Gaussian EDA from 0.288 to 0.534. The analysis thus predicts that the skewness matters when it is the only asymmetry of the sampler and adds little to AMS; Section \ref{sec:expts} tests both predictions.

\subsection{Plateau and Global Convergence}
\label{sec:plateau}
\begin{proposition}[Drift on a plateau]
\label{prop:drift}
If in generation $t$ all solutions have equal fitness and elitism is off, then for \algname{}, as for EMNA, $\mathbb{E}[\boldsymbol{\mu}_{t+1}]=\boldsymbol{\mu}_{t}$, $\mathbb{E}[\Sigma_{t+1}]=s\,(1-1/M)\,\Sigma_{t}$ and $\mathrm{Cov}(\boldsymbol{\mu}_{t+1}-\boldsymbol{\mu}_{t})=s\Sigma_{t}/M$, where the expectations are conditional on the state of generation $t$.
\end{proposition}
Indeed, the selected solutions are then $M$ independent samples of a distribution with mean $\boldsymbol{\mu}_{t}$ and covariance $s\Sigma_{t}$ (Theorem \ref{thm:mom}); their mean has expectation $\boldsymbol{\mu}_{t}$ and covariance $s\Sigma_{t}/M$, and their maximum-likelihood covariance has expectation $(1-1/M)s\Sigma_{t}$. Only the first two moments of the sampler enter, and the skewness changes neither. It therefore does not move a population that selection does not move and does not change the rate at which the variance of such a population decays, which is why it can stay on. It also gives \algname{} no mechanism for crossing a plateau that EMNA lacks: the improvement-triggered AVS does not compensate the loss of variance, since nothing improves there (Section \ref{sec:diag}).

\begin{theorem}[Global convergence with a variance floor]
\label{thm:glob}
Let $\Omega\subset\mathbb{R}^{D}$ be a box with nonempty interior and $f$ continuous on $\Omega$ with minimum $f^{*}$. Consider Algorithm \ref{alg:sn} with samples clipped to $\Omega$, elitism, a fixed variance floor $\sigma_{\min}>0$ added to the covariance ($\Sigma\leftarrow\Sigma+\sigma_{\min}^{2}I$), a bounded multiplier $\eta\le s\le s_{\max}$ with $\eta>0$, and $\rho_{\max}<1$. Then $f(\mathbf{x}_{bs,t})\to f^{*}$ almost surely as $t\to\infty$.
\end{theorem}
The proof is in supplementary Section \ref{S-sec:proof}, where the theorem is stated for a sampler that may also shift the mean. It conditions on the half-normal variable: given $|z_{0}|=r$ a sample is normal, with a mean within a fixed distance of $\Omega$ and a covariance whose eigenvalues lie between $\eta\sigma_{\min}^{2}(1-\rho_{\max}^{2})$ and a fixed upper bound, so the sampling density has a positive lower bound on $\Omega$ that is uniform in the history, and the standard argument for elitist algorithms applies \cite{rudolph1996}. The theorem needs no unique minimizer: its point is that the skewness does not destroy the positive density that drives the argument. It does not cover the runs reported below, which use the relative ridge $\epsilon I$ (not a fixed floor), no clipping and an unbounded domain; and, as for Gaussian EDAs, the bound is qualitative and decreases exponentially with $D$.

\section{Experiments}
\label{sec:expts}

The experiments ask what the skewness is worth. After the setup (Section \ref{sec:setup}), Section \ref{sec:diag} measures the direction inside the algorithm, Section \ref{sec:train} explores on a training set how the effect of the skewness depends on its magnitude, on a gate and on the other corrections, and Section \ref{sec:test} tests the resulting hypotheses on fresh instances under a protocol fixed in advance; Sections \ref{sec:why} and \ref{sec:disc} interpret the results.

\subsection{Setup}
\label{sec:setup}

\emph{Problems and budget.} All comparisons use the 24 noiseless BBOB functions of COCO \cite{hansen2009,hansen2021}. The search space is unconstrained ($\mathbb{R}^{D}$, the COCO convention) for all algorithms. Two starting conditions are used: the \emph{standard start}, a population uniform in $[-5,5]^{D}$ for the EDAs and a mean uniform in $[-4,4]^{D}$ with step size 2 for CMA-ES; and a \emph{distant start}, $\mathcal{N}(50\cdot\mathbf{1},I)$ for the EDAs and mean $50\cdot\mathbf{1}$ with step size 1 for CMA-ES. Since the optimum lies in $[-4,4]^{D}$, the distant start is about 50 initial standard deviations away from it in every coordinate, that is $50\sqrt{D}$ in Euclidean distance. It is not part of the standard BBOB protocol; we include it because a long approach to a far optimum is the regime that corrections along the direction of movement are meant for, and because a start centered on the domain can favor an algorithm that prefers that region (structural bias \cite{vanstein2024}). The budget is $10^{4}D$ evaluations, counted as in Section \ref{sec:alg}. An error below $10^{-8}$ counts as solved: runs stop there and smaller errors are not distinguished. This floor was fixed before the runs; differences below it have no practical meaning.

\emph{Algorithms.} All algorithms are one program with different parameters and share selection, sorting, evaluation counting and the random-number generator; within a (function, instance, start, dimension) cell all of them use the same random stream. The EDAs use $N=\lfloor17+3D^{1.5}\rfloor$, $\tau=0.35$ \cite{bosman2009} and elitism, and do not restart. \emph{EMNA+AVS} is EMNA with the AVS of Section \ref{sec:alg}, and \emph{\algname{}} is Algorithm \ref{alg:sn}, which differs from it in the sampler only. \emph{\algname{}-g} is the gated form with which the protocols were written: the size of its skewness is $\iota_{t}d_{\max}$, with the immaturity $\iota_{t}$ of supplementary Section \ref{S-sec:gate}, and $\rho_{\max}=0.95$ unless another value is given. \emph{EMNA+AVS+AMS} adds AMS as in \cite{bosman2013}, with $\lfloor M/2\rfloor$ shifted offspring and the raw difference of the means. \emph{AMaLGaM} is our reimplementation of the single-population algorithm of \cite{bosman2013}: SDR-AVS ($\theta_{\mathrm{SDR}}=1$, $\mathrm{NIS}_{\max}=25+D$, decay 0.9, multiplier allowed below one), AMS with $\delta_{\mathrm{AMS}}=2$ and the shift vector smoothed with $\eta_{\mathrm{shift}}$, and the covariance smoothed with $\eta_{\Sigma}$, both by the formulas of \cite{bosman2013}. It is not the original code and does not include the restart scheme of the parameter-free version. \emph{AMS+SN} and \emph{AMaLGaM+SN} are these two algorithms with their Gaussian sampler replaced by the gated skew-normal sampler; nothing else is changed. In these two configurations the gate ran with the thresholds $(2,6)$ of the earlier version instead of the null quantiles, a deviation from the protocol; Section \ref{sec:test} also reports their repetition with the null quantiles. \emph{CMA-ES} is our implementation of \cite{hansen2016} with the default parameters, including negative recombination weights, and \emph{IPOP-CMA-ES} restarts it with a doubled population whenever a stopping criterion fires \cite{auger2005}. For $D=10$, on the ten of the functions f1--f14 that both implementations solve in at least half of the runs, our CMA-ES needs 5\% more evaluations to $10^{-8}$ than the reference implementation pycma \cite{hansen2019} in the median, and at most 16\% more (supplementary Section \ref{S-sec:checks}). All algorithms with the AVS of Section \ref{sec:alg} use the decay 0.8, so that EMNA+AVS, \algname{}, \algname{}-g and EMNA+AVS+AMS differ in the sampler only. The protocols also contained a variant with a mean shift along the path (MS, SN+MS); its definition and all its results are in supplementary Section \ref{S-sec:gate}, and Section \ref{sec:test} summarizes them.

\emph{Stages.} The work proceeded in stages on disjoint sets of instances, each with a protocol that was written and stored before its runs (supplementary Section \ref{S-sec:prot}). Instances 1--5 with $D=10$ form the \emph{training set}: every choice of a parameter was made there. Instances 31--45, with $D=10$, 20 and 40, form the \emph{confirmatory set}; no program had been run on them when the protocol of the confirmatory test was frozen. Instances 6--30 had been used by the earlier version of this work, whose sampler also elongated the covariance along the path (supplementary Section \ref{S-sec:elong}; Section \ref{sec:why}).

\emph{Statistics.} The response is the \emph{anytime performance} $y_{\mathrm{any}}$, the mean over the $10^{4}D$ evaluations of $\log_{10}\max(\text{best error so far},10^{-8})$, which rewards both speed and final precision. All comparisons are paired. Within a setting (dimension $\times$ start) the blocks are the 24 functions: the paired difference of two algorithms is averaged over the instances of a function, and the 24 block differences are tested with the Wilcoxon signed-rank test (zero differences by Pratt's method \cite{pratt1959}, mid-ranks for ties). The effect size is the Hodges--Lehmann estimate (HL) of the block differences \cite{hodges1963} with a bootstrap interval over functions. The confirmatory comparisons form six families fixed in advance (G1--G6), and Holm's correction \cite{holm1979} is applied within each family over all its tests (comparisons and settings). Families G1, G3 and G5 and the first comparison of G4 concern the skewness alone (Table \ref{tab:fam}); G2, G6 and the other two comparisons of G4 concern a variant with a mean shift (supplementary Table \ref{S-tab:famms}). The dependence on the magnitude is tested with Page's trend test \cite{page1963} (normal approximation) over four levels of $\rho_{\max}$, with the functions as blocks. Because a non-significant difference is not evidence of equivalence \cite{lakens2017}, two algorithms are called \emph{practically equivalent} in a setting only if the 90\% interval of HL lies within $\pm0.25$, a margin fixed in advance. Successes are compared with the exact McNemar test on the paired runs. Everything on the training set, the per-function tests and the comparisons outside the six families are descriptive.

\subsection{The Direction Inside the Algorithm}
\label{sec:diag}

\emph{Direction.} Theorem \ref{thm:slope} asserts alignment only for an infinite population. With the population sizes used, on a linear function, the whitened direction $\mathbf{w}_{t}$ deviates from the whitened downhill direction by a median angle of 30, 27 and 23 degrees for $D=10$, 20 and 40 (supplementary Section \ref{S-sec:checks}). The angle is the price of estimating the direction from $M$ solutions; an angle of $30^{\circ}$ leaves $\cos^{3}30^{\circ}=0.65$ of the third cumulant along the true gradient.

\emph{Plateau.} A population started on a plateau (a constant function outside a ball of radius 1 around the optimum, sphere inside; start at distance 4, $D=10$, 15 runs) never reached the ball within the budget with any of EMNA+AVS, \algname{}, \algname{}-g, AMaLGaM, CMA-ES and IPOP-CMA-ES, in agreement with Proposition \ref{prop:drift} (supplementary Section \ref{S-sec:checks}).

\subsection{Exploration on the Training Set}
\label{sec:train}

All results of this subsection are descriptive and their $p$ values are uncorrected. Each algorithm was run on the 24 functions and five instances of the training set from both starts, which gives 48 blocks and 240 runs per algorithm.
\begin{table}[!t]
\caption{Training set (BBOB f1--f24, $D=10$, instances 1--5, both starts; 48 blocks): mean $y_{\mathrm{any}}$ of \algname{}-g as a function of the magnitude of the skewness, and the retained variance on a slope predicted by Theorem \ref{thm:slope}}
\label{tab:train}
\centering \footnotesize
\begin{tabular}{@{}cccc@{}}
\toprule
$\rho_{\max}$ & $c(\delta)$ & standard / distant start & both \\
\midrule
0 & 0.288 & $-$3.21 / $-$0.33 & $-$1.77 \\
0.25 & 0.292 & $-$3.20 / $-$0.26 & $-$1.73 \\
0.5 & 0.324 & $-$3.46 / $-$0.50 & $-$1.98 \\
0.75 & 0.413 & $-$3.45 / $-$1.18 & $-$2.32 \\
0.9 & 0.501 & $-$3.36 / $-$1.33 & $-$2.35 \\
0.95 & 0.534 & $-$3.46 / $-$1.58 & $-$2.52 \\
0.99 & 0.562 & $-$3.45 / $-$1.75 & $-$2.60 \\
\midrule
0.95, no gate (\algname{}) & 0.534 & $-$3.54 / $-$1.76 & $-$2.65 \\
\bottomrule
\end{tabular}
\par\vspace{2pt}\begin{minipage}{\columnwidth}\scriptsize Gated skewness with AVS (decay 0.8); $\rho_{\max}=0$ is EMNA+AVS. The last row is the ungated algorithm.\end{minipage}
\end{table}

\emph{Magnitude.} Table \ref{tab:train} gives the anytime performance of \algname{}-g as a function of $\rho_{\max}$. The skewness improves EMNA+AVS (HL $-$0.45, $p=6\times10^{-5}$ at $\rho_{\max}=0.95$), mostly with the distant start, and from $\rho_{\max}=0.5$ on the improvement grows monotonically, as the retained variance $c(\delta)$ of Theorem \ref{thm:slope} does; a weak skewness ($\rho_{\max}=0.25$, for which $c(\delta)$ has barely moved) changes nothing. Since $\rho_{\max}=0.99$ was not significantly better than 0.95, we kept 0.95, the value of the earlier version of this work (supplementary Table \ref{S-tab:train}).

\emph{Gate.} Removing the gate from the skewness makes no significant difference (mean $y_{\mathrm{any}}$ $-$2.65 against $-$2.52; HL $-$0.01, $p=0.64$). A skewness that fades on a fixed schedule, $\iota_{t}=e^{-t/\tau_{g}}$ with $\tau_{g}=25$ or 100 generations, is worse than the gated one (HL +0.14 and +0.13, $p=0.03$ and $0.01$): the skewness is useful whenever the population moves, not only at the start of a run.

\emph{On top of AMS and AMaLGaM.} Replacing the Gaussian sampler of EMNA+AVS+AMS or of AMaLGaM by the gated skew-normal sampler makes no difference: the mean $y_{\mathrm{any}}$ is $-$3.66 against $-$3.63 for AMS (HL +0.01, $p=0.59$) and $-$3.49 against $-$3.38 for AMaLGaM (HL $-$0.01, $p=0.61$). For AMS itself, eight settings ($\delta_{\mathrm{AMS}}\in\{1,1.5,2,3\}$, decay 0.8 or 0.9) had been run on the same blocks; the published $\delta_{\mathrm{AMS}}=2$ with the decay 0.8 was the best and is the EMNA+AVS+AMS used here. \algname{}-g remains behind both algorithms: its mean $y_{\mathrm{any}}$ is $-$2.52, against $-$3.63 for EMNA+AVS+AMS (HL +0.46, $p=1\times10^{-8}$) and $-$3.38 for AMaLGaM.

\emph{Hypotheses.} These observations became families of the confirmatory test, each with the outcome expected in the protocol: the skewness improves EMNA+AVS (G1), which had also been seen on instances 16--30 with the earlier version (HL $-$1.23 and $-$3.49 for $D=10$ and 20 with the distant start; supplementary Section \ref{S-sec:elong}); the improvement grows with $\rho_{\max}$ (G3); removing the gate makes no practical difference (G4); and adding the skewness to AMS or to AMaLGaM makes no practical difference (G5). Families G2 and G6 concerned the variant with a mean shift (supplementary Section \ref{S-sec:gate}).

\subsection{Confirmatory Test on Fresh Instances}
\label{sec:test}
\begin{table*}[!t]
\caption{Confirmatory test on BBOB f1--f24, instances 31--45 (15 runs per function, budget $10^{4}D$): number of functions solved to $10^{-8}$ in all 15 runs and mean anytime performance $y_{\mathrm{any}}$ over the 24 functions (lower is better); mean rank of $y_{\mathrm{any}}$ over the function--setting pairs}
\label{tab:test}
\centering \scriptsize
\begin{tabular}{@{}l@{\hspace{5pt}}rr@{\hspace{7pt}}rr@{\hspace{7pt}}rr@{\hspace{7pt}}rr@{\hspace{7pt}}rr@{\hspace{7pt}}rr@{\hspace{7pt}}cc@{}}
\toprule
 & \multicolumn{2}{c}{$D=10$, std.} & \multicolumn{2}{c}{$D=10$, dist.} & \multicolumn{2}{c}{$D=20$, std.} & \multicolumn{2}{c}{$D=20$, dist.} & \multicolumn{2}{c}{$D=40$, std.} & \multicolumn{2}{c}{$D=40$, dist.} & \multicolumn{2}{c}{mean rank} \\
Algorithm & sol. & $y$ & sol. & $y$ & sol. & $y$ & sol. & $y$ & sol. & $y$ & sol. & $y$ & $D\le20$ & all \\
\midrule
EMNA+AVS & 8 & $-$3.16 & 7 & $-$0.14 & 9 & $-$1.95 & 4 & 3.81 & 5 & $-$0.44 & 0 & 7.36 & 12.02 & 10.87 \\
\algname{}-g, $\rho_{\max}=0.5$ & 8 & $-$3.33 & 8 & $-$0.32 & 9 & $-$1.92 & 6 & 3.26 & -- & -- & -- & -- & 11.38 & -- \\
\algname{}-g, $\rho_{\max}=0.75$ & 10 & $-$3.45 & 7 & $-$1.07 & 8 & $-$2.14 & 7 & 1.72 & -- & -- & -- & -- & 10.55 & -- \\
\algname{}-g & 11 & $-$3.44 & 9 & $-$1.65 & 9 & $-$2.45 & 8 & 0.33 & 6 & $-$0.85 & 2 & 3.90 & 9.26 & 9.12 \\
\algname{} & 10 & $-$3.47 & 8 & $-$1.67 & 9 & $-$2.46 & 7 & 0.17 & 6 & $-$0.91 & 2 & 3.86 & 9.24 & 8.92 \\
\midrule
EMNA+AVS+AMS & 12 & $-$3.91 & 10 & $-$3.23 & 11 & $-$3.04 & 9 & $-$1.66 & 7 & $-$1.14 & 7 & 1.04 & 5.64 & 5.59 \\
AMS+SN & 10 & $-$3.87 & 10 & $-$3.23 & 11 & $-$3.00 & 11 & $-$1.72 & 7 & $-$1.14 & 5 & 1.01 & 5.82 & 5.64 \\
AMaLGaM & 10 & $-$3.74 & 9 & $-$3.06 & 12 & $-$3.20 & 8 & $-$2.08 & 8 & $-$1.82 & 8 & $-$0.04 & 5.52 & 4.62 \\
AMaLGaM+SN & 10 & $-$3.82 & 9 & $-$3.08 & 12 & $-$3.11 & 10 & $-$2.08 & 8 & $-$1.81 & 8 & $-$0.01 & 5.33 & 4.60 \\
CMA-ES & 8 & $-$3.33 & 8 & $-$3.48 & 8 & $-$2.96 & 8 & $-$2.86 & 9 & $-$2.65 & 9 & $-$2.54 & 6.12 & 5.12 \\
IPOP-CMA-ES & 14 & $-$4.88 & 14 & $-$4.71 & 14 & $-$4.25 & 12 & $-$4.11 & 12 & $-$3.80 & 11 & $-$3.61 & 1.76 & 1.53 \\
\bottomrule
\end{tabular}
\par\vspace{2pt}\begin{minipage}{\textwidth}\scriptsize std., dist.: standard and distant start; algorithms as defined in Section \ref{sec:setup}. The two lower levels of $\rho_{\max}$ were not run for $D=40$. The mean ranks are among all algorithms of the confirmatory test, which include four configurations with a mean shift (supplementary Table \ref{S-tab:testms}): the rank over $D\le20$ is among 15 algorithms (96 function--setting pairs), the one over all settings among the 13 run everywhere (144 pairs).\end{minipage}
\end{table*}

\begin{table*}[!t]
\caption{Confirmatory comparisons: Hodges--Lehmann estimate of the difference in $y_{\mathrm{any}}$ (first minus second; negative means that the first algorithm is better) over the 24 functions of each setting}
\label{tab:fam}
\centering \footnotesize \setlength{\tabcolsep}{5pt}
\begin{tabular}{@{}llcccccc@{}}
\toprule
 & & \multicolumn{2}{c}{$D=10$} & \multicolumn{2}{c}{$D=20$} & \multicolumn{2}{c}{$D=40$} \\
\cmidrule(lr){3-4}\cmidrule(lr){5-6}\cmidrule(l){7-8}
Family & Comparison & std. & dist. & std. & dist. & std. & dist. \\
\midrule
G1 (skewness alone) & \algname{}-g vs. EMNA+AVS & $-$0.03$^{\approx}$ & \textbf{$-$1.03} & \textbf{$-$0.09}$^{\approx}$ & \textbf{$-$3.33} & \textbf{$-$0.05}$^{\approx}$ & \textbf{$-$3.31} \\
G3 (dose) & trend over $\rho_{\max}$ & \textbf{\boldmath$p=0.007$} & \textbf{\boldmath$p=1\times10^{-10}$} & \textbf{\boldmath$p=0.007$} & \textbf{\boldmath$p=3\times10^{-16}$} & -- & -- \\
G4 (gate) & \algname{} vs. \algname{}-g & $-$0.01$^{\approx}$ & $-$0.06$^{\approx}$ & +0.01$^{\approx}$ & $-$0.11$^{\approx}$ & $-$0.02$^{\approx}$ & $-$0.03$^{\approx}$ \\
G5 (on top of others) & AMS+SN vs. EMNA+AVS+AMS & +0.01$^{\approx}$ & $-$0.01$^{\approx}$ & +0.02$^{\approx}$ & $-$0.02$^{\approx}$ & $-$0.01$^{\approx}$ & $-$0.01$^{\approx}$ \\
 & AMaLGaM+SN vs. AMaLGaM & $-$0.05$^{\approx}$ & +0.01$^{\approx}$ & +0.05$^{\approx}$ & +0.01$^{\approx}$ & +0.01$^{\approx}$ & +0.03$^{\approx}$ \\
\midrule
descriptive & \algname{} vs. EMNA+AVS & $-$0.03 & \textbf{$-$1.16} & $-$0.06 & \textbf{$-$3.49} & \textbf{$-$0.06}$^{\approx}$ & \textbf{$-$3.35} \\
 & \algname{} vs. EMNA+AVS+AMS & \textbf{+0.18} & \textbf{+0.77} & \textbf{+0.35} & \textbf{+1.41} & \textbf{+0.07}$^{\approx}$ & \textbf{+2.41} \\
 & \algname{} vs. AMaLGaM & +0.18 & \textbf{+0.72} & \textbf{+0.52} & \textbf{+1.65} & \textbf{+0.74} & \textbf{+3.68} \\
 & \algname{} vs. CMA-ES & +0.01 & \textbf{+0.86} & +0.33 & \textbf{+2.00} & +1.29 & \textbf{+5.77} \\
 & \algname{} vs. IPOP-CMA-ES & \textbf{+1.02} & \textbf{+2.39} & \textbf{+1.33} & \textbf{+3.19} & \textbf{+2.21} & \textbf{+6.75} \\
 & EMNA+AVS+AMS vs. AMaLGaM & $-$0.02 & +0.01 & +0.18 & \textbf{+0.37} & \textbf{+0.56} & \textbf{+0.96} \\
\bottomrule
\end{tabular}
\par\vspace{2pt}\begin{minipage}{\textwidth}\scriptsize Bold: significant at $\alpha=0.05$ (Wilcoxon signed-rank test over the 24 functions; in G1, G4 and G5 after Holm's correction over all tests of the family, in the descriptive rows uncorrected). $^{\approx}$: practically equivalent (90\% interval of HL within $\pm0.25$). G3: Holm-corrected $p$ of Page's trend test (normal approximation) for \algname{}-g over the four levels $\rho_{\max}=0$, 0.5, 0.75 and 0.95. The families G2 and G6 and the other two comparisons of G4, which concern the variant with a mean shift, are in supplementary Table \ref{S-tab:famms}. Intervals, per-function tests and paired success tables are in supplementary Section \ref{S-sec:test}.\end{minipage}
\end{table*}

\begin{figure*}[!t]
\centering
\includegraphics[width=0.80\textwidth]{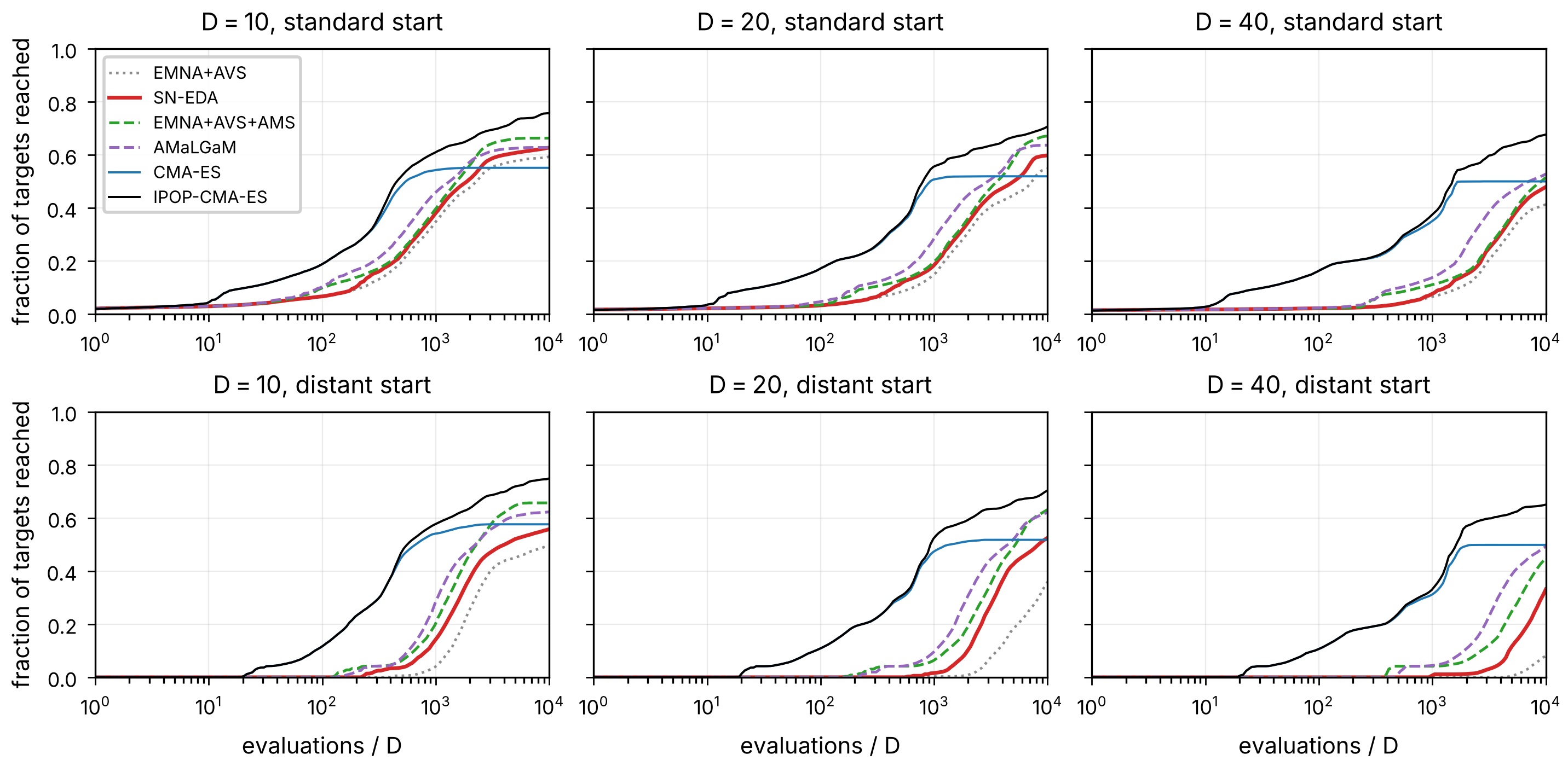}
\caption{Running-time distributions on the confirmatory set: fraction of the (function, instance, target) triples reached within a given number of evaluations, for the 51 targets $10^{2},10^{1.8},\ldots,10^{-8}$ on the 24 functions and 15 instances of each setting. Top: standard start; bottom: distant start.}
\label{fig:ecdf}
\end{figure*}

Table \ref{tab:test} summarizes the confirmatory test, which comprised 30{,}960 runs including the variant of the supplement, Table \ref{tab:fam} gives the comparisons, and Fig. \ref{fig:ecdf} shows the running-time distributions \cite{hansen2022}.

\emph{G1, the skewness against the Gaussian sampler.} \algname{}-g is better than EMNA+AVS in all 6 settings and significantly so in 5 after Holm's correction. With the distant start the effect is large (HL $-$1.03, $-$3.33 and $-$3.31 for $D=10$, 20 and 40; Holm-corrected $p\le4\times10^{-5}$): it solves 30, 56 and 50 runs that EMNA+AVS does not, against 11, 2 and 3 in the other direction. With the standard start the estimates are small (HL $-$0.027, $-$0.094 and $-$0.052; not significant for $D=10$, $p=0.09$) and inside the equivalence margin in all 3 settings. The gain is concentrated there on the linear slope f5, the benchmark's instance of the problem analysed in Section \ref{sec:slope}: on f5 the difference in $y_{\mathrm{any}}$ is $-$4.63, $-$9.24 and $-$7.55, and EMNA+AVS solves 10, 0 and 0 of the 15 runs against 15, 14 and 12 for \algname{}-g, whereas the mean difference over the other 23 functions is $-$0.10, $-$0.13 and $-$0.09.
\begin{figure}[!t]
\centering
\includegraphics[width=\columnwidth]{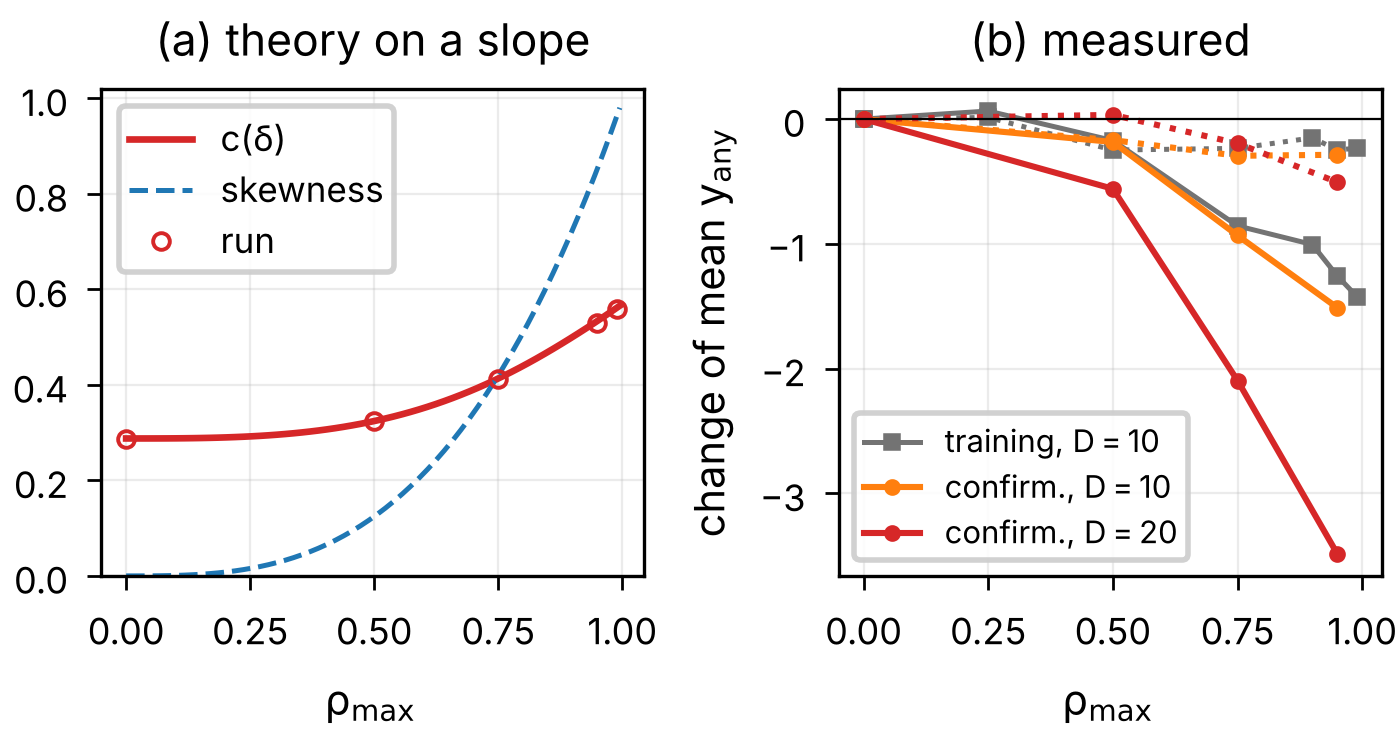}
\caption{Dose--response of the skewness. (a) Theory on a slope: fraction $c(\delta)$ of the variance retained by selection and skewness of the sampler as functions of $\rho_{\max}$; circles are runs with $N=10^{5}$. (b) Measured change of the mean $y_{\mathrm{any}}$ of \algname{}-g relative to $\rho_{\max}=0$ (EMNA+AVS); solid lines: distant start, dotted lines: standard start.}
\label{fig:dose}
\end{figure}

\emph{G3, dose and response.} The trend of \algname{}-g over the four levels $\rho_{\max}=0$, 0.5, 0.75, 0.95 is significant in all 4 settings in which it was tested (Page's test, Holm-corrected $p\le0.007$; Fig. \ref{fig:dose}(b)). For $D=20$ with the distant start the mean $y_{\mathrm{any}}$ falls from +3.81 to +3.26, +1.72 and +0.33. The shape follows the retained variance of Table \ref{tab:slope}: with the distant start only 12\% ($D=10$) and 16\% ($D=20$) of the gain is reached at $\rho_{\max}=0.5$, where $c(\delta)$ has moved from 0.288 to 0.324 only, and the rest comes above it.

\emph{G4, the gate.} As the protocol expected, the ungated skewness is practically equivalent to the gated one in all 6 settings (HL between $-$0.111 and +0.007, negative, that is in favor of the ungated form, in 5 of the 6). \algname{} is the ungated form. Measured directly against EMNA+AVS, in a comparison outside the six families and therefore descriptive, it gives HL $-$1.16, $-$3.49 and $-$3.35 with the distant start (uncorrected $p\le2\times10^{-7}$), where it solves 25, 61 and 56 runs that EMNA+AVS does not, against 11, 4 and 3 in the other direction, and HL $-$0.033, $-$0.062 and $-$0.055 with the standard start.

\emph{G5, on top of AMS and AMaLGaM.} Added to EMNA+AVS+AMS the gated skewness changes nothing: practically equivalent in all 6 settings (HL between $-$0.025 and +0.022). Added to AMaLGaM it changes nothing either: practically equivalent in all 6 settings (HL between $-$0.046 and +0.051), and no test of the family is significant after Holm's correction. The runs of this family were made with the gate thresholds $(2,6)$ of the earlier version, not with the null quantiles that the protocol prescribes, a deviation that we found after the test. Repeated with the prescribed thresholds, as a descriptive check, the skewness is practically equivalent in all 6 settings on top of EMNA+AVS+AMS (HL between $-$0.044 and +0.068) and in all 6 settings on top of AMaLGaM (HL between $-$0.030 and +0.044), and no test is significant after Holm's correction (supplementary Section \ref{S-sec:prot}, item 15).

\emph{Standing among the established algorithms (descriptive).} \algname{} does not reach the algorithms that also correct the second moment. Against EMNA+AVS+AMS it is worse in every setting, by HL +0.77, +1.41 and +2.41 with the distant start and by +0.18, +0.35 and +0.07 with the standard start (uncorrected $p\le0.009$). AMaLGaM and CMA-ES without restarts are further ahead as the dimension grows (HL +3.68 and +5.77 for $D=40$ with the distant start), although for $D=10$ and 20 with the standard start \algname{} is not significantly different from CMA-ES (HL +0.01 and +0.33; $p\ge0.47$). IPOP-CMA-ES is the best algorithm of the comparison in every setting. For $D=40$ with the distant start, the number of runs solved on the functions f1--f14 (210 runs) is 17 for EMNA+AVS, 68 for \algname{}, 105 for EMNA+AVS+AMS, 120 for AMaLGaM and 148 for CMA-ES. The mean ranks of $y_{\mathrm{any}}$ over the 144 function--setting pairs, among the 13 algorithms of the test, are 1.53 (IPOP-CMA-ES), 4.62 (AMaLGaM), 5.12 (CMA-ES), 5.59 (EMNA+AVS+AMS), 8.92 (\algname{}), 9.12 (\algname{}-g) and 10.87 (EMNA+AVS).

\emph{The variant with a mean shift.} The protocol also tested the skew-normal sampler together with a gated mean shift of one standard deviation along the path (SN+MS; supplementary Section \ref{S-sec:gate}). On top of this mean shift the skewness is again significant in all 3 settings with the distant start (G2: HL $-$0.29, $-$0.62 and $-$1.63) and practically equivalent with the standard start. SN+MS is better than \algname{} (mean rank 7.29 against 8.92) but worse than EMNA+AVS+AMS in every setting, significantly so in the 3 with the distant start (G6: HL +0.39, +0.71 and +0.75). Its mean shift needs a correction of the path for the shift it injects, without which the variant is significantly worse in all 6 settings (G4). Because the mean shift changes the first moment and requires this correction, it is not part of \algname{}.

\emph{Higher dimension (descriptive).} In supplementary runs with $D=80$ on the large-scale BBOB suite \cite{varelas2020} (five instances; supplementary Section \ref{S-sec:d80}), \algname{}-g differs from EMNA+AVS by HL $-$0.04 and $-$1.05 with the standard and the distant start ($p=0.16$ and $2\times10^{-5}$, uncorrected). At this dimension it is far behind AMaLGaM and CMA-ES: its mean $y_{\mathrm{any}}$ is +0.79 and +7.11, against $-$0.70 and +2.02 for AMaLGaM and $-$2.31 and $-$2.05 for CMA-ES. Algorithms designed for larger dimension \cite{glasmachers2025,qiu2026} were not part of the comparison.

\subsection{The Skewness Next to the Other Corrections}
\label{sec:why}

The pattern of the results is the one that the analysis of Section \ref{sec:theory} leads one to expect. Where the skewness is the only asymmetry of the sampler, it raises the retained variance on a slope from 0.288 to 0.534, and the measured effect depends on the magnitude as predicted (G1 and G3). AMS is different. By Proposition \ref{prop:ams} its offspring distribution is itself skewed along the direction of movement, with a skewness of 0.91 on a slope, and its variance along that direction is enlarged so much that selection no longer reduces it (ratio 1.89); replacing its Gaussian component by the skew-normal sampler changes the ratio by 4\%, and no effect is found (G5). The same reasoning identifies what \algname{} lacks relative to AMS. It is not the asymmetry, which it has, but the second moment: a sampler that preserves the estimated covariance cannot make the variance along the gradient grow and has to leave that to the multiplier, which the AVS of Section \ref{sec:alg} raises by the factor $1/0.8$ per generation and to at most 10.

The earlier version of this work supports this reading. Its sampler had, besides the skewness, a gated mean shift and a gated elongation that multiplies the variance along the path by $(1+e\iota_{t})^{2}$. In a confirmatory test on instances 16--30 (supplementary Section \ref{S-sec:elong}), the gated mean shift with this elongation ($\sigma_{1}=2$, $e=2$) was practically equivalent to EMNA+AVS+AMS in all 6 settings, and adding the skewness to an elongated sampler made no significant difference in any of 10 comparisons. The elongation, however, could not be left on: with $\iota_{t}\equiv1$ the configuration $\sigma_{1}=1$, $e=4$ solved 14 of the 240 training runs instead of 127, and on the confirmatory instances the gated configurations were significantly better than a tuned ungated one in 6 of 12 comparisons and never worse. Taken together, a correction that changes the estimated covariance is powerful and needs a gate; the skewness, which changes neither estimated moment, is weaker and needs none.

\subsection{Discussion}
\label{sec:disc}
\begin{itemize}
\item \emph{What is established.} A sampler that is skewed along the direction of movement and has exactly the mean and covariance estimated by EMNA; its one-generation constants on a slope; and, on fresh instances, that this skewness improves EMNA+AVS when the population starts far from the optimum, with an effect that depends on the magnitude as the theory predicts and is larger for $D=20$ and 40 than for $D=10$, and that it needs no gate.
\item \emph{What is not.} A practically relevant effect with the standard start beyond the linear slope: the estimates there are significant in two of three settings but lie inside the equivalence margin. Any effect of the skewness on top of AMS, AMaLGaM or an elongation. And any advantage of \algname{} in anytime performance over EMNA+AVS+AMS, AMaLGaM, CMA-ES or IPOP-CMA-ES.
\item \emph{Status of the evidence.} The confirmatory families were tested with the gated form of the sampler. For the ungated form, which is the algorithm of this paper, the comparison with EMNA+AVS is descriptive; that it carries over rests on the pre-registered equivalence of the two forms (G4).
\item \emph{Reproducibility.} Following \cite{lopez2021}, the supplementary package contains the source code, the frozen protocols with their hashes, the raw output of every run, the scripts that compute every number and table of this article from it, and the unit tests of the sampler; runs repeated on a second computer gave identical output (supplementary Section \ref{S-sec:prot}).
\end{itemize}

\section{Conclusion}
\label{sec:conc}

We proposed \algname{}, a Gaussian EDA that corrects the third moment of its sampling distribution. Its sampler is a skew-normal distribution with exactly the mean and the covariance estimated by EMNA and a closed-form third cumulant, skewed along a cumulative path of mean shifts with a constant magnitude. The algorithm changes one sampling step of EMNA and needs no gate, because the skewness biases neither estimated moment.

The skewness is an effective correction. On a linear function it raises the fraction of the variance that survives truncation from 0.288 to 0.534 and the progress that is possible without amplification from 2.28 to 4.12 standard deviations. On fresh benchmark instances it improved EMNA with variance scaling significantly in five of six settings; with a distant start it changed the anytime performance by $-$1.0, $-$3.3 and $-$3.3 for $D=10$, 20 and 40 (lower is better) with the gate of the protocol and by $-$1.2, $-$3.5 and $-$3.3 without it, with the predicted dependence on its magnitude. Its strength is bounded. It adds nothing detectable to AMS or to AMaLGaM, whose offspring distribution is already skewed and whose variance along the direction of movement is enlarged in addition (Proposition \ref{prop:ams}), and \algname{} stays behind the algorithms that also correct the second moment: against EMNA+AVS+AMS it is worse in every setting (HL +0.77, +1.41 and +2.41 with the distant start), AMaLGaM and CMA-ES are ahead for $D=40$, and IPOP-CMA-ES is the best algorithm of the comparison in every setting; the mean ranks over the function--setting pairs, among the 13 algorithms of the test, are 1.53 (IPOP-CMA-ES), 4.62 (AMaLGaM), 5.12 (CMA-ES), 5.59 (EMNA+AVS+AMS), 8.92 (\algname{}), 9.12 (\algname{}-g) and 10.87 (EMNA+AVS). What it lacks is a correction of the second moment, which a sampler that preserves the estimated covariance cannot provide.

The limits are those of the evidence. The slope analysis holds for an infinite population; with the population sizes used the direction is correct to within about $30^{\circ}$. The confirmatory families were tested with a gated form of the sampler, which the same test showed to be practically equivalent to \algname{}. The results hold for $D\le40$ (descriptively for $D=80$), noiseless functions, an unconstrained search space and the budget used; on the multimodal functions f15--f24 no EDA of the comparison solved more than 20\% of the runs. Two questions follow. The first is whether an asymmetric sampler outside the skew-normal family could do more: the skewness of that family is bounded by 0.9953, whereas the solutions selected on a slope have a skewness of 1.13, and the construction (\ref{eq:conv}) keeps the mean and the covariance for any scalar variable with a finite variance in place of the half-normal one. The second concerns the scale: in the earlier version of this work, a gated mean shift and elongation under the variance scaling of AMaLGaM were never significantly different from AMaLGaM (supplementary Section \ref{S-sec:elong}), which suggests that the difference between AMaLGaM and the algorithms with the simple AVS lies in the control of the scale rather than in the shape of the sampler; \algname{} under that variance scaling, without AMS, is the test to make next.
\section*{Acknowledgment}
The original idea of this work was proposed by Professor ZHANG Jun. Under Professor Zhang's direction, an AI system, Claude (Anthropic), was used to implement the algorithms and the experiments, and help complete the manuscript. The other authors and students reviewed and proofread the code, the results and the text. The authors take full responsibility for the content.

\clearpage
\onecolumn
\setcounter{section}{0}\setcounter{table}{0}\setcounter{figure}{0}\setcounter{theorem}{0}\setcounter{corollary}{0}\setcounter{equation}{0}\setcounter{algorithm}{0}
\renewcommand{\thesection}{S\arabic{section}}\renewcommand{\thetable}{S\arabic{table}}\renewcommand{\thefigure}{S\arabic{figure}}\renewcommand{\thetheorem}{S\arabic{theorem}}\renewcommand{\thecorollary}{S\arabic{corollary}}\renewcommand{\theequation}{S\arabic{equation}}\renewcommand{\thealgorithm}{S\arabic{algorithm}}
\renewcommand{\theHsection}{S\arabic{section}}\renewcommand{\theHtable}{S\arabic{table}}\renewcommand{\theHfigure}{S\arabic{figure}}\renewcommand{\theHtheorem}{S\arabic{theorem}}\renewcommand{\theHcorollary}{S\arabic{corollary}}\renewcommand{\theHequation}{S\arabic{equation}}\def\theHalgorithm{S\arabic{algorithm}}
\setlength{\parskip}{3pt}
\begin{center}{\Large Supplementary Material}\end{center}
\noindent This document contains the sections referred to in the manuscript. Section \ref{S-sec:gate} documents the gated form of the sampler and the variant with a mean shift, with which the protocols of this work were written. Sections \ref{S-sec:gate}, \ref{S-sec:test}, \ref{S-sec:checks} and \ref{S-sec:prot} and the tables in them are computed by programs from the raw results of the present work (one line per run in the accompanying data package). Section \ref{S-sec:elong} documents the variant with a covariance elongation that was the subject of the earlier version of this work; its experiments were run with the same program as those of the manuscript. Sections \ref{S-sec:param}, \ref{S-sec:heldout} and \ref{S-sec:pred} report studies made with the implementation of the earlier version; they concern algorithms without elongation, or without skewness, for which the sampler of that version coincides with the gated sampler of Section \ref{S-sec:gate}, and they are marked as such. Equation, table and theorem numbers without the prefix S refer to the manuscript. Reference numbers in square brackets refer to the reference list of the manuscript, except [S1]--[S6], which are listed at the end of this document.

\emph{Names.} The tables of this document keep the short names of the protocols. \emph{SN-on} is the algorithm of the manuscript, \algname{}: the skew-normal sampler without a gate. \emph{SN} is the gated skewness, called \algname{}-g in the manuscript; SN(0.5) and SN(0.75) are SN with $\rho_{\max}=0.5$ and 0.75. \emph{MS} is the gated mean shift ($\rho_{\max}=0$, $\sigma_{1}=1$) and \emph{SN+MS} the gated sampler with both terms. The suffix \emph{-on} denotes a configuration without the gate ($\iota_{t}\equiv1$ from the second generation on; the direction and the feedback correction are unchanged), and \emph{-nofb} the gated configuration without the feedback correction.

\section{The gated variant and the variant with a mean shift}
\label{S-sec:gate}

The protocols of this work were written for a form of the sampler in which a \emph{gate} scales the corrections along the direction of movement by an immaturity $\iota_{t}\in[0,1]$, and they included a second gated correction, a mean shift. The confirmatory test showed the gate to be dispensable for the skewness (family G4 in Table \ref{tab:fam} of the manuscript), and the algorithm of the manuscript, \algname{}, is the ungated sampler with the skewness alone. This section documents the gated form and the mean shift: the movement statistic and its exact null law, the gated sampling equation, the feedback correction that the mean shift needs, and all results of the variant with a mean shift, including the families G2 and G6 and the remaining comparisons of G4.

\subsection{The movement statistic}
\label{S-sec:stat}

Let $\Delta\boldsymbol{\mu}_{t}=\boldsymbol{\mu}_{t}-\boldsymbol{\mu}_{t-1}$. We whiten it with the current Cholesky factor and accumulate it with exponential smoothing:
\begin{equation}
\mathbf{z}_{t}=\sqrt{M}\,L_{t}^{-1}\Delta\boldsymbol{\mu}_{t}
\label{S-eq:zt}
\end{equation}
\begin{equation}
\mathbf{p}_{t}=(1-c_{p})\,\mathbf{p}_{t-1}+\sqrt{c_{p}(2-c_{p})}\,\mathbf{z}_{t}
\label{S-eq:path}
\end{equation}
\noindent The normalization $\sqrt{c_{p}(2-c_{p})}$ gives $\mathbf{p}_{t}$ the same stationary second moment as $\mathbf{z}_{t}$, because $c_{p}(2-c_{p})\sum_{k\ge0}(1-c_{p})^{2k}=1$, as in the cumulative step-size adaptation of CMA-ES \cite{hansen2001}. If the mean only fluctuates, successive $\mathbf{z}_{t}$ point in random directions and cancel; if it keeps moving, they add up. The movement statistic and the immaturity are
\begin{equation}
m_{t}=\frac{M-D-2}{M}\cdot\frac{\|\mathbf{p}_{t}\|^{2}}{D},\qquad \iota_{t}=\mathrm{clip}\left(\frac{m_{t}-\kappa_{0}}{\kappa_{1}-\kappa_{0}},0,1\right)
\label{S-eq:mstat}
\end{equation}
\noindent and the maturity is $1-\iota_{t}$. The factor $(M-D-2)/M$ removes the inverse-Wishart bias: because $L_{t}$ is estimated from $M$ solutions, $\mathbb{E}\|\mathbf{z}_{t}\|^{2}=MD/(M-D-2)$ rather than $D$ when selection causes no systematic shift (Theorem \ref{S-thm:calib}). After the correction $\mathbb{E}m_{t}=1$ in that case, whereas $m_{t}$ grows linearly with $M/D$ on a slope. The skewness and the mean shift of Section \ref{S-sec:gsampler} are off for $m_{t}\le\kappa_{0}$ and act with full strength for $m_{t}\ge\kappa_{1}$; the thresholds are set from the null distribution in Section \ref{S-sec:null}.

The direction $\mathbf{w}_{t}$ is that of (\ref{eq:Pt}) of the manuscript; in the first generation, and whenever $\mathbf{u}_{t}=\mathbf{0}$, $\mathbf{w}_{t}=\mathbf{0}$ and $\iota_{t}=0$. The increments $\mathbf{z}_{t}$ are whitened with the Cholesky factor of their own generation, so the path adds vectors expressed in frames that change from one generation to the next, and the statistic is not invariant under a rotation of the problem. Under the null model this choice is exactly what makes the law of $m_{t}$ free of the covariance (Theorem \ref{S-thm:calib}(b)); away from it the statistic is a heuristic measure of persistent movement, and Section \ref{S-sec:inside} quantifies the dependence on the coordinate system with paired rotations and compares the Cholesky factor with the symmetric square root.

\subsection{Calibration: null law and thresholds}
\label{S-sec:null}
The thresholds are set from the law of $m_{t}$ under an \emph{ideal null model}: in every generation the $M$ selected solutions are independent draws from the sampling distribution $\mathcal{N}(\boldsymbol{\mu}_{t-1},s\Sigma_{t-1})$. Selection produces such draws when it does not prefer any region (a neutral function without elitism), with the corrections off, no clipping, no regularization and $M>D+2$.
\begin{theorem}[Null law of the movement statistic]
\label{S-thm:calib}
Under the ideal null model:
(a) $\|\mathbf{z}_{t}\|^{2}\sim\frac{MD}{M-D}F_{D,M-D}$, hence $\mathbb{E}\|\mathbf{z}_{t}\|^{2}=MD/(M-D-2)$.
(b) $\mathbf{z}_{t}=\sqrt{M}\,T_{t}^{-1}\mathbf{g}_{t}$, where $\mathbf{g}_{t}\sim\mathcal{N}(\mathbf{0},I)$ and $T_{t}$ is a lower-triangular matrix independent of $\mathbf{g}_{t}$ with $T_{ii}^{2}\sim\chi^{2}_{M-i}$ and standard normal entries below the diagonal. The $\mathbf{z}_{t}$ of different generations are independent and identically distributed, and their law depends on $D$ and $M$ only: not on $\boldsymbol{\mu}$, $\Sigma$, $s$ or the history.
(c) After $k$ updates of a path started at $\mathbf{p}=\mathbf{0}$, $\mathbb{E}m=1-(1-c_{p})^{2k}$, which tends to one.
(d) The law of $m_{t}$ is that of a quadratic form in the accumulated path; it is not an $F$ or $\chi^{2}$ law, and its quantiles are obtained by sampling the recursion (\ref{S-eq:path}) with the $\mathbf{z}_{t}$ of (b).
\end{theorem}
\emph{Sketch of proof} (details in Section \ref{S-sec:proof}). Write the selected solutions as $\mathbf{x}_{j}=\boldsymbol{\mu}_{t-1}+\sqrt{s}L_{t-1}\mathbf{g}_{j}$ with independent $\mathbf{g}_{j}\sim\mathcal{N}(\mathbf{0},I)$, and let $\bar{\mathbf{g}}$ and $S$ be the mean and the maximum-likelihood covariance of the $\mathbf{g}_{j}$. Then $\Delta\boldsymbol{\mu}_{t}=\sqrt{s}L_{t-1}\bar{\mathbf{g}}$ and, by uniqueness of the Cholesky factor, $L_{t}=\sqrt{s}L_{t-1}\mathrm{chol}(S)$. Hence
\begin{equation}
\mathbf{z}_{t}=\sqrt{M}L_{t}^{-1}\Delta\boldsymbol{\mu}_{t}=\sqrt{M}\,\mathrm{chol}(S)^{-1}\bar{\mathbf{g}}
\label{S-eq:znull}
\end{equation}
\noindent contains neither $L_{t-1}$ nor $s$. Parts (a) and (b) follow from Hotelling's $T^{2}$ distribution and from Bartlett's decomposition of $MS$ \cite{muirhead1982}, and parts (c) and (d) from the independence of the $\mathbf{z}_{t}$ and the recursion (\ref{S-eq:path}).

Part (b) is specific to whitening with the Cholesky factor of a sampler of the form $\boldsymbol{\mu}+\sqrt{s}L\mathbf{g}$: the triangular factors telescope, and the null law of the statistic becomes a fixed distribution that can be sampled in $O(D^{2})$ operations per step without running the algorithm. We computed its stationary quantiles with $10^{6}$ steps ($2\times10^{5}$ for $D=80$) of (\ref{S-eq:path}) and (\ref{S-eq:znull}), with $c_{p}=0.3$ and the population sizes of Table \ref{S-tab:paramsg}: the 99\% quantile is 3.54, 2.53, 1.95, 1.74, 1.62 and 1.42 for $D=5$, 10, 20, 30, 40 and $80$ (Table \ref{S-tab:null} gives the other quantiles and the agreement of the samples with (a) and (c)). We set $\kappa_{0}$ to the 99\% quantile for the dimension at hand and $\kappa_{1}=\kappa_{0}+4$.

\emph{What the threshold does and does not guarantee.} The 1\% is a per-generation probability under the ideal null model. It is not a run-wise error rate: consecutive values of $m_{t}$ are correlated through the path and a run has hundreds of generations, so a population that sits at its optimum will open the gate now and then (with probability 0.92 at least once in 300 generations for $D=10$). Such an opening has a small effect: the immaturity averages 0.0009 for $D=10$ and 0.0003 for $D=40$ under the null model, and the feedback correction keeps an injected shift from sustaining itself. The implemented algorithm also departs from the ideal model in four ways, which Section \ref{S-sec:inside} measures one at a time: the elitist row, selection itself (near an optimum the selected solutions are not a Gaussian sample), the relative ridge, and the skewness and the mean shift once the gate has opened. The statistic is therefore used as a \emph{gate calibrated on a null model}, not as an exact test for the algorithm, and it measures persistent movement of the mean, not distance to the global optimum: a population that has stopped moving on a plateau or in a local optimum is mature in this sense.

\begin{corollary}[Conditional one-generation reduction to EMNA]
\label{S-cor:emna}
If $\iota_{t}=0$ in generation $t$, then both terms of Section \ref{S-sec:gsampler} vanish and, given the same $\boldsymbol{\mu}_{t}$, $\Sigma_{t}$, $s_{t}$ and Gaussian variates, the new population is that of EMNA. The statement concerns one generation: the gate need not stay closed, and the trajectory need not equal that of an EMNA run from the same initial population, since earlier generations may have been sampled with an open gate.
\end{corollary}
\subsection{The gated sampling equation: skewness and mean shift}
\label{S-sec:gsampler}

Let $\mathbf{z}_{0}\in\mathbb{R}^{N}$ and $Z\in\mathbb{R}^{N\times D}$ be standard normal, $s$ the variance multiplier, $\mathbf{w}$ the whitened unit direction and $\iota$ the immaturity of the current generation, and write
\begin{equation}
\delta=\iota\,d_{\max},\qquad \beta=1-\sqrt{1-(1-b^{2})\delta^{2}}
\label{S-eq:adb}
\end{equation}
\noindent for the size of the skewness and the contraction of Section \ref{sec:skew} of the manuscript. The gated variant samples
\begin{equation}
\begin{gathered}
X=\mathbf{1}\boldsymbol{\mu}^{T}+\sqrt{s}\,YL^{T},\qquad Y=Z+\mathbf{h}\mathbf{w}^{T}+\sigma_{1}\iota\,\mathbf{1}\mathbf{w}^{T}\\
\mathbf{h}=\delta\,(|\mathbf{z}_{0}|-b\mathbf{1})-\beta Z\mathbf{w}
\end{gathered}
\label{S-eq:mg}
\end{equation}
\noindent The rows of $Z+\mathbf{h}\mathbf{w}^{T}$ are the standardized skew-normal vectors of (\ref{eq:conv}) of the manuscript with $\mathbf{d}=\delta\mathbf{w}$: the population is skewed along the direction of movement by an amount that grows with the immaturity. The last term is a \emph{mean shift} of $\sigma_{1}\iota$ standard deviations along the same direction, a first-moment correction in the spirit of AMS that uses the same direction and the same gate; $\sigma_{1}=0$ gives the skewness alone. Both are rank-one terms of one matrix equation.
\begin{theorem}[Moments of the gated sampler]
\label{S-thm:joint}
For $0\le\iota\le1$, $\sigma_{1}\ge0$ and $\rho_{\max}<1$, a row $\mathbf{x}$ of (\ref{S-eq:mg}) has
\begin{align*}
\mathbb{E}\mathbf{x}&=\boldsymbol{\mu}+\sqrt{s}\,\sigma_{1}\iota\,L\mathbf{w},\\
\mathrm{Cov}\,\mathbf{x}&=s\,LL^{T},
\end{align*}
and third cumulant $s^{3/2}K_{3}\delta^{3}\,(L\mathbf{w})\otimes(L\mathbf{w})\otimes(L\mathbf{w})$. With $\iota=0$, $\mathbf{x}\sim\mathcal{N}(\boldsymbol{\mu},sLL^{T})$.
\end{theorem}
\begin{proof}
By Theorem \ref{thm:mom} of the manuscript with $\mathbf{d}=\delta\mathbf{w}$, a row of $Z+\mathbf{h}\mathbf{w}^{T}$ has mean zero, covariance $I$ and third cumulant $K_{3}\delta^{3}\mathbf{w}^{\otimes3}$. The shift adds $\sigma_{1}\iota\mathbf{w}$ to the mean only, and cumulants transform multilinearly under the affine map $\mathbf{x}=\boldsymbol{\mu}+\sqrt{s}L\mathbf{y}$. For $\iota=0$, $\delta=\beta=0$ and $\mathbf{y}=\mathbf{z}$.
\end{proof}

\emph{The skewness changes neither the mean nor the covariance of the sampling distribution, whatever its magnitude}: it can be weak, strong or permanently on without biasing the two moments that EMNA estimates. A mean shift, by contrast, moves the distribution by construction, and an anticipated mean shift or a covariance elongation changes its second moment as well. One consequence, confirmed in Section \ref{sec:expts} of the manuscript, is that the skewness does not need to be switched off at convergence, whereas the mean shift needs the correction described next. A unit test with $2\times10^{6}$ rows, a random non-diagonal $L$ and an arbitrary $\mathbf{w}$ reproduces the three moments of Theorem \ref{S-thm:joint} within sampling error (Section \ref{S-sec:checks}).

\emph{Feedback correction of the mean shift.} The mean shift changes the quantity the statistic measures. If the new population is shifted by $\mathbf{s}_{t}=\sqrt{s}\,\sigma_{1}\iota_{t}L\mathbf{w}_{t}$, the next selected mean moves by about $\mathbf{s}_{t}$ even under neutral selection, and the statistic registers its own shift as movement. A constant whitened increment $\mathbf{z}$ gives the path the stationary length $\sqrt{(2-c_{p})/c_{p}}\,\|\mathbf{z}\|$, and the injected shift has the whitened length $\sqrt{M}\sigma_{1}\iota$, so that by itself it contributes
\begin{equation}
m^{\mathrm{inj}}=\frac{2-c_{p}}{c_{p}}\cdot\frac{M-D-2}{D}\cdot\sigma_{1}^{2}\iota^{2}
\label{S-eq:minj}
\end{equation}
\noindent to the statistic. For $D=10$, $\sigma_{1}=1$ and $\iota=1$ this is 14.7, against $\kappa_{1}=6.53$: the shift can hold its own gate open and keep pushing the population past the optimum. We therefore remove the injected part before updating the paths: $\Delta\boldsymbol{\mu}_{t}\leftarrow\boldsymbol{\mu}_{t}-\boldsymbol{\mu}_{t-1}-\mathbf{s}_{t-1}$ in (\ref{S-eq:zt}) and (\ref{eq:Pt}) of the manuscript. The subtraction is exact when the shift is a common translation of the whole population and selection is translation invariant; in the algorithm it is an approximation, because $\mathbf{s}_{t-1}$ is the nominal shift whereas the shift that survives selection and the elitist row may differ from it. The skewness needs no such correction, because it does not move the mean.

\begin{algorithm}[htbp]
\caption{The gated variant with skewness and mean shift (one run)}
\label{S-alg:mg}
\begin{algorithmic}[1]
\REQUIRE matrix objective $F$; $N,M,D$; budget $E$; $c_{p}$; $\kappa_{0}<\kappa_{1}$; $\rho_{\max}<1$; $\sigma_{1}\ge0$; $\eta$
\ENSURE best solution $\mathbf{x}_{bs}$ and $f(\mathbf{x}_{bs})$
\STATE Sample $X$ (uniform in a box, or $\mathcal{N}(\mathbf{c}_{0},\sigma_{0}^{2}I)$); $\mathbf{f}\leftarrow F(X)$; $\mathbf{x}_{bs}\leftarrow$ best row; $\mathbf{P},\mathbf{p},\mathbf{s}_{0},\mathbf{w}\leftarrow\mathbf{0}$; $c_{\mathrm{mult}}\leftarrow1$
\FOR[while $N-1$ evaluations remain]{$t=1,2,\ldots$}
\STATE $X_{s}\leftarrow$ rows of $X$ with the $M$ smallest $\mathbf{f}$; $\boldsymbol{\mu},\Sigma$ by (\ref{eq:emna}) of the manuscript; $\Sigma\leftarrow\Sigma+\epsilon I$; $L\leftarrow\mathrm{chol}(\Sigma)$
\IF{$t=1$}
\STATE $\iota_{t}\leftarrow0$; $\mathbf{w}\leftarrow\mathbf{0}$
\ELSE
\STATE $\Delta\boldsymbol{\mu}\leftarrow\boldsymbol{\mu}-\boldsymbol{\mu}_{t-1}-\mathbf{s}_{t-1}$; update $\mathbf{p},m_{t},\iota_{t}$ by (\ref{S-eq:zt})--(\ref{S-eq:mstat}) and $\mathbf{P},\mathbf{w}$ by (\ref{eq:Pt}) of the manuscript
\ENDIF
\STATE draw $Z,\mathbf{z}_{0}$; $X\leftarrow$ (\ref{S-eq:adb}), (\ref{S-eq:mg}) with $s=\eta c_{\mathrm{mult}}$ \COMMENT{EMNA if $\iota_{t}=0$}
\STATE $\mathbf{s}_{t}\leftarrow\sqrt{s}\,\sigma_{1}\iota_{t}L\mathbf{w}$
\STATE first row $\leftarrow\mathbf{x}_{bs}$ (not re-evaluated); evaluate the other $N-1$ rows; update $\mathbf{x}_{bs}$
\STATE AVS: $c_{\mathrm{mult}}\leftarrow\min(c_{\mathrm{mult}}/0.8,10)$ if improved, else $\max(0.8c_{\mathrm{mult}},1)$
\ENDFOR
\end{algorithmic}
\end{algorithm}

Algorithm \ref{S-alg:mg} gives one run; Table \ref{S-tab:paramsg} lists the parameters with the values used. Per generation the work is that of EMNA, $\Theta(NC_{F}+N\log N+(M+N)D^{2}+D^{3})$, where $C_{F}$ is the cost of one function evaluation, plus $O(ND)$ for the two rank-one terms and two triangular solves of $O(D^{2})$ for the paths; the memory is that of EMNA, $O(ND+D^{2})$. \emph{Budget:} the initial population costs $N$ evaluations and every generation $N-1$, because the elitist row keeps its stored value, so a budget of $E$ gives $\lfloor(E-N)/(N-1)\rfloor$ generations; a last partial generation is not run. \emph{Ties:} solutions are sorted by value and, among equal values, by row index, so the elitist row precedes its ties. \emph{Regularization:} $\epsilon=10^{-14}\mathrm{tr}(\Sigma)/D$; if the Cholesky factorization still fails, $\epsilon$ is multiplied by 10 until it succeeds (this never happened in the experiments reported). \emph{Domain:} $M>D+2$ is required (the statistic uses $M-D-2$).

\begin{table}[htbp]
\caption{Parameters of the gated variant and the values used}
\label{S-tab:paramsg}
\centering \footnotesize
\begin{tabular}{@{}lll@{}}
\toprule
Parameter & Value & Remark \\
\midrule
$N$; $\tau$ & $\lfloor17+3D^{1.5}\rfloor$; 0.35 & population size; truncation ratio \\
$\rho_{\max}$ & 0.95 & magnitude of the skewness, which is $0.853\,\iota_{t}^{3}$ \\
$\sigma_{1}$ & 1; 0 & mean shift in standard deviations: 1 in SN+MS and MS, 0 in SN \\
$c_{p}$ & 0.3 & learning rate of the paths \\
$\kappa_{0}$; $\kappa_{1}$ & Section \ref{S-sec:null}; $\kappa_{0}+4$ & 99\% null quantile of $m_{t}$: 2.53, 1.95, 1.62 for $D=10$, 20, 40 \\
AVS & $\times1/0.8$, $\times0.8$ & multiplier in [1, 10] \\
\bottomrule
\end{tabular}
\end{table}

\subsection{On a slope and on a plateau}
\label{S-sec:gslope}

With a mean shift, Theorem \ref{thm:slope} of the manuscript holds with $\lambda(\delta)$ replaced by $\Lambda=\sigma_{1}\iota+\lambda(\delta)$ in the map of the mean, and Corollary \ref{cor:prog} of the manuscript with the same replacement; the proof in Section \ref{S-sec:proof} is written for this case. Table \ref{S-tab:slopems} gives the constants. The mean shift does not change the retained variance but adds $\sigma_{1}$ to the step: it is a progress term, not a remedy for the variance loss.
\begin{table}[htbp]
\caption{One-generation constants on a linear function with a mean shift ($\tau=0.35$, $s=1$, $\iota=1$): theory and a run with $D=10$, $N=10^{5}$, no elitism (mean of generations 5--13); columns as in Table \ref{tab:slope} of the manuscript}
\label{S-tab:slopems}
\centering \footnotesize
\begin{tabular}{@{}lccccc@{}}
\toprule
 & \multicolumn{2}{c}{variance ratio} & \multicolumn{2}{c}{step} & \\
\cmidrule(lr){2-3}\cmidrule(lr){4-5}
Sampler & theory & run & theory & run & $1/c$ \\
\midrule
Gaussian (EMNA) & 0.288 & 0.287 & 1.058 & 1.059 & 3.47 \\
mean shift, $\sigma_{1}=1$ (MS) & 0.288 & 0.287 & 2.058 & 2.059 & 3.47 \\
skewness 0.95 (SN) & 0.534 & 0.531 & 1.109 & 1.108 & 1.87 \\
skewness 0.95 + mean shift (SN+MS) & 0.534 & 0.531 & 2.109 & 2.108 & 1.87 \\
\bottomrule
\end{tabular}
\end{table}

With the feedback correction the whitened increment on a slope is $\|\mathbf{z}_{t}\|^{2}=M\lambda(\delta)^{2}/c(\delta)$ whatever $\sigma_{1}$ is, because the injected shift is removed; a component of constant direction accumulates in the path with the factor $(2-c_{p})/c_{p}$, so the stationary value of the statistic is
\begin{equation}
m_{t}\approx\frac{2-c_{p}}{c_{p}}\cdot\frac{M-D-2}{D}\cdot\frac{\lambda(\delta)^{2}}{c(\delta)}
\label{S-eq:mslope}
\end{equation}
\noindent which is 57 for the Gaussian sampler and 34 at $\rho_{\max}=0.95$ with $D=10$, $M=38$, and grows with $M/D$: on a slope the gate is fully open, $\iota_{t}=1$.

\begin{theorem}[Drift under a closed gate]
\label{S-thm:drift}
If in generation $t$ all solutions have equal fitness, elitism is off and $\iota_{t}=0$, then $\mathbb{E}[\Sigma_{t+1}\mid\Sigma_{t}]=s\,(1-1/M)\,\Sigma_{t}$ and $\mathrm{Cov}(\boldsymbol{\mu}_{t+1}-\boldsymbol{\mu}_{t}\mid\Sigma_{t})=s\Sigma_{t}/M$.
\end{theorem}
Indeed, with $\iota_{t}=0$ the sampler (\ref{S-eq:mg}) is Gaussian with covariance $s\Sigma_{t}$ (Theorem \ref{S-thm:joint}) and the selected solutions are $M$ independent samples of it, whose maximum-likelihood covariance has expectation $(1-1/M)s\Sigma_{t}$ and whose mean has covariance $s\Sigma_{t}/M$. The theorem is conditional on the gate being closed in that generation. On a plateau the ideal null model of Theorem \ref{S-thm:calib} applies, so the gate is closed in 99\% of the generations but not in all, and in the generations in which it opens the corrections act along a direction that is pure noise. The improvement-triggered AVS does not compensate the loss of variance on a plateau, since nothing improves there. The gated variant therefore has no mechanism for crossing a plateau that EMNA lacks; Section \ref{S-sec:inside} measures the escape from a plateau and confirms this.

\subsection{The statistic inside the algorithm}
\label{S-sec:inside}

The tables and the figure behind this subsection are in Section \ref{S-sec:checks}.

\emph{Departures from the null model.} With the corrections off, the algorithm reproduces the null law on a neutral function: without elitism the gate is open in 1.0\% of the generations for $D=10$ and 1.2\% for $D=20$ (1.2\% and 1.3\% with the elitist row), against the nominal 1\%. Selection matters more: on a sphere centered on the initial mean the rate is 4.5\% and 1.9\%, because the solutions kept by selection near an optimum are not a Gaussian sample; the mean immaturity nevertheless stays at 0.010 and 0.001, so the corrections are weak when they act there. Run-wise, the gate opens at least once within 55 generations of the neutral function in 37\% of the runs for $D=10$, which confirms the caveat of Section \ref{S-sec:null}.

\emph{Feedback correction.} On the neutral function (elitism, 300 generations) an opening of the gate lasts 1.2 generations on average for SN, MS and SN+MS with the feedback correction, in every dimension tried, and the immaturity never reaches one. Without the correction the behavior follows (\ref{S-eq:minj}): an opening of SN+MS lasts 13.5, 7.0 and 4.1 generations on average for $D=10$, 20 and 40, the gate is fully open in 5.8\%, 3.3\% and 2.0\% of all generations, and single openings last up to 294 of the 295 generations evaluated.

\emph{Coordinate system.} The dependence on the coordinate system noted in Section \ref{S-sec:stat} is small. In paired runs of SN+MS on a sphere, an ellipsoid with condition number $10^{6}$ and the Rosenbrock function, with and without a random rotation (15 pairs, $D=10$ and 20), the median number of evaluations to $10^{-8}$ changed by at most 13\% under rotation, and by at most 11\% when the Cholesky factor was replaced by the symmetric square root of the covariance. We keep the Cholesky factor, for which the null law is exact.

\emph{When the corrections act.} With the distant start the median immaturity of SN+MS is one from generation 2 to generation 12 on the sphere and on the Rosenbrock function ($D=10$ and 20; Fig. \ref{S-fig:mat}): every generation improves, the multiplier grows, and the distribution widens until it covers the optimum while the mean hardly moves. The immaturity is back at zero in generation 15 for $D=10$ and in generation 16 (sphere) and 18 (Rosenbrock) for $D=20$. Afterwards the gate reopens in episodes: in 6\% of the generations on the sphere with $D=10$ (mean immaturity 0.01), once more between generations 22 and 46 for $D=20$ while the mean travels to the optimum, and intermittently along the valley of the Rosenbrock function (32\% and 55\% of the generations for $D=10$ and 20). The two terms thus act at full strength for about ten generations at the start of a distant approach and in episodes afterwards.

\subsection{Results of the variant with a mean shift}
\label{S-sec:msres}

\emph{Training set (descriptive).} Table \ref{S-tab:trainms} gives the mean $y_{\mathrm{any}}$ on the training set with and without the gated mean shift. The skewness improves the gated mean shift as it improves EMNA+AVS (HL $-$0.14, $p=0.002$ for SN+MS against MS). Removing the gate from SN+MS makes no significant difference (mean $y_{\mathrm{any}}$ $-$3.01 against $-$3.14; HL +0.03, $p=0.48$). A mean shift of 0.5, 1, 1.5, 2 and 3 standard deviations gives the means $-$2.89, $-$3.14, $-$3.12, $-$3.11 and $-$3.23; values above one were not significantly better, and $\sigma_{1}=1$ was kept. SN+MS remains behind the established algorithms (HL +0.19, $p=2\times10^{-5}$ against EMNA+AVS+AMS and HL +0.22, $p=0.02$ against AMaLGaM).
\begin{table}[htbp]
\caption{Training set (BBOB f1--f24, $D=10$, instances 1--5, both starts; 48 blocks): mean $y_{\mathrm{any}}$ as a function of the magnitude of the skewness, without and with the gated mean shift ($\sigma_{1}=1$, with feedback correction)}
\label{S-tab:trainms}
\centering \footnotesize
\begin{tabular}{@{}ccc@{}}
\toprule
$\rho_{\max}$ & skewness alone & with mean shift \\
\midrule
0 & $-$1.77 & $-$2.64 \\
0.5 & $-$1.98 & $-$2.92 \\
0.75 & $-$2.32 & $-$2.95 \\
0.95 & $-$2.52 & $-$3.14 \\
0.99 & $-$2.60 & $-$3.23 \\
\bottomrule
\end{tabular}
\par\vspace{2pt}\begin{minipage}{0.5\textwidth}\scriptsize $\rho_{\max}=0$ is EMNA+AVS (left) and MS (right); 0.95 is SN and SN+MS.\end{minipage}
\end{table}

\begin{figure}[htbp]
\centering
\includegraphics[width=0.9\textwidth]{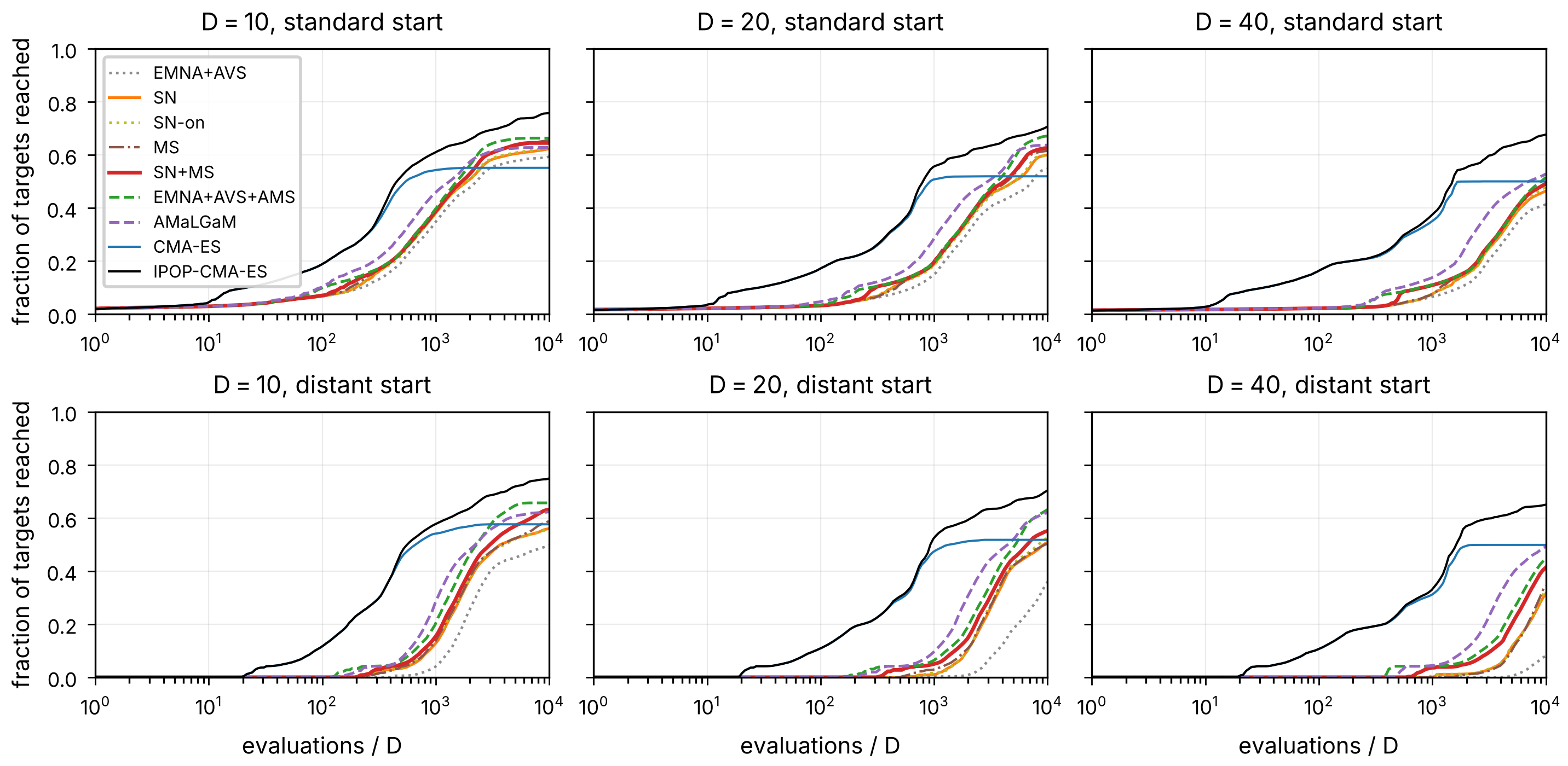}
\caption{Running-time distributions on the confirmatory set, including the configurations with a mean shift (names of the protocol): fraction of the (function, instance, target) triples reached within a given number of evaluations; targets and layout as in Fig. \ref{fig:ecdf} of the manuscript.}
\label{S-fig:ecdfms}
\end{figure}

\emph{Confirmatory test.} Table \ref{S-tab:testms} gives the results of the configurations with a mean shift on the confirmatory instances 31--45, next to EMNA+AVS and the skewness alone, and Table \ref{S-tab:famms} the comparisons of the families G2, G4 and G6 and the descriptive comparisons; the mean ranks are those of Table \ref{tab:test} of the manuscript. Fig. \ref{S-fig:ecdfms} shows the running-time distributions of these configurations.
\begin{table}[htbp]
\caption{Confirmatory test, configurations with a mean shift: number of functions solved to $10^{-8}$ in all 15 runs and mean $y_{\mathrm{any}}$ over the 24 functions (lower is better); mean rank of $y_{\mathrm{any}}$ among all algorithms of the test, as in Table \ref{tab:test} of the manuscript}
\label{S-tab:testms}
\centering \scriptsize
\begin{tabular}{@{}l@{\hspace{5pt}}rr@{\hspace{7pt}}rr@{\hspace{7pt}}rr@{\hspace{7pt}}rr@{\hspace{7pt}}rr@{\hspace{7pt}}rr@{\hspace{7pt}}cc@{}}
\toprule
 & \multicolumn{2}{c}{$D=10$, std.} & \multicolumn{2}{c}{$D=10$, dist.} & \multicolumn{2}{c}{$D=20$, std.} & \multicolumn{2}{c}{$D=20$, dist.} & \multicolumn{2}{c}{$D=40$, std.} & \multicolumn{2}{c}{$D=40$, dist.} & \multicolumn{2}{c}{mean rank} \\
Configuration & sol. & $y$ & sol. & $y$ & sol. & $y$ & sol. & $y$ & sol. & $y$ & sol. & $y$ & $D\le20$ & all \\
\midrule
EMNA+AVS & 8 & $-$3.16 & 7 & $-$0.14 & 9 & $-$1.95 & 4 & 3.81 & 5 & $-$0.44 & 0 & 7.36 & 12.02 & 10.87 \\
SN & 11 & $-$3.44 & 9 & $-$1.65 & 9 & $-$2.45 & 8 & 0.33 & 6 & $-$0.85 & 2 & 3.90 & 9.26 & 9.12 \\
SN-on & 10 & $-$3.47 & 8 & $-$1.67 & 9 & $-$2.46 & 7 & 0.17 & 6 & $-$0.91 & 2 & 3.86 & 9.24 & 8.92 \\
MS & 11 & $-$3.70 & 8 & $-$1.92 & 11 & $-$2.65 & 8 & 0.27 & 7 & $-$1.03 & 2 & 4.00 & 8.59 & 8.52 \\
SN+MS & 11 & $-$3.70 & 8 & $-$2.48 & 11 & $-$2.73 & 8 & $-$0.51 & 7 & $-$1.06 & 4 & 2.16 & 7.34 & 7.29 \\
SN+MS-on & 10 & $-$3.69 & 8 & $-$2.50 & 11 & $-$2.87 & 8 & $-$0.62 & 7 & $-$1.06 & 3 & 2.18 & 8.47 & 7.82 \\
SN+MS-nofb & 3 & $-$2.02 & 2 & 0.82 & 1 & $-$0.29 & 1 & 2.80 & 1 & 0.94 & 1 & 4.29 & 12.95 & 11.35 \\
\bottomrule
\end{tabular}
\end{table}
\begin{table}[htbp]
\caption{Confirmatory comparisons that involve the mean shift: Hodges--Lehmann estimate of the difference in $y_{\mathrm{any}}$ (first minus second; negative means that the first configuration is better) over the 24 functions of each setting}
\label{S-tab:famms}
\centering \footnotesize
\begin{tabular}{@{}llcccccc@{}}
\toprule
 & & \multicolumn{2}{c}{$D=10$} & \multicolumn{2}{c}{$D=20$} & \multicolumn{2}{c}{$D=40$} \\
\cmidrule(lr){3-4}\cmidrule(lr){5-6}\cmidrule(l){7-8}
Family & Comparison & std. & dist. & std. & dist. & std. & dist. \\
\midrule
G2 (with mean shift) & SN+MS vs. MS & +0.01$^{\approx}$ & \textbf{$-$0.29} & $-$0.03$^{\approx}$ & \textbf{$-$0.62} & 0.00$^{\approx}$ & \textbf{$-$1.63} \\
\midrule
G4 (gate, feedback) & SN+MS-on vs. SN+MS & +0.02$^{\approx}$ & +0.10$^{\approx}$ & $-$0.02 & $-$0.01$^{\approx}$ & 0.00$^{\approx}$ & 0.00$^{\approx}$ \\
 & SN+MS-nofb vs. SN+MS & \textbf{+1.46} & \textbf{+3.30} & \textbf{+2.38} & \textbf{+3.12} & \textbf{+1.79} & \textbf{+1.69} \\
\midrule
G6 (distance) & SN+MS vs. EMNA+AVS+AMS & +0.13 & \textbf{+0.39} & +0.14 & \textbf{+0.71} & +0.04$^{\approx}$ & \textbf{+0.75} \\
 & SN+MS vs. AMaLGaM & +0.05$^{\approx}$ & +0.34 & +0.32 & \textbf{+1.04} & \textbf{+0.63} & \textbf{+2.22} \\
\midrule
descriptive & MS vs. EMNA+AVS & \textbf{$-$0.26} & \textbf{$-$1.29} & \textbf{$-$0.21} & \textbf{$-$3.31} & \textbf{$-$0.11}$^{\approx}$ & \textbf{$-$3.27} \\
 & SN+MS vs. SN & $-$0.05$^{\approx}$ & \textbf{$-$0.55} & $-$0.09 & \textbf{$-$0.79} & \textbf{$-$0.07}$^{\approx}$ & \textbf{$-$1.25} \\
 & SN+MS vs. EMNA+AVS & \textbf{$-$0.09} & \textbf{$-$1.65} & \textbf{$-$0.22} & \textbf{$-$4.20} & \textbf{$-$0.10} & \textbf{$-$5.20} \\
 & SN+MS vs. CMA-ES & $-$0.09 & +0.70 & +0.18 & \textbf{+1.53} & +1.18 & \textbf{+4.08} \\
 & SN+MS vs. IPOP-CMA-ES & \textbf{+1.08} & \textbf{+1.99} & \textbf{+1.11} & \textbf{+2.61} & \textbf{+2.06} & \textbf{+5.11} \\
\bottomrule
\end{tabular}
\par\vspace{2pt}\begin{minipage}{0.86\textwidth}\scriptsize Bold: significant at $\alpha=0.05$ (Wilcoxon signed-rank test over the 24 functions; in G2, G4 and G6 after Holm's correction over all tests of the family, the family G4 including the comparison of SN-on with SN of Table \ref{tab:fam} of the manuscript; in the descriptive rows uncorrected). $^{\approx}$: practically equivalent (90\% interval of HL within $\pm0.25$).\end{minipage}
\end{table}

\emph{G2, the skewness with a mean shift.} On top of the gated mean shift the skewness is again significant in all 3 settings with the distant start, with an effect that grows with the dimension (HL $-$0.29, $-$0.62 and $-$1.63; Holm-corrected $p\le0.0004$), and makes no difference with the standard start (practically equivalent in all 3, HL between $-$0.034 and +0.006).

\emph{G4, gate and feedback correction.} As expected, the skewness without the gate ($\iota_{t}\equiv1$) is practically equivalent to the gated skewness in all 6 settings (HL between $-$0.111 and +0.007, negative, that is in favor of the ungated version, in 5 of the 6). SN+MS without the gate is practically equivalent to SN+MS in 5 of the six settings and not significantly different in any (HL between $-$0.024 and +0.098). The expectation of the protocol was wrong, however, for the feedback correction: without it SN+MS is significantly worse in all 6 settings, by +1.46 to +3.30, and worse even than EMNA+AVS in the mean rank (11.35 against 10.87). The two results are compatible because the ungated variants still remove the injected shift from the path: what the mean shift needs is not a gate but a path that does not contain its own shift (on the wrong expectation of the protocol see Section \ref{S-sec:prot}).

\emph{G6, distance to the established EDAs.} SN+MS does not reach the established EDAs. Against EMNA+AVS+AMS it is worse in every setting, significantly in the 3 with the distant start (HL +0.39, +0.71 and +0.75) and by +0.04 to +0.14 with the standard start (not significant after correction; practically equivalent for $D=40$). Against AMaLGaM it is practically equivalent only for $D=10$ with the standard start (HL +0.05) and significantly worse in 3 settings, by up to +2.22 for $D=40$ with the distant start.

\emph{Descriptive comparisons.} The gated mean shift alone is about as effective as the skewness alone (HL of MS against EMNA+AVS with the distant start: $-$1.29, $-$3.31 and $-$3.27), and the two are complementary: together they change the performance of EMNA+AVS by $-$1.65, $-$4.20 and $-$5.20. For $D=40$ with the distant start, the number of runs solved on the functions f1--f14 (210 runs) is 17 for EMNA+AVS, 62 for SN, 69 for MS, 99 for SN+MS, 105 for EMNA+AVS+AMS, 120 for AMaLGaM and 148 for CMA-ES. IPOP-CMA-ES is the best algorithm of the comparison in every setting (mean ranks in Table \ref{tab:test} of the manuscript). Against CMA-ES without restarts, SN+MS is not significantly different for $D=10$ and for $D=20$ with the standard start (HL $-$0.09, +0.70 and +0.18; uncorrected $p\ge0.06$); for $D=10$ with the standard start it solves 39 runs that CMA-ES does not, against 1 in the other direction, mainly on the step ellipsoid f7 and on the multimodal functions f15--f24, so that the two algorithms have different footprints on the benchmark \cite{nikolikj2025}. For $D=40$ CMA-ES is better (HL +1.18 and +4.08; 124 such runs against 34).

\emph{Higher dimension (descriptive).} In supplementary runs with $D=80$ on the large-scale BBOB suite \cite{varelas2020} (five instances; Section \ref{S-sec:d80}), SN differs from EMNA+AVS by HL $-$0.04 and $-$1.05 with the standard and the distant start ($p=0.16$ and $2\times10^{-5}$, uncorrected), SN+MS from MS by 0.00 and $-$2.62 ($p=0.94$ and $1\times10^{-7}$), and SN+MS from EMNA+AVS+AMS by +0.02 and +0.96. At this dimension SN+MS is behind AMaLGaM (HL +1.43 and +2.60) and CMA-ES (+2.98 and +6.15). Algorithms designed for larger dimension \cite{glasmachers2025,qiu2026} were not part of the comparison.

\section{Proofs of Theorem \ref{S-thm:calib} and of Theorems \ref{thm:slope} and \ref{thm:glob} of the manuscript}
\label{S-sec:proof}

\emph{Theorem \ref{S-thm:calib} (null law of the movement statistic).} The notation is that of Sections \ref{S-sec:stat} and \ref{S-sec:null}, which give the first step of the proof.

\begin{proof}
Write the selected solutions as $\mathbf{x}_{j}=\boldsymbol{\mu}_{t-1}+\sqrt{s}L_{t-1}\mathbf{g}_{j}$ with independent $\mathbf{g}_{j}\sim\mathcal{N}(\mathbf{0},I)$, and let $\bar{\mathbf{g}}$ and $S=\frac{1}{M}\sum_{j}(\mathbf{g}_{j}-\bar{\mathbf{g}})(\mathbf{g}_{j}-\bar{\mathbf{g}})^{T}$ be their mean and maximum-likelihood covariance. Then $\Delta\boldsymbol{\mu}_{t}=\sqrt{s}L_{t-1}\bar{\mathbf{g}}$ and $\Sigma_{t}=sL_{t-1}SL_{t-1}^{T}$. The product of two lower-triangular matrices with positive diagonals is lower triangular with a positive diagonal, so by uniqueness of the Cholesky factor $L_{t}=\sqrt{s}L_{t-1}\mathrm{chol}(S)$, which gives (\ref{S-eq:znull}); its right-hand side contains neither $L_{t-1}$ nor $s$. For normal samples $\bar{\mathbf{g}}\sim\mathcal{N}(\mathbf{0},I/M)$ is independent of $MS\sim W_{D}(I,M-1)$, and by Bartlett's decomposition $MS=TT^{T}$ with $T$ as stated \cite{muirhead1982}; substituting $\mathrm{chol}(S)=T/\sqrt{M}$ and $\bar{\mathbf{g}}=\mathbf{g}/\sqrt{M}$ gives (b). Each generation uses fresh samples, so the $\mathbf{z}_{t}$ are independent. For (a), $\|\mathbf{z}_{t}\|^{2}=M\bar{\mathbf{g}}^{T}S^{-1}\bar{\mathbf{g}}=\frac{M}{M-1}H$, where Hotelling's statistic $H=M\bar{\mathbf{g}}^{T}S_{u}^{-1}\bar{\mathbf{g}}$ with the unbiased covariance $S_{u}=\frac{M}{M-1}S$ has the law $\frac{(M-1)D}{M-D}F_{D,M-D}$ \cite{muirhead1982}, and $\mathbb{E}F_{D,M-D}=(M-D)/(M-D-2)$. For (c), the $\mathbf{z}_{t}$ are independent with mean zero (the law of $\bar{\mathbf{g}}$ is symmetric), so the cross terms of $\mathbb{E}\|\mathbf{p}\|^{2}$ vanish and $\mathbb{E}\|\mathbf{p}\|^{2}=c_{p}(2-c_{p})\sum_{j<k}(1-c_{p})^{2j}\,\mathbb{E}\|\mathbf{z}\|^{2}=(1-(1-c_{p})^{2k})MD/(M-D-2)$; multiplying by $(M-D-2)/(MD)$ gives the claim. (d) follows from (b) and (\ref{S-eq:path}).
\end{proof}

\emph{Theorem \ref{thm:slope} of the manuscript (one generation on a linear function).} The notation is that of Section \ref{sec:slope} of the manuscript. The proof is written for the gated sampler (\ref{S-eq:mg}) with a mean shift, for which the map (\ref{eq:slopemap}) holds with $\lambda(\delta)$ replaced by $\Lambda=\sigma_{1}\iota+\lambda(\delta)$, where $\delta=\iota d_{\max}$; the theorem of the manuscript is the case $\sigma_{1}=0$.

\begin{proof}
A whitened sample is $\mathbf{y}=\mathbf{y}_{\mathrm{SN}}+\sigma_{1}\iota\mathbf{w}$ with $\mathbf{y}_{\mathrm{SN}}=Q\mathbf{z}+\delta(|z_{0}|-b)\mathbf{w}$, and $f(\mathbf{x})=\mathbf{v}^{T}\boldsymbol{\mu}_{t}+\sqrt{s}\,\mathbf{v}^{T}L_{t}\mathbf{y}=\mathbf{v}^{T}\boldsymbol{\mu}_{t}-\sqrt{s}\,\sigma_{\mathbf{v},t}\,\mathbf{w}^{T}\mathbf{y}$, so selection keeps the fraction $\tau$ with the largest $\mathbf{w}^{T}\mathbf{y}=Y_{\delta}+\sigma_{1}\iota$, where $Y_{\delta}=\mathbf{w}^{T}\mathbf{y}_{\mathrm{SN}}=\delta(|z_{0}|-b)+(1-\beta)\mathbf{w}^{T}\mathbf{z}$ has the stated law because $1-\beta=\sqrt{1-(1-b^{2})\delta^{2}}$. The part of $\mathbf{y}$ orthogonal to $\mathbf{w}$ is $(I-\mathbf{w}\mathbf{w}^{T})\mathbf{z}$, which is independent of $(z_{0},\mathbf{w}^{T}\mathbf{z})$ and unaffected by selection. The selected $\mathbf{y}$ therefore have mean $\Lambda\mathbf{w}$ and covariance $I-(1-c(\delta))\mathbf{w}\mathbf{w}^{T}$; transforming back with $L_{t}\mathbf{w}=-\Sigma_{t}\mathbf{v}/\sigma_{\mathbf{v},t}$ gives (\ref{eq:slopemap}), and $\mathbf{v}^{T}\Sigma_{t+1}\mathbf{v}=s\,c(\delta)\sigma_{\mathbf{v},t}^{2}$, $\Sigma_{t+1}\mathbf{v}=s\,c(\delta)\Sigma_{t}\mathbf{v}$. For the direction: in a Gaussian first generation the mean moves along $-\Sigma_{1}\mathbf{v}$. If the direction is downhill up to generation $t$, every mean shift so far and every nominal shift $\sqrt{s}\sigma_{1}\iota L\mathbf{w}$ is parallel to $-\Sigma_{1}\mathbf{v}$, and so is $\Sigma_{t+1}\mathbf{v}$ up to sign; hence $\mathbf{P}_{t+1}\parallel-\Sigma_{t+1}\mathbf{v}$ and $\mathbf{u}_{t+1}=L_{t+1}^{-1}\mathbf{P}_{t+1}\parallel-L_{t+1}^{T}\mathbf{v}\parallel\mathbf{w}_{t+1}$.
\end{proof}

\emph{Theorem \ref{thm:glob} of the manuscript (global convergence with a variance floor).} Theorem \ref{thm:glob} of the manuscript is the case $\sigma_{1}=0$, $\iota_{t}\equiv1$ of the following statement: Algorithm \ref{S-alg:mg} with samples clipped to a box $\Omega$, elitism, a fixed variance floor $\sigma_{\min}>0$, a bounded multiplier $\eta\le s\le s_{\max}$, $\rho_{\max}<1$ and bounded $\sigma_{1}\ge0$ converges almost surely to the minimum of a continuous $f$ on $\Omega$.

\begin{proof}
Condition on the history and on $|z_{0}|=r$. By (\ref{S-eq:mg}), $\mathbf{y}=Q\mathbf{z}+[\delta(r-b)+\sigma_{1}\iota]\mathbf{w}$, so an unclipped sample is normal,
\begin{equation*}
\mathbf{x}\mid r\sim\mathcal{N}\big(\boldsymbol{\mu}+\sqrt{s}\,[\delta(r-b)+\sigma_{1}\iota]L\mathbf{w},\;sLQQ^{T}L^{T}\big).
\end{equation*}
The matrix $QQ^{T}$ has the eigenvalue $1-(1-b^{2})\delta^{2}\ge1-\rho_{\max}^{2}$ along $\mathbf{w}$ and one elsewhere. The estimate $\Sigma$ is the covariance of points of $\Omega$ plus the floor, so its eigenvalues lie in $[\sigma_{\min}^{2},\,\mathrm{diam}(\Omega)^{2}/4+\sigma_{\min}^{2}]$ \cite{bhatia2000}. Hence the conditional covariance has all eigenvalues in $[\underline{v},\bar{v}]$ with $\underline{v}=\eta\sigma_{\min}^{2}(1-\rho_{\max}^{2})>0$ and $\bar{v}=s_{\max}(\mathrm{diam}(\Omega)^{2}/4+\sigma_{\min}^{2})$. For $r\in[0,1]$, $|\delta(r-b)+\sigma_{1}\iota|\le D_{\lim}+\sigma_{1}$, and $\boldsymbol{\mu}\in\Omega$ as an average of points of $\Omega$, so the conditional mean lies within a fixed distance $R$ of $\Omega$. A normal density with these bounds is at least $c_{0}=(2\pi\bar{v})^{-D/2}\exp(-(\mathrm{diam}(\Omega)+R)^{2}/(2\underline{v}))$ at every point of $\Omega$. Integrating over $r\in[0,1]$, which has probability $2\Phi(1)-1$, the unconditional density of an unclipped sample is at least $c=(2\Phi(1)-1)c_{0}>0$ on $\Omega$, uniformly in the history. Clipping moves only samples outside $\Omega$ and can only add probability to subsets of $\Omega$.

Let $\varepsilon>0$ and $A_{\varepsilon}=\{\mathbf{x}\in\Omega:f(\mathbf{x})<f^{*}+\varepsilon\}$. By continuity $A_{\varepsilon}$ is a nonempty relatively open subset of $\Omega$ and therefore has positive Lebesgue measure $|A_{\varepsilon}|$. Given the history, the $N-1$ new samples of a generation are independent, so the probability that none of them falls in $A_{\varepsilon}$ is at most $(1-c|A_{\varepsilon}|)^{N-1}<1$; multiplying the conditional bounds, the probability of no hit in $T$ generations is at most $(1-c|A_{\varepsilon}|)^{(N-1)T}\to0$. With elitism the best value never increases, so almost surely $f(\mathbf{x}_{bs,t})<f^{*}+\varepsilon$ for all large $t$; taking $\varepsilon=1/k$, $k=1,2,\ldots$, gives the claim \cite{rudolph1996}.
\end{proof}

\section{Noise of the third-moment vector}
\label{S-sec:mardia}

This section gives the statement, proof and measurement behind the claim of the manuscript that the skewness cannot be estimated from the selected solutions at the population sizes used. The whitened variable $\mathbf{y}$ is that of the mean- and covariance-preserving sampler, (\ref{eq:conv}) of the manuscript; $\mathbf{g}=\frac{1}{M}\sum_{k}\|\mathbf{y}_{k}\|^{2}\mathbf{y}_{k}$ is the third-moment vector of the $M$ whitened selected solutions, $K_{3}$ the third cumulant of the standardized half-normal variable and $d_{\max}$ the largest skewness magnitude used.
\begin{theorem}[Third-moment vector]
\label{S-thm:mardia}
(a) For the whitened variable $\mathbf{y}$ of the mean- and covariance-preserving sampler, (\ref{eq:conv}) of the manuscript, $\mathbb{E}[\|\mathbf{y}\|^{2}\mathbf{y}]=K_{3}\|\mathbf{d}\|^{2}\mathbf{d}$. (b) If the selected solutions are $M$ independent samples of $\mathcal{N}(\boldsymbol{\mu},\Sigma)$ and $\mathbf{g}$ is computed after centering with the sample mean and whitening with the sample covariance, then $M\|\mathbf{g}\|^{2}$ converges in distribution to $2(D+2)\chi_{D}^{2}$ as $M\to\infty$, so that $\mathbb{E}\|\mathbf{g}\|^{2}\approx2D(D+2)/M$.
\end{theorem}
\begin{proof}
(a) Because $\mathbb{E}\mathbf{y}=\mathbf{0}$, third moments equal third cumulants, $\mathbb{E}[y_{i}y_{j}y_{k}]=K_{3}d_{i}d_{j}d_{k}$; summing over $j=k$ gives the claim. (b) Centering contributes a first-order correction, $\mathbf{g}\approx\frac{1}{M}\sum_{k}(\|\mathbf{y}_{k}\|^{2}-D-2)\mathbf{y}_{k}$ with $\mathbf{y}_{k}$ whitened by the true parameters; the first-order correction from whitening with the sample covariance involves odd moments of the normal distribution and vanishes. Hence $\sqrt{M}\mathbf{g}$ is asymptotically normal with covariance $\frac{1}{D}\mathbb{E}[(\|\mathbf{y}\|^{2}-D-2)^{2}\|\mathbf{y}\|^{2}]I=2(D+2)I$, computed from the first three moments of $\chi_{D}^{2}$. $M\|\mathbf{g}\|^{2}$ is $M$ times the multivariate skewness of Móri, Rohatgi and Székely \cite{mori1993}, whose limit law under normality is the same multiple of $\chi_{D}^{2}$ \cite{henze1997,klar2002}.
\end{proof}

At $d_{\max}$ the signal is $\|\mathbb{E}\mathbf{g}\|^{2}=(K_{3}d_{\max}^{3})^{2}=0.728$, whereas the noise grows with $D^{2}/M$. Table \ref{S-tab:mardia} confirms that, with the population sizes of the algorithm, the noise exceeds the signal already for $D=5$ (by a factor of 2.9) and is more than ten times larger from $D=30$ on; the measured values are below the asymptotic formula because $M$ is not large compared with $D$. In a direct check with $2\times10^{6}$ skew-normal samples ($D=5$, $\|\mathbf{d}\|=0.9D_{\lim}$), the cosine between $\mathbf{g}$ and $\mathbf{d}$ was 0.999997 and $\|\mathbf{g}\|/(K_{3}\|\mathbf{d}\|^{3})=1.010$; the sample mean deviated by $9.5\times10^{-4}$ standard deviations, the largest relative error of the covariance was $9.9\times10^{-4}$, and the third cumulant along $L\mathbf{d}$ was 25.53 against 25.38 predicted by Theorem \ref{thm:mom} of the manuscript.

\begin{table}[htbp]\centering\small
\caption{Noise of the third-moment vector under normality}
\label{S-tab:mardia}
\begin{tabular}{@{}cccccc@{}}
\toprule
$D$ & $M$ & Measured $\mathbb{E}\|\mathbf{g}\|^{2}$ (s.e.) & Asymptotic $2D(D+2)/M$ & Signal at $d_{\max}$ & Noise/signal \\
\midrule
5 & 17 & 2.146 (0.004) & 4.118 & 0.728 & 2.9 \\
10 & 38 & 4.052 (0.006) & 6.316 & 0.728 & 5.6 \\
20 & 99 & 6.724 (0.007) & 8.889 & 0.728 & 9.2 \\
30 & 178 & 8.717 (0.007) & 10.787 & 0.728 & 12.0 \\
40 & 271 & 10.360 (0.007) & 12.399 & 0.728 & 14.2 \\
\bottomrule
\end{tabular}
\par\vspace{2pt}\begin{minipage}{0.8\textwidth}\scriptsize 100{,}000 repetitions per row (program \nolinkurl{gnoise.py}), standard error of the mean in parentheses; $M=\lfloor0.35N\rfloor$ is the number of selected solutions for $N=\lfloor17+3D^{1.5}\rfloor$. The solutions are centered with the sample mean and whitened with the sample covariance. Noise/signal is the ratio of the measured $\mathbb{E}\|\mathbf{g}\|^{2}$ to the signal $(K_{3}d_{\max}^{3})^{2}$.\end{minipage}
\end{table}

\section{Earlier parameter study of the skewness-controlled algorithm}
\label{S-sec:param}

This section reports the parameter study of the earlier version on the 48 training blocks (24 BBOB functions, $D=10$, instances 1--5, standard and distant start). It was run with the earlier implementation, which differed from the one of the manuscript in three conventions: samples were clipped to the box ($[-5,5]^{D}$ for the standard start), the elitist row was re-evaluated in every generation (900 generations, 100{,}011 evaluations), and the thresholds were fixed numbers. None of the configurations reported here uses the elongation, so the sampler is the gated sampler (\ref{S-eq:mg}).

\begin{figure*}[!htbp]
\centering
\includegraphics[width=0.85\textwidth]{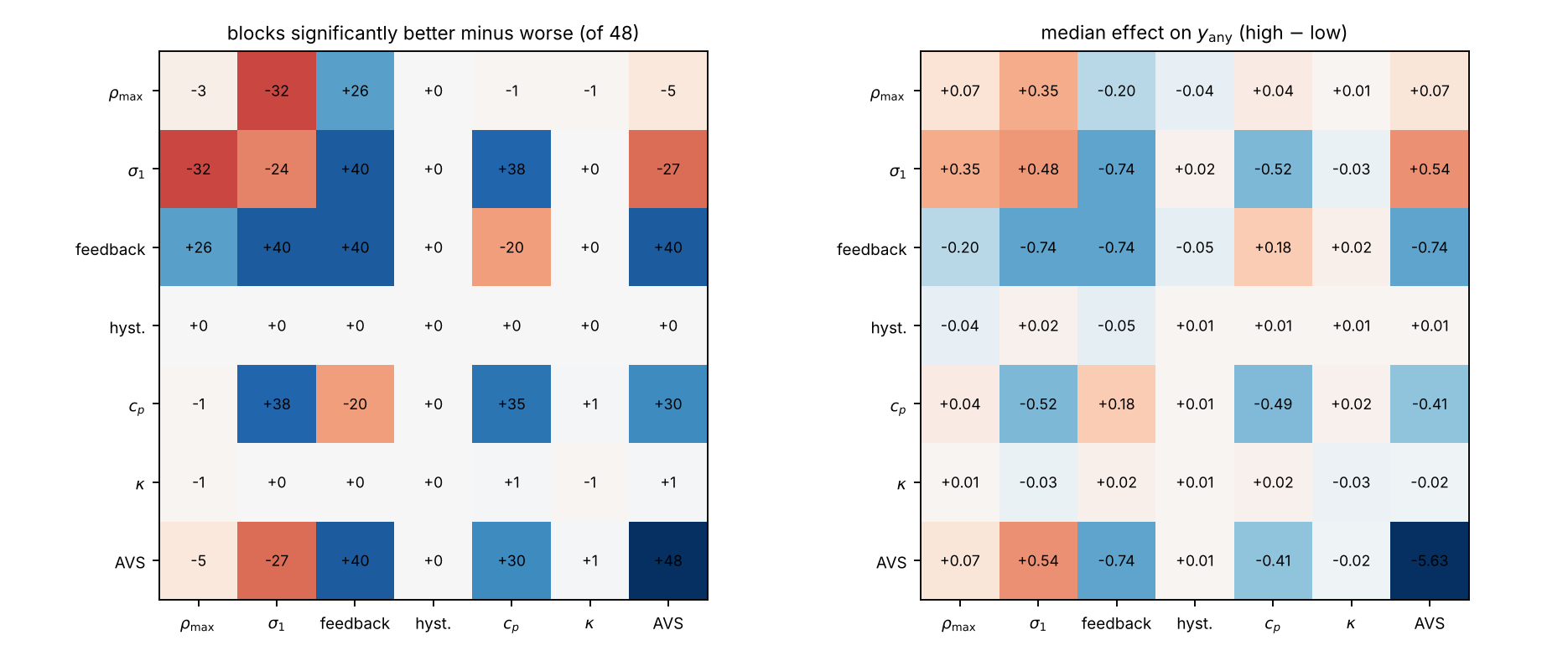}
\caption{Main effects (diagonal) and two-factor interactions (off-diagonal) of the earlier $2^{7}$ study on $y_{\mathrm{any}}$, all 48 blocks, blocked analysis. Left: number of blocks in which the term is significantly better minus significantly worse; right: median effect (high $-$ low). Blue means that raising the factor improves performance.}
\label{S-fig:heat}
\end{figure*}

\emph{Full factorial design.} Seven parameters of the skewness-controlled algorithm were varied at two levels each ($2^{7}=128$ configurations $\times$ 48 blocks $\times$ 5 instances): the skewness magnitude $\rho_{\max}$, the mean shift $\sigma_{1}$, the feedback correction, a hysteresis of $\iota_{t}$, $c_{p}$, the thresholds and AVS. The protocol was written before the runs. All configurations of a block share the random numbers of an instance, so the five instances are repeated measures; the analysis of the earlier version, which treated the 640 observations of a block as independent, is replaced here by two analyses that respect this. (i) Within each block, a randomized complete block model with the instance as block factor, seven main effects and 21 two-factor interactions is fitted, and each term is tested with Holm's correction over the 28 terms; Table \ref{S-tab:factfull} counts the blocks in which a term is significant in either direction. (ii) Across the 48 blocks, the effect of each term (high minus low level, averaged over the five instances) is tested with the Wilcoxon signed-rank test, with Holm's correction over the 28 terms. The counts of (i) differ from those of the earlier analysis by at most two blocks, and no conclusion changes. AVS is significant in all 48 blocks with a median effect of $-$5.63, an order of magnitude larger than any other: the skewness and the mean shift refine a scale control and do not replace it. The skewness magnitude ($\rho_{\max}=0.95$ against 0.5) has no significant main effect in this design (significantly better in 9 blocks and worse in 12; Holm-corrected $p=0.91$ across the blocks), because its effect depends on the other factors, half of whose combinations are poor algorithms: it is favorable with the feedback correction (interaction significant in 26 blocks and in none in the opposite direction) and unfavorable together with the mean shift, half of whose configurations lack that correction (interaction significant against in 34 blocks). The mean shift is harmful as a main effect (30 blocks) only because half of its configurations lack the feedback correction: the interaction of the two is significant in 40 blocks and in none in the opposite direction. $c_{p}=0.3$ is better than 0.1 (35 blocks), mainly when the mean shift is on. The hysteresis is significant in no block and the thresholds in one. Fig. \ref{S-fig:heat} shows the same results as a matrix.

\begin{figure*}[!htbp]
\centering
\includegraphics[width=\textwidth]{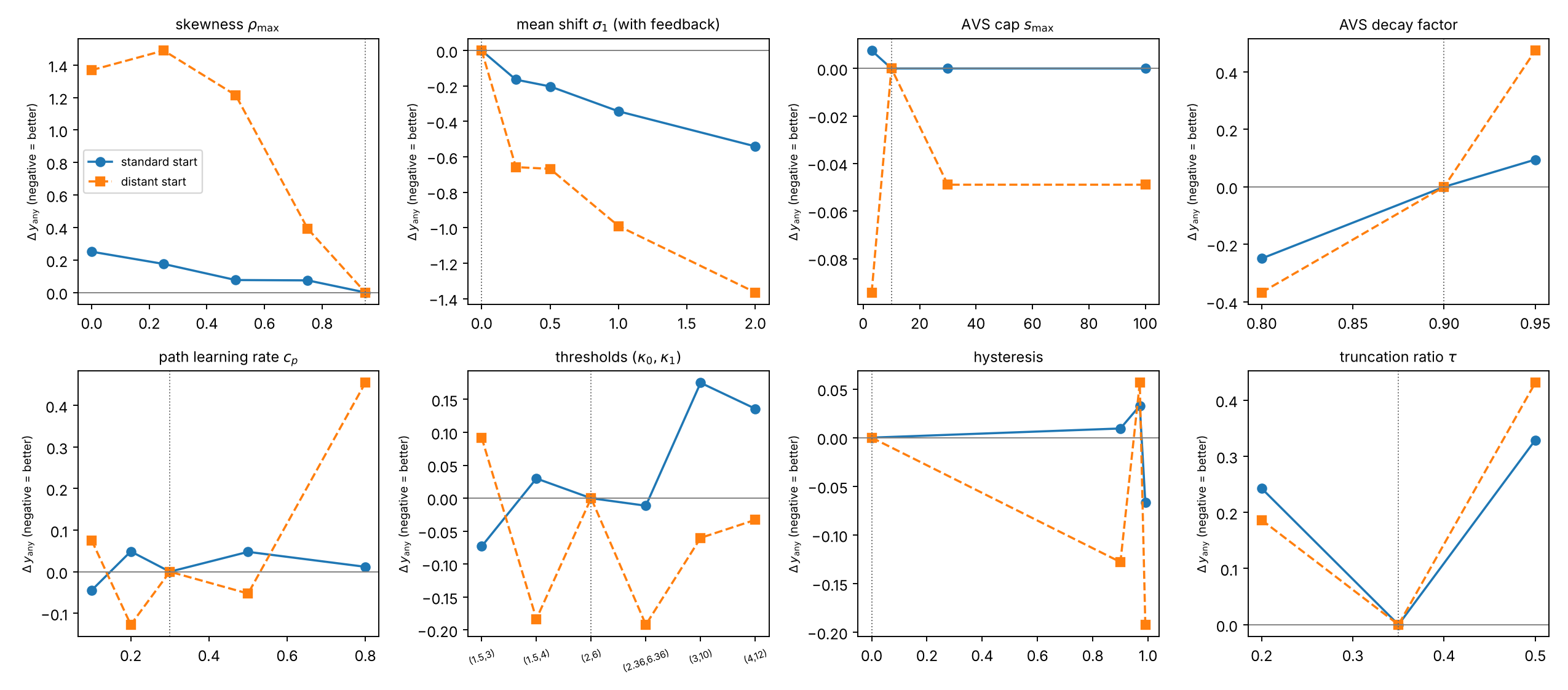}
\caption{Earlier one-at-a-time sensitivity around ASN+AVS on the 48 training blocks (earlier implementation). Change of the block mean of $y_{\mathrm{any}}$ relative to ASN+AVS (negative = better) for the standard start (circles) and the distant start (squares); the dotted vertical line marks the value of ASN+AVS. The panel on the elongation used the joint sampler of the earlier version and is superseded by Section \ref{S-sec:elong}.}
\label{S-fig:oat}
\end{figure*}

\emph{One-at-a-time sweeps.} Fig. \ref{S-fig:oat} shows sweeps of one parameter at a time around ASN+AVS (the gated configuration SN with decay 0.9 and thresholds $(2,6)$); each point is the mean over the blocks of the change of $y_{\mathrm{any}}$ relative to ASN+AVS. Removing the skewness costs +0.81 (8/38 blocks better/worse), and the sweep of its magnitude shows the dose--response relation of Table \ref{tab:train} of the manuscript: relative to $\rho_{\max}=0.95$, the mean $y_{\mathrm{any}}$ is worse by 0.81, 0.83, 0.65 and 0.23 for $\rho_{\max}=0$, 0.25, 0.5 and 0.75. The mean shift with feedback improves up to $\sigma_{1}=2$, $-$0.95 (35/10 blocks better/worse), and without feedback it is harmful, +2.05 (5/42 blocks better/worse). Switching AVS off costs +6.36; a floor of 0.5 instead of 1 on the multiplier costs +2.58; the decay 0.8 instead of 0.9 gives $-$0.31 (uncorrected $p=0.10$). The three rules that let the movement statistic drive the scale are all worse than the improvement-triggered rule: the cumulative-step-size rule $\ln s\leftarrow\ln s+c(\sqrt{m_{t}}-1)$ by +2.91 at its best coefficient, the threshold rule by +0.44 and AVS with the additional trigger $m_{t}>\kappa_{0}$ by +0.39. Rules of this kind enlarge the scale too fast on a slope and release it too slowly near the optimum, where $m_{t}$ falls only gradually; the improvement-triggered rule reacts within one generation. Replacing the gate by a fixed maximal skewness changes $y_{\mathrm{any}}$ by $-$0.13 (uncorrected $p=0.56$): for the skewness alone the gate is immaterial, which the confirmatory test of the manuscript establishes as family G4.

\begin{table}[htbp]\centering\scriptsize
\caption{Earlier full factorial $2^{7}$ study on the training set ($D=10$, 48 blocks), re-analysed with the five instances as blocks: number of blocks in which the term is significant (randomized complete block model, Holm over the 28 terms of a block, $\alpha=0.05$) with the high level better or worse; median over the blocks of the effect on $y_{\mathrm{any}}$ (high $-$ low; negative is better); number of blocks with a negative and a positive effect; Holm-corrected $p$ of the Wilcoxon signed-rank test of the effect over the 48 blocks}
\label{S-tab:factfull}
\begin{tabular}{@{}llcccccc@{}}
\toprule
Term & Factor (low/high) or interaction & sig. better & sig. worse & median effect & blocks $<0$ & blocks $>0$ & $p$ (Holm) \\
\midrule
A & $\rho_{\max}$ (0.5/0.95) & 9 & 12 & +0.07 & 16 & 32 & $0.91$ \\
B & mean shift $\sigma_{1}$ (0/1) & 6 & 30 & +0.48 & 9 & 39 & $6\times10^{-5}$ \\
C & feedback correction fb (off/on) & 40 & 0 & $-$0.74 & 47 & 1 & $4\times10^{-13}$ \\
D & hysteresis (0/0.97) & 0 & 0 & +0.01 & 20 & 28 & $0.89$ \\
E & $c_{p}$ (0.1/0.3) & 35 & 0 & $-$0.49 & 44 & 4 & $3\times10^{-10}$ \\
F & $(\kappa_{0},\kappa_{1})$ ((1.5,4)/(2,6)) & 0 & 1 & $-$0.03 & 31 & 17 & $0.45$ \\
G & AVS (off/on) & 48 & 0 & $-$5.63 & 48 & 0 & $2\times10^{-13}$ \\
\midrule
AB & $\rho_{\max}\times \sigma_{1}$ & 2 & 34 & +0.35 & 3 & 45 & $9\times10^{-10}$ \\
AC & $\rho_{\max}\times \mathrm{fb}$ & 26 & 0 & $-$0.20 & 46 & 2 & $2\times10^{-11}$ \\
AD & $\rho_{\max}\times \mathrm{hyst}$ & 0 & 0 & $-$0.04 & 39 & 9 & $6\times10^{-5}$ \\
AE & $\rho_{\max}\times c_{p}$ & 2 & 3 & +0.04 & 21 & 27 & $0.19$ \\
AF & $\rho_{\max}\times \kappa$ & 0 & 1 & +0.01 & 20 & 28 & $0.91$ \\
AG & $\rho_{\max}\times \mathrm{AVS}$ & 8 & 13 & +0.07 & 17 & 31 & $0.89$ \\
BC & $\sigma_{1}\times \mathrm{fb}$ & 40 & 0 & $-$0.74 & 47 & 1 & $4\times10^{-13}$ \\
BD & $\sigma_{1}\times \mathrm{hyst}$ & 0 & 0 & +0.02 & 15 & 33 & $0.007$ \\
BE & $\sigma_{1}\times c_{p}$ & 38 & 0 & $-$0.52 & 46 & 2 & $1\times10^{-11}$ \\
BF & $\sigma_{1}\times \kappa$ & 0 & 0 & $-$0.03 & 37 & 11 & $0.007$ \\
BG & $\sigma_{1}\times \mathrm{AVS}$ & 4 & 31 & +0.54 & 6 & 42 & $4\times10^{-6}$ \\
CD & $\mathrm{fb}\times \mathrm{hyst}$ & 0 & 0 & $-$0.05 & 39 & 9 & $1\times10^{-6}$ \\
CE & $\mathrm{fb}\times c_{p}$ & 0 & 20 & +0.18 & 12 & 36 & $3\times10^{-6}$ \\
CF & $\mathrm{fb}\times \kappa$ & 0 & 0 & +0.02 & 11 & 37 & $0.0002$ \\
CG & $\mathrm{fb}\times \mathrm{AVS}$ & 40 & 0 & $-$0.74 & 46 & 2 & $2\times10^{-12}$ \\
DE & $\mathrm{hyst}\times c_{p}$ & 0 & 0 & +0.01 & 22 & 26 & $0.91$ \\
DF & $\mathrm{hyst}\times \kappa$ & 0 & 0 & +0.01 & 20 & 28 & $0.91$ \\
DG & $\mathrm{hyst}\times \mathrm{AVS}$ & 0 & 0 & +0.01 & 22 & 26 & $0.89$ \\
EF & $c_{p}\times \kappa$ & 1 & 0 & +0.02 & 18 & 30 & $0.90$ \\
EG & $c_{p}\times \mathrm{AVS}$ & 31 & 1 & $-$0.41 & 43 & 5 & $5\times10^{-10}$ \\
FG & $\kappa\times \mathrm{AVS}$ & 1 & 0 & $-$0.02 & 30 & 18 & $0.62$ \\
\bottomrule
\end{tabular}
\end{table}

\section{Details of the training-set study and of the confirmatory test}
\label{S-sec:test}

\subsection{Training set}
Table \ref{S-tab:train} lists every configuration that was run on the training set for the skewness study (protocols A and A7; BBOB f1--f24, $D=10$, instances 1--5, both starts; 48 blocks and 240 runs per configuration), with the paired comparisons quoted in Section \ref{sec:train} of the manuscript. All of it is descriptive.

\begin{table}[htbp]\centering\scriptsize
\caption{Training set: mean $y_{\mathrm{any}}$ over all 48 blocks and over the blocks of each start, runs solved to $10^{-8}$ (of 240), and paired comparison with a reference configuration (Hodges--Lehmann estimate of the difference over the 48 blocks with 95\% bootstrap interval, uncorrected Wilcoxon $p$, blocks better/worse)}
\label{S-tab:train}
\begin{tabular}{@{}llrrrrlccc@{}}
\toprule
 & & \multicolumn{3}{c}{mean $y_{\mathrm{any}}$} & & \multicolumn{4}{c}{comparison} \\
\cmidrule(lr){3-5}\cmidrule(l){7-10}
Configuration & description & all & std. & dist. & solved & reference & HL [95\% interval] & $p$ & b/w \\
\midrule
EMNA+AVS & no correction & $-$1.77 & $-$3.21 & $-$0.33 & 110 & -- & -- & -- & -- \\
SN(0.25) & gated skewness, $\rho_{\max}=0.25$ & $-$1.73 & $-$3.20 & $-$0.26 & 109 & -- & -- & -- & -- \\
SN(0.5) & gated skewness, $\rho_{\max}=0.5$ & $-$1.98 & $-$3.46 & $-$0.50 & 111 & -- & -- & -- & -- \\
SN(0.75) & gated skewness, $\rho_{\max}=0.75$ & $-$2.32 & $-$3.45 & $-$1.18 & 110 & -- & -- & -- & -- \\
SN(0.9) & gated skewness, $\rho_{\max}=0.9$ & $-$2.35 & $-$3.36 & $-$1.33 & 107 & -- & -- & -- & -- \\
SN & gated skewness, $\rho_{\max}=0.95$ & $-$2.52 & $-$3.46 & $-$1.58 & 115 & EMNA+AVS & $-$0.45 [$-$1.07, $-$0.13] & $6\times10^{-5}$ & 38/10 \\
SN(0.99) & gated skewness, $\rho_{\max}=0.99$ & $-$2.60 & $-$3.45 & $-$1.75 & 113 & SN & 0.00 [$-$0.05, +0.06] & $0.92$ & 26/22 \\
SN-on & skewness without gate ($\iota_{t}\equiv1$) & $-$2.65 & $-$3.54 & $-$1.76 & 114 & SN & $-$0.01 [$-$0.07, +0.04] & $0.64$ & 26/22 \\
SN-decay25 & scheduled skewness, $\iota_{t}=e^{-t/25}$ & $-$2.13 & $-$3.55 & $-$0.71 & 115 & SN & +0.14 [+0.02, +0.49] & $0.03$ & 20/28 \\
SN-decay100 & scheduled skewness, $\iota_{t}=e^{-t/100}$ & $-$2.22 & $-$3.38 & $-$1.06 & 106 & SN & +0.13 [+0.03, +0.50] & $0.01$ & 17/31 \\
\midrule
MS & gated mean shift, $\sigma_{1}=1$ & $-$2.64 & $-$3.52 & $-$1.77 & 114 & EMNA+AVS & $-$0.48 [$-$0.91, $-$0.19] & $4\times10^{-6}$ & 41/7 \\
MS(2) & gated mean shift, $\sigma_{1}=2$ & $-$3.02 & $-$3.80 & $-$2.24 & 127 & -- & -- & -- & -- \\
MS+SN(0.5) & $\sigma_{1}=1$, $\rho_{\max}=0.5$ & $-$2.92 & $-$3.61 & $-$2.23 & 124 & -- & -- & -- & -- \\
MS+SN(0.75) & $\sigma_{1}=1$, $\rho_{\max}=0.75$ & $-$2.95 & $-$3.72 & $-$2.18 & 121 & -- & -- & -- & -- \\
SN+MS & $\sigma_{1}=1$, $\rho_{\max}=0.95$ & $-$3.14 & $-$3.76 & $-$2.53 & 131 & MS & $-$0.14 [$-$0.75, $-$0.03] & $0.002$ & 33/15 \\
MS+SN(0.99) & $\sigma_{1}=1$, $\rho_{\max}=0.99$ & $-$3.23 & $-$3.73 & $-$2.74 & 135 & SN+MS & $-$0.03 [$-$0.11, +0.01] & $0.10$ & 32/16 \\
SN+MS(0.5) & $\sigma_{1}=0.5$, $\rho_{\max}=0.95$ & $-$2.89 & $-$3.50 & $-$2.28 & 123 & SN+MS & +0.05 [0.00, +0.19] & $0.04$ & 18/30 \\
SN+MS(1.5) & $\sigma_{1}=1.5$, $\rho_{\max}=0.95$ & $-$3.12 & $-$3.72 & $-$2.51 & 127 & SN+MS & $-$0.02 [$-$0.07, +0.07] & $0.48$ & 30/18 \\
SN+MS(2) & $\sigma_{1}=2$, $\rho_{\max}=0.95$ & $-$3.11 & $-$3.68 & $-$2.55 & 126 & SN+MS & $-$0.02 [$-$0.08, +0.02] & $0.33$ & 27/21 \\
SN+MS(3) & $\sigma_{1}=3$, $\rho_{\max}=0.95$ & $-$3.23 & $-$3.66 & $-$2.79 & 129 & SN+MS & $-$0.04 [$-$0.10, +0.01] & $0.15$ & 30/18 \\
SN+MS-on & $\sigma_{1}=1$, $\rho_{\max}=0.95$, no gate & $-$3.01 & $-$3.66 & $-$2.37 & 123 & SN+MS & +0.03 [$-$0.06, +0.12] & $0.48$ & 23/25 \\
\midrule
EMNA+AVS+AMS & AMS & $-$3.63 & $-$3.91 & $-$3.36 & 136 & -- & -- & -- & -- \\
AMS+SN & AMS with gated skewness & $-$3.66 & $-$4.07 & $-$3.26 & 137 & EMNA+AVS+AMS & +0.01 [$-$0.05, +0.06] & $0.59$ & 20/28 \\
AMaLGaM & AMaLGaM & $-$3.38 & $-$3.67 & $-$3.09 & 121 & -- & -- & -- & -- \\
AMaLGaM+SN & AMaLGaM with gated skewness & $-$3.49 & $-$3.77 & $-$3.21 & 125 & AMaLGaM & $-$0.01 [$-$0.08, +0.04] & $0.61$ & 21/27 \\
\bottomrule
\end{tabular}
\par\vspace{2pt}\begin{minipage}{0.97\textwidth}\scriptsize All configurations above the second rule use the AVS of Section \ref{sec:alg} of the manuscript with the decay 0.8 and, where gated, the thresholds of Table \ref{S-tab:null}; the mean shift has the feedback correction. AMS+SN and AMaLGaM+SN ran with the gate thresholds $(2,6)$ (Section \ref{S-sec:prot}, item 15). Further comparisons: SN+MS against SN, HL $-$0.13 [$-$0.50, $-$0.05], $p=0.0002$; SN+MS against EMNA+AVS+AMS, +0.19 [+0.09, +0.41], $p=2\times10^{-5}$; SN+MS against AMaLGaM, +0.22 [+0.04, +0.38], $p=0.02$; MS+SN(0.99) against MS, $-$0.21, $p=5\times10^{-5}$.\end{minipage}
\end{table}

\subsection{Confirmatory test}
Table \ref{S-tab:famfull} gives, for every comparison of Table \ref{tab:fam} of the manuscript and of Table \ref{S-tab:famms} and every setting, the mean and the Hodges--Lehmann estimate of the block differences with its 95\% bootstrap interval, the Wilcoxon $p$ value before and after Holm's correction over all tests of the family, the verdict on practical equivalence, the number of functions on which the first algorithm has the better and the worse mean $y_{\mathrm{any}}$, the number of functions on which the difference is significant in a paired Wilcoxon test over the 15 instances (Holm over the 24 functions), and the paired success table: runs solved by the first algorithm only and by the second only, with the exact McNemar $p$. Table \ref{S-tab:page} gives Page's trend test of family G3. Wilcoxon $p$ values are computed from the exact null distribution of the signed-rank statistic when a comparison has neither zero nor tied differences, and from the normal approximation with tie correction (zeros by Pratt's method, no continuity correction) otherwise. Among the 54 block-level tests of families G1, G2 and G4--G6 the latter occurs once (SN+MS-on against SN+MS for $D=40$ with the standard start, where all 15 runs on the linear slope f5 are identical because the gate stays fully open until the target is reached), and among their 1296 per-function tests in 29; a zero difference means that the two runs of a pair are identical. Tables \ref{S-tab:pf0}--\ref{S-tab:pf5} give the mean $y_{\mathrm{any}}$ of every algorithm on every function, Tables \ref{S-tab:ert0}--\ref{S-tab:ert5} the expected running times to $10^{-8}$, and Table \ref{S-tab:gatestat} the behavior of the gate by function group.

{\scriptsize
\setlength{\tabcolsep}{3pt}
\begin{longtable}{@{}llrlcccccccccc@{}}
\caption{Confirmatory comparisons in full (difference first minus second in $y_{\mathrm{any}}$; negative: the first algorithm is better)}\label{S-tab:famfull}\\
\toprule
Fam. & Comparison & $D$ & start & mean & HL & 95\% interval & $p$ & $p$ (Holm) & equiv. & fn. b/w & sig. b/w & solved only 1st/2nd & McNemar $p$ \\
\midrule
\endfirsthead
\toprule
Fam. & Comparison & $D$ & start & mean & HL & 95\% interval & $p$ & $p$ (Holm) & equiv. & fn. b/w & sig. b/w & solved only 1st/2nd & McNemar $p$ \\
\midrule
\endhead
G1 & SN vs. EMNA+AVS & 10 & std. & $-$0.29 & $-$0.03 & [$-$0.24, 0.00] & $0.09$ & $0.09$ & yes & 18/6 & 2/0 & 13/6 & $0.17$ \\
G1 & SN vs. EMNA+AVS & 10 & dist. & $-$1.51 & $-$1.03 & [$-$1.98, $-$0.66] & $1\times10^{-5}$ & $4\times10^{-5}$ & no & 22/2 & 15/0 & 30/11 & $0.004$ \\
G1 & SN vs. EMNA+AVS & 20 & std. & $-$0.51 & $-$0.09 & [$-$0.24, $-$0.01] & $0.01$ & $0.03$ & yes & 18/6 & 2/0 & 28/5 & $7\times10^{-5}$ \\
G1 & SN vs. EMNA+AVS & 20 & dist. & $-$3.48 & $-$3.33 & [$-$4.30, $-$2.40] & $1\times10^{-7}$ & $7\times10^{-7}$ & no & 24/0 & 23/0 & 56/2 & $1\times10^{-14}$ \\
G1 & SN vs. EMNA+AVS & 40 & std. & $-$0.40 & $-$0.05 & [$-$0.21, $-$0.01] & $0.01$ & $0.03$ & yes & 19/5 & 5/0 & 13/1 & $0.002$ \\
G1 & SN vs. EMNA+AVS & 40 & dist. & $-$3.47 & $-$3.31 & [$-$4.11, $-$2.66] & $1\times10^{-7}$ & $7\times10^{-7}$ & no & 24/0 & 22/0 & 50/3 & $6\times10^{-12}$ \\
G2 & SN+MS vs. MS & 10 & std. & 0.00 & +0.01 & [$-$0.03, +0.19] & $0.81$ & $1.00$ & yes & 12/12 & 2/0 & 14/11 & $0.69$ \\
G2 & SN+MS vs. MS & 10 & dist. & $-$0.56 & $-$0.29 & [$-$0.74, $-$0.08] & $0.0001$ & $0.0004$ & no & 21/3 & 4/0 & 38/13 & $0.0006$ \\
G2 & SN+MS vs. MS & 20 & std. & $-$0.08 & $-$0.03 & [$-$0.10, +0.01] & $0.19$ & $0.56$ & yes & 15/9 & 2/0 & 9/5 & $0.42$ \\
G2 & SN+MS vs. MS & 20 & dist. & $-$0.78 & $-$0.62 & [$-$1.13, $-$0.36] & $4\times10^{-7}$ & $2\times10^{-6}$ & no & 23/1 & 14/0 & 22/5 & $0.002$ \\
G2 & SN+MS vs. MS & 40 & std. & $-$0.03 & 0.00 & [$-$0.04, +0.03] & $0.90$ & $1.00$ & yes & 9/15 & 1/0 & 2/2 & $1.00$ \\
G2 & SN+MS vs. MS & 40 & dist. & $-$1.85 & $-$1.63 & [$-$2.45, $-$1.08] & $1\times10^{-7}$ & $7\times10^{-7}$ & no & 24/0 & 20/0 & 33/3 & $2\times10^{-7}$ \\
G4 & SN-on vs. SN & 10 & std. & $-$0.03 & $-$0.01 & [$-$0.08, +0.06] & $0.58$ & $1.00$ & yes & 14/10 & 0/0 & 8/8 & $1.00$ \\
G4 & SN-on vs. SN & 10 & dist. & $-$0.02 & $-$0.06 & [$-$0.14, +0.03] & $0.15$ & $1.00$ & yes & 16/8 & 1/0 & 10/15 & $0.42$ \\
G4 & SN-on vs. SN & 20 & std. & 0.00 & +0.01 & [$-$0.08, +0.05] & $0.70$ & $1.00$ & yes & 9/15 & 0/0 & 4/6 & $0.75$ \\
G4 & SN-on vs. SN & 20 & dist. & $-$0.16 & $-$0.11 & [$-$0.25, $-$0.04] & $0.009$ & $0.10$ & yes & 20/4 & 0/0 & 15/12 & $0.70$ \\
G4 & SN-on vs. SN & 40 & std. & $-$0.06 & $-$0.02 & [$-$0.07, +0.01] & $0.14$ & $1.00$ & yes & 15/9 & 0/0 & 0/0 & $1.00$ \\
G4 & SN-on vs. SN & 40 & dist. & $-$0.03 & $-$0.03 & [$-$0.08, +0.02] & $0.29$ & $1.00$ & yes & 15/9 & 2/0 & 10/4 & $0.18$ \\
G4 & SN+MS-on vs. SN+MS & 10 & std. & +0.01 & +0.02 & [$-$0.09, +0.11] & $0.77$ & $1.00$ & yes & 11/13 & 0/1 & 9/13 & $0.52$ \\
G4 & SN+MS-on vs. SN+MS & 10 & dist. & $-$0.02 & +0.10 & [$-$0.21, +0.26] & $0.41$ & $1.00$ & yes & 8/16 & 2/1 & 18/23 & $0.53$ \\
G4 & SN+MS-on vs. SN+MS & 20 & std. & $-$0.14 & $-$0.02 & [$-$0.36, +0.08] & $0.77$ & $1.00$ & no & 11/13 & 5/2 & 12/13 & $1.00$ \\
G4 & SN+MS-on vs. SN+MS & 20 & dist. & $-$0.11 & $-$0.01 & [$-$0.18, +0.08] & $0.88$ & $1.00$ & yes & 11/13 & 2/0 & 15/13 & $0.85$ \\
G4 & SN+MS-on vs. SN+MS & 40 & std. & 0.00 & 0.00 & [$-$0.08, +0.09] & $0.92$ & $1.00$ & yes & 12/11 & 6/1 & 0/2 & $0.50$ \\
G4 & SN+MS-on vs. SN+MS & 40 & dist. & +0.02 & 0.00 & [$-$0.08, +0.09] & $0.90$ & $1.00$ & yes & 13/11 & 1/1 & 3/15 & $0.008$ \\
G4 & SN+MS-nofb vs. SN+MS & 10 & std. & +1.68 & +1.46 & [+0.68, +2.41] & $6\times10^{-7}$ & $1\times10^{-5}$ & no & 2/22 & 0/16 & 0/56 & $3\times10^{-17}$ \\
G4 & SN+MS-nofb vs. SN+MS & 10 & dist. & +3.30 & +3.30 & [+2.17, +4.35] & $2\times10^{-7}$ & $4\times10^{-6}$ & no & 1/23 & 1/23 & 3/110 & $5\times10^{-29}$ \\
G4 & SN+MS-nofb vs. SN+MS & 20 & std. & +2.44 & +2.38 & [+1.10, +3.33] & $5\times10^{-5}$ & $0.0007$ & no & 5/19 & 1/16 & 0/143 & $2\times10^{-43}$ \\
G4 & SN+MS-nofb vs. SN+MS & 20 & dist. & +3.31 & +3.12 & [+1.53, +4.55] & $2\times10^{-7}$ & $4\times10^{-6}$ & no & 1/23 & 1/22 & 1/145 & $3\times10^{-42}$ \\
G4 & SN+MS-nofb vs. SN+MS & 40 & std. & +1.99 & +1.79 & [+0.44, +2.92] & $3\times10^{-5}$ & $0.0004$ & no & 5/19 & 0/17 & 0/93 & $2\times10^{-28}$ \\
G4 & SN+MS-nofb vs. SN+MS & 40 & dist. & +2.14 & +1.69 & [+0.98, +3.10] & $3\times10^{-6}$ & $4\times10^{-5}$ & no & 3/21 & 0/19 & 0/85 & $5\times10^{-26}$ \\
G5 & AMS+SN vs. EMNA+AVS+AMS & 10 & std. & +0.04 & +0.01 & [$-$0.02, +0.16] & $0.17$ & $1.00$ & yes & 7/17 & 1/0 & 11/12 & $1.00$ \\
G5 & AMS+SN vs. EMNA+AVS+AMS & 10 & dist. & 0.00 & $-$0.01 & [$-$0.08, +0.05] & $0.62$ & $1.00$ & yes & 14/10 & 0/0 & 17/16 & $1.00$ \\
G5 & AMS+SN vs. EMNA+AVS+AMS & 20 & std. & +0.04 & +0.02 & [$-$0.02, +0.07] & $0.34$ & $1.00$ & yes & 11/13 & 0/0 & 7/12 & $0.36$ \\
G5 & AMS+SN vs. EMNA+AVS+AMS & 20 & dist. & $-$0.06 & $-$0.02 & [$-$0.11, +0.03] & $0.41$ & $1.00$ & yes & 13/11 & 0/0 & 14/4 & $0.03$ \\
G5 & AMS+SN vs. EMNA+AVS+AMS & 40 & std. & 0.00 & $-$0.01 & [$-$0.04, +0.02] & $0.42$ & $1.00$ & yes & 15/9 & 0/0 & 1/2 & $1.00$ \\
G5 & AMS+SN vs. EMNA+AVS+AMS & 40 & dist. & $-$0.02 & $-$0.01 & [$-$0.06, +0.02] & $0.41$ & $1.00$ & yes & 14/10 & 0/0 & 0/4 & $0.12$ \\
G5 & AMaLGaM+SN vs. AMaLGaM & 10 & std. & $-$0.09 & $-$0.05 & [$-$0.18, $-$0.01] & $0.01$ & $0.13$ & yes & 19/5 & 0/0 & 13/7 & $0.26$ \\
G5 & AMaLGaM+SN vs. AMaLGaM & 10 & dist. & $-$0.03 & +0.01 & [$-$0.09, +0.14] & $0.51$ & $1.00$ & yes & 9/15 & 0/0 & 9/9 & $1.00$ \\
G5 & AMaLGaM+SN vs. AMaLGaM & 20 & std. & +0.08 & +0.05 & [+0.01, +0.11] & $0.02$ & $0.17$ & yes & 7/17 & 0/0 & 4/9 & $0.27$ \\
G5 & AMaLGaM+SN vs. AMaLGaM & 20 & dist. & $-$0.01 & +0.01 & [$-$0.09, +0.06] & $0.92$ & $1.00$ & yes & 11/13 & 0/0 & 8/7 & $1.00$ \\
G5 & AMaLGaM+SN vs. AMaLGaM & 40 & std. & 0.00 & +0.01 & [$-$0.02, +0.05] & $0.49$ & $1.00$ & yes & 10/14 & 0/0 & 8/4 & $0.39$ \\
G5 & AMaLGaM+SN vs. AMaLGaM & 40 & dist. & +0.03 & +0.03 & [$-$0.02, +0.09] & $0.30$ & $1.00$ & yes & 9/15 & 0/0 & 2/6 & $0.29$ \\
G6 & SN+MS vs. EMNA+AVS+AMS & 10 & std. & +0.21 & +0.13 & [+0.01, +0.33] & $0.03$ & $0.13$ & no & 9/15 & 0/4 & 9/15 & $0.31$ \\
G6 & SN+MS vs. EMNA+AVS+AMS & 10 & dist. & +0.75 & +0.39 & [+0.20, +1.03] & $0.0001$ & $0.0009$ & no & 3/21 & 0/7 & 18/29 & $0.14$ \\
G6 & SN+MS vs. EMNA+AVS+AMS & 20 & std. & +0.31 & +0.14 & [+0.01, +0.34] & $0.03$ & $0.13$ & no & 9/15 & 0/4 & 4/29 & $1\times10^{-5}$ \\
G6 & SN+MS vs. EMNA+AVS+AMS & 20 & dist. & +1.15 & +0.71 & [+0.29, +1.36] & $7\times10^{-6}$ & $6\times10^{-5}$ & no & 3/21 & 0/16 & 4/22 & $0.0005$ \\
G6 & SN+MS vs. EMNA+AVS+AMS & 40 & std. & +0.08 & +0.04 & [0.00, +0.10] & $0.08$ & $0.16$ & yes & 7/17 & 0/4 & 2/2 & $1.00$ \\
G6 & SN+MS vs. EMNA+AVS+AMS & 40 & dist. & +1.12 & +0.75 & [+0.42, +1.56] & $1\times10^{-7}$ & $1\times10^{-6}$ & no & 0/24 & 0/22 & 0/7 & $0.02$ \\
G6 & SN+MS vs. AMaLGaM & 10 & std. & +0.04 & +0.05 & [$-$0.12, +0.21] & $0.60$ & $0.60$ & yes & 11/13 & 0/9 & 19/6 & $0.01$ \\
G6 & SN+MS vs. AMaLGaM & 10 & dist. & +0.58 & +0.34 & [+0.05, +0.76] & $0.04$ & $0.13$ & no & 6/18 & 0/12 & 20/13 & $0.30$ \\
G6 & SN+MS vs. AMaLGaM & 20 & std. & +0.46 & +0.32 & [+0.08, +0.58] & $0.01$ & $0.07$ & no & 8/16 & 0/14 & 12/20 & $0.22$ \\
G6 & SN+MS vs. AMaLGaM & 20 & dist. & +1.56 & +1.04 & [+0.60, +1.65] & $4\times10^{-7}$ & $4\times10^{-6}$ & no & 1/23 & 0/13 & 10/23 & $0.04$ \\
G6 & SN+MS vs. AMaLGaM & 40 & std. & +0.76 & +0.63 & [+0.20, +1.16] & $8\times10^{-6}$ & $7\times10^{-5}$ & no & 2/22 & 0/18 & 0/24 & $1\times10^{-7}$ \\
G6 & SN+MS vs. AMaLGaM & 40 & dist. & +2.20 & +2.22 & [+1.14, +2.92] & $1\times10^{-7}$ & $1\times10^{-6}$ & no & 0/24 & 0/24 & 0/27 & $1\times10^{-8}$ \\
desc. & SN+MS vs. SN & 10 & std. & $-$0.26 & $-$0.05 & [$-$0.26, 0.00] & $0.08$ & -- & yes & 15/9 & 2/0 & 28/5 & $7\times10^{-5}$ \\
desc. & SN+MS vs. SN & 10 & dist. & $-$0.83 & $-$0.55 & [$-$1.13, $-$0.12] & $0.0006$ & -- & no & 21/3 & 6/0 & 38/11 & $0.0001$ \\
desc. & SN+MS vs. SN & 20 & std. & $-$0.28 & $-$0.09 & [$-$0.51, 0.00] & $0.08$ & -- & no & 13/11 & 5/0 & 12/3 & $0.04$ \\
desc. & SN+MS vs. SN & 20 & dist. & $-$0.84 & $-$0.79 & [$-$1.23, $-$0.36] & $9\times10^{-5}$ & -- & no & 22/2 & 15/0 & 21/5 & $0.002$ \\
desc. & SN+MS vs. SN & 40 & std. & $-$0.21 & $-$0.07 & [$-$0.16, $-$0.03] & $0.001$ & -- & yes & 18/6 & 4/0 & 5/0 & $0.06$ \\
desc. & SN+MS vs. SN & 40 & dist. & $-$1.74 & $-$1.25 & [$-$2.27, $-$0.85] & $1\times10^{-7}$ & -- & no & 24/0 & 19/0 & 40/4 & $2\times10^{-8}$ \\
desc. & MS vs. EMNA+AVS & 10 & std. & $-$0.55 & $-$0.26 & [$-$0.52, $-$0.06] & $0.002$ & -- & no & 18/6 & 2/0 & 31/4 & $3\times10^{-6}$ \\
desc. & MS vs. EMNA+AVS & 10 & dist. & $-$1.78 & $-$1.29 & [$-$2.25, $-$0.71] & $5\times10^{-6}$ & -- & no & 22/2 & 15/0 & 30/9 & $0.001$ \\
desc. & MS vs. EMNA+AVS & 20 & std. & $-$0.70 & $-$0.21 & [$-$0.66, $-$0.03] & $0.006$ & -- & no & 16/8 & 6/0 & 33/5 & $4\times10^{-6}$ \\
desc. & MS vs. EMNA+AVS & 20 & dist. & $-$3.55 & $-$3.31 & [$-$4.35, $-$2.63] & $1\times10^{-7}$ & -- & no & 24/0 & 24/0 & 55/2 & $2\times10^{-14}$ \\
desc. & MS vs. EMNA+AVS & 40 & std. & $-$0.59 & $-$0.11 & [$-$0.27, $-$0.05] & $0.0001$ & -- & yes & 20/4 & 7/0 & 18/1 & $8\times10^{-5}$ \\
desc. & MS vs. EMNA+AVS & 40 & dist. & $-$3.36 & $-$3.27 & [$-$3.93, $-$2.66] & $1\times10^{-7}$ & -- & no & 24/0 & 23/0 & 55/2 & $2\times10^{-14}$ \\
desc. & SN+MS vs. EMNA+AVS & 10 & std. & $-$0.54 & $-$0.09 & [$-$0.43, $-$0.02] & $0.007$ & -- & no & 19/5 & 3/0 & 35/5 & $1\times10^{-6}$ \\
desc. & SN+MS vs. EMNA+AVS & 10 & dist. & $-$2.34 & $-$1.65 & [$-$2.87, $-$0.84] & $6\times10^{-7}$ & -- & no & 22/2 & 18/0 & 55/9 & $4\times10^{-9}$ \\
desc. & SN+MS vs. EMNA+AVS & 20 & std. & $-$0.78 & $-$0.22 & [$-$0.78, $-$0.02] & $0.002$ & -- & no & 18/6 & 7/0 & 36/4 & $2\times10^{-7}$ \\
desc. & SN+MS vs. EMNA+AVS & 20 & dist. & $-$4.33 & $-$4.20 & [$-$5.36, $-$3.12] & $1\times10^{-7}$ & -- & no & 24/0 & 23/0 & 72/2 & $3\times10^{-19}$ \\
desc. & SN+MS vs. EMNA+AVS & 40 & std. & $-$0.61 & $-$0.10 & [$-$0.28, $-$0.03] & $0.001$ & -- & no & 18/6 & 7/0 & 18/1 & $8\times10^{-5}$ \\
desc. & SN+MS vs. EMNA+AVS & 40 & dist. & $-$5.21 & $-$5.20 & [$-$6.23, $-$4.10] & $1\times10^{-7}$ & -- & no & 24/0 & 23/0 & 85/2 & $5\times10^{-23}$ \\
desc. & SN vs. EMNA+AVS+AMS & 10 & std. & +0.47 & +0.12 & [+0.01, +0.74] & $0.04$ & -- & no & 7/17 & 0/3 & 5/34 & $2\times10^{-6}$ \\
desc. & SN vs. EMNA+AVS+AMS & 10 & dist. & +1.58 & +0.82 & [+0.39, +1.54] & $1\times10^{-7}$ & -- & no & 0/24 & 0/16 & 10/48 & $5\times10^{-7}$ \\
desc. & SN vs. EMNA+AVS+AMS & 20 & std. & +0.59 & +0.35 & [+0.03, +0.81] & $0.02$ & -- & no & 8/16 & 0/8 & 4/38 & $6\times10^{-8}$ \\
desc. & SN vs. EMNA+AVS+AMS & 20 & dist. & +1.99 & +1.59 & [+0.64, +2.48] & $1\times10^{-5}$ & -- & no & 1/23 & 0/19 & 6/40 & $3\times10^{-7}$ \\
desc. & SN vs. EMNA+AVS+AMS & 40 & std. & +0.29 & +0.08 & [+0.02, +0.36] & $0.005$ & -- & no & 5/19 & 0/8 & 0/5 & $0.06$ \\
desc. & SN vs. EMNA+AVS+AMS & 40 & dist. & +2.86 & +2.40 & [+1.52, +3.82] & $1\times10^{-7}$ & -- & no & 0/24 & 0/24 & 1/44 & $3\times10^{-12}$ \\
desc. & SN+MS vs. CMA-ES & 10 & std. & $-$0.37 & $-$0.09 & [$-$0.78, +0.34] & $0.73$ & -- & no & 10/14 & 5/8 & 39/1 & $7\times10^{-11}$ \\
desc. & SN+MS vs. CMA-ES & 10 & dist. & +1.01 & +0.70 & [$-$0.01, +1.59] & $0.06$ & -- & no & 8/16 & 0/13 & 33/11 & $0.001$ \\
desc. & SN+MS vs. CMA-ES & 20 & std. & +0.22 & +0.18 & [$-$0.67, +1.23] & $0.60$ & -- & no & 11/13 & 8/11 & 49/15 & $2\times10^{-5}$ \\
desc. & SN+MS vs. CMA-ES & 20 & dist. & +2.35 & +1.53 & [+0.45, +2.98] & $0.01$ & -- & no & 6/18 & 2/16 & 44/33 & $0.25$ \\
desc. & SN+MS vs. CMA-ES & 40 & std. & +1.59 & +1.18 & [$-$0.06, +3.35] & $0.09$ & -- & no & 9/15 & 6/12 & 18/59 & $3\times10^{-6}$ \\
desc. & SN+MS vs. CMA-ES & 40 & dist. & +4.70 & +4.08 & [+2.35, +6.11] & $1\times10^{-5}$ & -- & no & 1/23 & 1/21 & 16/65 & $4\times10^{-8}$ \\
desc. & SN+MS vs. IPOP-CMA-ES & 10 & std. & +1.18 & +1.08 & [+0.46, +1.62] & $2\times10^{-6}$ & -- & no & 1/23 & 0/19 & 5/55 & $1\times10^{-11}$ \\
desc. & SN+MS vs. IPOP-CMA-ES & 10 & dist. & +2.23 & +1.99 & [+1.11, +2.87] & $2\times10^{-7}$ & -- & no & 1/23 & 0/23 & 5/60 & $5\times10^{-13}$ \\
desc. & SN+MS vs. IPOP-CMA-ES & 20 & std. & +1.52 & +1.11 & [+0.61, +2.20] & $2\times10^{-5}$ & -- & no & 2/22 & 1/19 & 3/45 & $1\times10^{-10}$ \\
desc. & SN+MS vs. IPOP-CMA-ES & 20 & dist. & +3.60 & +2.61 & [+1.92, +3.84] & $1\times10^{-7}$ & -- & no & 0/24 & 0/24 & 1/64 & $4\times10^{-18}$ \\
desc. & SN+MS vs. IPOP-CMA-ES & 40 & std. & +2.75 & +2.06 & [+1.16, +4.18] & $1\times10^{-7}$ & -- & no & 0/24 & 0/22 & 0/101 & $8\times10^{-31}$ \\
desc. & SN+MS vs. IPOP-CMA-ES & 40 & dist. & +5.77 & +5.11 & [+3.62, +7.13] & $1\times10^{-7}$ & -- & no & 0/24 & 0/24 & 0/90 & $2\times10^{-27}$ \\
desc. & EMNA+AVS+AMS vs. AMaLGaM & 10 & std. & $-$0.17 & $-$0.02 & [$-$0.43, +0.14] & $0.83$ & -- & no & 12/12 & 1/6 & 23/4 & $0.0003$ \\
desc. & EMNA+AVS+AMS vs. AMaLGaM & 10 & dist. & $-$0.17 & +0.01 & [$-$0.46, +0.21] & $0.92$ & -- & no & 9/15 & 1/7 & 27/9 & $0.004$ \\
desc. & EMNA+AVS+AMS vs. AMaLGaM & 20 & std. & +0.16 & +0.18 & [$-$0.05, +0.41] & $0.16$ & -- & no & 8/16 & 0/11 & 22/5 & $0.002$ \\
desc. & EMNA+AVS+AMS vs. AMaLGaM & 20 & dist. & +0.42 & +0.37 & [+0.08, +0.63] & $0.005$ & -- & no & 6/18 & 1/8 & 18/13 & $0.47$ \\
desc. & EMNA+AVS+AMS vs. AMaLGaM & 40 & std. & +0.68 & +0.56 & [+0.12, +1.11] & $4\times10^{-5}$ & -- & no & 3/21 & 1/16 & 2/26 & $3\times10^{-6}$ \\
desc. & EMNA+AVS+AMS vs. AMaLGaM & 40 & dist. & +1.08 & +0.96 & [+0.46, +1.59] & $2\times10^{-6}$ & -- & no & 2/22 & 1/19 & 1/21 & $1\times10^{-5}$ \\
desc. & AMaLGaM vs. CMA-ES & 10 & std. & $-$0.41 & $-$0.14 & [$-$0.76, +0.16] & $0.47$ & -- & no & 12/12 & 6/8 & 27/2 & $2\times10^{-6}$ \\
desc. & AMaLGaM vs. CMA-ES & 10 & dist. & +0.43 & +0.41 & [+0.01, +0.82] & $0.06$ & -- & no & 7/17 & 0/10 & 22/7 & $0.008$ \\
desc. & AMaLGaM vs. CMA-ES & 20 & std. & $-$0.24 & +0.05 & [$-$0.95, +0.62] & $0.92$ & -- & no & 11/13 & 7/10 & 42/0 & $5\times10^{-13}$ \\
desc. & AMaLGaM vs. CMA-ES & 20 & dist. & +0.78 & +0.52 & [$-$0.18, +1.63] & $0.15$ & -- & no & 7/17 & 3/14 & 42/18 & $0.003$ \\
desc. & AMaLGaM vs. CMA-ES & 40 & std. & +0.83 & +0.52 & [$-$0.40, +1.99] & $0.32$ & -- & no & 10/14 & 9/12 & 40/57 & $0.10$ \\
desc. & AMaLGaM vs. CMA-ES & 40 & dist. & +2.50 & +1.63 & [+0.54, +4.18] & $0.002$ & -- & no & 5/19 & 3/18 & 37/59 & $0.03$ \\
desc. & SN(0.5) vs. EMNA+AVS & 10 & std. & $-$0.17 & $-$0.06 & [$-$0.27, 0.00] & $0.02$ & -- & no & 15/9 & 0/0 & 21/10 & $0.07$ \\
desc. & SN(0.5) vs. EMNA+AVS & 10 & dist. & $-$0.18 & $-$0.20 & [$-$0.33, $-$0.04] & $0.03$ & -- & no & 19/5 & 1/0 & 14/17 & $0.72$ \\
desc. & SN(0.5) vs. EMNA+AVS & 20 & std. & +0.03 & +0.01 & [$-$0.01, +0.03] & $0.56$ & -- & yes & 11/13 & 0/0 & 9/8 & $1.00$ \\
desc. & SN(0.5) vs. EMNA+AVS & 20 & dist. & $-$0.56 & $-$0.50 & [$-$0.73, $-$0.32] & $3\times10^{-5}$ & -- & no & 22/2 & 8/0 & 17/3 & $0.003$ \\
desc. & SN(0.75) vs. SN(0.5) & 10 & std. & $-$0.13 & $-$0.01 & [$-$0.13, +0.09] & $0.81$ & -- & yes & 14/10 & 2/0 & 14/12 & $0.85$ \\
desc. & SN(0.75) vs. SN(0.5) & 10 & dist. & $-$0.75 & $-$0.54 & [$-$0.89, $-$0.32] & $1\times10^{-5}$ & -- & no & 22/2 & 5/0 & 23/10 & $0.04$ \\
desc. & SN(0.75) vs. SN(0.5) & 20 & std. & $-$0.23 & $-$0.06 & [$-$0.16, 0.00] & $0.11$ & -- & yes & 14/10 & 0/0 & 14/4 & $0.03$ \\
desc. & SN(0.75) vs. SN(0.5) & 20 & dist. & $-$1.54 & $-$1.22 & [$-$1.97, $-$0.82] & $1\times10^{-7}$ & -- & no & 24/0 & 14/0 & 25/2 & $6\times10^{-6}$ \\
desc. & SN vs. SN(0.75) & 10 & std. & +0.01 & 0.00 & [$-$0.05, +0.08] & $0.94$ & -- & yes & 11/13 & 2/0 & 4/10 & $0.18$ \\
desc. & SN vs. SN(0.75) & 10 & dist. & $-$0.58 & $-$0.35 & [$-$0.90, $-$0.13] & $0.005$ & -- & no & 19/5 & 5/0 & 25/16 & $0.21$ \\
desc. & SN vs. SN(0.75) & 20 & std. & $-$0.31 & $-$0.06 & [$-$0.13, $-$0.02] & $0.002$ & -- & yes & 19/5 & 1/0 & 18/6 & $0.02$ \\
desc. & SN vs. SN(0.75) & 20 & dist. & $-$1.39 & $-$1.28 & [$-$1.85, $-$0.83] & $2\times10^{-7}$ & -- & no & 23/1 & 16/0 & 20/3 & $0.0005$ \\
\bottomrule
\end{longtable}
}

\begin{table}[htbp]\centering\footnotesize
\caption{Family G3: Page's trend test over the levels $\rho_{\max}=0$, 0.5, 0.75, 0.95 (EMNA+AVS, SN(0.5), SN(0.75), SN), with the 24 functions as blocks and the alternative that $y_{\mathrm{any}}$ improves monotonically with $\rho_{\max}$; mean $y_{\mathrm{any}}$ at the four levels}
\label{S-tab:page}
\begin{tabular}{@{}llcccccc@{}}
\toprule
$D$ & start & $L$ & $p$ & $p$ (Holm) & exact $p$ (Holm) & \multicolumn{2}{c}{mean $y_{\mathrm{any}}$ at $\rho_{\max}=0$, 0.5, 0.75, 0.95} \\
\midrule
10 & std. & 636 & $0.005$ & $0.007$ & $0.007$ & \multicolumn{2}{c}{$-$3.16, $-$3.33, $-$3.45, $-$3.44} \\
10 & dist. & 692 & $4\times10^{-11}$ & $1\times10^{-10}$ & $7\times10^{-13}$ & \multicolumn{2}{c}{$-$0.14, $-$0.32, $-$1.07, $-$1.65} \\
20 & std. & 638 & $0.004$ & $0.007$ & $0.007$ & \multicolumn{2}{c}{$-$1.95, $-$1.92, $-$2.14, $-$2.45} \\
20 & dist. & 717 & $7\times10^{-17}$ & $3\times10^{-16}$ & $2\times10^{-28}$ & \multicolumn{2}{c}{+3.81, +3.26, +1.72, +0.33} \\
\bottomrule
\end{tabular}
\par\vspace{2pt}\begin{minipage}{0.75\textwidth}\scriptsize $L$ is Page's statistic for four treatments and 24 blocks. $p$ and $p$ (Holm) are from its normal approximation (scipy.stats.page\_trend\_test with its default method), which is what the analysis of the protocol used and what the manuscript reports; the last $p$ column is from the exact null distribution of $L$ (Holm-corrected), which gives the same decisions with smaller values in the two settings with the distant start. The two lower levels were not run for $D=40$.\end{minipage}
\end{table}

\begin{table}[htbp]\centering\scriptsize
\setlength{\tabcolsep}{3.2pt}
\caption{Confirmatory test, $D=10$, standard start: mean anytime performance $y_{\mathrm{any}}$ over the 15 instances of each function (lower is better; best of the row in bold)}
\label{S-tab:pf0}
\begin{tabular}{@{}lrrrrrrrrrrrrrrr@{}}
\toprule
 & EMNA & SN(.5) & SN(.75) & SN & SN-on & MS & SN+MS & S+M-on & S+M-nofb & AMS & AMS+SN & AMaL. & AMaL.+SN & CMA & IPOP \\
\midrule
f1 & $-$7.43 & $-$7.43 & $-$7.39 & $-$7.44 & $-$7.47 & $-$7.43 & $-$7.45 & $-$7.46 & $-$6.20 & $-$7.47 & $-$7.46 & $-$7.66 & $-$7.67 & \textbf{$-$7.93} & \textbf{$-$7.93} \\
f2 & $-$6.99 & $-$7.01 & $-$7.04 & $-$7.01 & $-$6.95 & $-$7.00 & $-$7.00 & $-$6.92 & $-$5.62 & $-$6.97 & $-$6.97 & $-$7.32 & $-$7.31 & \textbf{$-$7.63} & \textbf{$-$7.63} \\
f3 & 0.65 & 0.63 & 0.74 & 0.78 & 0.74 & \textbf{0.31} & 0.74 & 0.95 & 1.32 & 0.84 & 0.65 & 0.84 & 0.71 & 1.20 & 0.51 \\
f4 & 1.40 & 1.36 & 1.34 & 1.30 & 1.24 & 1.27 & 1.35 & 1.21 & 1.40 & 1.10 & 1.14 & 1.16 & 1.11 & 1.22 & \textbf{0.89} \\
f5 & $-$3.09 & $-$3.61 & $-$7.52 & $-$7.72 & $-$7.75 & $-$7.79 & $-$7.84 & $-$7.86 & $-$7.85 & $-$7.92 & $-$7.91 & $-$7.90 & $-$7.89 & \textbf{$-$7.99} & \textbf{$-$7.99} \\
\midrule
f6 & 0.01 & $-$0.13 & $-$0.73 & $-$2.06 & $-$2.34 & $-$3.35 & $-$5.01 & $-$5.37 & $-$3.94 & $-$6.34 & $-$6.33 & $-$6.51 & $-$6.54 & \textbf{$-$7.78} & \textbf{$-$7.78} \\
f7 & $-$6.99 & $-$7.55 & $-$7.56 & $-$7.59 & $-$6.68 & $-$7.57 & $-$7.56 & $-$7.17 & $-$1.96 & $-$7.56 & $-$7.22 & $-$6.02 & $-$6.89 & $-$0.66 & \textbf{$-$7.69} \\
f8 & $-$6.00 & $-$5.60 & $-$6.09 & $-$6.07 & $-$5.95 & $-$6.14 & $-$5.72 & $-$5.88 & $-$2.76 & $-$6.28 & $-$5.77 & $-$6.50 & $-$6.56 & $-$5.95 & \textbf{$-$7.46} \\
f9 & $-$5.54 & $-$5.52 & $-$5.57 & $-$6.01 & $-$6.00 & $-$6.14 & $-$6.12 & $-$6.36 & $-$1.91 & $-$6.27 & $-$6.27 & $-$5.59 & $-$6.00 & $-$7.04 & \textbf{$-$7.55} \\
\midrule
f10 & $-$6.96 & $-$6.94 & $-$6.99 & $-$6.97 & $-$6.98 & $-$6.96 & $-$6.96 & $-$6.93 & $-$4.10 & $-$6.95 & $-$6.94 & $-$7.33 & $-$7.26 & \textbf{$-$7.63} & \textbf{$-$7.63} \\
f11 & $-$7.36 & $-$7.36 & $-$7.35 & $-$7.33 & $-$7.34 & $-$7.28 & $-$7.31 & $-$7.24 & $-$6.05 & $-$7.35 & $-$7.33 & $-$7.49 & $-$7.50 & \textbf{$-$7.76} & \textbf{$-$7.76} \\
f12 & $-$5.28 & $-$5.17 & $-$5.37 & $-$5.30 & $-$5.32 & $-$5.43 & $-$5.41 & $-$5.45 & $-$0.17 & $-$5.72 & $-$5.48 & $-$5.89 & $-$6.19 & \textbf{$-$7.34} & \textbf{$-$7.34} \\
f13 & $-$6.60 & $-$6.58 & $-$6.59 & $-$6.63 & $-$6.63 & $-$6.65 & $-$6.62 & $-$6.54 & $-$3.03 & $-$6.59 & $-$6.62 & $-$7.08 & $-$7.10 & $-$7.14 & \textbf{$-$7.56} \\
f14 & $-$7.40 & $-$7.41 & $-$7.41 & $-$7.43 & $-$7.40 & $-$7.41 & $-$7.43 & $-$7.38 & $-$6.05 & $-$7.43 & $-$7.43 & $-$7.60 & $-$7.61 & \textbf{$-$7.84} & \textbf{$-$7.84} \\
\midrule
f15 & 0.19 & 0.15 & 0.64 & 0.28 & 0.56 & 0.65 & 0.54 & 0.91 & 1.27 & 0.71 & 0.57 & 0.64 & 0.13 & 1.17 & \textbf{$-$3.01} \\
f16 & $-$0.63 & $-$0.75 & $-$1.28 & $-$0.89 & $-$1.37 & $-$0.98 & $-$0.93 & $-$1.22 & $-$0.37 & $-$1.44 & $-$0.77 & $-$0.37 & $-$0.73 & $-$0.00 & \textbf{$-$5.43} \\
f17 & $-$3.86 & $-$3.85 & $-$3.33 & $-$2.75 & $-$3.67 & $-$4.32 & $-$4.67 & $-$4.94 & $-$0.53 & $-$5.27 & $-$5.41 & $-$4.25 & $-$4.33 & $-$2.01 & \textbf{$-$6.79} \\
f18 & $-$2.03 & $-$3.23 & $-$2.31 & $-$2.37 & $-$2.46 & $-$3.65 & $-$3.25 & $-$3.12 & $-$0.94 & $-$4.42 & $-$5.34 & $-$3.68 & $-$2.91 & $-$1.28 & \textbf{$-$6.05} \\
f19 & $-$0.54 & \textbf{$-$0.64} & $-$0.56 & $-$0.55 & $-$0.51 & $-$0.52 & $-$0.58 & $-$0.39 & $-$0.51 & $-$0.57 & $-$0.62 & $-$0.56 & $-$0.62 & $-$0.33 & $-$0.59 \\
\midrule
f20 & 0.22 & 0.09 & 0.19 & 0.21 & 0.20 & 0.23 & 0.20 & 0.12 & 0.17 & 0.17 & 0.19 & 0.20 & 0.18 & 0.19 & \textbf{$-$0.26} \\
f21 & $-$1.70 & $-$2.22 & $-$2.25 & $-$1.68 & $-$2.78 & $-$2.28 & $-$1.79 & $-$1.20 & $-$1.54 & $-$1.21 & $-$1.21 & $-$1.62 & $-$1.65 & 0.82 & \textbf{$-$4.17} \\
f22 & $-$0.30 & $-$0.30 & $-$0.27 & $-$0.30 & 0.35 & $-$0.27 & $-$0.30 & $-$0.15 & $-$0.14 & \textbf{$-$0.32} & $-$0.31 & $-$0.20 & $-$0.22 & 0.76 & 0.10 \\
f23 & $-$0.95 & \textbf{$-$2.15} & $-$1.49 & $-$0.48 & $-$0.16 & $-$1.50 & $-$0.99 & $-$1.66 & $-$0.37 & $-$1.95 & $-$1.50 & $-$0.37 & $-$0.22 & $-$0.36 & $-$1.07 \\
f24 & 1.42 & 1.42 & 1.37 & 1.40 & 1.32 & 1.35 & 1.32 & 1.47 & 1.36 & 1.29 & 1.36 & 1.36 & 1.28 & 1.30 & \textbf{1.03} \\
\midrule
mean & $-$3.16 & $-$3.33 & $-$3.45 & $-$3.44 & $-$3.47 & $-$3.70 & $-$3.70 & $-$3.69 & $-$2.02 & $-$3.91 & $-$3.87 & $-$3.74 & $-$3.82 & $-$3.33 & $-$4.88 \\
\bottomrule
\end{tabular}
\end{table}
\begin{table}[htbp]\centering\scriptsize
\setlength{\tabcolsep}{3.2pt}
\caption{Confirmatory test, $D=10$, distant start: mean anytime performance $y_{\mathrm{any}}$ over the 15 instances of each function (lower is better; best of the row in bold)}
\label{S-tab:pf1}
\begin{tabular}{@{}lrrrrrrrrrrrrrrr@{}}
\toprule
 & EMNA & SN(.5) & SN(.75) & SN & SN-on & MS & SN+MS & S+M-on & S+M-nofb & AMS & AMS+SN & AMaL. & AMaL.+SN & CMA & IPOP \\
\midrule
f1 & $-$5.85 & $-$6.17 & $-$6.42 & $-$6.64 & $-$6.65 & $-$6.64 & $-$6.77 & $-$6.71 & $-$3.18 & $-$6.86 & $-$6.84 & $-$7.08 & $-$7.07 & \textbf{$-$7.87} & \textbf{$-$7.87} \\
f2 & $-$5.29 & $-$5.49 & $-$5.56 & $-$5.66 & $-$5.59 & $-$5.67 & $-$5.75 & $-$5.57 & $-$0.01 & $-$6.10 & $-$6.14 & $-$6.63 & $-$6.53 & \textbf{$-$7.49} & \textbf{$-$7.49} \\
f3 & 4.63 & 4.44 & 3.23 & 2.28 & 2.13 & 2.03 & 1.05 & 1.58 & 2.37 & 1.17 & 1.22 & 1.05 & 1.07 & 1.10 & \textbf{0.59} \\
f4 & 2.08 & 1.90 & 1.85 & 1.73 & 1.72 & 1.63 & 1.62 & 1.40 & 2.51 & 1.44 & 1.40 & 1.39 & 1.41 & 1.36 & \textbf{0.94} \\
f5 & $-$6.08 & $-$6.41 & $-$7.10 & $-$7.49 & $-$7.61 & $-$7.53 & $-$7.69 & $-$7.72 & $-$7.73 & $-$7.83 & $-$7.82 & $-$7.79 & $-$7.78 & \textbf{$-$7.97} & \textbf{$-$7.97} \\
\midrule
f6 & 6.03 & 5.50 & 4.65 & 2.95 & 3.03 & 1.80 & $-$1.07 & $-$1.66 & 0.34 & $-$4.41 & $-$4.54 & $-$4.81 & $-$4.95 & \textbf{$-$7.65} & \textbf{$-$7.65} \\
f7 & $-$6.33 & $-$6.69 & $-$6.34 & $-$6.77 & $-$6.85 & $-$6.82 & $-$6.83 & $-$5.64 & 1.84 & $-$6.95 & $-$6.96 & $-$6.71 & $-$6.76 & $-$3.92 & \textbf{$-$7.72} \\
f8 & 2.41 & 2.33 & $-$0.29 & $-$2.24 & $-$1.00 & $-$1.37 & $-$3.26 & $-$2.56 & 3.28 & $-$4.15 & $-$3.57 & $-$4.11 & $-$4.39 & $-$5.67 & \textbf{$-$7.12} \\
f9 & 1.06 & 1.90 & $-$1.68 & $-$4.31 & $-$3.00 & $-$2.70 & $-$2.50 & $-$3.43 & 2.69 & $-$4.64 & $-$5.16 & $-$5.43 & $-$4.91 & $-$6.84 & \textbf{$-$7.34} \\
\midrule
f10 & $-$5.22 & $-$5.44 & $-$5.65 & $-$5.72 & $-$5.58 & $-$5.69 & $-$5.68 & $-$5.60 & 1.17 & $-$6.21 & $-$6.16 & $-$6.56 & $-$6.57 & \textbf{$-$7.49} & \textbf{$-$7.49} \\
f11 & $-$4.89 & $-$5.26 & $-$5.56 & $-$5.96 & $-$5.88 & $-$5.85 & $-$6.08 & $-$6.00 & $-$0.67 & $-$6.76 & $-$6.77 & $-$7.02 & $-$7.00 & \textbf{$-$7.66} & \textbf{$-$7.66} \\
f12 & 16.00 & 15.17 & 12.59 & 8.57 & 9.59 & 5.36 & 1.67 & $-$0.74 & 5.21 & $-$3.61 & $-$3.36 & $-$2.72 & $-$3.71 & \textbf{$-$6.54} & \textbf{$-$6.54} \\
f13 & $-$4.74 & $-$5.15 & $-$5.61 & $-$5.76 & $-$5.86 & $-$5.75 & $-$5.88 & $-$5.75 & $-$0.17 & $-$5.93 & $-$5.93 & $-$6.46 & $-$6.48 & $-$7.50 & \textbf{$-$7.52} \\
f14 & $-$5.53 & $-$5.66 & $-$6.10 & $-$6.41 & $-$6.50 & $-$6.46 & $-$6.47 & $-$6.35 & $-$2.09 & $-$6.76 & $-$6.76 & $-$7.00 & $-$7.01 & \textbf{$-$7.71} & \textbf{$-$7.71} \\
\midrule
f15 & 5.90 & 6.09 & 4.76 & 3.42 & 2.39 & 1.88 & 1.37 & 1.56 & 2.53 & 1.00 & 1.05 & 0.98 & 0.98 & 1.12 & \textbf{$-$2.39} \\
f16 & $-$0.29 & 0.14 & $-$0.25 & $-$0.84 & $-$1.15 & $-$0.37 & $-$0.28 & $-$1.64 & 0.94 & $-$2.29 & $-$0.80 & $-$0.13 & 0.16 & $-$0.09 & \textbf{$-$4.66} \\
f17 & $-$1.47 & $-$2.27 & $-$2.83 & $-$1.91 & $-$2.37 & $-$3.23 & $-$4.25 & $-$3.74 & 1.30 & $-$4.81 & $-$4.97 & $-$3.60 & $-$3.33 & $-$2.30 & \textbf{$-$6.63} \\
f18 & $-$0.09 & 0.18 & $-$0.61 & $-$1.14 & $-$1.16 & $-$2.18 & $-$2.52 & $-$3.05 & 1.40 & $-$3.41 & $-$4.59 & $-$1.99 & $-$2.46 & $-$0.98 & \textbf{$-$5.95} \\
f19 & 1.90 & 1.46 & 0.64 & 0.43 & 0.34 & 0.29 & 0.21 & 0.42 & 1.62 & $-$0.00 & $-$0.04 & $-$0.09 & 0.17 & $-$0.12 & \textbf{$-$0.22} \\
\midrule
f20 & 1.12 & 1.08 & 0.79 & 0.65 & 0.66 & 0.64 & 0.59 & 0.49 & 1.08 & 0.54 & 0.51 & 0.52 & 0.52 & 0.32 & \textbf{0.00} \\
f21 & $-$1.42 & $-$0.97 & $-$1.26 & $-$0.40 & $-$1.86 & $-$0.88 & $-$1.85 & $-$0.80 & 1.15 & $-$1.37 & $-$1.49 & $-$0.49 & $-$1.50 & $-$1.57 & \textbf{$-$4.37} \\
f22 & 0.36 & 0.20 & 0.30 & 0.62 & 0.49 & 0.42 & 0.42 & 0.72 & 0.81 & 0.37 & 0.57 & 0.04 & 0.40 & 0.77 & \textbf{$-$0.36} \\
f23 & 0.17 & $-$0.69 & $-$1.06 & $-$0.85 & $-$1.06 & $-$0.57 & $-$1.11 & $-$0.88 & 0.70 & $-$1.53 & \textbf{$-$1.83} & $-$0.21 & 0.12 & $-$0.28 & $-$1.06 \\
f24 & 2.19 & 2.05 & 1.77 & 1.85 & 1.73 & 1.68 & 1.60 & 1.78 & 2.63 & 1.68 & 1.56 & 1.51 & 1.60 & 1.37 & \textbf{1.17} \\
\midrule
mean & $-$0.14 & $-$0.32 & $-$1.07 & $-$1.65 & $-$1.67 & $-$1.92 & $-$2.48 & $-$2.50 & 0.82 & $-$3.23 & $-$3.23 & $-$3.06 & $-$3.08 & $-$3.48 & $-$4.71 \\
\bottomrule
\end{tabular}
\end{table}
\begin{table}[htbp]\centering\scriptsize
\setlength{\tabcolsep}{3.2pt}
\caption{Confirmatory test, $D=20$, standard start: mean anytime performance $y_{\mathrm{any}}$ over the 15 instances of each function (lower is better; best of the row in bold)}
\label{S-tab:pf2}
\begin{tabular}{@{}lrrrrrrrrrrrrrrr@{}}
\toprule
 & EMNA & SN(.5) & SN(.75) & SN & SN-on & MS & SN+MS & S+M-on & S+M-nofb & AMS & AMS+SN & AMaL. & AMaL.+SN & CMA & IPOP \\
\midrule
f1 & $-$6.80 & $-$6.79 & $-$6.77 & $-$6.77 & $-$6.77 & $-$6.83 & $-$6.77 & $-$6.75 & $-$2.67 & $-$6.82 & $-$6.77 & $-$7.21 & $-$7.20 & \textbf{$-$7.93} & \textbf{$-$7.93} \\
f2 & $-$5.82 & $-$5.81 & $-$5.84 & $-$5.87 & $-$5.71 & $-$5.78 & $-$5.83 & $-$5.76 & 0.35 & $-$5.79 & $-$5.74 & $-$6.58 & $-$6.54 & \textbf{$-$7.35} & \textbf{$-$7.35} \\
f3 & 0.98 & 1.03 & 1.08 & 1.08 & 0.97 & 1.01 & 0.96 & 1.24 & 2.06 & 1.00 & 1.03 & 1.10 & 1.08 & 1.47 & \textbf{0.92} \\
f4 & 1.90 & 1.89 & 1.83 & 1.77 & 1.70 & 1.64 & 1.68 & 1.54 & 1.78 & 1.50 & 1.49 & 1.49 & 1.52 & 1.68 & \textbf{1.25} \\
f5 & 2.35 & 2.35 & $-$1.54 & $-$6.89 & $-$6.89 & $-$7.51 & $-$7.74 & $-$7.74 & $-$7.76 & $-$7.84 & $-$7.85 & $-$7.83 & $-$7.83 & \textbf{$-$7.99} & \textbf{$-$7.99} \\
\midrule
f6 & 1.95 & 1.84 & 1.69 & 1.50 & 1.55 & 1.29 & 0.32 & 0.44 & 0.04 & $-$2.90 & $-$2.82 & $-$3.90 & $-$3.77 & \textbf{$-$7.71} & \textbf{$-$7.71} \\
f7 & $-$7.01 & $-$7.07 & $-$6.91 & $-$7.03 & $-$7.02 & $-$7.09 & $-$7.04 & $-$6.97 & 0.94 & $-$7.03 & $-$7.05 & \textbf{$-$7.32} & $-$7.30 & 0.77 & $-$7.26 \\
f8 & $-$1.60 & $-$1.61 & $-$1.88 & $-$1.89 & $-$2.09 & $-$2.86 & $-$2.90 & $-$3.78 & 0.23 & $-$3.52 & $-$3.39 & $-$3.91 & $-$3.81 & $-$6.81 & \textbf{$-$7.28} \\
f9 & $-$1.51 & $-$1.57 & $-$1.86 & $-$2.01 & $-$2.25 & $-$2.98 & $-$2.97 & $-$3.87 & $-$0.75 & $-$3.62 & $-$3.53 & $-$4.22 & $-$4.16 & $-$6.21 & \textbf{$-$7.13} \\
\midrule
f10 & $-$5.80 & $-$5.74 & $-$5.75 & $-$5.78 & $-$5.70 & $-$5.75 & $-$5.81 & $-$5.65 & 0.18 & $-$5.81 & $-$5.73 & $-$6.49 & $-$6.55 & \textbf{$-$7.35} & \textbf{$-$7.35} \\
f11 & $-$6.65 & $-$6.66 & $-$6.66 & $-$6.68 & $-$6.67 & $-$6.66 & $-$6.59 & $-$6.55 & $-$3.15 & $-$6.64 & $-$6.72 & $-$7.15 & $-$7.08 & \textbf{$-$7.67} & \textbf{$-$7.67} \\
f12 & $-$2.34 & $-$2.29 & $-$2.63 & $-$2.64 & $-$2.94 & $-$3.15 & $-$3.19 & $-$3.80 & 2.06 & $-$3.43 & $-$3.50 & $-$4.41 & $-$4.15 & \textbf{$-$7.30} & \textbf{$-$7.30} \\
f13 & $-$5.04 & $-$4.99 & $-$5.05 & $-$5.09 & $-$5.07 & $-$5.00 & $-$5.05 & $-$4.91 & 0.86 & $-$4.88 & $-$5.00 & $-$6.12 & $-$6.02 & $-$3.54 & \textbf{$-$7.04} \\
f14 & $-$6.79 & $-$6.78 & $-$6.72 & $-$6.75 & $-$6.76 & $-$6.76 & $-$6.78 & $-$6.70 & $-$3.88 & $-$6.71 & $-$6.82 & $-$7.15 & $-$7.15 & \textbf{$-$7.77} & \textbf{$-$7.77} \\
\midrule
f15 & 0.96 & 0.95 & 0.98 & 0.87 & 0.96 & 0.83 & 0.82 & 1.10 & 1.96 & 0.89 & 0.88 & 0.80 & 0.94 & 1.56 & \textbf{0.02} \\
f16 & 0.03 & 0.03 & 0.05 & $-$0.36 & $-$0.13 & $-$0.07 & $-$0.25 & $-$1.14 & $-$0.29 & $-$0.64 & $-$0.39 & $-$0.16 & $-$0.08 & 0.38 & \textbf{$-$3.14} \\
f17 & $-$3.07 & $-$2.71 & $-$2.93 & $-$4.14 & $-$4.51 & $-$4.70 & $-$5.20 & $-$5.02 & $-$0.26 & $-$5.65 & $-$5.72 & $-$4.90 & $-$4.23 & $-$1.54 & \textbf{$-$6.50} \\
f18 & $-$2.76 & $-$2.30 & $-$2.83 & $-$2.28 & $-$2.45 & $-$3.23 & $-$3.44 & $-$4.71 & 0.18 & \textbf{$-$4.90} & $-$4.77 & $-$3.84 & $-$3.29 & $-$0.58 & $-$4.84 \\
f19 & $-$0.46 & $-$0.38 & $-$0.49 & $-$0.46 & $-$0.44 & $-$0.40 & $-$0.45 & $-$0.35 & $-$0.49 & $-$0.40 & $-$0.52 & $-$0.43 & \textbf{$-$0.61} & $-$0.09 & $-$0.16 \\
\midrule
f20 & 0.39 & 0.44 & 0.40 & 0.43 & 0.42 & 0.40 & 0.46 & 0.24 & 0.27 & 0.39 & 0.40 & 0.37 & 0.35 & 0.26 & \textbf{0.14} \\
f21 & $-$1.67 & $-$1.69 & $-$1.68 & $-$1.68 & $-$1.15 & \textbf{$-$1.73} & $-$1.67 & $-$1.10 & $-$1.13 & $-$1.18 & $-$1.19 & $-$1.23 & $-$1.22 & 0.82 & $-$0.87 \\
f22 & 0.43 & 0.43 & 0.43 & 0.43 & 0.51 & 0.41 & 0.43 & 0.43 & 0.50 & 0.42 & 0.43 & 0.52 & 0.52 & 0.53 & \textbf{0.16} \\
f23 & $-$0.16 & $-$0.24 & $-$0.09 & $-$0.31 & $-$0.20 & $-$0.49 & $-$0.26 & $-$0.80 & 0.03 & \textbf{$-$0.96} & $-$0.41 & $-$0.01 & 0.15 & $-$0.26 & $-$0.63 \\
f24 & 1.70 & 1.73 & 1.74 & 1.66 & 1.69 & 1.74 & 1.68 & 1.67 & 1.87 & 1.56 & 1.64 & 1.84 & 1.71 & 1.67 & \textbf{1.37} \\
\midrule
mean & $-$1.95 & $-$1.92 & $-$2.14 & $-$2.45 & $-$2.46 & $-$2.65 & $-$2.73 & $-$2.87 & $-$0.29 & $-$3.04 & $-$3.00 & $-$3.20 & $-$3.11 & $-$2.96 & $-$4.25 \\
\bottomrule
\end{tabular}
\end{table}
\begin{table}[htbp]\centering\scriptsize
\setlength{\tabcolsep}{3.2pt}
\caption{Confirmatory test, $D=20$, distant start: mean anytime performance $y_{\mathrm{any}}$ over the 15 instances of each function (lower is better; best of the row in bold)}
\label{S-tab:pf3}
\begin{tabular}{@{}lrrrrrrrrrrrrrrr@{}}
\toprule
 & EMNA & SN(.5) & SN(.75) & SN & SN-on & MS & SN+MS & S+M-on & S+M-nofb & AMS & AMS+SN & AMaL. & AMaL.+SN & CMA & IPOP \\
\midrule
f1 & 3.92 & 2.87 & $-$3.30 & $-$4.98 & $-$5.11 & $-$4.98 & $-$5.56 & $-$5.53 & 0.22 & $-$5.71 & $-$5.80 & $-$6.45 & $-$6.31 & \textbf{$-$7.88} & \textbf{$-$7.88} \\
f2 & $-$3.04 & $-$3.29 & $-$3.51 & $-$3.46 & $-$3.48 & $-$3.53 & $-$3.55 & $-$3.39 & 4.81 & $-$4.17 & $-$4.13 & $-$5.31 & $-$5.36 & \textbf{$-$7.14} & \textbf{$-$7.14} \\
f3 & 8.22 & 8.02 & 6.60 & 5.04 & 4.50 & 5.10 & 3.90 & 3.96 & 4.95 & 2.21 & 2.28 & 1.98 & 1.96 & 1.58 & \textbf{1.03} \\
f4 & 3.56 & 3.46 & 3.01 & 2.44 & 2.33 & 2.42 & 2.17 & 1.99 & 2.86 & 1.91 & 1.91 & 1.91 & 1.90 & 1.72 & \textbf{1.28} \\
f5 & $-$1.36 & $-$1.75 & $-$2.58 & $-$5.53 & $-$5.54 & $-$6.79 & $-$7.50 & $-$7.53 & $-$7.54 & $-$7.73 & $-$7.74 & $-$7.69 & $-$7.68 & \textbf{$-$7.98} & \textbf{$-$7.98} \\
\midrule
f6 & 7.92 & 7.68 & 6.88 & 5.76 & 5.78 & 5.74 & 4.61 & 4.80 & 4.79 & 1.17 & 1.28 & 0.36 & $-$0.01 & \textbf{$-$7.60} & \textbf{$-$7.60} \\
f7 & $-$2.63 & $-$4.59 & $-$5.42 & $-$5.62 & $-$5.25 & $-$5.54 & $-$5.66 & $-$5.55 & 3.14 & $-$5.90 & $-$5.94 & $-$5.98 & $-$6.31 & 0.59 & \textbf{$-$7.28} \\
f8 & 6.77 & 6.00 & 4.13 & 0.47 & 1.01 & 1.28 & $-$0.92 & $-$0.49 & 4.05 & $-$2.00 & $-$1.88 & $-$2.05 & $-$2.14 & $-$6.06 & \textbf{$-$6.91} \\
f9 & 9.02 & 8.56 & 4.66 & 2.67 & 2.32 & 3.08 & $-$0.04 & $-$1.06 & 4.47 & $-$1.50 & $-$1.69 & $-$2.20 & $-$2.49 & $-$6.59 & \textbf{$-$7.04} \\
\midrule
f10 & $-$2.33 & $-$2.71 & $-$2.88 & $-$3.40 & $-$3.46 & $-$3.66 & $-$3.60 & $-$3.34 & 4.72 & $-$4.29 & $-$4.27 & $-$5.37 & $-$5.32 & \textbf{$-$7.18} & \textbf{$-$7.18} \\
f11 & $-$0.13 & $-$0.34 & $-$1.15 & $-$3.60 & $-$3.68 & $-$3.41 & $-$4.27 & $-$4.14 & 3.00 & $-$5.56 & $-$5.60 & $-$6.29 & $-$6.22 & \textbf{$-$7.57} & \textbf{$-$7.57} \\
f12 & 19.05 & 19.79 & 18.58 & 14.76 & 14.37 & 14.19 & 12.03 & 11.58 & 13.71 & 2.64 & 1.93 & $-$0.03 & 0.11 & \textbf{$-$6.66} & \textbf{$-$6.66} \\
f13 & 3.93 & 2.15 & $-$1.47 & $-$3.49 & $-$3.51 & $-$3.32 & $-$3.76 & $-$3.67 & 2.46 & $-$3.86 & $-$3.85 & $-$5.29 & $-$5.24 & $-$3.15 & \textbf{$-$6.94} \\
f14 & 1.05 & 0.50 & $-$2.34 & $-$4.66 & $-$4.78 & $-$4.93 & $-$5.14 & $-$4.98 & $-$0.26 & $-$5.58 & $-$5.47 & $-$6.16 & $-$6.14 & \textbf{$-$7.61} & \textbf{$-$7.61} \\
\midrule
f15 & 9.47 & 9.05 & 8.55 & 6.57 & 6.37 & 6.62 & 4.63 & 4.67 & 5.46 & 1.67 & 1.60 & 1.53 & 1.62 & 1.50 & \textbf{0.50} \\
f16 & 0.69 & 0.65 & 0.36 & 0.27 & 0.06 & 0.16 & $-$0.21 & $-$0.29 & 1.50 & $-$0.13 & $-$0.25 & $-$0.11 & $-$0.07 & 0.48 & \textbf{$-$2.14} \\
f17 & 1.41 & 1.43 & 0.32 & $-$0.66 & $-$1.62 & $-$1.56 & $-$2.74 & $-$3.22 & 1.42 & $-$3.89 & $-$4.14 & $-$3.47 & $-$3.69 & $-$1.39 & \textbf{$-$6.03} \\
f18 & 2.77 & 2.09 & 1.17 & 0.77 & $-$0.43 & $-$0.73 & $-$1.22 & $-$2.74 & 1.61 & $-$2.95 & $-$3.18 & $-$2.18 & $-$1.96 & $-$0.25 & \textbf{$-$4.33} \\
f19 & 3.25 & 2.72 & 1.93 & 1.01 & 0.86 & 0.99 & 0.48 & 0.52 & 1.86 & 0.21 & 0.19 & 0.24 & 0.06 & 0.13 & \textbf{$-$0.03} \\
\midrule
f20 & 6.26 & 5.10 & 2.34 & 1.38 & 1.35 & 1.39 & 1.12 & 0.97 & 2.46 & 0.95 & 0.96 & 0.85 & 0.83 & 0.31 & \textbf{0.17} \\
f21 & 3.76 & 2.90 & 0.42 & $-$1.29 & $-$1.70 & 0.06 & $-$0.48 & $-$0.60 & 1.45 & $-$0.25 & $-$0.70 & $-$1.24 & $-$0.81 & $-$0.09 & \textbf{$-$2.41} \\
f22 & 3.27 & 2.77 & 1.44 & 0.91 & 0.81 & 0.85 & 0.70 & 0.89 & 1.57 & 0.77 & 0.83 & 0.58 & 0.68 & 0.65 & \textbf{0.38} \\
f23 & 1.28 & 0.77 & 0.40 & 0.04 & 0.27 & 0.47 & 0.40 & 0.01 & 1.06 & 0.03 & 0.18 & 0.33 & 0.48 & $-$0.17 & \textbf{$-$0.70} \\
f24 & 5.45 & 4.29 & 3.11 & 2.56 & 2.53 & 2.53 & 2.33 & 2.30 & 3.34 & 2.19 & 2.24 & 2.19 & 2.10 & 1.73 & \textbf{1.45} \\
\midrule
mean & 3.81 & 3.26 & 1.72 & 0.33 & 0.17 & 0.27 & $-$0.51 & $-$0.62 & 2.80 & $-$1.66 & $-$1.72 & $-$2.08 & $-$2.08 & $-$2.86 & $-$4.11 \\
\bottomrule
\end{tabular}
\end{table}
\begin{table}[htbp]\centering\scriptsize
\setlength{\tabcolsep}{3.2pt}
\caption{Confirmatory test, $D=40$, standard start: mean anytime performance $y_{\mathrm{any}}$ over the 15 instances of each function (lower is better; best of the row in bold)}
\label{S-tab:pf4}
\begin{tabular}{@{}lrrrrrrrrrrrrr@{}}
\toprule
 & EMNA & SN & SN-on & MS & SN+MS & S+M-on & S+M-nofb & AMS & AMS+SN & AMaL. & AMaL.+SN & CMA & IPOP \\
\midrule
f1 & $-$5.23 & $-$5.24 & $-$5.24 & $-$5.24 & $-$5.27 & $-$5.18 & $-$0.32 & $-$5.24 & $-$5.33 & $-$6.43 & $-$6.40 & \textbf{$-$7.93} & \textbf{$-$7.93} \\
f2 & $-$2.86 & $-$2.90 & $-$2.86 & $-$2.83 & $-$2.98 & $-$2.58 & 2.90 & $-$2.68 & $-$2.68 & $-$5.11 & $-$4.92 & \textbf{$-$6.76} & \textbf{$-$6.76} \\
f3 & 1.56 & 1.48 & 1.52 & 1.45 & 1.52 & 1.67 & 2.48 & 1.49 & 1.48 & 1.45 & 1.44 & 1.93 & \textbf{1.31} \\
f4 & 2.39 & 2.21 & 2.10 & 2.12 & 2.13 & 1.91 & 2.10 & 1.93 & 1.93 & 1.92 & 1.89 & 2.05 & \textbf{1.59} \\
f5 & 2.71 & $-$4.84 & $-$4.86 & $-$6.75 & $-$7.49 & $-$7.49 & $-$7.51 & $-$7.70 & $-$7.71 & $-$7.67 & $-$7.66 & \textbf{$-$7.99} & \textbf{$-$7.99} \\
\midrule
f6 & 3.14 & 3.30 & 3.11 & 3.05 & 3.17 & 3.26 & 2.99 & 2.00 & 2.29 & 1.01 & 1.08 & \textbf{$-$7.61} & \textbf{$-$7.61} \\
f7 & $-$5.24 & $-$5.65 & $-$5.60 & $-$5.70 & $-$5.74 & $-$5.56 & 1.62 & $-$5.68 & $-$5.63 & $-$6.40 & \textbf{$-$6.50} & 1.15 & $-$5.84 \\
f8 & 1.80 & 1.76 & 1.74 & 1.68 & 1.69 & 1.54 & 2.19 & 1.64 & 1.63 & 1.49 & 1.49 & $-$5.65 & \textbf{$-$6.39} \\
f9 & 1.76 & 1.70 & 1.70 & 1.66 & 1.66 & 1.55 & 1.72 & 1.60 & 1.62 & 1.45 & 1.46 & \textbf{$-$6.60} & \textbf{$-$6.60} \\
\midrule
f10 & $-$2.80 & $-$2.88 & $-$2.79 & $-$2.89 & $-$2.72 & $-$2.55 & 2.85 & $-$2.71 & $-$2.83 & $-$4.95 & $-$5.00 & \textbf{$-$6.76} & \textbf{$-$6.76} \\
f11 & $-$4.98 & $-$5.09 & $-$5.03 & $-$5.16 & $-$5.07 & $-$4.88 & $-$0.00 & $-$5.13 & $-$5.06 & $-$6.22 & $-$6.09 & \textbf{$-$7.52} & \textbf{$-$7.52} \\
f12 & 1.12 & 1.00 & 0.97 & 0.83 & 0.70 & 0.21 & 4.15 & 0.53 & 0.52 & $-$1.56 & $-$1.34 & \textbf{$-$7.29} & \textbf{$-$7.29} \\
f13 & $-$1.62 & $-$1.53 & $-$1.66 & $-$1.67 & $-$1.60 & $-$1.40 & 1.93 & $-$1.68 & $-$1.43 & $-$4.24 & $-$4.37 & $-$0.55 & \textbf{$-$5.05} \\
f14 & $-$5.08 & $-$5.11 & $-$5.15 & $-$5.25 & $-$5.19 & $-$5.08 & $-$1.72 & $-$5.07 & $-$5.21 & $-$6.32 & $-$6.24 & \textbf{$-$7.64} & \textbf{$-$7.64} \\
\midrule
f15 & 1.14 & 1.33 & 1.32 & 1.29 & 1.24 & 1.48 & 2.51 & 1.23 & 1.23 & 1.20 & 1.19 & 1.86 & \textbf{0.42} \\
f16 & 0.11 & 0.10 & 0.13 & 0.03 & $-$0.08 & $-$0.38 & $-$0.02 & 0.01 & $-$0.17 & 0.08 & 0.09 & 0.56 & \textbf{$-$1.22} \\
f17 & $-$1.60 & $-$2.05 & $-$2.82 & $-$2.95 & $-$2.93 & $-$3.04 & $-$0.31 & $-$3.14 & $-$3.16 & $-$4.04 & $-$3.84 & $-$1.33 & \textbf{$-$6.06} \\
f18 & $-$0.92 & $-$1.86 & $-$2.10 & $-$2.21 & $-$2.27 & $-$2.66 & 0.04 & $-$2.54 & $-$2.61 & $-$2.77 & $-$3.25 & $-$0.16 & \textbf{$-$3.82} \\
f19 & 0.02 & 0.02 & 0.04 & 0.00 & 0.02 & 0.01 & 0.01 & 0.01 & $-$0.02 & $-$0.19 & \textbf{$-$0.21} & 0.05 & $-$0.04 \\
\midrule
f20 & 0.60 & 0.61 & 0.59 & 0.61 & 0.59 & 0.45 & 0.73 & 0.57 & 0.56 & 0.48 & 0.50 & 0.28 & \textbf{0.24} \\
f21 & 0.21 & 0.21 & 0.16 & 0.19 & 0.20 & 0.18 & 0.61 & 0.21 & 0.19 & 0.13 & 0.13 & 0.12 & \textbf{$-$1.25} \\
f22 & 0.48 & 0.47 & 0.40 & 0.46 & 0.47 & 0.48 & 0.67 & 0.46 & 0.47 & 0.42 & 0.39 & 0.55 & \textbf{0.19} \\
f23 & 0.50 & 0.50 & 0.51 & 0.50 & 0.50 & 0.49 & 0.50 & 0.50 & 0.52 & 0.47 & 0.55 & $-$0.38 & \textbf{$-$1.00} \\
f24 & 2.14 & 2.11 & 2.08 & 2.09 & 2.10 & 2.09 & 2.36 & 2.08 & 2.08 & 2.16 & 2.06 & 2.01 & \textbf{1.72} \\
\midrule
mean & $-$0.44 & $-$0.85 & $-$0.91 & $-$1.03 & $-$1.06 & $-$1.06 & 0.94 & $-$1.14 & $-$1.14 & $-$1.82 & $-$1.81 & $-$2.65 & $-$3.80 \\
\bottomrule
\end{tabular}
\end{table}
\begin{table}[htbp]\centering\scriptsize
\setlength{\tabcolsep}{3.2pt}
\caption{Confirmatory test, $D=40$, distant start: mean anytime performance $y_{\mathrm{any}}$ over the 15 instances of each function (lower is better; best of the row in bold)}
\label{S-tab:pf5}
\begin{tabular}{@{}lrrrrrrrrrrrrr@{}}
\toprule
 & EMNA & SN & SN-on & MS & SN+MS & S+M-on & S+M-nofb & AMS & AMS+SN & AMaL. & AMaL.+SN & CMA & IPOP \\
\midrule
f1 & 4.99 & $-$1.05 & $-$0.96 & $-$0.32 & $-$2.78 & $-$2.79 & 2.10 & $-$3.45 & $-$3.20 & $-$5.04 & $-$5.11 & \textbf{$-$7.88} & \textbf{$-$7.88} \\
f2 & 1.50 & 0.97 & 0.91 & 0.99 & 0.78 & 1.33 & 6.88 & 0.25 & 0.24 & $-$3.08 & $-$2.98 & \textbf{$-$6.45} & \textbf{$-$6.45} \\
f3 & 9.75 & 7.22 & 7.16 & 7.37 & 6.68 & 6.83 & 6.93 & 4.56 & 4.45 & 3.87 & 3.69 & 1.91 & \textbf{1.42} \\
f4 & 7.24 & 3.99 & 3.87 & 4.01 & 3.02 & 2.88 & 3.84 & 2.78 & 2.77 & 2.59 & 2.55 & 2.07 & \textbf{1.62} \\
f5 & 3.40 & 0.85 & 0.60 & $-$1.33 & $-$7.01 & $-$7.09 & $-$7.03 & $-$7.54 & $-$7.54 & $-$7.42 & $-$7.44 & \textbf{$-$7.97} & \textbf{$-$7.97} \\
\midrule
f6 & 8.32 & 7.72 & 7.70 & 7.72 & 6.61 & 6.70 & 6.46 & 5.34 & 5.35 & 4.67 & 4.78 & \textbf{$-$7.46} & \textbf{$-$7.46} \\
f7 & 5.43 & $-$3.20 & $-$3.23 & $-$2.93 & $-$3.23 & $-$3.27 & 3.58 & $-$3.68 & $-$3.65 & $-$4.82 & $-$4.91 & 1.17 & \textbf{$-$6.33} \\
f8 & 10.39 & 4.22 & 4.11 & 5.11 & 3.24 & 3.16 & 5.41 & 2.86 & 2.90 & 2.55 & 2.56 & $-$5.39 & \textbf{$-$6.13} \\
f9 & 10.82 & 5.58 & 5.44 & 5.99 & 3.58 & 3.32 & 5.50 & 2.99 & 2.93 & 2.58 & 2.63 & \textbf{$-$6.50} & \textbf{$-$6.50} \\
\midrule
f10 & 3.73 & 1.33 & 1.30 & 1.60 & 1.11 & 1.43 & 6.92 & 0.14 & 0.12 & $-$2.90 & $-$2.85 & \textbf{$-$6.47} & \textbf{$-$6.47} \\
f11 & 7.92 & 3.25 & 3.32 & 3.26 & $-$0.52 & $-$0.27 & 4.91 & $-$3.20 & $-$3.06 & $-$4.88 & $-$4.92 & \textbf{$-$7.40} & \textbf{$-$7.40} \\
f12 & 22.97 & 20.79 & 21.09 & 20.37 & 15.72 & 15.88 & 15.52 & 11.92 & 11.57 & 9.46 & 9.31 & \textbf{$-$6.81} & \textbf{$-$6.81} \\
f13 & 4.85 & 2.09 & 2.17 & 2.38 & 0.45 & 0.52 & 3.43 & 0.17 & 0.15 & $-$2.94 & $-$2.83 & $-$0.36 & \textbf{$-$5.15} \\
f14 & 5.65 & 3.77 & 3.46 & 2.70 & $-$1.83 & $-$1.50 & 2.46 & $-$2.35 & $-$2.59 & $-$4.36 & $-$4.35 & \textbf{$-$7.45} & \textbf{$-$7.45} \\
\midrule
f15 & 10.36 & 9.21 & 9.29 & 9.47 & 8.49 & 8.55 & 8.56 & 3.98 & 3.92 & 3.19 & 3.18 & 1.90 & \textbf{1.27} \\
f16 & 4.24 & 1.03 & 1.12 & 1.10 & 0.77 & 0.54 & 1.91 & 0.64 & 0.68 & 0.50 & 0.67 & 0.73 & \textbf{$-$0.82} \\
f17 & 8.23 & 4.61 & 4.69 & 5.69 & 2.98 & 3.05 & 4.14 & 0.58 & 0.18 & $-$1.40 & $-$1.04 & $-$1.17 & \textbf{$-$5.24} \\
f18 & 8.59 & 5.96 & 5.66 & 6.61 & 3.77 & 3.50 & 4.91 & 0.91 & 0.97 & $-$0.56 & $-$0.19 & $-$0.05 & \textbf{$-$2.83} \\
f19 & 6.62 & 2.63 & 2.64 & 2.92 & 1.33 & 1.21 & 2.16 & 0.89 & 0.90 & 0.67 & 0.70 & 0.16 & \textbf{0.05} \\
\midrule
f20 & 8.20 & 3.67 & 3.68 & 3.65 & 1.96 & 1.83 & 4.13 & 1.66 & 1.64 & 1.36 & 1.37 & 0.29 & \textbf{0.28} \\
f21 & 4.90 & 1.18 & 1.12 & 1.59 & 0.98 & 0.90 & 1.94 & 0.41 & 0.71 & 0.60 & 0.19 & 0.10 & \textbf{$-$1.68} \\
f22 & 4.90 & 1.95 & 1.91 & 1.96 & 1.14 & 1.22 & 1.99 & 1.06 & 0.93 & 0.90 & 0.91 & 0.49 & \textbf{0.21} \\
f23 & 4.85 & 1.59 & 1.55 & 1.73 & 1.17 & 1.10 & 1.78 & 0.97 & 0.95 & 0.71 & 0.92 & $-$0.46 & \textbf{$-$0.78} \\
f24 & 8.90 & 4.18 & 4.12 & 4.48 & 3.29 & 3.21 & 4.56 & 2.99 & 2.95 & 2.75 & 2.90 & 2.02 & \textbf{1.81} \\
\midrule
mean & 7.36 & 3.90 & 3.86 & 4.00 & 2.16 & 2.18 & 4.29 & 1.04 & 1.01 & $-$0.04 & $-$0.01 & $-$2.54 & $-$3.61 \\
\bottomrule
\end{tabular}
\end{table}
\noindent Column headings of Tables \ref{S-tab:pf0}--\ref{S-tab:ert5}: EMNA = EMNA+AVS, SN(.5) and SN(.75) = SN with $\rho_{\max}=0.5$ and 0.75, S+M = SN+MS, AMS = EMNA+AVS+AMS, AMaL. = AMaLGaM, CMA = CMA-ES, IPOP = IPOP-CMA-ES.

\begin{table}[htbp]\centering\scriptsize
\setlength{\tabcolsep}{3.5pt}
\caption{Confirmatory test, $D=10$, standard start: expected running time to $10^{-8}$ divided by $D$ (evaluations spent by all 15 runs divided by the number of successful runs), with the number of successful runs in parentheses; -- if no run succeeded}
\label{S-tab:ert0}
\begin{tabular}{@{}lrrrrrrrrrrr@{}}
\toprule
 & EMNA & SN & MS & SN+MS & S+M-nofb & AMS & AMS+SN & AMaL. & AMaL.+SN & CMA & IPOP \\
\midrule
f1 & 1168 (15) & 1159 (15) & 1164 (15) & 1142 (15) & 3957 (15) & 1135 (15) & 1134 (15) & 735 (15) & 742 (15) & 151 (15) & 151 (15) \\
f2 & 1534 (15) & 1511 (15) & 1575 (15) & 1538 (15) & 4314 (14) & 1621 (15) & 1580 (15) & 1003 (15) & 1026 (15) & 433 (15) & 433 (15) \\
f3 & -- (0) & -- (0) & 141858 (1) & -- (0) & -- (0) & -- (0) & -- (0) & -- (0) & -- (0) & -- (0) & -- (0) \\
f4 & -- (0) & -- (0) & -- (0) & -- (0) & -- (0) & -- (0) & -- (0) & -- (0) & -- (0) & -- (0) & -- (0) \\
f5 & 7399 (10) & 286 (15) & 212 (15) & 166 (15) & 154 (15) & 87 (15) & 89 (15) & 110 (15) & 111 (15) & 12 (15) & 12 (15) \\
\midrule
f6 & -- (0) & -- (0) & 12407 (11) & 6029 (15) & 9317 (13) & 3241 (15) & 3259 (15) & 2907 (15) & 2869 (15) & 415 (15) & 415 (15) \\
f7 & 2119 (13) & 559 (15) & 598 (15) & 603 (15) & 28078 (4) & 624 (15) & 1282 (14) & 4117 (11) & 1960 (13) & 2331 (1) & 458 (15) \\
f8 & 2828 (15) & 2801 (15) & 2635 (15) & 3317 (14) & 10709 (11) & 2474 (15) & 3239 (14) & 2085 (15) & 2054 (15) & 696 (12) & 700 (15) \\
f9 & 3594 (14) & 2843 (15) & 2637 (15) & 2723 (15) & 15774 (8) & 2499 (15) & 2488 (15) & 3580 (13) & 2839 (14) & 593 (14) & 594 (15) \\
\midrule
f10 & 1594 (15) & 1620 (15) & 1597 (15) & 1628 (15) & 7704 (13) & 1631 (15) & 1595 (15) & 1011 (15) & 1141 (15) & 430 (15) & 430 (15) \\
f11 & 1261 (15) & 1270 (15) & 1368 (15) & 1326 (15) & 3648 (14) & 1252 (15) & 1294 (15) & 928 (15) & 923 (15) & 323 (15) & 323 (15) \\
f12 & 4563 (15) & 4395 (15) & 4176 (15) & 4252 (15) & 22922 (6) & 3710 (15) & 3927 (15) & 3667 (15) & 3250 (15) & 1055 (15) & 1055 (15) \\
f13 & 2572 (15) & 2491 (15) & 2465 (15) & 2526 (15) & 17956 (7) & 2569 (15) & 2541 (15) & 1704 (15) & 1650 (15) & 872 (13) & 847 (15) \\
f14 & 1306 (15) & 1277 (15) & 1294 (15) & 1254 (15) & 4584 (15) & 1262 (15) & 1260 (15) & 885 (15) & 885 (15) & 416 (15) & 416 (15) \\
\midrule
f15 & 141689 (1) & 141463 (1) & -- (0) & -- (0) & -- (0) & -- (0) & -- (0) & -- (0) & 141380 (1) & -- (0) & 6817 (14) \\
f16 & -- (0) & 141638 (1) & 141716 (1) & 141968 (1) & -- (0) & 66562 (2) & -- (0) & -- (0) & 141399 (1) & -- (0) & 4542 (13) \\
f17 & 67588 (2) & 143081 (1) & 22725 (5) & 11189 (8) & -- (0) & 17416 (6) & 9062 (9) & 41806 (3) & 29156 (4) & -- (0) & 2803 (15) \\
f18 & -- (0) & 142587 (1) & 67624 (2) & 67539 (2) & 144841 (1) & 17606 (6) & 11430 (8) & 67122 (2) & 66813 (2) & -- (0) & 3685 (15) \\
f19 & -- (0) & -- (0) & -- (0) & -- (0) & -- (0) & -- (0) & -- (0) & -- (0) & -- (0) & -- (0) & -- (0) \\
\midrule
f20 & -- (0) & -- (0) & -- (0) & -- (0) & -- (0) & -- (0) & -- (0) & -- (0) & -- (0) & -- (0) & -- (0) \\
f21 & 28204 (4) & 28170 (4) & 20705 (5) & 28159 (4) & 29633 (4) & 40603 (3) & 40590 (3) & 28687 (4) & 28550 (4) & -- (0) & 7850 (9) \\
f22 & 141102 (1) & 141036 (1) & 141369 (1) & 141116 (1) & 143207 (1) & 140750 (1) & 140924 (1) & 140648 (1) & 140474 (1) & -- (0) & 147406 (1) \\
f23 & 45295 (3) & 144684 (1) & 31744 (4) & 44422 (3) & 148485 (1) & 20010 (6) & 32186 (4) & 146317 (1) & 146402 (1) & -- (0) & 144620 (1) \\
f24 & -- (0) & -- (0) & -- (0) & -- (0) & -- (0) & -- (0) & -- (0) & -- (0) & -- (0) & -- (0) & -- (0) \\
\bottomrule
\end{tabular}
\end{table}
\begin{table}[htbp]\centering\scriptsize
\setlength{\tabcolsep}{3.5pt}
\caption{Confirmatory test, $D=10$, distant start: expected running time to $10^{-8}$ divided by $D$ (evaluations spent by all 15 runs divided by the number of successful runs), with the number of successful runs in parentheses; -- if no run succeeded}
\label{S-tab:ert1}
\begin{tabular}{@{}lrrrrrrrrrrr@{}}
\toprule
 & EMNA & SN & MS & SN+MS & S+M-nofb & AMS & AMS+SN & AMaL. & AMaL.+SN & CMA & IPOP \\
\midrule
f1 & 2484 (15) & 1834 (15) & 1812 (15) & 1716 (15) & 7327 (15) & 1620 (15) & 1712 (15) & 1207 (15) & 1242 (15) & 197 (15) & 197 (15) \\
f2 & 2660 (15) & 2376 (15) & 2413 (15) & 2328 (15) & 12162 (11) & 2186 (15) & 2149 (15) & 1461 (15) & 1590 (15) & 490 (15) & 490 (15) \\
f3 & -- (0) & -- (0) & -- (0) & 143865 (1) & -- (0) & -- (0) & -- (0) & -- (0) & -- (0) & -- (0) & -- (0) \\
f4 & -- (0) & -- (0) & -- (0) & -- (0) & -- (0) & -- (0) & -- (0) & -- (0) & -- (0) & -- (0) & -- (0) \\
f5 & 1741 (15) & 467 (15) & 429 (15) & 281 (15) & 246 (15) & 159 (15) & 168 (15) & 199 (15) & 205 (15) & 25 (15) & 25 (15) \\
\midrule
f6 & -- (0) & -- (0) & -- (0) & 9686 (14) & 74164 (2) & 4905 (15) & 4750 (15) & 4301 (15) & 4150 (15) & 510 (15) & 510 (15) \\
f7 & 2001 (14) & 1236 (15) & 1201 (15) & 1200 (15) & -- (0) & 1100 (15) & 1096 (15) & 1648 (14) & 1587 (14) & 526 (6) & 359 (15) \\
f8 & 28025 (5) & 6951 (13) & 8387 (12) & 5965 (13) & 73611 (2) & 5012 (13) & 6144 (12) & 4451 (14) & 4184 (14) & 1028 (12) & 1003 (15) \\
f9 & 13027 (9) & 4005 (15) & 5986 (13) & 6611 (13) & 74429 (2) & 4107 (14) & 3247 (15) & 2826 (15) & 3662 (14) & 763 (14) & 765 (15) \\
\midrule
f10 & 2601 (15) & 2364 (15) & 2393 (15) & 2399 (15) & 16303 (8) & 2047 (15) & 2125 (15) & 1528 (15) & 1566 (15) & 489 (15) & 489 (15) \\
f11 & 2881 (15) & 2211 (15) & 2263 (15) & 2197 (15) & 10838 (12) & 1737 (15) & 1737 (15) & 1224 (15) & 1269 (15) & 372 (15) & 372 (15) \\
f12 & 147321 (1) & 34926 (4) & 15814 (8) & 10320 (11) & 74847 (2) & 4828 (15) & 5493 (14) & 6498 (13) & 4943 (15) & 1769 (15) & 1769 (15) \\
f13 & 4058 (15) & 3228 (15) & 3246 (15) & 3134 (15) & -- (0) & 3092 (15) & 3119 (15) & 2188 (15) & 2152 (15) & 922 (13) & 906 (15) \\
f14 & 2808 (15) & 2092 (15) & 2047 (15) & 2072 (15) & 10215 (13) & 1837 (15) & 1857 (15) & 1410 (15) & 1414 (15) & 544 (15) & 544 (15) \\
\midrule
f15 & -- (0) & -- (0) & -- (0) & -- (0) & -- (0) & -- (0) & -- (0) & -- (0) & -- (0) & -- (0) & 10397 (10) \\
f16 & 142506 (1) & 42797 (3) & 142563 (1) & 142862 (1) & -- (0) & 22134 (5) & 142228 (1) & -- (0) & -- (0) & -- (0) & 5418 (14) \\
f17 & 43329 (3) & 43613 (3) & 43542 (3) & 14505 (7) & -- (0) & 14522 (7) & 11719 (8) & 67559 (2) & 67454 (2) & -- (0) & 3401 (15) \\
f18 & 143237 (1) & 143875 (1) & 43525 (3) & 68604 (2) & -- (0) & 30805 (4) & 11929 (8) & 67902 (2) & 142948 (1) & -- (0) & 3882 (15) \\
f19 & -- (0) & -- (0) & -- (0) & -- (0) & -- (0) & -- (0) & -- (0) & -- (0) & -- (0) & -- (0) & -- (0) \\
\midrule
f20 & -- (0) & -- (0) & -- (0) & -- (0) & 147832 (1) & -- (0) & -- (0) & -- (0) & -- (0) & -- (0) & -- (0) \\
f21 & 22136 (5) & 66343 (2) & 41388 (3) & 21284 (5) & -- (0) & 28831 (4) & 28765 (4) & 65840 (2) & 28604 (4) & 1051 (4) & 6643 (10) \\
f22 & 142066 (1) & -- (0) & -- (0) & -- (0) & 145598 (1) & -- (0) & -- (0) & 141058 (1) & -- (0) & -- (0) & 140202 (1) \\
f23 & -- (0) & 45130 (3) & 46543 (3) & 33510 (4) & -- (0) & 25691 (5) & 20056 (6) & 144480 (1) & -- (0) & -- (0) & 144635 (1) \\
f24 & -- (0) & -- (0) & -- (0) & -- (0) & -- (0) & -- (0) & -- (0) & -- (0) & -- (0) & -- (0) & -- (0) \\
\bottomrule
\end{tabular}
\end{table}
\begin{table}[htbp]\centering\scriptsize
\setlength{\tabcolsep}{3.5pt}
\caption{Confirmatory test, $D=20$, standard start: expected running time to $10^{-8}$ divided by $D$ (evaluations spent by all 15 runs divided by the number of successful runs), with the number of successful runs in parentheses; -- if no run succeeded}
\label{S-tab:ert2}
\begin{tabular}{@{}lrrrrrrrrrrr@{}}
\toprule
 & EMNA & SN & MS & SN+MS & S+M-nofb & AMS & AMS+SN & AMaL. & AMaL.+SN & CMA & IPOP \\
\midrule
f1 & 2369 (15) & 2418 (15) & 2356 (15) & 2421 (15) & 36498 (4) & 2359 (15) & 2461 (15) & 1535 (15) & 1577 (15) & 142 (15) & 142 (15) \\
f2 & 3188 (15) & 3117 (15) & 3270 (15) & 3197 (15) & -- (0) & 3316 (15) & 3317 (15) & 1975 (15) & 2111 (15) & 703 (15) & 703 (15) \\
f3 & -- (0) & -- (0) & -- (0) & -- (0) & -- (0) & -- (0) & -- (0) & -- (0) & -- (0) & -- (0) & -- (0) \\
f4 & -- (0) & -- (0) & -- (0) & -- (0) & -- (0) & -- (0) & -- (0) & -- (0) & -- (0) & -- (0) & -- (0) \\
f5 & -- (0) & 1163 (14) & 494 (15) & 262 (15) & 241 (15) & 166 (15) & 157 (15) & 172 (15) & 169 (15) & 13 (15) & 13 (15) \\
\midrule
f6 & -- (0) & -- (0) & -- (0) & -- (0) & -- (0) & 9699 (14) & 10734 (13) & 7687 (15) & 7798 (15) & 509 (15) & 509 (15) \\
f7 & 1263 (15) & 1239 (15) & 1177 (15) & 1217 (15) & -- (0) & 1244 (15) & 1232 (15) & 892 (15) & 915 (15) & -- (0) & 918 (15) \\
f8 & 8250 (15) & 7876 (15) & 6761 (15) & 6746 (15) & -- (0) & 5955 (15) & 6113 (15) & 5269 (15) & 5404 (15) & 876 (14) & 892 (15) \\
f9 & 8351 (15) & 7807 (15) & 6617 (15) & 6654 (15) & -- (0) & 5957 (15) & 6011 (15) & 4861 (15) & 5041 (15) & 1045 (13) & 1059 (15) \\
\midrule
f10 & 3248 (15) & 3192 (15) & 3262 (15) & 3248 (15) & -- (0) & 3226 (15) & 3289 (15) & 2159 (15) & 2037 (15) & 704 (15) & 704 (15) \\
f11 & 2507 (15) & 2491 (15) & 2548 (15) & 2626 (15) & 33580 (4) & 2600 (15) & 2460 (15) & 1565 (15) & 1698 (15) & 404 (15) & 404 (15) \\
f12 & 23551 (6) & 11229 (12) & 7851 (15) & 7852 (15) & -- (0) & 6992 (15) & 6975 (15) & 5950 (15) & 6141 (15) & 1050 (15) & 1050 (15) \\
f13 & 5213 (15) & 5159 (15) & 5329 (15) & 5307 (15) & -- (0) & 5486 (15) & 5308 (15) & 3306 (15) & 3489 (15) & 3585 (3) & 2291 (15) \\
f14 & 2549 (15) & 2631 (15) & 2649 (15) & 2550 (15) & 9361 (14) & 2724 (15) & 2499 (15) & 1794 (15) & 1817 (15) & 660 (15) & 660 (15) \\
\midrule
f15 & -- (0) & -- (0) & -- (0) & -- (0) & -- (0) & -- (0) & -- (0) & -- (0) & -- (0) & -- (0) & 70946 (2) \\
f16 & -- (0) & -- (0) & -- (0) & -- (0) & -- (0) & -- (0) & -- (0) & -- (0) & -- (0) & -- (0) & 12748 (9) \\
f17 & 32739 (4) & 13843 (8) & 16280 (7) & 9943 (10) & -- (0) & 6308 (13) & 6362 (13) & 31283 (4) & 68949 (2) & -- (0) & 3862 (15) \\
f18 & 70473 (2) & -- (0) & 45443 (3) & 70281 (2) & -- (0) & 11848 (9) & 11995 (9) & 31713 (4) & 69401 (2) & -- (0) & 6812 (15) \\
f19 & -- (0) & -- (0) & -- (0) & -- (0) & -- (0) & -- (0) & -- (0) & -- (0) & -- (0) & -- (0) & -- (0) \\
\midrule
f20 & -- (0) & -- (0) & -- (0) & -- (0) & -- (0) & -- (0) & -- (0) & -- (0) & -- (0) & -- (0) & -- (0) \\
f21 & 28995 (4) & 28978 (4) & 28793 (4) & 28992 (4) & 32626 (4) & 41420 (3) & 41359 (3) & 40725 (3) & 40989 (3) & -- (0) & 26199 (5) \\
f22 & -- (0) & -- (0) & -- (0) & -- (0) & -- (0) & -- (0) & -- (0) & -- (0) & -- (0) & -- (0) & -- (0) \\
f23 & 148590 (1) & 74381 (2) & 148055 (1) & 49306 (3) & -- (0) & 29165 (5) & 148915 (1) & 145345 (1) & -- (0) & -- (0) & -- (0) \\
f24 & -- (0) & -- (0) & -- (0) & -- (0) & -- (0) & -- (0) & -- (0) & -- (0) & -- (0) & -- (0) & -- (0) \\
\bottomrule
\end{tabular}
\end{table}
\begin{table}[htbp]\centering\scriptsize
\setlength{\tabcolsep}{3.5pt}
\caption{Confirmatory test, $D=20$, distant start: expected running time to $10^{-8}$ divided by $D$ (evaluations spent by all 15 runs divided by the number of successful runs), with the number of successful runs in parentheses; -- if no run succeeded}
\label{S-tab:ert3}
\begin{tabular}{@{}lrrrrrrrrrrr@{}}
\toprule
 & EMNA & SN & MS & SN+MS & S+M-nofb & AMS & AMS+SN & AMaL. & AMaL.+SN & CMA & IPOP \\
\midrule
f1 & -- (0) & 3923 (15) & 3919 (15) & 3398 (15) & -- (0) & 3372 (15) & 3215 (15) & 2149 (15) & 2293 (15) & 188 (15) & 188 (15) \\
f2 & 4873 (15) & 4648 (15) & 4623 (15) & 4620 (15) & -- (0) & 4286 (15) & 4359 (15) & 2941 (15) & 2911 (15) & 768 (15) & 768 (15) \\
f3 & -- (0) & -- (0) & -- (0) & -- (0) & -- (0) & -- (0) & -- (0) & -- (0) & -- (0) & -- (0) & -- (0) \\
f4 & -- (0) & -- (0) & -- (0) & -- (0) & -- (0) & -- (0) & -- (0) & -- (0) & -- (0) & -- (0) & -- (0) \\
f5 & 5878 (15) & 2190 (15) & 1075 (15) & 443 (15) & 408 (15) & 248 (15) & 239 (15) & 292 (15) & 299 (15) & 22 (15) & 22 (15) \\
\midrule
f6 & -- (0) & -- (0) & -- (0) & -- (0) & -- (0) & -- (0) & -- (0) & 149160 (1) & 149769 (1) & 595 (15) & 595 (15) \\
f7 & 5514 (12) & 2346 (15) & 2417 (15) & 2326 (15) & -- (0) & 2168 (15) & 2133 (15) & 2403 (14) & 1685 (15) & -- (0) & 871 (15) \\
f8 & -- (0) & 11081 (12) & 16915 (8) & 8995 (13) & -- (0) & 6918 (15) & 6977 (15) & 6522 (15) & 6957 (14) & 1204 (13) & 1227 (15) \\
f9 & -- (0) & 36284 (4) & 48478 (3) & 9811 (13) & -- (0) & 7785 (14) & 7085 (15) & 7154 (13) & 6052 (15) & 1074 (14) & 1088 (15) \\
\midrule
f10 & 5334 (15) & 4669 (15) & 4610 (15) & 4558 (15) & -- (0) & 4184 (15) & 4232 (15) & 2797 (15) & 2912 (15) & 745 (15) & 745 (15) \\
f11 & 6579 (15) & 4530 (15) & 4749 (15) & 4177 (15) & -- (0) & 3482 (15) & 3356 (15) & 2233 (15) & 2300 (15) & 441 (15) & 441 (15) \\
f12 & -- (0) & -- (0) & 149548 (1) & 149803 (1) & -- (0) & 19841 (7) & 15429 (9) & 8671 (14) & 8503 (15) & 1576 (15) & 1576 (15) \\
f13 & -- (0) & 6496 (15) & 6585 (15) & 6244 (15) & -- (0) & 6172 (15) & 6295 (15) & 3924 (15) & 4111 (15) & 10540 (1) & 2642 (15) \\
f14 & 9492 (14) & 4324 (15) & 4107 (15) & 3984 (15) & -- (0) & 3691 (15) & 3805 (15) & 2636 (15) & 2682 (15) & 848 (15) & 848 (15) \\
\midrule
f15 & -- (0) & -- (0) & -- (0) & -- (0) & -- (0) & -- (0) & -- (0) & -- (0) & -- (0) & -- (0) & 49051 (3) \\
f16 & -- (0) & -- (0) & -- (0) & 144609 (1) & -- (0) & -- (0) & -- (0) & -- (0) & -- (0) & -- (0) & 14170 (9) \\
f17 & 48455 (3) & 72031 (2) & 21434 (6) & 15356 (8) & -- (0) & 9939 (11) & 6229 (15) & 25024 (5) & 20176 (6) & -- (0) & 5335 (14) \\
f18 & -- (0) & -- (0) & 47105 (3) & 147261 (1) & -- (0) & 15433 (8) & 10385 (11) & 145954 (1) & -- (0) & -- (0) & 10377 (11) \\
f19 & -- (0) & -- (0) & -- (0) & -- (0) & -- (0) & -- (0) & -- (0) & -- (0) & -- (0) & -- (0) & -- (0) \\
\midrule
f20 & -- (0) & -- (0) & -- (0) & -- (0) & -- (0) & -- (0) & -- (0) & -- (0) & -- (0) & -- (0) & -- (0) \\
f21 & 149836 (1) & 22824 (5) & 67529 (2) & 42443 (3) & 148960 (1) & 67373 (2) & 42185 (3) & 29364 (4) & 41647 (3) & 4344 (1) & 17055 (6) \\
f22 & -- (0) & -- (0) & -- (0) & -- (0) & -- (0) & -- (0) & -- (0) & -- (0) & -- (0) & -- (0) & -- (0) \\
f23 & -- (0) & 149521 (1) & -- (0) & -- (0) & -- (0) & 149301 (1) & -- (0) & 147150 (1) & -- (0) & -- (0) & -- (0) \\
f24 & -- (0) & -- (0) & -- (0) & -- (0) & -- (0) & -- (0) & -- (0) & -- (0) & -- (0) & -- (0) & -- (0) \\
\bottomrule
\end{tabular}
\end{table}
\begin{table}[htbp]\centering\scriptsize
\setlength{\tabcolsep}{3.5pt}
\caption{Confirmatory test, $D=40$, standard start: expected running time to $10^{-8}$ divided by $D$ (evaluations spent by all 15 runs divided by the number of successful runs), with the number of successful runs in parentheses; -- if no run succeeded}
\label{S-tab:ert4}
\begin{tabular}{@{}lrrrrrrrrrrr@{}}
\toprule
 & EMNA & SN & MS & SN+MS & S+M-nofb & AMS & AMS+SN & AMaL. & AMaL.+SN & CMA & IPOP \\
\midrule
f1 & 5276 (15) & 5302 (15) & 5270 (15) & 5232 (15) & -- (0) & 5321 (15) & 5173 (15) & 2841 (15) & 3038 (15) & 132 (15) & 132 (15) \\
f2 & 7064 (15) & 7181 (15) & 7141 (15) & 7085 (15) & -- (0) & 7263 (15) & 7287 (15) & 3960 (15) & 4185 (15) & 1326 (15) & 1326 (15) \\
f3 & -- (0) & -- (0) & -- (0) & -- (0) & -- (0) & -- (0) & -- (0) & -- (0) & -- (0) & -- (0) & -- (0) \\
f4 & -- (0) & -- (0) & -- (0) & -- (0) & -- (0) & -- (0) & -- (0) & -- (0) & -- (0) & -- (0) & -- (0) \\
f5 & -- (0) & 3724 (12) & 1213 (15) & 495 (15) & 471 (15) & 290 (15) & 282 (15) & 321 (15) & 329 (15) & 14 (15) & 14 (15) \\
\midrule
f6 & -- (0) & -- (0) & -- (0) & -- (0) & -- (0) & -- (0) & -- (0) & -- (0) & -- (0) & 677 (15) & 677 (15) \\
f7 & 3563 (14) & 2804 (15) & 2782 (15) & 2717 (15) & -- (0) & 2815 (15) & 2851 (15) & 1931 (15) & 1824 (15) & -- (0) & 2617 (15) \\
f8 & -- (0) & -- (0) & -- (0) & -- (0) & -- (0) & -- (0) & -- (0) & -- (0) & -- (0) & 1851 (13) & 1860 (15) \\
f9 & -- (0) & -- (0) & -- (0) & -- (0) & -- (0) & -- (0) & -- (0) & -- (0) & -- (0) & 1607 (15) & 1607 (15) \\
\midrule
f10 & 7153 (15) & 7236 (15) & 7171 (15) & 7216 (15) & -- (0) & 7377 (15) & 7288 (15) & 4090 (15) & 4057 (15) & 1321 (15) & 1321 (15) \\
f11 & 5607 (15) & 5426 (15) & 5311 (15) & 5351 (15) & -- (0) & 5379 (15) & 5417 (15) & 3210 (15) & 3418 (15) & 541 (15) & 541 (15) \\
f12 & -- (0) & -- (0) & -- (0) & -- (0) & -- (0) & -- (0) & -- (0) & 74838 (2) & 149520 (1) & 1067 (15) & 1067 (15) \\
f13 & -- (0) & -- (0) & 149732 (1) & -- (0) & -- (0) & -- (0) & -- (0) & 6498 (15) & 6353 (15) & -- (0) & 8174 (13) \\
f14 & 5953 (15) & 5881 (15) & 5714 (15) & 5770 (15) & -- (0) & 5952 (15) & 5676 (15) & 3510 (15) & 3582 (15) & 1221 (15) & 1221 (15) \\
\midrule
f15 & 147345 (1) & -- (0) & -- (0) & -- (0) & -- (0) & -- (0) & -- (0) & -- (0) & -- (0) & -- (0) & 71128 (2) \\
f16 & -- (0) & -- (0) & -- (0) & -- (0) & -- (0) & -- (0) & -- (0) & -- (0) & -- (0) & -- (0) & 73537 (2) \\
f17 & -- (0) & -- (0) & 149713 (1) & 74818 (2) & -- (0) & 74807 (2) & 149549 (1) & 19046 (7) & 16569 (8) & -- (0) & 5437 (15) \\
f18 & -- (0) & -- (0) & -- (0) & -- (0) & -- (0) & -- (0) & -- (0) & 72847 (2) & 23230 (6) & -- (0) & 14665 (9) \\
f19 & -- (0) & -- (0) & -- (0) & -- (0) & -- (0) & -- (0) & -- (0) & -- (0) & -- (0) & -- (0) & -- (0) \\
\midrule
f20 & -- (0) & -- (0) & -- (0) & -- (0) & -- (0) & -- (0) & -- (0) & -- (0) & -- (0) & -- (0) & -- (0) \\
f21 & 142947 (1) & 142811 (1) & 142343 (1) & 142823 (1) & -- (0) & 143055 (1) & 142332 (1) & 141668 (1) & 141696 (1) & 3282 (1) & 40395 (3) \\
f22 & -- (0) & -- (0) & -- (0) & -- (0) & -- (0) & -- (0) & -- (0) & -- (0) & -- (0) & -- (0) & -- (0) \\
f23 & -- (0) & -- (0) & -- (0) & -- (0) & -- (0) & -- (0) & -- (0) & -- (0) & -- (0) & -- (0) & -- (0) \\
f24 & -- (0) & -- (0) & -- (0) & -- (0) & -- (0) & -- (0) & -- (0) & -- (0) & -- (0) & -- (0) & -- (0) \\
\bottomrule
\end{tabular}
\end{table}
\begin{table}[htbp]\centering\scriptsize
\setlength{\tabcolsep}{3.5pt}
\caption{Confirmatory test, $D=40$, distant start: expected running time to $10^{-8}$ divided by $D$ (evaluations spent by all 15 runs divided by the number of successful runs), with the number of successful runs in parentheses; -- if no run succeeded}
\label{S-tab:ert5}
\begin{tabular}{@{}lrrrrrrrrrrr@{}}
\toprule
 & EMNA & SN & MS & SN+MS & S+M-nofb & AMS & AMS+SN & AMaL. & AMaL.+SN & CMA & IPOP \\
\midrule
f1 & -- (0) & 8512 (15) & 10483 (13) & 7250 (15) & -- (0) & 6743 (15) & 7062 (15) & 4131 (15) & 3950 (15) & 174 (15) & 174 (15) \\
f2 & 13016 (11) & 11806 (12) & 11743 (12) & 10043 (14) & -- (0) & 9153 (15) & 10526 (13) & 5287 (15) & 5194 (15) & 1378 (15) & 1378 (15) \\
f3 & -- (0) & -- (0) & -- (0) & -- (0) & -- (0) & -- (0) & -- (0) & -- (0) & -- (0) & -- (0) & -- (0) \\
f4 & -- (0) & -- (0) & -- (0) & -- (0) & -- (0) & -- (0) & -- (0) & -- (0) & -- (0) & -- (0) & -- (0) \\
f5 & 147006 (1) & 28517 (4) & 5760 (15) & 863 (15) & 848 (15) & 411 (15) & 414 (15) & 521 (15) & 506 (15) & 23 (15) & 23 (15) \\
\midrule
f6 & -- (0) & -- (0) & -- (0) & -- (0) & -- (0) & -- (0) & -- (0) & -- (0) & -- (0) & 802 (15) & 802 (15) \\
f7 & 147682 (1) & 4683 (15) & 4873 (15) & 4715 (15) & -- (0) & 4398 (15) & 4376 (15) & 3177 (15) & 3130 (15) & -- (0) & 1962 (15) \\
f8 & -- (0) & -- (0) & -- (0) & -- (0) & -- (0) & -- (0) & -- (0) & -- (0) & -- (0) & 2032 (13) & 2050 (15) \\
f9 & -- (0) & -- (0) & -- (0) & -- (0) & -- (0) & -- (0) & -- (0) & -- (0) & -- (0) & 1652 (15) & 1652 (15) \\
\midrule
f10 & 36998 (4) & 12994 (11) & 24357 (6) & 11860 (12) & -- (0) & 9025 (15) & 9623 (14) & 5524 (15) & 5434 (15) & 1368 (15) & 1368 (15) \\
f11 & -- (0) & 28811 (5) & 28937 (5) & 9858 (13) & -- (0) & 6938 (15) & 7107 (15) & 4069 (15) & 4051 (15) & 583 (15) & 583 (15) \\
f12 & -- (0) & -- (0) & -- (0) & -- (0) & -- (0) & -- (0) & -- (0) & -- (0) & -- (0) & 1467 (15) & 1467 (15) \\
f13 & -- (0) & -- (0) & -- (0) & -- (0) & -- (0) & -- (0) & -- (0) & 7410 (15) & 7501 (15) & -- (0) & 10932 (10) \\
f14 & -- (0) & -- (0) & 49529 (3) & 8528 (15) & -- (0) & 8278 (15) & 8047 (15) & 5047 (15) & 5058 (15) & 1487 (15) & 1487 (15) \\
\midrule
f15 & -- (0) & -- (0) & -- (0) & -- (0) & -- (0) & -- (0) & -- (0) & -- (0) & -- (0) & -- (0) & -- (0) \\
f16 & -- (0) & -- (0) & -- (0) & -- (0) & -- (0) & -- (0) & -- (0) & -- (0) & -- (0) & -- (0) & 149490 (1) \\
f17 & -- (0) & -- (0) & -- (0) & -- (0) & -- (0) & -- (0) & -- (0) & 24206 (6) & 149406 (1) & -- (0) & 10928 (10) \\
f18 & -- (0) & -- (0) & -- (0) & -- (0) & -- (0) & -- (0) & -- (0) & -- (0) & -- (0) & -- (0) & -- (0) \\
f19 & -- (0) & -- (0) & -- (0) & -- (0) & -- (0) & -- (0) & -- (0) & -- (0) & -- (0) & -- (0) & -- (0) \\
\midrule
f20 & -- (0) & -- (0) & -- (0) & -- (0) & -- (0) & -- (0) & -- (0) & -- (0) & -- (0) & -- (0) & -- (0) \\
f21 & -- (0) & 70870 (2) & 147221 (1) & 144686 (1) & -- (0) & 69523 (2) & 144341 (1) & 143021 (1) & 67969 (2) & 3932 (1) & 28161 (4) \\
f22 & -- (0) & -- (0) & -- (0) & -- (0) & -- (0) & -- (0) & -- (0) & -- (0) & -- (0) & -- (0) & -- (0) \\
f23 & -- (0) & -- (0) & -- (0) & -- (0) & -- (0) & -- (0) & -- (0) & -- (0) & -- (0) & -- (0) & -- (0) \\
f24 & -- (0) & -- (0) & -- (0) & -- (0) & -- (0) & -- (0) & -- (0) & -- (0) & -- (0) & -- (0) & -- (0) \\
\bottomrule
\end{tabular}
\end{table}
\begin{table}[htbp]\centering\scriptsize
\caption{The gate of SN and of SN+MS on the confirmatory set, by function group: fraction of generations with the gate open ($\iota_{t}>0$) and mean immaturity, averaged over the runs of a group; rows ``open after $10^{-6}$'': mean number of generations with the gate open after the error first fell below $10^{-6}$, over the runs that reached $10^{-8}$}
\label{S-tab:gatestat}
\begin{tabular}{@{}llcccccccccccc@{}}
\toprule
 & & \multicolumn{2}{c}{$D=10$, std.} & \multicolumn{2}{c}{$D=10$, dist.} & \multicolumn{2}{c}{$D=20$, std.} & \multicolumn{2}{c}{$D=20$, dist.} & \multicolumn{2}{c}{$D=40$, std.} & \multicolumn{2}{c}{$D=40$, dist.} \\
 & Group & open & $\iota$ & open & $\iota$ & open & $\iota$ & open & $\iota$ & open & $\iota$ & open & $\iota$ \\
\midrule
SN & f1--f5 separable & 0.35 & 0.22 & 0.41 & 0.27 & 0.39 & 0.27 & 0.50 & 0.36 & 0.36 & 0.26 & 0.61 & 0.46 \\
 & f6--f9 moderate & 0.43 & 0.26 & 0.48 & 0.31 & 0.59 & 0.36 & 0.65 & 0.41 & 0.69 & 0.53 & 0.74 & 0.56 \\
 & f10--f14 ill-conditioned & 0.17 & 0.06 & 0.21 & 0.14 & 0.22 & 0.10 & 0.29 & 0.20 & 0.20 & 0.11 & 0.48 & 0.39 \\
 & f15--f19 multimodal & 0.39 & 0.21 & 0.39 & 0.23 & 0.44 & 0.19 & 0.54 & 0.30 & 0.53 & 0.20 & 0.76 & 0.49 \\
 & f20--f24 weak structure & 0.20 & 0.06 & 0.23 & 0.09 & 0.25 & 0.10 & 0.29 & 0.15 & 0.19 & 0.10 & 0.43 & 0.32 \\
 & open after $10^{-6}$ & \multicolumn{2}{c}{2.5} & \multicolumn{2}{c}{2.1} & \multicolumn{2}{c}{6.5} & \multicolumn{2}{c}{0.9} & \multicolumn{2}{c}{0.3} & \multicolumn{2}{c}{0.3} \\
\midrule
SN+MS & f1--f5 separable & 0.33 & 0.20 & 0.36 & 0.22 & 0.35 & 0.25 & 0.40 & 0.26 & 0.31 & 0.21 & 0.48 & 0.29 \\
 & f6--f9 moderate & 0.34 & 0.17 & 0.39 & 0.22 & 0.46 & 0.19 & 0.48 & 0.21 & 0.60 & 0.25 & 0.60 & 0.25 \\
 & f10--f14 ill-conditioned & 0.13 & 0.04 & 0.18 & 0.10 & 0.16 & 0.04 & 0.24 & 0.15 & 0.17 & 0.05 & 0.25 & 0.14 \\
 & f15--f19 multimodal & 0.34 & 0.16 & 0.35 & 0.19 & 0.35 & 0.13 & 0.40 & 0.17 & 0.35 & 0.07 & 0.58 & 0.24 \\
 & f20--f24 weak structure & 0.20 & 0.05 & 0.23 & 0.08 & 0.22 & 0.08 & 0.23 & 0.08 & 0.16 & 0.04 & 0.27 & 0.10 \\
 & open after $10^{-6}$ & \multicolumn{2}{c}{7.9} & \multicolumn{2}{c}{7.6} & \multicolumn{2}{c}{5.3} & \multicolumn{2}{c}{1.3} & \multicolumn{2}{c}{0.8} & \multicolumn{2}{c}{0.3} \\
\bottomrule
\end{tabular}
\end{table}

\subsection{Supplementary runs with $D=80$}
\label{S-sec:d80}
The functions of the standard BBOB suite cannot be built for $D=80$, because the code of COCO that generates their rotation matrices is limited to $D^{2}<2000$; the runs with $D=80$ therefore use the large-scale BBOB suite of COCO (\texttt{bbob-largescale}) \cite{varelas2020}, whose 24 functions correspond to those of the standard suite with the full rotations replaced by permuted block-diagonal matrices, with instances 1--5. All other settings are those of the confirmatory test (two starting conditions, budget $10^{4}D$, unconstrained, exact counting, floor $10^{-8}$); $N=2163$, $M=757$ and $\kappa_{0}=1.42$. Protocol B4 added EMNA+AVS, SN, MS and SN+MS to the three baselines that had been run on the same instances and with the same random streams for the earlier version (Section \ref{S-sec:elong}). The results are descriptive: five instances, no correction for multiplicity, and an instance set that had been used before, although not for any choice concerning these algorithms.

Table \ref{S-tab:d80a} gives the mean performance and the numbers of solved runs and functions, and Table \ref{S-tab:d80b} the paired comparisons. With the distant start the skewness improves the anytime performance of EMNA+AVS (HL $-$1.05, better on 22 of the 24 functions) and that of the algorithm with the mean shift (SN+MS against MS: $-$2.62, better on all 24 functions). With the standard start the two differences are $-$0.04 and 0.00, and their intervals contain zero. SN+MS is behind EMNA+AVS+AMS with the distant start (+0.96, worse on all 24 functions), whereas with the standard start the interval of their difference contains zero; it is behind AMaLGaM and CMA-ES with both starts. At this dimension the five algorithms that estimate a full covariance matrix from the current generation (EMNA+AVS, SN, MS, SN+MS and EMNA+AVS+AMS) solve at most 11 of the 120 runs of a setting, against 31 or more for AMaLGaM and 48 or more for CMA-ES.

\begin{table}[htbp]\centering\footnotesize
\caption{$D=80$, large-scale BBOB suite, instances 1--5: mean $y_{\mathrm{any}}$ over the 24 functions, runs solved to $10^{-8}$ (of 120) and functions solved in all five runs}
\label{S-tab:d80a}
\begin{tabular}{@{}lcccccc@{}}
\toprule
 & \multicolumn{3}{c}{standard start} & \multicolumn{3}{c}{distant start} \\
\cmidrule(lr){2-4}\cmidrule(l){5-7}
Algorithm & $y_{\mathrm{any}}$ & runs & functions & $y_{\mathrm{any}}$ & runs & functions \\
\midrule
EMNA+AVS & 1.12 & 6 & 1 & 8.29 & 0 & 0 \\
SN & 0.79 & 10 & 1 & 7.11 & 1 & 0 \\
MS & 0.77 & 10 & 1 & 7.42 & 0 & 0 \\
SN+MS & 0.62 & 11 & 2 & 4.67 & 6 & 0 \\
EMNA+AVS+AMS & 0.60 & 11 & 2 & 3.45 & 8 & 1 \\
AMaLGaM & $-$0.70 & 36 & 7 & 2.02 & 31 & 4 \\
CMA-ES & $-$2.31 & 50 & 9 & $-$2.05 & 48 & 8 \\
\bottomrule
\end{tabular}
\end{table}

\begin{table}[htbp]\centering\footnotesize
\caption{$D=80$: Hodges--Lehmann estimate of the difference in $y_{\mathrm{any}}$ (first minus second) over the 24 functions, with 95\% bootstrap interval, uncorrected Wilcoxon $p$, and functions on which the first algorithm is better/worse}
\label{S-tab:d80b}
\begin{tabular}{@{}lcc@{}}
\toprule
Comparison & standard start & distant start \\
\midrule
SN vs. EMNA+AVS & $-$0.04 [$-$0.13, +0.02], $p=0.16$, 14/10 & $-$1.05 [$-$1.62, $-$0.68], $p=2\times10^{-5}$, 22/0 \\
MS vs. EMNA+AVS & $-$0.10 [$-$0.18, $-$0.03], $p=0.007$, 16/8 & $-$0.86 [$-$1.19, $-$0.32], $p=1\times10^{-7}$, 24/0 \\
SN+MS vs. MS & 0.00 [$-$0.04, +0.03], $p=0.94$, 12/12 & $-$2.62 [$-$3.36, $-$1.95], $p=1\times10^{-7}$, 24/0 \\
SN+MS vs. SN & $-$0.03 [$-$0.08, 0.00], $p=0.07$, 17/7 & $-$2.20 [$-$3.08, $-$1.56], $p=1\times10^{-7}$, 24/0 \\
SN+MS vs. EMNA+AVS & $-$0.07 [$-$0.22, 0.00], $p=0.06$, 15/9 & $-$3.43 [$-$4.14, $-$2.90], $p=1\times10^{-7}$, 24/0 \\
SN+MS vs. EMNA+AVS+AMS & +0.02 [$-$0.03, +0.09], $p=0.47$, 12/12 & +0.96 [+0.58, +1.67], $p=1\times10^{-7}$, 0/24 \\
SN+MS vs. AMaLGaM & +1.43 [+0.36, +1.80], $p=2\times10^{-7}$, 1/23 & +2.60 [+1.64, +3.32], $p=1\times10^{-7}$, 0/24 \\
SN+MS vs. CMA-ES & +2.98 [+0.53, +4.27], $p=0.0001$, 3/21 & +6.15 [+4.25, +8.37], $p=1\times10^{-7}$, 0/24 \\
\bottomrule
\end{tabular}
\end{table}

\section{The held-out test of the earlier version, re-analysed}
\label{S-sec:heldout}

The earlier version of this work tested its algorithms on BBOB instances 6--15 ($D=10$ and 20, ten instances, the two starting conditions) with a protocol written before the runs. The runs used the earlier implementation (samples clipped to the box, elitist row re-evaluated, fixed thresholds, decay 0.9 for the baselines, AMaLGaM without the smoothing of the shift and of the covariance). ASN+AVS is the gated configuration SN in that implementation (thresholds $(2,6)$, decay 0.9); MG-A and MG-B are the configurations with a gated mean shift and a gated elongation of Section \ref{S-sec:elong} (thresholds $(2.36,6.36)$, decay 0.8), which use no skewness. The analysis of the earlier version used unpaired tests although the runs of a function share their random numbers; Tables \ref{S-tab:old1} and \ref{S-tab:old2} give the paired analysis of Section \ref{sec:setup} of the manuscript (blocks = functions; descriptive, uncorrected). The first row of Table \ref{S-tab:old2} is the earliest held-out evidence on the skewness alone: with the distant start ASN+AVS is better than EMNA+AVS by HL $-$1.40 for $D=10$ and $-$3.02 for $D=20$, and with the standard start the difference is small, as on the two later sets of instances (16--30, Section \ref{S-sec:elong}; 31--45, the confirmatory test of the manuscript).

\begin{table}[htbp]\centering\footnotesize
\caption{Held-out test of the earlier version (instances 6--15): functions solved to $10^{-8}$ in all ten runs, mean $y_{\mathrm{any}}$ over the 24 functions, and mean rank over the 96 function--setting pairs}
\label{S-tab:old1}
\begin{tabular}{@{}lrrrrrrrrc@{}}
\toprule
 & \multicolumn{2}{c}{$D=10$, std.} & \multicolumn{2}{c}{$D=10$, dist.} & \multicolumn{2}{c}{$D=20$, std.} & \multicolumn{2}{c}{$D=20$, dist.} & \\
Algorithm & sol. & $y$ & sol. & $y$ & sol. & $y$ & sol. & $y$ & rank \\
\midrule
EMNA+AVS & 9 & $-$2.90 & 7 & 0.04 & 9 & $-$1.92 & 4 & 3.90 & 5.46 \\
ASN+AVS & 10 & $-$2.97 & 8 & $-$1.48 & 9 & $-$2.03 & 7 & 0.68 & 4.81 \\
EMNA+AVS+AMS & 12 & $-$3.87 & 10 & $-$3.17 & 11 & $-$2.76 & 10 & $-$1.78 & 2.96 \\
AMaLGaM & 12 & $-$3.72 & 11 & $-$3.20 & 11 & $-$2.98 & 11 & $-$2.10 & 2.23 \\
MG-A & 12 & $-$3.84 & 11 & $-$3.06 & 11 & $-$2.89 & 10 & $-$1.57 & 2.86 \\
MG-B & 11 & $-$3.87 & 11 & $-$3.13 & 12 & $-$2.96 & 10 & $-$1.80 & 2.68 \\
\bottomrule
\end{tabular}
\end{table}

\begin{table}[htbp]\centering\footnotesize
\caption{Held-out test of the earlier version, paired comparisons: Hodges--Lehmann estimate of the difference in $y_{\mathrm{any}}$ (first minus second) over the 24 functions, with the number of functions on which the first algorithm is better/worse; bold: Wilcoxon signed-rank $p<0.05$ (uncorrected)}
\label{S-tab:old2}
\begin{tabular}{@{}lcccc@{}}
\toprule
Comparison & $D=10$, std. & $D=10$, dist. & $D=20$, std. & $D=20$, dist. \\
\midrule
ASN+AVS vs. EMNA+AVS & $-$0.01 (13/11) & \textbf{$-$1.40} (22/2) & \textbf{$-$0.09} (18/6) & \textbf{$-$3.02} (24/0) \\
MG-A vs. ASN+AVS & \textbf{$-$0.10} (23/1) & \textbf{$-$1.07} (24/0) & \textbf{$-$0.48} (19/5) & \textbf{$-$2.01} (23/1) \\
MG-A vs. EMNA+AVS & \textbf{$-$0.26} (21/3) & \textbf{$-$2.46} (23/1) & \textbf{$-$0.55} (21/3) & \textbf{$-$5.34} (24/0) \\
MG-B vs. EMNA+AVS & \textbf{$-$0.48} (21/3) & \textbf{$-$2.37} (23/1) & \textbf{$-$0.72} (18/6) & \textbf{$-$5.50} (24/0) \\
MG-A vs. EMNA+AVS+AMS & +0.06 (10/14) & +0.01 (14/10) & +0.01 (12/12) & $-$0.01 (15/9) \\
MG-B vs. EMNA+AVS+AMS & +0.01 (12/12) & $-$0.07 (17/7) & $-$0.01 (12/12) & $-$0.02 (14/10) \\
MG-A vs. AMaLGaM & +0.08 (9/15) & +0.18 (7/17) & \textbf{+0.34} (7/17) & \textbf{+0.39} (6/18) \\
MG-B vs. AMaLGaM & $-$0.02 (11/13) & +0.09 (10/14) & +0.21 (7/17) & \textbf{+0.33} (9/15) \\
EMNA+AVS+AMS vs. AMaLGaM & +0.03 (10/14) & +0.13 (8/15) & +0.21 (7/16) & \textbf{+0.34} (6/18) \\
\bottomrule
\end{tabular}
\end{table}

\section{Checks of the implementation and diagnostics}
\label{S-sec:checks}

\emph{Moments of the sampler.} The program contains a test (mode \texttt{unit}) that draws $2\times10^{6}$ rows of the sampler for a random non-diagonal lower-triangular $L$ and a random unit vector $\mathbf{w}$, with $s=2$ and $\sigma_{1}=1$, and compares the sample mean, the full sample covariance matrix and the third cumulant along $\mathbf{w}$ with the theory. Table \ref{S-tab:unit} gives the results for $D=5$ and $D=12$ (different seeds). The first six cases are the sampler (\ref{S-eq:mg}), which for $\sigma_{1}=0$ and $\iota=1$ is the sampler (\ref{eq:sample}) of the manuscript, including the special cases $\iota=0$ and $\rho_{\max}=0$, and agree with Theorem \ref{S-thm:joint} within sampling error. The next five are the sampler with elongation of Section \ref{S-sec:elong}, and the last one is the joint sampler printed in the earlier version, whose covariance was not the one claimed for it.

\begin{table}[htbp]\centering\scriptsize
\caption{Moments of the sampler along $\mathbf{w}$ in whitened coordinates: theory and measurement with $2\times10^{6}$ samples ($D=5$ and $D=12$); largest error of the mean (in standard deviations) and of the covariance matrix (relative to $\sqrt{C_{ii}C_{jj}}$) over all entries}
\label{S-tab:unit}
\begin{tabular}{@{}ccclcccccccc@{}}
\toprule
 & & & & \multicolumn{3}{c}{variance along $\mathbf{w}$} & \multicolumn{3}{c}{third cumulant along $\mathbf{w}$} & \multicolumn{2}{c}{largest error} \\
\cmidrule(lr){5-7}\cmidrule(lr){8-10}\cmidrule(l){11-12}
$\iota$ & $e$ & $\rho_{\max}$ & sampler & theory & $D=5$ & $D=12$ & theory & $D=5$ & $D=12$ & mean & covariance \\
\midrule
0.00 & 0 & 0.00 & manuscript & 1.0000 & 1.0011 & 0.9970 & 0.0000 & $-$0.0007 & 0.0004 & 0.0015 & 0.0020 \\
1.00 & 0 & 0.95 & manuscript & 1.0000 & 0.9998 & 0.9995 & 0.8533 & 0.8549 & 0.8515 & 0.0012 & 0.0020 \\
0.50 & 0 & 0.95 & manuscript & 1.0000 & 1.0001 & 1.0008 & 0.1067 & 0.1088 & 0.1062 & 0.0008 & 0.0018 \\
1.00 & 0 & 0.50 & manuscript & 1.0000 & 1.0005 & 1.0011 & 0.1244 & 0.1234 & 0.1245 & 0.0020 & 0.0020 \\
0.30 & 0 & 0.80 & manuscript & 1.0000 & 1.0005 & 1.0000 & 0.0138 & 0.0115 & 0.0107 & 0.0018 & 0.0019 \\
1.00 & 0 & 0.99 & manuscript & 1.0000 & 1.0017 & 0.9996 & 0.9657 & 0.9680 & 0.9610 & 0.0014 & 0.0018 \\
\midrule
1.00 & 4 & 0.00 & Section \ref{S-sec:elong} & 25.0000 & 24.9959 & 25.0130 & 0.0000 & $-$0.0039 & 0.0722 & 0.0015 & 0.0017 \\
1.00 & 4 & 0.95 & Section \ref{S-sec:elong} & 25.0000 & 25.0327 & 25.0149 & 106.6651 & 106.8886 & 106.7079 & 0.0034 & 0.0019 \\
0.50 & 4 & 0.95 & Section \ref{S-sec:elong} & 9.0000 & 8.9818 & 8.9938 & 2.8800 & 2.7893 & 2.8912 & 0.0041 & 0.0022 \\
1.00 & 2 & 0.50 & Section \ref{S-sec:elong} & 9.0000 & 9.0103 & 8.9871 & 3.3590 & 3.3171 & 3.3336 & 0.0029 & 0.0014 \\
0.30 & 8 & 0.80 & Section \ref{S-sec:elong} & 11.5600 & 11.5589 & 11.5622 & 0.5408 & 0.5219 & 0.4470 & 0.0031 & 0.0025 \\
\midrule
1.00 & 4 & 0.95 & earlier version & 3.3400 & 3.3355 & 3.3362 & 0.8533 & 0.8450 & 0.8563 & 0.0019 & 0.8552 \\
\bottomrule
\end{tabular}
\par\vspace{2pt}\begin{minipage}{0.92\textwidth}\scriptsize $e$ is the elongation of Section \ref{S-sec:elong}; $e=0$ is the sampler (\ref{S-eq:mg}). Theory: variance $a^{2}$ and third cumulant $a^{3}K_{3}\delta^{3}$ with $a=1+e\iota$ and $\delta=\iota\rho_{\max}D_{\lim}$; for the sampler of the earlier version, variance $a^{2}(1-(1-b^{2})\delta^{2})+(1-b^{2})\delta^{2}$ and third cumulant $K_{3}\delta^{3}$, and its covariance error is taken against the covariance claimed for it.\end{minipage}
\end{table}

\emph{Slope constants for other truncation ratios.} Table \ref{S-tab:const} extends Table \ref{tab:slope} of the manuscript to $\tau=0.3$ and 0.5 (numerical integration of (\ref{eq:lamc}), program \nolinkurl{slope_theory.py}). The last column is the skewness of the solutions selected from the sampler; for the Gaussian sampler its magnitude already exceeds the supremum 0.9953 of the skew-normal family.

\begin{table}[htbp]\centering\footnotesize
\caption{One-generation constants of the skew-normal sampler on a linear function for three truncation ratios}
\label{S-tab:const}
\begin{tabular}{@{}cccccccc@{}}
\toprule
$\tau$ & $\rho_{\max}$ & skewness of the sampler & $\lambda$ & $c$ & $\lambda/(1-\sqrt{c})$ & $1/c$ & skewness of the selected solutions \\
\midrule
0.3 & 0 & 0.000 & 1.1590 & 0.2645 & 2.386 & 3.780 & 1.177 \\
0.3 & 0.5 & 0.124 & 1.1700 & 0.3009 & 2.592 & 3.323 & 1.251 \\
0.3 & 0.75 & 0.420 & 1.1961 & 0.3885 & 3.175 & 2.574 & 1.319 \\
0.3 & 0.95 & 0.853 & 1.2398 & 0.5030 & 4.264 & 1.988 & 1.326 \\
0.3 & 0.99 & 0.966 & 1.2518 & 0.5295 & 4.597 & 1.888 & 1.326 \\
\midrule
0.35 & 0 & 0.000 & 1.0583 & 0.2878 & 2.283 & 3.474 & 1.131 \\
0.35 & 0.5 & 0.124 & 1.0650 & 0.3243 & 2.474 & 3.083 & 1.208 \\
0.35 & 0.75 & 0.420 & 1.0805 & 0.4134 & 3.027 & 2.419 & 1.288 \\
0.35 & 0.95 & 0.853 & 1.1091 & 0.5341 & 4.120 & 1.872 & 1.298 \\
0.35 & 0.99 & 0.966 & 1.1177 & 0.5622 & 4.467 & 1.779 & 1.298 \\
\midrule
0.5 & 0 & 0.000 & 0.7979 & 0.3634 & 2.009 & 2.752 & 0.995 \\
0.5 & 0.5 & 0.124 & 0.7962 & 0.3993 & 2.163 & 2.505 & 1.077 \\
0.5 & 0.75 & 0.420 & 0.7906 & 0.4893 & 2.631 & 2.044 & 1.192 \\
0.5 & 0.95 & 0.853 & 0.7847 & 0.6237 & 3.732 & 1.603 & 1.223 \\
0.5 & 0.99 & 0.966 & 0.7849 & 0.6565 & 4.137 & 1.523 & 1.223 \\
\bottomrule
\end{tabular}
\par\vspace{2pt}\begin{minipage}{0.85\textwidth}\scriptsize The skewness of the selected solutions is given in the direction of improvement.\end{minipage}
\end{table}

\emph{AMS on a slope.} The fixed point of Section \ref{sec:amsmix} of the manuscript (program \nolinkurl{ams_theory.py}) is $d=3.6327$, $\Lambda_{\mathrm{A}}=2.5000$, $\rho_{\mathrm{A}}=1.8945$ for the Gaussian component and $d=3.7526$, $\Lambda_{\mathrm{A}}=2.6349$, $\rho_{\mathrm{A}}=1.9720$ with the skew-normal component at $\rho_{\max}=0.95$; the offspring distribution then has skewness 0.908 and 1.100 and its variance along the direction is 2.905 and 3.033 times that of the model. The moments of the mixture and the two constants were checked against $4\times10^{6}$ samples of the mixture, and the constants against runs of the program (Table \ref{tab:slope}).

\emph{Null law of the movement statistic.} Table \ref{S-tab:null} gives the stationary quantiles of $m_{t}$ under the ideal null model of Theorem \ref{S-thm:calib}; the thresholds of the gate are its 99\% column. The quantiles were computed with $10^{6}$ steps ($2\times10^{5}$ for $D=80$) of the recursion (\ref{S-eq:path}) with the increments (\ref{S-eq:znull}), for $c_{p}=0.3$, $N=\lfloor17+3D^{1.5}\rfloor$ and $M=\lfloor0.35N\rfloor$. The sampled mean of $m_{t}$ is 1.000--1.002, in agreement with part (c) of the theorem, and the sampled $\mathbb{E}\|\mathbf{z}\|^{2}$ agrees with part (a) to within 0.1\%.

\begin{table}[htbp]\centering\small
\caption{Stationary quantiles of $m_{t}$ under the ideal null model (exact sampling, $c_{p}=0.3$)}
\label{S-tab:null}
\begin{tabular}{@{}ccccccc@{}}
\toprule
$D$ & $M$ & mean & 90\% & 95\% & 99\% ($\kappa_{0}$) & 99.9\% \\
\midrule
5 & 17 & 1.000 & 1.92 & 2.38 & 3.54 & 5.53 \\
10 & 38 & 1.001 & 1.65 & 1.92 & 2.53 & 3.40 \\
20 & 99 & 1.000 & 1.44 & 1.61 & 1.95 & 2.41 \\
30 & 178 & 1.000 & 1.36 & 1.48 & 1.74 & 2.07 \\
40 & 271 & 1.000 & 1.30 & 1.41 & 1.62 & 1.88 \\
80 & 757 & 1.002 & 1.22 & 1.28 & 1.42 & 1.58 \\
\bottomrule
\end{tabular}
\end{table}

\emph{The statistic inside the algorithm.} Table \ref{S-tab:diag} compares the ideal null model with the algorithm, with the corrections off. On a neutral function without elitism the algorithm reproduces the null law: the gate is open in 1.0\% of the generations for $D=10$ and 1.2\% for $D=20$, against the nominal 1\%. The elitist row raises this slightly (1.2\% and 1.3\%). Selection matters more: on a sphere centered on the initial mean, where the population converges in place, the rate is 4.5\% for $D=10$ and 1.9\% for $D=20$ (5.3\% and 1.7\% without elitism), because the solutions kept by selection near an optimum are not a Gaussian sample. The mean immaturity nevertheless stays at 0.010 and 0.001, so the corrections are weak when they act there. Within 55 generations of the neutral function the gate opens at least once in 37\% of the runs for $D=10$.

\begin{table}[htbp]\centering\footnotesize
\caption{The statistic inside the algorithm (generations 6--60, 300 runs): mean of $m_{t}$, fraction of generations with the gate open, mean immaturity}
\label{S-tab:diag}
\begin{tabular}{@{}l@{\hspace{3pt}}ccc@{\hspace{6pt}}ccc@{}}
\toprule
 & \multicolumn{3}{c}{$D=10$} & \multicolumn{3}{c}{$D=20$} \\
\cmidrule(lr){2-4}\cmidrule(l){5-7}
Case & $m$ & open & $\iota$ & $m$ & open & $\iota$ \\
\midrule
ideal null model (exact) & 1.00 & 1.0\% & 0.0009 & 1.00 & 1.0\% & 0.0005 \\
neutral function, no elitism & 0.99 & 1.0\% & 0.0009 & 1.01 & 1.2\% & 0.0006 \\
neutral function, elitism & 1.01 & 1.2\% & 0.0012 & 1.01 & 1.3\% & 0.0006 \\
centered sphere, no elitism & 1.18 & 5.3\% & 0.0127 & 0.99 & 1.7\% & 0.0012 \\
centered sphere, elitism & 1.16 & 4.5\% & 0.0101 & 0.99 & 1.9\% & 0.0013 \\
\bottomrule
\end{tabular}
\par\vspace{2pt}\begin{minipage}{0.62\textwidth}\scriptsize Corrections off (EMNA; on the sphere with AVS), start $\mathcal{N}(\mathbf{0},I)$, thresholds of Table \ref{S-tab:null}. On the neutral function all values are equal and ties are broken by row index, so with elitism the elitist row is always selected.\end{minipage}
\end{table}

\emph{The relative ridge.} In a long neutral run the mean of $m_{t}$ sinks, to 0.87 in generations 201--300 for $D=10$. The cause is not the statistic: under neutral selection the condition number of the covariance grows geometrically, and once it approaches $10^{14}$ the relative ridge $\epsilon I$ inflates the smallest variances and shortens the whitened increments. With $\epsilon=0$ the mean stays at 1.04 in the same window. The sinking mean matters for calibration, since quantiles taken from such a window are too low, and is the reason why Table \ref{S-tab:null} is computed from the exact law and not from runs of the algorithm; it does not occur on functions with selection pressure, where the condition number stays bounded.

\emph{CMA-ES against the reference implementation.} Table \ref{S-tab:cma} compares our implementation of CMA-ES with the reference implementation pycma 4.6.0 on BBOB instances 16--30 for $D=10$, under the same rules: unconstrained, initial mean uniform in $[-4,4]^{D}$, initial step size 2, budget $10^{4}D$, target $10^{-8}$, no restarts. The two programs use different random numbers, so only distributions can be compared; each was therefore run with five seeds per instance, 75 runs per function, and the evaluations are counted up to the first one at which the target is reached. For our implementation the first seed is the random stream of protocol B, and its 210 runs are identical to the stored ones. On the ten functions that both solve in at least half of the runs, our implementation needs more evaluations than pycma on nine; the difference of the medians is +5\% in the median and at most +16\% (f12). The comparison is a repetition made for the present version (Section \ref{S-sec:prot}, item 14).

\begin{table}[htbp]\centering\footnotesize
\caption{CMA-ES: runs solved (of 75) and median evaluations to $10^{-8}$ of the solved runs, our implementation and pycma; the last column is the difference of our median from that of pycma, for the functions that both solve in at least half of the runs}
\label{S-tab:cma}
\begin{tabular}{@{}lccccc@{}}
\toprule
 & \multicolumn{2}{c}{our implementation} & \multicolumn{2}{c}{pycma 4.6.0} & \\
Function & solved & median evaluations & solved & median evaluations & difference \\
\midrule
f1 & 75 & 1487 & 75 & 1474 & +1\% \\
f2 & 75 & 4336 & 75 & 4123 & +5\% \\
f3 & 0 & -- & 0 & -- & -- \\
f5 & 75 & 131 & 75 & 132 & $-$1\% \\
f6 & 75 & 4132 & 75 & 4093 & +1\% \\
f7 & 7 & 1366 & 1 & 1594 & -- \\
f8 & 68 & 5527 & 68 & 5202 & +6\% \\
f10 & 75 & 4373 & 75 & 4186 & +4\% \\
f11 & 75 & 3217 & 75 & 3053 & +5\% \\
f12 & 75 & 11750 & 75 & 10126 & +16\% \\
f13 & 51 & 7758 & 47 & 7356 & +5\% \\
f14 & 75 & 4111 & 75 & 3906 & +5\% \\
f15 & 0 & -- & 0 & -- & -- \\
f20 & 0 & -- & 0 & -- & -- \\
\bottomrule
\end{tabular}
\end{table}

\emph{Feedback correction on a neutral function.} Table \ref{S-tab:fbn} gives the experiment of Section \ref{S-sec:inside}: all function values are equal, ties are broken by row index, elitism is on, the start is $\mathcal{N}(\mathbf{0},I)$, and 300 generations are run (200 runs for $D=10$ and 20, 100 for $D=40$; the first five generations are discarded).

\begin{table}[htbp]\centering\footnotesize
\caption{Neutral function: fraction of generations with the gate open ($\iota_{t}>0$) and fully open ($\iota_{t}=1$), mean length of an opening in generations, and the longest opening observed (of 295 generations)}
\label{S-tab:fbn}
\begin{tabular}{@{}lcccccccccccc@{}}
\toprule
 & \multicolumn{4}{c}{$D=10$} & \multicolumn{4}{c}{$D=20$} & \multicolumn{4}{c}{$D=40$} \\
\cmidrule(lr){2-5}\cmidrule(lr){6-9}\cmidrule(l){10-13}
Algorithm & open & full & mean & longest & open & full & mean & longest & open & full & mean & longest \\
\midrule
SN & 0.64\% & 0.00\% & 1.2 & 6 & 0.72\% & 0.00\% & 1.2 & 5 & 0.82\% & 0.00\% & 1.1 & 3 \\
MS & 0.64\% & 0.00\% & 1.2 & 4 & 0.72\% & 0.00\% & 1.2 & 5 & 0.83\% & 0.00\% & 1.2 & 4 \\
SN+MS & 0.63\% & 0.00\% & 1.2 & 6 & 0.73\% & 0.00\% & 1.2 & 5 & 0.82\% & 0.00\% & 1.1 & 3 \\
MS without feedback correction & 6.26\% & 5.45\% & 12.8 & 294 & 3.78\% & 2.92\% & 6.4 & 291 & 3.00\% & 1.99\% & 4.2 & 248 \\
SN+MS without feedback correction & 6.68\% & 5.84\% & 13.5 & 294 & 4.20\% & 3.33\% & 7.0 & 291 & 3.00\% & 2.00\% & 4.1 & 248 \\
\bottomrule
\end{tabular}
\end{table}

\emph{Paired rotations and the whitening factor.} Table \ref{S-tab:rot} gives the experiment of Section \ref{S-sec:inside}: SN+MS with the Cholesky factor and with the symmetric square root of the covariance, on three functions with and without a random rotation (a new rotation for every pair, the same random numbers within a pair), start $\mathcal{N}(\mathbf{0},I)$, optimum at $30\cdot\mathbf{1}$, unconstrained, budget $10^{4}D$. The sphere is invariant under rotation, so its two columns coincide.

\begin{table}[htbp]\centering\footnotesize
\caption{Paired rotations: median evaluations to $10^{-8}$ over the solved runs (runs solved of 15)}
\label{S-tab:rot}
\begin{tabular}{@{}lccccc@{}}
\toprule
 & & \multicolumn{2}{c}{Cholesky factor} & \multicolumn{2}{c}{symmetric square root} \\
Function & $D$ & unrotated & rotated & unrotated & rotated \\
\midrule
sphere & 10 & 16697 (15) & 16697 (15) & 16369 (15) & 16369 (15) \\
sphere & 20 & 63251 (15) & 63251 (15) & 68380 (15) & 68380 (15) \\
ellipsoid & 10 & 22343 (15) & 22465 (15) & 23098 (15) & 21033 (15) \\
ellipsoid & 20 & 88783 (15) & 89384 (15) & 89869 (15) & 88651 (15) \\
rosenbrock & 10 & 40216 (12) & 35113 (15) & 35805 (13) & 35479 (13) \\
rosenbrock & 20 & 153798 (15) & 153716 (13) & 158026 (14) & 151523 (15) \\
\bottomrule
\end{tabular}
\end{table}

\emph{Direction on a slope.} Table \ref{S-tab:dir} gives the angle between the whitened direction $\mathbf{w}_{t}$ and the whitened downhill direction $-A_{t}^{T}\mathbf{v}/\|A_{t}^{T}\mathbf{v}\|$ on the linear function $f(\mathbf{x})=x_{1}$, where $A_{t}$ is the factor used for whitening, in generations 3--12 of 20 runs with the population size of the algorithm, AVS on, no elitism, start $\mathcal{N}(\mathbf{0},I)$; in all of these generations $\iota_{t}=1$. The first row is the ungated algorithm of the manuscript; for $D=20$ and 40 its runs are identical to those of SN, whose gate is fully open from the second generation on.

\begin{table}[htbp]\centering\footnotesize
\caption{Angle in degrees between the whitened direction and the whitened downhill direction on a linear function: median (10\%--90\% quantiles) over 200 generations}
\label{S-tab:dir}
\begin{tabular}{@{}lccc@{}}
\toprule
Configuration & $D=10$, $N=111$ & $D=20$, $N=285$ & $D=40$, $N=775$ \\
\midrule
SN-on (Cholesky factor) & 29.8 (21.2--38.5) & 27.0 (21.9--32.1) & 22.9 (19.4--26.1) \\
SN (Cholesky factor) & 29.7 (21.2--38.5) & 27.0 (21.9--32.1) & 22.9 (19.4--26.1) \\
SN+MS (Cholesky factor) & 29.7 (21.5--38.9) & 27.0 (21.8--31.5) & 22.7 (19.9--26.1) \\
SN+MS (symmetric square root) & 29.8 (21.0--37.0) & 27.1 (22.1--32.0) & 22.9 (20.1--26.3) \\
\bottomrule
\end{tabular}
\end{table}

\emph{Plateau.} Table \ref{S-tab:plat} gives the plateau experiment of Section \ref{sec:diag} of the manuscript: $f(\mathbf{x})=\|\mathbf{x}\|^{2}$ for $\|\mathbf{x}\|\le1$ and $f(\mathbf{x})=2$ otherwise, $D=10$, start $\mathcal{N}(c\,\mathbf{e}_{1},0.25I)$ for the EDAs and mean $c\,\mathbf{e}_{1}$ with step size 0.5 for CMA-ES, unconstrained, budget $10^{5}$, 15 runs. Without restarts no algorithm crosses the plateau reliably; CMA-ES stops when all values are equal, and with restarts it reaches the ball from $c=2$ through its growing population.

\begin{table}[htbp]\centering\footnotesize
\caption{Plateau: runs that reached the ball / runs that reached $10^{-8}$ (of 15); in parentheses the fraction of generations with the gate open (gated configurations)}
\label{S-tab:plat}
\begin{tabular}{@{}lcc@{}}
\toprule
Algorithm & $c=2$ & $c=4$ \\
\midrule
EMNA+AVS & 1 / 1 & 0 / 0 \\
SN-on & 1 / 1 & 0 / 0 \\
SN & 1 / 1 (4.6\%) & 0 / 0 (2.9\%) \\
SN+MS & 1 / 1 (3.8\%) & 0 / 0 (2.9\%) \\
AMaLGaM & 1 / 0 & 0 / 0 \\
CMA-ES & 0 / 0 & 0 / 0 \\
IPOP-CMA-ES & 15 / 10 & 0 / 0 \\
\bottomrule
\end{tabular}
\end{table}

\begin{figure}[htbp]
\centering
\includegraphics[width=0.9\textwidth]{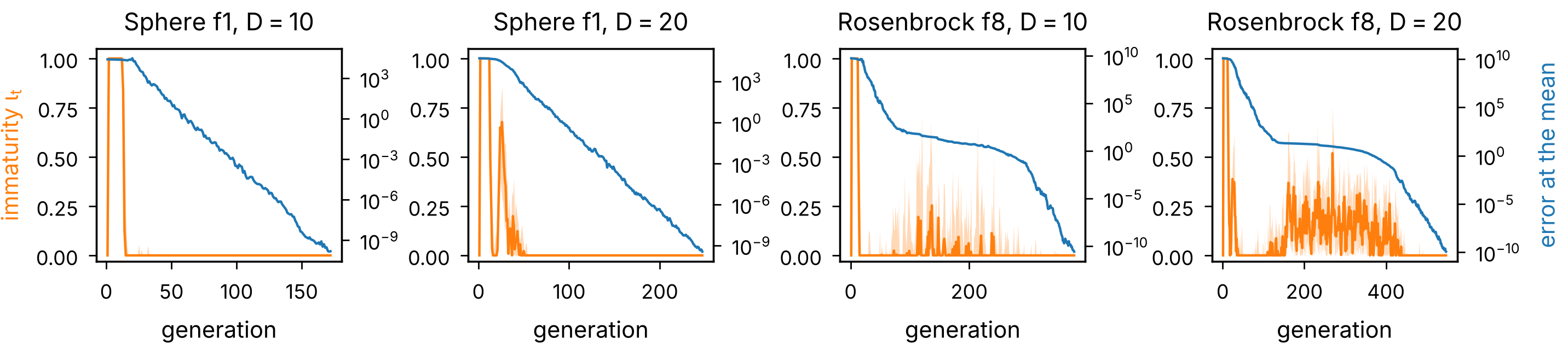}
\caption{SN+MS with the distant start, instances 31--45: median immaturity $\iota_{t}$ (orange, with interquartile band) and median error at the mean (blue, right axis) on the sphere and the Rosenbrock function; generations numbered as in Algorithm \ref{S-alg:mg}.}
\label{S-fig:mat}
\end{figure}

\emph{When the corrections act.} Fig. \ref{S-fig:mat} shows the median immaturity of SN+MS with the distant start. In all four cases it is one from generation 2 to generation 12. In this phase every generation improves, the multiplier grows (its median reaches 7.5 on the sphere with $D=10$), and the distribution widens until it covers the optimum, $50\sqrt{D}$ initial standard deviations away, while the mean hardly moves (the error at the mean falls only from $2.5\times10^{4}$ to $2.2\times10^{4}$). The immaturity is back at zero in generation 15 for $D=10$ and in generation 16 (sphere) and 18 (Rosenbrock) for $D=20$. On the sphere with $D=10$ it stays low until the target is reached in generation 148: the gate is open in 6\% of these generations with a mean immaturity of 0.01, so the algorithm is essentially EMNA+AVS. For $D=20$ the gate opens once more between generations 22 and 46 (median up to 0.68), while the mean travels to the optimum. On the Rosenbrock function the gate opens intermittently while the distribution follows the curved valley floor, in 32\% of the generations for $D=10$ and 55\% for $D=20$ (mean immaturity 0.10 and 0.16). The two terms thus act at full strength for about ten generations at the start of a distant approach and in episodes afterwards.

\section{Protocols, computers, deviations and corrections}
\label{S-sec:prot}

\emph{Protocols.} Each stage was run from a protocol file that was written and stored before its runs; the files are in the data package, and Table \ref{S-tab:prot} lists the first 16 hexadecimal digits of their SHA-256 hashes. The stored files carry modification times that precede the first result file of their stage. The hash was entered in the work log before the runs only for protocol B4 and for the plan of the post hoc check C1 (item 15 below); for protocols A, A7 and B3 it was entered within about an hour after the runs had started, and for protocols B and B2 on the following day, from the stored files, as part of a correction of the log that is itself recorded there. Protocols A and A7 concern the training set, protocol B3 the confirmatory test of the manuscript and protocol B4 its supplementary runs with $D=80$; protocols B and B2 are the confirmatory test and the runs with $D=80$ of the variant with elongation (Section \ref{S-sec:elong}).

\begin{table}[htbp]\centering\footnotesize
\caption{Protocol and program files with the first 16 hexadecimal digits of their SHA-256 hashes}
\label{S-tab:prot}
\begin{tabular}{@{}lll@{}}
\toprule
File & Hash & Remark \\
\midrule
Protocol A, training set & \texttt{18260b7f3e4c8b78} & as first written \\
Protocol A with amendment & \texttt{a034902e3fc258b0} & thresholds changed to the null quantiles before any result (item 1 below) \\
Protocol A7, training set, skewness & \texttt{67b2ed90af071f45} & exploratory; written after protocols A, B and B2 \\
Protocol B, instances 16--30 & \texttt{0d55780c8e47964b} & variant with elongation; written before any run on these instances (but see item 14) \\
Protocol B2, $D=80$ & \texttt{eca5cef4c8b85921} & variant with elongation; as amended before the runs on the large-scale suite (Section \ref{S-sec:elong}) \\
Protocol B2 before the amendment & \texttt{dd8d043306206d2e} & text restored afterwards from the amended file \\
Protocol B3, instances 31--45 & \texttt{2581049b9f32750f} & confirmatory test of the manuscript; written before any run on these instances \\
Protocol B4, $D=80$, skewness & \texttt{2b6b33f095f0b931} & descriptive \\
Plan of the post hoc check C1 & \texttt{43e37ee55a4ebad7} & descriptive; thresholds of AMS+SN and AMaLGaM+SN (item 15 below) \\
\midrule
Source code, version 2 & \texttt{1dc536d0db53b8a0} & used for protocols A and B \\
Source code, version 3 & \texttt{cb574a4f1be83135} & used for protocols B2, A7, B3 and B4; adds diagnostic modes and the large-scale suite \\
Source code, version 4 & \texttt{58c781ab4a138ae9} & used for the slope figure and the diagnostics; adds an option of the slope mode \\
Source code, version 5 & \texttt{db9c6f2502349d8a} & used for the unit test; adds four unit cases without elongation and Makefile comments \\
\bottomrule
\end{tabular}
\par\vspace{2pt}\begin{minipage}{0.95\textwidth}\scriptsize The files that implement the algorithms are byte-identical in versions 3, 4 and 5; versions 3 to 5 change the driver program only (relative to version 2, version 3 also adds an optional recording variable to the file of the EDAs; see item 6).\end{minipage}
\end{table}

\emph{Computers.} Two computers were used, both x86-64 Linux with different versions of the compiler (A and B; the versions and the options are listed in the description of the code package). The program is single-threaded and uses no external numerical libraries; the COCO functions are compiled from the source of that library, version 2.8.3. All algorithms of a (function, instance, start, dimension) cell of a protocol ran on the same computer. Protocol A: standard start on A, distant start on B; the tuning runs on B, together with four algorithms for both starts, so that 480 standard-start runs exist from both computers; all 480 are identical in final error, evaluation count and every recorded hitting time. Protocol A7: on B. Protocol B: $D=10$ and 20 on B; $D=40$, f1--f18 on A and f19--f24 on B. Protocol B2: f1--f12 and f17--f21 on A, f13--f16 and f22--f24 on B; ten runs of f21 were made on both computers and are identical. Protocol B3: $D=10$ and 20 on B; $D=40$, f19--f23 on B and f1--f18 and f24 on A. Protocol B4 and the diagnostics of Section \ref{S-sec:checks}: on A; the baselines with which protocol B4 is compared come from protocol B2, that is partly from computer B. The runs added in the present version (items 10 and 14 below) were made on a machine of the configuration of A, on which stored results were first reproduced exactly: the direction file of SN for $D=10$ and the 210 runs of CMA-ES of protocol B for $D=10$. The runs of item 15 were made on a machine of the same configuration, on which 24 stored runs of protocol B3 ($D=10$, 20 and 40) were first reproduced exactly; 88 of the new runs were repeated on computer B and are identical.

\emph{Deviations, incidents and corrections.} (1) The first runs of protocol A were started with the threshold 2.36 of the earlier version and stopped after about two minutes, without their results being inspected, when the exact null law of Theorem \ref{S-thm:calib} became available; the protocol was amended to use its 99\% quantiles and the runs were restarted. (2) The tuning of the alternatives to the gate (twelve ungated settings and three further time constants of the scheduled gate) was added to protocol A after the results of the gate ablation were known, in order to give the alternatives their own tuning; these runs, like everything on the training set, are descriptive. (3) Protocols B and B3 prescribe Holm's correction within each family; it was applied over all tests of a family (settings and comparisons), the stricter of the two readings of the text. (4) Three algorithms of protocol B and the two lower skewness levels of protocol B3 were not run for $D=40$, as stated in the protocols. (5) For the amendment of protocol B2 see Section \ref{S-sec:elong}. (6) After the runs of protocols A and B the program was extended by diagnostic modes and by the option for the large-scale suite; the files that implement the algorithms were not changed, apart from an optional recording variable. As a regression check, 27 runs of protocol B (f8, $D=10$, distant start, instances 16--18, nine algorithms) were repeated with version 3 on computer A and are identical to the stored results from computer B. (7) In protocol B3 the function f24 for $D=40$ was moved from computer B to computer A to save time; all its algorithms ran on A. (8) During protocol B3 the scratch volume of computer B, which is shared with other work, became full, and one half-written line was left at the end of each of the result files being written. These lines were discarded, the four affected commands (each one function, instance and start with all algorithms) were re-run, and the number of runs of every completed command was checked; no result was lost or altered. (9) Protocol B3 justifies family G4 with the remark that on the training set none of its three comparisons had shown a significant difference. For the third comparison, SN+MS with and without the feedback correction, this remark was an error of ours: that pair had not been run on the training set; what had been seen there was that the variant with elongation loses without the correction (Table \ref{S-tab:gate}) and that a mean shift without it is harmful in the earlier factorial study (Table \ref{S-tab:factfull}). The tests were carried out as specified, and the result contradicts the stated expectation: without the feedback correction SN+MS is significantly worse in all six settings. (10) The trace runs behind Fig. \ref{S-fig:mat}, the slope runs and the diagnostics of Section \ref{S-sec:checks} for SN and SN+MS were made after the confirmatory test; they are not part of a protocol. The direction and plateau runs of SN-on (Tables \ref{S-tab:dir} and \ref{S-tab:plat}) were added in the present version, after the decision of item 13. (11) No run of any protocol was excluded. (12) The earlier studies reproduced in Sections \ref{S-sec:param}, \ref{S-sec:heldout} and \ref{S-sec:pred} belong to the earlier version and were re-run on another computer at the time; that comparison is part of the data package of that version. (13) The algorithm of the manuscript is the configuration SN-on of protocol B3, the skewness without the gate. It was adopted after the confirmatory test and because of it: family G4 found it practically equivalent to the gated configuration SN in all six settings. The confirmatory statements about the skewness (G1, G3 and G5) were tested with the gated sampler, as the protocol prescribes; the comparisons of SN-on with EMNA+AVS and with the other algorithms are descriptive. (14) The comparison of our CMA-ES with pycma (Table \ref{S-tab:cma}) was first made before protocol B was written, on instances 16--30 with $D=10$, the standard start and one seed per instance, so that these instances had been seen by the CMA-ES baseline, and by it alone, when that protocol was frozen. The raw results of that first comparison were not kept. It was repeated for the present version with five seeds per instance; the figures of the first comparison (medians of 15 runs that differed by 2\% in the median and by at most 10\%) are replaced by those of the repetition.
(15) In the configurations AMS+SN and AMaLGaM+SN of protocols A7 and B3 the gate ran with the thresholds $\kappa_{0}=2$ and $\kappa_{1}=6$ of the earlier version, not with the null quantiles that protocol B3 prescribes for its gated configurations ($\kappa_{0}=2.53$, 1.95 and 1.62 for $D=10$, 20 and 40, with $\kappa_{1}=\kappa_{0}+4$): the program sets the null quantiles by the name of a configuration, and these two names were not covered. We found the deviation in the final check of the present version, after the confirmatory test, by repeating stored runs with and without an explicit threshold. The results of family G5 in Table \ref{tab:fam} of the manuscript and in Table \ref{S-tab:famfull}, and the rows of these two configurations in Tables \ref{tab:test} of the manuscript and \ref{S-tab:train}, are those of the runs as they were made. Under a plan written before its runs (Table \ref{S-tab:prot}), both configurations were then run again with the prescribed thresholds: 4{,}320 runs on the confirmatory instances, with the random streams of protocol B3, and 480 runs on the training set. The analysis is that of family G5; it is post hoc and therefore descriptive. Table \ref{S-tab:c1} gives the result. With the prescribed thresholds the skewness added to EMNA+AVS+AMS is practically equivalent in all 6 settings (HL between $-$0.044 and +0.068), and added to AMaLGaM in all 6 settings (HL between $-$0.030 and +0.044). None of the 12 tests is significant after Holm's correction over the 12. The conclusion of family G5 is therefore the same with either pair of thresholds. The repeated configurations differ from the configurations as run by HL between $-$0.030 and +0.043, and the two are practically equivalent in all 12 settings; none of these differences is significant after Holm's correction. On the training set the repeated configurations give HL +0.03 against EMNA+AVS+AMS ($p=0.24$, uncorrected; as run +0.01) and $-$0.05 against AMaLGaM ($p=0.08$; as run $-$0.01); their mean $y_{\mathrm{any}}$ over the 48 blocks is $-$3.56 and $-$3.53, against $-$3.66 and $-$3.49 as run.

\begin{table}[htbp]\centering\scriptsize
\setlength{\tabcolsep}{3pt}
\caption{Post hoc check C1 (item 15): AMS+SN and AMaLGaM+SN on the confirmatory instances with the gate thresholds as run, $(2,6)$, and repeated with the thresholds prescribed by the protocol (difference first minus second in $y_{\mathrm{any}}$ over the 24 functions; negative: the first algorithm is better). Descriptive}
\label{S-tab:c1}
\begin{tabular}{@{}lrlcccccccccccc@{}}
\toprule
 & & & \multicolumn{2}{c}{as run} & \multicolumn{8}{c}{repeated with the prescribed thresholds} & \multicolumn{2}{c}{repeated vs. as run} \\
\cmidrule(lr){4-5}\cmidrule(lr){6-13}\cmidrule(l){14-15}
Comparison & $D$ & start & HL & equiv. & mean & HL & 95\% interval & $p$ & $p$ (Holm) & equiv. & fn. b/w & solved only 1st/2nd & HL & 95\% interval \\
\midrule
AMS+SN vs. EMNA+AVS+AMS & 10 & std. & +0.01 & yes & +0.02 & +0.03 & [0.00, +0.07] & $0.02$ & $0.20$ & yes & 5/19 & 11/13 & +0.01 & [$-$0.02, +0.04] \\
AMS+SN vs. EMNA+AVS+AMS & 10 & dist. & $-$0.01 & yes & +0.18 & +0.07 & [$-$0.03, +0.23] & $0.39$ & $1.00$ & yes & 13/11 & 12/20 & +0.03 & [$-$0.03, +0.25] \\
AMS+SN vs. EMNA+AVS+AMS & 20 & std. & +0.02 & yes & $-$0.01 & $-$0.02 & [$-$0.06, +0.01] & $0.30$ & $1.00$ & yes & 13/11 & 9/9 & $-$0.03 & [$-$0.05, $-$0.01] \\
AMS+SN vs. EMNA+AVS+AMS & 20 & dist. & $-$0.02 & yes & $-$0.06 & $-$0.04 & [$-$0.12, +0.01] & $0.14$ & $1.00$ & yes & 16/8 & 14/9 & $-$0.01 & [$-$0.07, +0.03] \\
AMS+SN vs. EMNA+AVS+AMS & 40 & std. & $-$0.01 & yes & 0.00 & +0.01 & [$-$0.03, +0.03] & $0.58$ & $1.00$ & yes & 8/16 & 3/2 & +0.01 & [$-$0.02, +0.02] \\
AMS+SN vs. EMNA+AVS+AMS & 40 & dist. & $-$0.01 & yes & $-$0.02 & $-$0.02 & [$-$0.05, +0.01] & $0.16$ & $1.00$ & yes & 15/9 & 0/3 & 0.00 & [$-$0.03, +0.03] \\
\midrule
AMaLGaM+SN vs. AMaLGaM & 10 & std. & $-$0.05 & yes & $-$0.06 & $-$0.03 & [$-$0.09, 0.00] & $0.06$ & $0.62$ & yes & 17/7 & 11/7 & +0.01 & [$-$0.02, +0.11] \\
AMaLGaM+SN vs. AMaLGaM & 10 & dist. & +0.01 & yes & $-$0.08 & 0.00 & [$-$0.18, +0.04] & $0.88$ & $1.00$ & yes & 11/13 & 8/8 & $-$0.03 & [$-$0.13, +0.01] \\
AMaLGaM+SN vs. AMaLGaM & 20 & std. & +0.05 & yes & +0.02 & +0.01 & [$-$0.03, +0.05] & $0.56$ & $1.00$ & yes & 10/14 & 7/7 & $-$0.03 & [$-$0.10, +0.01] \\
AMaLGaM+SN vs. AMaLGaM & 20 & dist. & +0.01 & yes & +0.08 & +0.04 & [$-$0.01, +0.14] & $0.23$ & $1.00$ & yes & 10/14 & 9/15 & +0.04 & [0.00, +0.10] \\
AMaLGaM+SN vs. AMaLGaM & 40 & std. & +0.01 & yes & +0.03 & +0.01 & [$-$0.02, +0.04] & $0.44$ & $1.00$ & yes & 11/13 & 7/10 & +0.01 & [$-$0.03, +0.06] \\
AMaLGaM+SN vs. AMaLGaM & 40 & dist. & +0.03 & yes & +0.03 & +0.04 & [$-$0.03, +0.09] & $0.22$ & $1.00$ & yes & 7/17 & 2/3 & $-$0.01 & [$-$0.06, +0.05] \\
\bottomrule
\end{tabular}
\par\vspace{2pt}\begin{minipage}{0.97\textwidth}\scriptsize Columns as in Table \ref{S-tab:famfull}; Holm's correction over the 12 tests of the table. equiv.: the 90\% interval of HL lies within $\pm0.25$. ``as run'' repeats family G5 of Table \ref{S-tab:famfull}. The last two columns compare each repeated configuration with the same configuration as run.\end{minipage}
\end{table}

\section{The variant with a covariance elongation (earlier version of this work)}
\label{S-sec:elong}

The earlier version of this work combined three gated terms along the direction of movement: the mean shift and the skewness of (\ref{S-eq:mg}) and an \emph{elongation} of the covariance. Its subject was the gate. This section documents that variant: the joint sampler in its corrected form, what the gate is worth for it, and its confirmatory test on instances 16--30. In the manuscript the elongation is not part of the algorithm, and these results are the basis of Section \ref{sec:why} there. All runs of this section were made with the program of the manuscript, under protocols A, B and B2 (Section \ref{S-sec:prot}). The configurations are named as in the earlier version: MG-A ($\sigma_{1}=1$, $e=4$) and MG-B ($\sigma_{1}=2$, $e=2$) use no skewness, MG-S ($\sigma_{1}=1$, $e=2$) and MG-A+skew use $\rho_{\max}=0.95$, and ASN+AVS is the gated configuration SN.

\subsection{The joint sampler}
With the elongation factor $a=1+e\iota$ and $\delta$, $\beta$ of (\ref{S-eq:adb}), the joint sampler is
\begin{equation}
X=\mathbf{1}\boldsymbol{\mu}^{T}+\sqrt{s}\,YL^{T},\qquad Y=Z'+\mathbf{h}\mathbf{w}^{T}+\sigma_{1}\iota\,\mathbf{1}\mathbf{w}^{T},\qquad Z'=Z+e\iota\,(Z\mathbf{w})\mathbf{w}^{T},\qquad \mathbf{h}=a\,\delta\,(|\mathbf{z}_{0}|-b\mathbf{1})-\beta Z'\mathbf{w}
\label{S-eq:mgE}
\end{equation}
\noindent In terms of the standardized skew-normal rows $\mathbf{y}_{\mathrm{SN}}=Q\mathbf{z}+\delta(|z_{0}|-b)\mathbf{w}$, a row of $Y$ is $\mathbf{y}=B\,\mathbf{y}_{\mathrm{SN}}+\sigma_{1}\iota\,\mathbf{w}$ with $B=I+(a-1)\mathbf{w}\mathbf{w}^{T}$: the sampler first draws a standardized skew-normal population along $\mathbf{w}$, then stretches its whole component along $\mathbf{w}$, then shifts it. With $e=0$ it is the sampler (\ref{S-eq:mg}).
\begin{theorem}[Moments of the joint sampler]
\label{S-thm:jointE}
For $0\le\iota\le1$, $e,\sigma_{1}\ge0$ and $\rho_{\max}<1$, a row $\mathbf{x}$ of (\ref{S-eq:mgE}) has mean $\boldsymbol{\mu}+\sqrt{s}\,\sigma_{1}\iota\,L\mathbf{w}$, covariance $s\,L\big(I+(a^{2}-1)\mathbf{w}\mathbf{w}^{T}\big)L^{T}$ and third cumulant $s^{3/2}a^{3}K_{3}\delta^{3}\,(L\mathbf{w})\otimes(L\mathbf{w})\otimes(L\mathbf{w})$.
\end{theorem}
\begin{proof}
By Theorem \ref{thm:mom}, $\mathbf{y}_{\mathrm{SN}}$ has mean zero, covariance $I$ and third cumulant $K_{3}\delta^{3}\mathbf{w}^{\otimes3}$. Since $B\mathbf{w}=a\mathbf{w}$ and $B$ is symmetric, $B\mathbf{y}_{\mathrm{SN}}$ has mean zero, covariance $B^{2}=I+(a^{2}-1)\mathbf{w}\mathbf{w}^{T}$ and third cumulant $K_{3}\delta^{3}(B\mathbf{w})^{\otimes3}=a^{3}K_{3}\delta^{3}\mathbf{w}^{\otimes3}$. The shift adds $\sigma_{1}\iota\mathbf{w}$ to the mean only, and the affine map $\mathbf{x}=\boldsymbol{\mu}+\sqrt{s}L\mathbf{y}$ gives the statement.
\end{proof}

\emph{Correction of the earlier version.} The sampler printed in the earlier version had $\mathbf{h}=\delta(|\mathbf{z}_{0}|-b\mathbf{1})-\beta Z'\mathbf{w}$, without the factor $a$ in front of the half-normal term. Its variance along $\mathbf{w}$ is $a^{2}(1-(1-b^{2})\delta^{2})+(1-b^{2})\delta^{2}$, which is 3.34 instead of $a^{2}=25$ for $e=4$, $\iota=1$ and $\rho_{\max}=0.95$, so it did not have the covariance claimed for it (last row of Table \ref{S-tab:unit}). The error affected only the configurations that combine elongation and skewness, and the statement of that version that the two corrections cancel each other on a slope was a consequence of it. That version also applied $\rho_{\max}$ twice in the definition of the skewness vector; $\delta=\iota d_{\max}$ with $d_{\max}=\rho_{\max}D_{\lim}$ is the definition used here and, with $\iota=1$, in the manuscript.

\emph{On a slope.} Theorem \ref{thm:slope} of the manuscript extends to (\ref{S-eq:mgE}) with the same proof: the selected $\mathbf{w}^{T}\mathbf{y}$ is $aY_{\delta}+\sigma_{1}\iota$, so the variance of the model along the gradient is multiplied by $s\,a^{2}c(\delta)$ per generation and the mean progresses by $\sqrt{s}\,[\sigma_{1}\iota+a\lambda(\delta)]$ standard deviations of the model. Table \ref{S-tab:const10} gives these constants and their measurement. With $e=2$ and 4 the ratio is above one: the variance along the gradient grows without any amplification, as it does with AMS. The skewness multiplies the ratio by a further $c(d_{\max})/c_{\tau}=1.86$. The contribution of the injected shift to the statistic, (\ref{S-eq:minj}), becomes $m^{\mathrm{inj}}/a^{2}$: for $D=10$ and $\iota=1$ it is 6.5 for MG-B and 0.6 for MG-A, against $\kappa_{1}=6.53$, so that a strong elongation hides the shift from the whitened statistic. On the neutral function, an opening of the gate of MG-B without the feedback correction lasts 4.0, 16.7 and 30.5 generations on average for $D=10$, 20 and 40, and one of MG-A 1.3 to 1.4, against 1.2 with the correction.

\emph{Global convergence.} Theorem \ref{thm:glob} of the manuscript also holds for (\ref{S-eq:mgE}) with a bounded $e\ge0$. Given $|z_{0}|=r$ a sample is normal with mean $\boldsymbol{\mu}+\sqrt{s}\,[a\delta(r-b)+\sigma_{1}\iota]L\mathbf{w}$ and covariance $sLBQQ^{T}BL^{T}$. The matrix $BQQ^{T}B$ has the eigenvalue $a^{2}(1-(1-b^{2})\delta^{2})$ along $\mathbf{w}$, which lies between $1-\rho_{\max}^{2}$ and $(1+e)^{2}$, and the eigenvalue one elsewhere; for $r\in[0,1]$ the factor of $L\mathbf{w}$ in the mean is at most $(1+e)D_{\lim}+\sigma_{1}$ in absolute value. The bounds of the proof in Section \ref{S-sec:proof} therefore hold with $\bar{v}$ multiplied by $(1+e)^{2}$ and a correspondingly larger $R$, and the rest of the proof is unchanged.

\begin{table}[htbp]\centering\footnotesize
\caption{One-generation constants of the joint sampler on a linear function ($\tau=0.35$, $s=1$, $\iota=1$): theory and a run with $D=10$, $N=10^{5}$, no elitism (mean of generations 5--13)}
\label{S-tab:const10}
\begin{tabular}{@{}lcccc@{}}
\toprule
 & \multicolumn{2}{c}{$a^{2}c(\delta)$} & \multicolumn{2}{c}{$\sigma_{1}+a\lambda(\delta)$} \\
\cmidrule(lr){2-3}\cmidrule(lr){4-5}
Sampler & theory & run & theory & run \\
\midrule
Gaussian & 0.288 & 0.287 & 1.058 & 1.059 \\
elongation $e=2$ & 2.590 & 2.582 & 3.175 & 3.176 \\
elongation $e=4$ & 7.195 & 7.171 & 5.291 & 5.294 \\
MG-A ($\sigma_{1}=1$, $e=4$) & 7.195 & 7.171 & 6.291 & 6.294 \\
MG-B ($\sigma_{1}=2$, $e=2$) & 2.590 & 2.582 & 5.175 & 5.176 \\
MG-A with skewness $\rho_{\max}=0.95$ & 13.352 & 13.265 & 6.545 & 6.540 \\
MG-S ($\sigma_{1}=1$, $e=2$, $\rho_{\max}=0.95$) & 4.807 & 4.776 & 4.327 & 4.324 \\
\bottomrule
\end{tabular}
\par\vspace{2pt}\begin{minipage}{0.7\textwidth}\scriptsize First pair of columns: ratio of the variance of the selected solutions along the gradient to that of the model from which they were sampled. Second pair: step of the mean in standard deviations of the model.\end{minipage}
\end{table}

\subsection{What the gate is worth}
The value of the gate was measured by changing the gate alone, under the same variance scaling, population, initialization and random numbers (Table \ref{S-tab:gate}; training set, descriptive). First, the corrections of MG-A cannot be left on: with $\iota_{t}\equiv1$ the same $\sigma_{1}=1$ and $e=4$ solve 14 of 240 runs against 127, because an elongation that multiplies the variance along the path by 7.2 per generation is only useful while there is a slope to follow. Second, an ungated algorithm with its own tuning is a respectable baseline but not as good: the best of twelve always-on settings ($\sigma_{1}=0.5$, $e=0.5$) trails MG-A by 0.08 and MG-B by 0.13 in $y_{\mathrm{any}}$ (HL; $p=0.07$ and $0.002$). Third, the kind of gate matters: a gate that closes on a fixed schedule, $\iota_{t}=e^{-t/\tau_{g}}$ with the best of six time constants, is worse by 0.12 ($p=0.008$); a gate that uses the same thresholds on the uncorrected statistic $\|\mathbf{p}_{t}\|^{2}/D$, without the inverse-Wishart factor, is worse by 0.06, which is at the edge of significance ($p=0.05$, uncorrected). Fourth, without the feedback correction MG-A loses 0.10 ($p=0.0002$). The two alternatives with their own tuning were carried to the confirmatory test as family F1 (Table \ref{S-tab:fam16}). There the gated configurations were significantly better than the ungated one in 6 of 12 comparisons after Holm's correction, in 4 of the six with the distant start and in 2 of the six with the standard start, the latter for $D=40$; they were never worse. MG-A was better than the scheduled gate in 3 of 4 comparisons. For a correction that enlarges the covariance, the benefit of the calibrated gate is thus established for the approach from afar, and it grows with the dimension (HL of MG-B against the ungated configuration with the distant start: $-$0.28, $-$0.44, $-$0.85 for $D=10$, 20 and 40).

\begin{table}[htbp]
\caption{Variant with elongation, what the gate is worth: variants of MG-A that differ in the gate or in one correction (training set: BBOB f1--f24, $D=10$, instances 1--5, standard and distant start; 48 blocks, 240 runs)}
\label{S-tab:gate}
\centering \footnotesize
\begin{tabular}{@{}lccccc@{}}
\toprule
Variant & mean $y_{\mathrm{any}}$ & runs solved & gate open & HL difference to MG-A [95\% interval] & $p$ \\
\midrule
MG-A (calibrated gate, $\sigma_{1}=1$, $e=4$) & $-$3.40 & 127 & 0.27 & \multicolumn{2}{c}{reference} \\
no gate, same strength ($\iota_{t}\equiv1$) & +1.07 & 14 & 0.99 & +4.38 [+3.56, +5.38] & $1\times10^{-14}$ \\
no gate, own tuning ($\sigma_{1}=0.5$, $e=0.5$; best of 12) & $-$3.29 & 131 & 0.99 & +0.08 [$-$0.00, +0.18] & $0.07$ \\
scheduled gate $\iota_{t}=e^{-t/25}$ (best of 6 time constants) & $-$2.97 & 118 & 0.99 & +0.12 [+0.03, +0.25] & $0.008$ \\
uncalibrated statistic (no inverse-Wishart factor) & $-$3.21 & 127 & 0.38 & +0.06 [+0.00, +0.22] & $0.05$ \\
no feedback correction & $-$2.98 & 119 & 0.31 & +0.10 [+0.05, +0.29] & $0.0002$ \\
symmetric square root instead of Cholesky factor & $-$3.39 & 128 & 0.26 & +0.01 [$-$0.05, +0.05] & $0.80$ \\
\midrule
elongation only ($e=4$) & $-$3.06 & 120 & 0.30 & +0.11 [+0.03, +0.23] & $0.002$ \\
mean shift only ($\sigma_{1}=1$) & $-$2.64 & 114 & 0.27 & +0.38 [+0.15, +0.70] & $3\times10^{-6}$ \\
no correction (EMNA+AVS) & $-$1.77 & 110 & 0.42 & +0.92 [+0.34, +1.59] & $2\times10^{-7}$ \\
\midrule
MG-B (calibrated gate, $\sigma_{1}=2$, $e=2$) & $-$3.48 & 133 & 0.26 & $-$0.04 [$-$0.10, +0.01] & $0.12$ \\
\bottomrule
\end{tabular}
\par\vspace{2pt}\begin{minipage}{\textwidth}\scriptsize Differences are variant minus MG-A in $y_{\mathrm{any}}$ (positive: the variant is worse); HL is the Hodges--Lehmann estimate over the 48 blocks, $p$ from the Wilcoxon signed-rank test (descriptive, uncorrected). ``Gate open'' is the fraction of generations with $\iota_{t}>0$. Against MG-B the tuned ungated variant differs by +0.13 [+0.06, +0.25], $p=0.002$.\end{minipage}
\end{table}

\subsection{Strength of the corrections}
Table \ref{S-tab:grid} gives the anytime performance on the training set for $\sigma_{1}\in\{0,1,2\}$ and $e\in\{0,2,4\}$, with and without the skewness. The best configurations combine a moderate shift with a moderate elongation. The skewness improves the configurations without elongation ($e=0$) and changes those with elongation by at most 0.12 in either direction: once the covariance is elongated, a third-moment correction has no detectable effect, which family F4 of the confirmatory test confirms. Under the SDR-AVS of AMaLGaM, the corrections of MG-B without AMS score $-$3.35 against $-$3.38 for AMaLGaM (HL +0.01, $p=0.67$).

\begin{table}[htbp]
\caption{Mean anytime performance on the training set for the mean shift $\sigma_{1}$ and the elongation $e$, without / with the skewness ($\rho_{\max}=0$ / $0.95$)}
\label{S-tab:grid}
\centering \footnotesize
\begin{tabular}{@{}cccc@{}}
\toprule
 & $e=0$ & $e=2$ & $e=4$ \\
\midrule
$\sigma_{1}=0$ & $-$1.77 / $-$2.52 & $-$3.29 / $-$3.24 & $-$3.06 / $-$3.09 \\
$\sigma_{1}=1$ & $-$2.64 / $-$3.14 & $-$3.37 / $-$3.48 & $-$3.40 / $-$3.36 \\
$\sigma_{1}=2$ & $-$3.02 / $-$3.11 & $-$3.48 / $-$3.41 & $-$3.41 / $-$3.37 \\
\bottomrule
\end{tabular}
\par\vspace{2pt}\begin{minipage}{0.6\textwidth}\scriptsize 48 blocks, AVS decay 0.8, calibrated gate. $(0,0)$ without skewness is EMNA+AVS and with skewness SN; $(1,0)$ is MS and, with skewness, SN+MS; $(1,4)$ and $(2,2)$ without skewness are MG-A and MG-B; $(1,2)$ with skewness is MG-S.\end{minipage}
\end{table}

\subsection{Confirmatory test on instances 16--30}
\begin{figure}[htbp]
\centering
\includegraphics[width=0.95\textwidth]{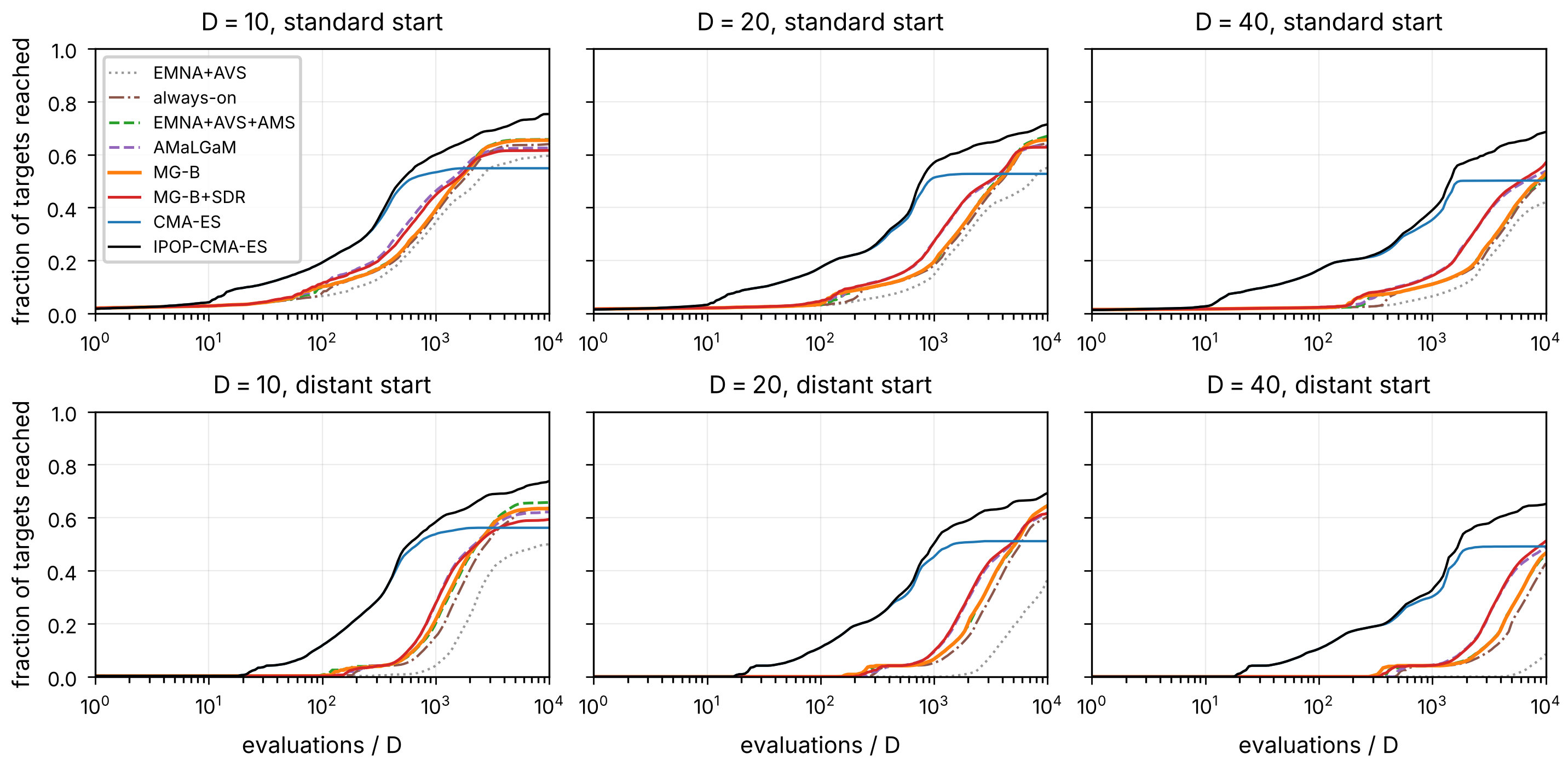}
\caption{Variant with elongation: running-time distributions of the confirmatory test on instances 16--30 (fraction of the (function, instance, target) triples reached within a given number of evaluations, 51 targets $10^{2},10^{1.8},\ldots,10^{-8}$). Top: standard start; bottom: distant start.}
\label{S-fig:ecdf16}
\end{figure}

Protocol B tested the variant on BBOB instances 16--30 with five families of comparisons fixed in advance (Tables \ref{S-tab:test16} and \ref{S-tab:fam16}; Fig. \ref{S-fig:ecdf16}; all comparisons in Table \ref{S-tab:famfull16}). These instances had not been used before, with one exception: our CMA-ES had been run on them for $D=10$ with the standard start when its implementation was first compared with pycma (Section \ref{S-sec:prot}, item 14), a check that concerned this baseline only. \emph{F2:} MG-A and MG-B are better than EMNA+AVS in all 12 comparisons (Holm-corrected $p\le0.02$); with the distant start the difference is large (HL from $-$1.96 to $-$6.43). \emph{F3:} MG-B is practically equivalent to EMNA+AVS+AMS in all 6 settings (HL between $-$0.033 and +0.028), and MG-S in 5 of six; neither differs significantly from AMS anywhere, and no configuration is significantly better than AMS in any setting. \emph{F4:} neither MG-A with skewness against MG-A nor MG-S against MG-B differs significantly in any setting, and they are practically equivalent in all 10 comparisons. \emph{F5:} with the SDR-AVS of AMaLGaM in place of the AVS of the manuscript, the corrections of MG-B without AMS are never significantly different from AMaLGaM and practically equivalent in 4 of six settings. For $D=40$ AMaLGaM is better than both MG-B (HL +0.53 and +0.73) and EMNA+AVS+AMS (+0.62 and +0.80), whereas with SDR-AVS the corrections of MG-B are as good as AMaLGaM there (mean $y_{\mathrm{any}}$ $-$1.98 against $-$1.88 with the standard start and $-$0.29 against $-$0.04 with the distant start): the simple AVS, whose multiplier is confined to $[1,10]$, is the weak part in higher dimension. \emph{Descriptive:} ASN+AVS, the skewness alone, is better than EMNA+AVS with the distant start (HL $-$1.23 and $-$3.49 for $D=10$ and 20) and slightly better with the standard start ($-$0.10 and $-$0.09); this is the observation that the confirmatory test of the manuscript replicates on instances 31--45 as family G1. The mean ranks over the 144 function--setting pairs are 1.37 (IPOP-CMA-ES), 4.55 (CMA-ES), 4.66 (AMaLGaM), 4.69 (MG-B+SDR), 5.47 (MG-B), 5.58 (EMNA+AVS+AMS), 5.69 (MG-S), 6.41 (MG-A), 7.83 (always-on), 8.75 (EMNA+AVS).

\begin{table}[htbp]
\caption{Variant with elongation: confirmatory test on BBOB f1--f24, instances 16--30 (15 runs per function, budget $10^{4}D$): number of functions solved to $10^{-8}$ in all 15 runs and mean anytime performance $y_{\mathrm{any}}$ over the 24 functions (lower is better); mean rank of $y_{\mathrm{any}}$ over the function--setting pairs}
\label{S-tab:test16}
\centering \scriptsize
\begin{tabular}{@{}l@{\hspace{5pt}}rr@{\hspace{7pt}}rr@{\hspace{7pt}}rr@{\hspace{7pt}}rr@{\hspace{7pt}}rr@{\hspace{7pt}}rr@{\hspace{7pt}}cc@{}}
\toprule
 & \multicolumn{2}{c}{$D=10$, std.} & \multicolumn{2}{c}{$D=10$, dist.} & \multicolumn{2}{c}{$D=20$, std.} & \multicolumn{2}{c}{$D=20$, dist.} & \multicolumn{2}{c}{$D=40$, std.} & \multicolumn{2}{c}{$D=40$, dist.} & \multicolumn{2}{c}{mean rank} \\
Algorithm & sol. & $y$ & sol. & $y$ & sol. & $y$ & sol. & $y$ & sol. & $y$ & sol. & $y$ & $D\le20$ & all \\
\midrule
EMNA+AVS & 9 & $-$3.16 & 8 & 0.02 & 8 & $-$1.95 & 3 & 3.81 & 6 & $-$0.48 & 0 & 7.30 & 10.86 & 8.75 \\
ASN+AVS & 8 & $-$3.48 & 7 & $-$1.49 & 10 & $-$2.45 & 8 & 0.27 & -- & -- & -- & -- & 9.73 & -- \\
EMNA+AVS+AMS & 12 & $-$3.87 & 9 & $-$3.20 & 11 & $-$3.03 & 10 & $-$1.69 & 7 & $-$1.18 & 7 & 0.89 & 5.86 & 5.58 \\
AMaLGaM & 10 & $-$3.74 & 10 & $-$3.07 & 11 & $-$3.13 & 9 & $-$1.97 & 8 & $-$1.88 & 8 & $-$0.04 & 5.78 & 4.66 \\
CMA-ES & 8 & $-$3.30 & 8 & $-$3.36 & 9 & $-$3.04 & 8 & $-$2.80 & 9 & $-$2.67 & 8 & $-$2.46 & 5.90 & 4.55 \\
IPOP-CMA-ES & 14 & $-$4.83 & 15 & $-$4.64 & 13 & $-$4.35 & 13 & $-$4.06 & 11 & $-$3.91 & 11 & $-$3.58 & 1.53 & 1.37 \\
\midrule
MG-A & 11 & $-$3.79 & 9 & $-$2.90 & 11 & $-$2.81 & 8 & $-$1.48 & 7 & $-$1.15 & 7 & 0.99 & 7.12 & 6.41 \\
MG-B & 12 & $-$3.84 & 11 & $-$3.10 & 12 & $-$2.97 & 9 & $-$1.76 & 7 & $-$1.23 & 7 & 0.79 & 5.96 & 5.47 \\
MG-S & 11 & $-$3.79 & 10 & $-$3.07 & 11 & $-$2.90 & 10 & $-$1.70 & 7 & $-$1.21 & 7 & 0.75 & 6.31 & 5.69 \\
MG-A+skew & 12 & $-$3.75 & 8 & $-$2.93 & 11 & $-$2.79 & 10 & $-$1.49 & -- & -- & -- & -- & 7.23 & -- \\
MG-B+SDR & 11 & $-$3.63 & 8 & $-$2.87 & 12 & $-$3.14 & 10 & $-$2.07 & 8 & $-$1.98 & 8 & $-$0.29 & 6.16 & 4.69 \\
\midrule
always-on & 12 & $-$3.67 & 8 & $-$2.73 & 11 & $-$2.83 & 10 & $-$1.05 & 7 & $-$1.05 & 4 & 1.81 & 8.97 & 7.83 \\
time-decay & 10 & $-$3.40 & 8 & $-$2.39 & 9 & $-$2.34 & 8 & $-$0.65 & -- & -- & -- & -- & 9.58 & -- \\
\bottomrule
\end{tabular}
\par\vspace{2pt}\begin{minipage}{\textwidth}\scriptsize std.: standard start; dist.: distant start. MG-A+skew: MG-A with $\rho_{\max}=0.95$. MG-B+SDR: the corrections of MG-B with the SDR-AVS of AMaLGaM, without AMS. always-on: no gate, $\sigma_{1}=e=0.5$; time-decay: $\iota_{t}=e^{-t/25}$ with the strength of MG-A (the two alternatives to the gate with their own tuning, Table \ref{S-tab:gate}). Three algorithms were not run for $D=40$; the rank over $D\le20$ is among all 13 algorithms (96 pairs), the rank over all settings among the ten run everywhere (144 pairs).\end{minipage}
\end{table}

\begin{table}[htbp]
\caption{Variant with elongation, confirmatory comparisons on instances 16--30: Hodges--Lehmann estimate of the difference in $y_{\mathrm{any}}$ (first minus second; negative means that the first algorithm is better) over the 24 functions of each setting}
\label{S-tab:fam16}
\centering \footnotesize
\begin{tabular}{@{}llcccccc@{}}
\toprule
 & & \multicolumn{2}{c}{$D=10$} & \multicolumn{2}{c}{$D=20$} & \multicolumn{2}{c}{$D=40$} \\
\cmidrule(lr){3-4}\cmidrule(lr){5-6}\cmidrule(l){7-8}
Family & Comparison & std. & dist. & std. & dist. & std. & dist. \\
\midrule
F1 (gate) & MG-A vs. always-on & $-$0.02$^{\approx}$ & $-$0.20 & $-$0.02$^{\approx}$ & $-$0.35 & \textbf{$-$0.10}$^{\approx}$ & \textbf{$-$0.71} \\
 & MG-B vs. always-on & $-$0.03$^{\approx}$ & \textbf{$-$0.28} & $-$0.06$^{\approx}$ & \textbf{$-$0.44} & \textbf{$-$0.15}$^{\approx}$ & \textbf{$-$0.85} \\
 & MG-A vs. time-decay & \textbf{$-$0.16} & $-$0.09$^{\approx}$ & \textbf{$-$0.28} & \textbf{$-$0.37} & -- & -- \\
\midrule
F2 (corrections) & MG-A vs. EMNA+AVS & \textbf{$-$0.15} & \textbf{$-$1.96} & \textbf{$-$0.24} & \textbf{$-$4.96} & \textbf{$-$0.16} & \textbf{$-$6.19} \\
 & MG-B vs. EMNA+AVS & \textbf{$-$0.22} & \textbf{$-$1.98} & \textbf{$-$0.26} & \textbf{$-$5.17} & \textbf{$-$0.18} & \textbf{$-$6.43} \\
\midrule
F3 (AMS) & MG-A vs. EMNA+AVS+AMS & +0.03$^{\approx}$ & +0.25 & \textbf{+0.08}$^{\approx}$ & +0.14 & +0.03$^{\approx}$ & +0.04$^{\approx}$ \\
 & MG-B vs. EMNA+AVS+AMS & +0.01$^{\approx}$ & $-$0.00$^{\approx}$ & +0.03$^{\approx}$ & $-$0.01$^{\approx}$ & $-$0.02$^{\approx}$ & $-$0.03$^{\approx}$ \\
 & MG-S vs. EMNA+AVS+AMS & +0.04$^{\approx}$ & +0.05 & +0.05$^{\approx}$ & $-$0.00$^{\approx}$ & $-$0.01$^{\approx}$ & $-$0.06$^{\approx}$ \\
\midrule
F4 (skewness) & MG-A+skew vs. MG-A & +0.01$^{\approx}$ & +0.00$^{\approx}$ & +0.02$^{\approx}$ & $-$0.01$^{\approx}$ & -- & -- \\
 & MG-S vs. MG-B & +0.03$^{\approx}$ & +0.05$^{\approx}$ & +0.02$^{\approx}$ & +0.02$^{\approx}$ & +0.02$^{\approx}$ & $-$0.04$^{\approx}$ \\
\midrule
F5 (SDR-AVS) & MG-B+SDR vs. AMaLGaM & +0.05$^{\approx}$ & +0.08 & +0.00$^{\approx}$ & $-$0.04$^{\approx}$ & $-$0.03$^{\approx}$ & $-$0.12 \\
\midrule
descriptive & MG-B vs. AMaLGaM & +0.02 & +0.00 & +0.15 & +0.22 & \textbf{+0.53} & \textbf{+0.73} \\
 & EMNA+AVS+AMS vs. AMaLGaM & +0.02 & $-$0.00 & +0.13 & +0.29 & \textbf{+0.62} & \textbf{+0.80} \\
 & MG-B vs. CMA-ES & $-$0.17 & +0.41 & +0.17 & \textbf{+0.90} & +1.02 & \textbf{+2.65} \\
 & MG-B vs. IPOP-CMA-ES & \textbf{+0.83} & \textbf{+1.41} & \textbf{+1.22} & \textbf{+1.82} & \textbf{+2.09} & \textbf{+3.64} \\
 & ASN+AVS vs. EMNA+AVS & $-$0.10 & \textbf{$-$1.23} & \textbf{$-$0.09} & \textbf{$-$3.49} & -- & -- \\
 & MG-S vs. MG-A & +0.00$^{\approx}$ & $-$0.07 & \textbf{$-$0.04}$^{\approx}$ & \textbf{$-$0.09} & $-$0.02$^{\approx}$ & \textbf{$-$0.09} \\
\bottomrule
\end{tabular}
\par\vspace{2pt}\begin{minipage}{\textwidth}\scriptsize Bold: significant at $\alpha=0.05$ (Wilcoxon signed-rank test over the 24 functions; in F1--F5 after Holm's correction over all tests of the family, in the descriptive rows uncorrected). $^{\approx}$: practically equivalent (90\% interval of HL within $\pm0.25$). Intervals, per-function tests and paired success tables are in Table \ref{S-tab:famfull16}.\end{minipage}
\end{table}

\clearpage
{\scriptsize
\setlength{\tabcolsep}{3pt}
\begin{longtable}{@{}llrlcccccccccc@{}}
\caption{Variant with elongation: confirmatory comparisons on instances 16--30 in full (difference first minus second in $y_{\mathrm{any}}$; negative: the first algorithm is better; columns as in Table \ref{S-tab:famfull})}\label{S-tab:famfull16}\\
\toprule
Fam. & Comparison & $D$ & start & mean & HL & 95\% interval & $p$ & $p$ (Holm) & equiv. & fn. b/w & sig. b/w & solved only 1st/2nd & McNemar $p$ \\
\midrule
\endfirsthead
\toprule
Fam. & Comparison & $D$ & start & mean & HL & 95\% interval & $p$ & $p$ (Holm) & equiv. & fn. b/w & sig. b/w & solved only 1st/2nd & McNemar $p$ \\
\midrule
\endhead
F1 & MG-A vs. always-on & 10 & std. & $-$0.12 & $-$0.02 & [$-$0.23, +0.03] & $0.25$ & $0.53$ & yes & 16/8 & 0/1 & 16/7 & $0.09$ \\
F1 & MG-A vs. always-on & 10 & dist. & $-$0.17 & $-$0.20 & [$-$0.33, $-$0.03] & $0.03$ & $0.17$ & no & 19/5 & 6/0 & 16/21 & $0.51$ \\
F1 & MG-A vs. always-on & 20 & std. & +0.02 & $-$0.02 & [$-$0.10, +0.14] & $0.73$ & $0.73$ & yes & 15/9 & 1/3 & 7/8 & $1.00$ \\
F1 & MG-A vs. always-on & 20 & dist. & $-$0.43 & $-$0.35 & [$-$0.59, $-$0.08] & $0.02$ & $0.13$ & no & 18/6 & 15/0 & 7/16 & $0.09$ \\
F1 & MG-A vs. always-on & 40 & std. & $-$0.10 & $-$0.10 & [$-$0.17, $-$0.04] & $0.005$ & $0.04$ & yes & 17/7 & 5/2 & 2/1 & $1.00$ \\
F1 & MG-A vs. always-on & 40 & dist. & $-$0.82 & $-$0.71 & [$-$1.17, $-$0.32] & $2\times10^{-5}$ & $0.0002$ & no & 23/1 & 18/0 & 23/2 & $2\times10^{-5}$ \\
F1 & MG-B vs. always-on & 10 & std. & $-$0.17 & $-$0.03 & [$-$0.24, $-$0.01] & $0.03$ & $0.17$ & yes & 17/7 & 2/0 & 14/7 & $0.19$ \\
F1 & MG-B vs. always-on & 10 & dist. & $-$0.37 & $-$0.28 & [$-$0.48, $-$0.14] & $0.002$ & $0.02$ & no & 20/4 & 10/0 & 19/19 & $1.00$ \\
F1 & MG-B vs. always-on & 20 & std. & $-$0.13 & $-$0.06 & [$-$0.16, +0.03] & $0.18$ & $0.53$ & yes & 15/9 & 3/0 & 24/9 & $0.01$ \\
F1 & MG-B vs. always-on & 20 & dist. & $-$0.71 & $-$0.44 & [$-$0.69, $-$0.23] & $0.0003$ & $0.004$ & no & 20/4 & 14/0 & 22/10 & $0.05$ \\
F1 & MG-B vs. always-on & 40 & std. & $-$0.18 & $-$0.15 & [$-$0.22, $-$0.09] & $4\times10^{-5}$ & $0.0006$ & yes & 20/4 & 7/0 & 4/1 & $0.38$ \\
F1 & MG-B vs. always-on & 40 & dist. & $-$1.02 & $-$0.85 & [$-$1.28, $-$0.39] & $1\times10^{-7}$ & $2\times10^{-6}$ & no & 24/0 & 17/0 & 26/1 & $4\times10^{-7}$ \\
F1 & MG-A vs. time-decay & 10 & std. & $-$0.39 & $-$0.16 & [$-$0.47, $-$0.05] & $0.005$ & $0.04$ & no & 19/5 & 7/0 & 33/16 & $0.02$ \\
F1 & MG-A vs. time-decay & 10 & dist. & $-$0.50 & $-$0.09 & [$-$0.29, 0.00] & $0.06$ & $0.26$ & yes & 18/6 & 2/0 & 29/8 & $0.0008$ \\
F1 & MG-A vs. time-decay & 20 & std. & $-$0.47 & $-$0.28 & [$-$0.78, $-$0.10] & $0.0004$ & $0.004$ & no & 19/5 & 6/0 & 24/5 & $0.0005$ \\
F1 & MG-A vs. time-decay & 20 & dist. & $-$0.83 & $-$0.37 & [$-$0.87, $-$0.17] & $1\times10^{-6}$ & $2\times10^{-5}$ & no & 22/2 & 9/0 & 16/5 & $0.03$ \\
F2 & MG-A vs. EMNA+AVS & 10 & std. & $-$0.62 & $-$0.15 & [$-$0.60, $-$0.02] & $0.005$ & $0.01$ & no & 19/5 & 2/0 & 35/7 & $2\times10^{-5}$ \\
F2 & MG-A vs. EMNA+AVS & 10 & dist. & $-$2.92 & $-$1.96 & [$-$3.46, $-$0.96] & $2\times10^{-6}$ & $1\times10^{-5}$ & no & 22/2 & 21/0 & 45/10 & $2\times10^{-6}$ \\
F2 & MG-A vs. EMNA+AVS & 20 & std. & $-$0.87 & $-$0.24 & [$-$1.03, $-$0.03] & $0.02$ & $0.02$ & no & 17/7 & 7/0 & 41/6 & $2\times10^{-7}$ \\
F2 & MG-A vs. EMNA+AVS & 20 & dist. & $-$5.29 & $-$4.96 & [$-$6.33, $-$3.44] & $1\times10^{-7}$ & $1\times10^{-6}$ & no & 24/0 & 24/0 & 73/1 & $8\times10^{-21}$ \\
F2 & MG-A vs. EMNA+AVS & 40 & std. & $-$0.67 & $-$0.16 & [$-$0.40, $-$0.07] & $2\times10^{-5}$ & $0.0001$ & no & 20/4 & 12/0 & 16/0 & $3\times10^{-5}$ \\
F2 & MG-A vs. EMNA+AVS & 40 & dist. & $-$6.31 & $-$6.19 & [$-$7.41, $-$5.18] & $1\times10^{-7}$ & $1\times10^{-6}$ & no & 24/0 & 24/0 & 90/0 & $2\times10^{-27}$ \\
F2 & MG-B vs. EMNA+AVS & 10 & std. & $-$0.68 & $-$0.22 & [$-$0.60, $-$0.06] & $0.0002$ & $0.0008$ & no & 19/5 & 2/0 & 34/8 & $7\times10^{-5}$ \\
F2 & MG-B vs. EMNA+AVS & 10 & dist. & $-$3.12 & $-$1.98 & [$-$3.56, $-$1.02] & $3\times10^{-6}$ & $2\times10^{-5}$ & no & 23/1 & 21/0 & 50/10 & $2\times10^{-7}$ \\
F2 & MG-B vs. EMNA+AVS & 20 & std. & $-$1.02 & $-$0.26 & [$-$1.11, $-$0.06] & $0.001$ & $0.004$ & no & 21/3 & 8/0 & 56/5 & $6\times10^{-12}$ \\
F2 & MG-B vs. EMNA+AVS & 20 & dist. & $-$5.57 & $-$5.17 & [$-$6.68, $-$3.75] & $1\times10^{-7}$ & $1\times10^{-6}$ & no & 24/0 & 24/0 & 95/2 & $6\times10^{-26}$ \\
F2 & MG-B vs. EMNA+AVS & 40 & std. & $-$0.75 & $-$0.18 & [$-$0.51, $-$0.09] & $4\times10^{-6}$ & $2\times10^{-5}$ & no & 23/1 & 9/0 & 18/0 & $8\times10^{-6}$ \\
F2 & MG-B vs. EMNA+AVS & 40 & dist. & $-$6.51 & $-$6.43 & [$-$7.59, $-$5.38] & $1\times10^{-7}$ & $1\times10^{-6}$ & no & 24/0 & 24/0 & 94/0 & $1\times10^{-28}$ \\
F3 & MG-A vs. EMNA+AVS+AMS & 10 & std. & +0.09 & +0.03 & [$-$0.02, +0.19] & $0.41$ & $1.00$ & yes & 11/13 & 0/2 & 15/14 & $1.00$ \\
F3 & MG-A vs. EMNA+AVS+AMS & 10 & dist. & +0.30 & +0.25 & [$-$0.01, +0.56] & $0.08$ & $1.00$ & no & 10/14 & 0/1 & 11/29 & $0.006$ \\
F3 & MG-A vs. EMNA+AVS+AMS & 20 & std. & +0.21 & +0.08 & [+0.03, +0.25] & $0.0004$ & $0.007$ & yes & 5/19 & 0/3 & 5/26 & $0.0002$ \\
F3 & MG-A vs. EMNA+AVS+AMS & 20 & dist. & +0.21 & +0.14 & [+0.02, +0.38] & $0.03$ & $0.50$ & no & 7/17 & 1/3 & 5/26 & $0.0002$ \\
F3 & MG-A vs. EMNA+AVS+AMS & 40 & std. & +0.03 & +0.03 & [$-$0.01, +0.06] & $0.15$ & $1.00$ & yes & 10/14 & 0/2 & 1/2 & $1.00$ \\
F3 & MG-A vs. EMNA+AVS+AMS & 40 & dist. & +0.10 & +0.04 & [$-$0.03, +0.16] & $0.34$ & $1.00$ & yes & 11/13 & 0/1 & 0/3 & $0.25$ \\
F3 & MG-B vs. EMNA+AVS+AMS & 10 & std. & +0.04 & +0.01 & [$-$0.03, +0.06] & $0.60$ & $1.00$ & yes & 12/12 & 1/0 & 13/14 & $1.00$ \\
F3 & MG-B vs. EMNA+AVS+AMS & 10 & dist. & +0.10 & 0.00 & [$-$0.11, +0.17] & $0.97$ & $1.00$ & yes & 12/12 & 2/0 & 15/28 & $0.07$ \\
F3 & MG-B vs. EMNA+AVS+AMS & 20 & std. & +0.06 & +0.03 & [$-$0.01, +0.12] & $0.14$ & $1.00$ & yes & 9/15 & 2/0 & 6/11 & $0.33$ \\
F3 & MG-B vs. EMNA+AVS+AMS & 20 & dist. & $-$0.07 & $-$0.01 & [$-$0.07, +0.06] & $0.77$ & $1.00$ & yes & 14/10 & 1/1 & 15/15 & $1.00$ \\
F3 & MG-B vs. EMNA+AVS+AMS & 40 & std. & $-$0.04 & $-$0.02 & [$-$0.05, +0.01] & $0.21$ & $1.00$ & yes & 15/9 & 3/0 & 3/2 & $1.00$ \\
F3 & MG-B vs. EMNA+AVS+AMS & 40 & dist. & $-$0.10 & $-$0.03 & [$-$0.21, +0.03] & $0.41$ & $1.00$ & yes & 13/11 & 3/0 & 1/0 & $1.00$ \\
F3 & MG-S vs. EMNA+AVS+AMS & 10 & std. & +0.08 & +0.04 & [$-$0.02, +0.22] & $0.34$ & $1.00$ & yes & 12/12 & 0/2 & 15/16 & $1.00$ \\
F3 & MG-S vs. EMNA+AVS+AMS & 10 & dist. & +0.13 & +0.05 & [$-$0.06, +0.35] & $0.34$ & $1.00$ & no & 10/14 & 1/0 & 11/20 & $0.15$ \\
F3 & MG-S vs. EMNA+AVS+AMS & 20 & std. & +0.13 & +0.05 & [+0.01, +0.16] & $0.03$ & $0.50$ & yes & 8/16 & 0/2 & 6/15 & $0.08$ \\
F3 & MG-S vs. EMNA+AVS+AMS & 20 & dist. & $-$0.01 & 0.00 & [$-$0.06, +0.06] & $0.90$ & $1.00$ & yes & 12/12 & 1/0 & 4/13 & $0.05$ \\
F3 & MG-S vs. EMNA+AVS+AMS & 40 & std. & $-$0.02 & $-$0.01 & [$-$0.03, +0.02] & $0.64$ & $1.00$ & yes & 12/12 & 1/0 & 1/2 & $1.00$ \\
F3 & MG-S vs. EMNA+AVS+AMS & 40 & dist. & $-$0.14 & $-$0.06 & [$-$0.21, 0.00] & $0.06$ & $0.91$ & yes & 17/7 & 2/0 & 2/1 & $1.00$ \\
F4 & MG-A+skew vs. MG-A & 10 & std. & +0.04 & +0.01 & [$-$0.03, +0.08] & $0.75$ & $1.00$ & yes & 12/12 & 0/0 & 7/16 & $0.09$ \\
F4 & MG-A+skew vs. MG-A & 10 & dist. & $-$0.04 & 0.00 & [$-$0.19, +0.06] & $0.99$ & $1.00$ & yes & 10/14 & 0/0 & 22/16 & $0.42$ \\
F4 & MG-A+skew vs. MG-A & 20 & std. & +0.03 & +0.02 & [$-$0.01, +0.06] & $0.25$ & $1.00$ & yes & 11/13 & 0/0 & 5/12 & $0.14$ \\
F4 & MG-A+skew vs. MG-A & 20 & dist. & $-$0.01 & $-$0.01 & [$-$0.05, +0.03] & $0.68$ & $1.00$ & yes & 14/10 & 0/0 & 12/9 & $0.66$ \\
F4 & MG-S vs. MG-B & 10 & std. & +0.05 & +0.03 & [$-$0.01, +0.08] & $0.10$ & $0.81$ & yes & 10/14 & 0/1 & 10/10 & $1.00$ \\
F4 & MG-S vs. MG-B & 10 & dist. & +0.03 & +0.05 & [$-$0.04, +0.26] & $0.21$ & $1.00$ & yes & 8/16 & 0/0 & 19/15 & $0.61$ \\
F4 & MG-S vs. MG-B & 20 & std. & +0.06 & +0.02 & [$-$0.01, +0.08] & $0.28$ & $1.00$ & yes & 10/14 & 0/2 & 9/13 & $0.52$ \\
F4 & MG-S vs. MG-B & 20 & dist. & +0.06 & +0.02 & [$-$0.03, +0.07] & $0.46$ & $1.00$ & yes & 10/14 & 0/1 & 13/22 & $0.18$ \\
F4 & MG-S vs. MG-B & 40 & std. & +0.02 & +0.02 & [0.00, +0.05] & $0.07$ & $0.69$ & yes & 7/17 & 0/2 & 2/4 & $0.69$ \\
F4 & MG-S vs. MG-B & 40 & dist. & $-$0.04 & $-$0.04 & [$-$0.10, +0.01] & $0.08$ & $0.71$ & yes & 15/9 & 1/1 & 1/1 & $1.00$ \\
F5 & MG-B+SDR vs. AMaLGaM & 10 & std. & +0.11 & +0.05 & [$-$0.01, +0.19] & $0.09$ & $0.45$ & yes & 8/16 & 0/0 & 3/5 & $0.73$ \\
F5 & MG-B+SDR vs. AMaLGaM & 10 & dist. & +0.20 & +0.08 & [$-$0.01, +0.39] & $0.12$ & $0.48$ & no & 9/15 & 0/0 & 7/20 & $0.02$ \\
F5 & MG-B+SDR vs. AMaLGaM & 20 & std. & $-$0.01 & 0.00 & [$-$0.04, +0.05] & $0.86$ & $0.86$ & yes & 13/11 & 2/0 & 4/6 & $0.75$ \\
F5 & MG-B+SDR vs. AMaLGaM & 20 & dist. & $-$0.10 & $-$0.04 & [$-$0.16, +0.06] & $0.37$ & $0.75$ & yes & 14/10 & 2/0 & 19/7 & $0.03$ \\
F5 & MG-B+SDR vs. AMaLGaM & 40 & std. & $-$0.10 & $-$0.03 & [$-$0.14, +0.02] & $0.21$ & $0.62$ & yes & 13/11 & 4/0 & 13/6 & $0.17$ \\
F5 & MG-B+SDR vs. AMaLGaM & 40 & dist. & $-$0.25 & $-$0.12 & [$-$0.35, $-$0.02] & $0.02$ & $0.14$ & no & 16/8 & 3/0 & 5/3 & $0.73$ \\
desc. & MG-B vs. AMaLGaM & 10 & std. & $-$0.10 & +0.02 & [$-$0.38, +0.18] & $0.86$ & -- & no & 9/15 & 1/7 & 20/4 & $0.002$ \\
desc. & MG-B vs. AMaLGaM & 10 & dist. & $-$0.03 & 0.00 & [$-$0.29, +0.21] & $0.99$ & -- & no & 11/13 & 0/5 & 16/10 & $0.33$ \\
desc. & MG-B vs. AMaLGaM & 20 & std. & +0.16 & +0.15 & [$-$0.04, +0.35] & $0.14$ & -- & no & 9/15 & 1/10 & 14/2 & $0.004$ \\
desc. & MG-B vs. AMaLGaM & 20 & dist. & +0.21 & +0.22 & [$-$0.03, +0.42] & $0.11$ & -- & no & 9/15 & 1/8 & 25/7 & $0.002$ \\
desc. & MG-B vs. AMaLGaM & 40 & std. & +0.66 & +0.53 & [+0.11, +1.05] & $0.0010$ & -- & no & 5/19 & 1/16 & 2/28 & $9\times10^{-7}$ \\
desc. & MG-B vs. AMaLGaM & 40 & dist. & +0.83 & +0.73 & [+0.25, +1.34] & $0.0001$ & -- & no & 3/21 & 1/17 & 4/21 & $0.0009$ \\
desc. & EMNA+AVS+AMS vs. AMaLGaM & 10 & std. & $-$0.14 & +0.02 & [$-$0.40, +0.18] & $0.86$ & -- & no & 10/14 & 1/6 & 20/3 & $0.0005$ \\
desc. & EMNA+AVS+AMS vs. AMaLGaM & 10 & dist. & $-$0.13 & 0.00 & [$-$0.37, +0.24] & $1.00$ & -- & no & 11/13 & 2/7 & 28/9 & $0.003$ \\
desc. & EMNA+AVS+AMS vs. AMaLGaM & 20 & std. & +0.10 & +0.13 & [$-$0.11, +0.36] & $0.53$ & -- & no & 12/12 & 2/10 & 22/5 & $0.002$ \\
desc. & EMNA+AVS+AMS vs. AMaLGaM & 20 & dist. & +0.27 & +0.29 & [0.00, +0.58] & $0.08$ & -- & no & 7/17 & 3/8 & 25/7 & $0.002$ \\
desc. & EMNA+AVS+AMS vs. AMaLGaM & 40 & std. & +0.70 & +0.62 & [+0.14, +1.06] & $0.0002$ & -- & no & 4/20 & 1/17 & 1/28 & $1\times10^{-7}$ \\
desc. & EMNA+AVS+AMS vs. AMaLGaM & 40 & dist. & +0.93 & +0.80 & [+0.36, +1.43] & $2\times10^{-5}$ & -- & no & 3/21 & 1/21 & 3/21 & $0.0003$ \\
desc. & MG-B vs. CMA-ES & 10 & std. & $-$0.54 & $-$0.17 & [$-$0.74, +0.22] & $0.42$ & -- & no & 13/11 & 5/8 & 45/2 & $2\times10^{-11}$ \\
desc. & MG-B vs. CMA-ES & 10 & dist. & +0.26 & +0.41 & [$-$0.14, +1.02] & $0.22$ & -- & no & 9/15 & 3/10 & 34/7 & $3\times10^{-5}$ \\
desc. & MG-B vs. CMA-ES & 20 & std. & +0.07 & +0.17 & [$-$0.50, +1.15] & $0.73$ & -- & no & 13/11 & 8/10 & 54/2 & $4\times10^{-14}$ \\
desc. & MG-B vs. CMA-ES & 20 & dist. & +1.04 & +0.90 & [+0.12, +2.00] & $0.03$ & -- & no & 5/19 & 3/14 & 57/17 & $3\times10^{-6}$ \\
desc. & MG-B vs. CMA-ES & 40 & std. & +1.45 & +1.02 & [$-$0.22, +3.30] & $0.18$ & -- & no & 11/13 & 7/12 & 20/59 & $1\times10^{-5}$ \\
desc. & MG-B vs. CMA-ES & 40 & dist. & +3.25 & +2.65 & [+0.90, +4.60] & $0.0002$ & -- & no & 2/22 & 1/19 & 21/57 & $6\times10^{-5}$ \\
desc. & MG-B vs. IPOP-CMA-ES & 10 & std. & +0.99 & +0.83 & [+0.44, +1.29] & $8\times10^{-6}$ & -- & no & 2/22 & 0/16 & 6/53 & $2\times10^{-10}$ \\
desc. & MG-B vs. IPOP-CMA-ES & 10 & dist. & +1.54 & +1.41 & [+0.88, +2.07] & $1\times10^{-7}$ & -- & no & 0/24 & 0/20 & 2/52 & $2\times10^{-13}$ \\
desc. & MG-B vs. IPOP-CMA-ES & 20 & std. & +1.39 & +1.22 & [+0.63, +2.01] & $4\times10^{-6}$ & -- & no & 1/23 & 1/19 & 2/30 & $2\times10^{-7}$ \\
desc. & MG-B vs. IPOP-CMA-ES & 20 & dist. & +2.30 & +1.82 & [+1.26, +2.94] & $1\times10^{-7}$ & -- & no & 0/24 & 0/24 & 3/37 & $2\times10^{-8}$ \\
desc. & MG-B vs. IPOP-CMA-ES & 40 & std. & +2.68 & +2.09 & [+1.27, +4.13] & $1\times10^{-7}$ & -- & no & 0/24 & 0/23 & 0/104 & $1\times10^{-31}$ \\
desc. & MG-B vs. IPOP-CMA-ES & 40 & dist. & +4.37 & +3.64 & [+2.33, +5.43] & $1\times10^{-7}$ & -- & no & 0/24 & 0/24 & 2/82 & $4\times10^{-22}$ \\
desc. & ASN+AVS vs. EMNA+AVS & 10 & std. & $-$0.32 & $-$0.10 & [$-$0.28, +0.01] & $0.13$ & -- & no & 15/9 & 2/0 & 17/13 & $0.58$ \\
desc. & ASN+AVS vs. EMNA+AVS & 10 & dist. & $-$1.51 & $-$1.23 & [$-$2.04, $-$0.63] & $7\times10^{-6}$ & -- & no & 22/2 & 10/0 & 18/11 & $0.26$ \\
desc. & ASN+AVS vs. EMNA+AVS & 20 & std. & $-$0.50 & $-$0.09 & [$-$0.28, $-$0.01] & $0.03$ & -- & no & 17/7 & 1/0 & 30/7 & $0.0002$ \\
desc. & ASN+AVS vs. EMNA+AVS & 20 & dist. & $-$3.53 & $-$3.49 & [$-$4.42, $-$2.42] & $1\times10^{-7}$ & -- & no & 24/0 & 21/0 & 56/3 & $1\times10^{-13}$ \\
desc. & MG-S vs. MG-A & 10 & std. & $-$0.01 & 0.00 & [$-$0.03, +0.04] & $0.94$ & -- & yes & 11/13 & 1/0 & 11/13 & $0.84$ \\
desc. & MG-S vs. MG-A & 10 & dist. & $-$0.17 & $-$0.07 & [$-$0.36, +0.02] & $0.19$ & -- & no & 15/9 & 0/0 & 20/11 & $0.15$ \\
desc. & MG-S vs. MG-A & 20 & std. & $-$0.09 & $-$0.04 & [$-$0.07, 0.00] & $0.05$ & -- & yes & 18/6 & 1/0 & 22/10 & $0.05$ \\
desc. & MG-S vs. MG-A & 20 & dist. & $-$0.22 & $-$0.09 & [$-$0.38, $-$0.01] & $0.05$ & -- & no & 18/6 & 2/0 & 16/4 & $0.01$ \\
desc. & MG-S vs. MG-A & 40 & std. & $-$0.06 & $-$0.02 & [$-$0.05, +0.01] & $0.20$ & -- & yes & 16/8 & 4/0 & 2/2 & $1.00$ \\
desc. & MG-S vs. MG-A & 40 & dist. & $-$0.23 & $-$0.09 & [$-$0.36, $-$0.02] & $0.04$ & -- & no & 18/6 & 3/0 & 4/0 & $0.12$ \\
\bottomrule
\end{longtable}
}

\subsection{Runs with $D=80$}
Protocol B2 had specified the standard BBOB functions for $D=80$. They cannot be built there, because the code of COCO that generates their rotation matrices is limited to $D^{2}<2000$; after the two functions that need no rotation had run and the others had stopped at this limit, the protocol was amended, before any comparison was looked at, to the large-scale BBOB suite of COCO (\texttt{bbob-largescale}) with instances 1--5. All other settings are those of the confirmatory test. Six algorithms were run, 1440 runs in all; the results are descriptive (Tables \ref{S-tab:d80old} and \ref{S-tab:d80oldb}). MG-B does not differ from EMNA+AVS+AMS (HL $-$0.01 and +0.02 for the two starts), is better than the tuned ungated configuration ($-$0.14 and $-$0.64), and is worse than AMaLGaM (+1.37 and +1.09) and CMA-ES (+3.03 and +5.18).

\begin{table}[htbp]\centering\footnotesize
\caption{Variant with elongation, $D=80$, large-scale BBOB suite, instances 1--5: mean $y_{\mathrm{any}}$ over the 24 functions and number of runs solved to $10^{-8}$ (of 120)}
\label{S-tab:d80old}
\begin{tabular}{@{}lcccc@{}}
\toprule
 & \multicolumn{2}{c}{standard start} & \multicolumn{2}{c}{distant start} \\
Algorithm & $y_{\mathrm{any}}$ & solved & $y_{\mathrm{any}}$ & solved \\
\midrule
EMNA+AVS+AMS & 0.60 & 11 & 3.45 & 8 \\
AMaLGaM & $-$0.70 & 36 & 2.02 & 31 \\
CMA-ES & $-$2.31 & 50 & $-$2.05 & 48 \\
MG-B & 0.57 & 11 & 3.40 & 7 \\
MG-S & 0.59 & 11 & 3.36 & 9 \\
always-on & 0.76 & 10 & 4.31 & 6 \\
\bottomrule
\end{tabular}
\par\vspace{2pt}\begin{minipage}{0.6\textwidth}\scriptsize always-on: no gate, $\sigma_{1}=e=0.5$.\end{minipage}
\end{table}

\begin{table}[htbp]\centering\footnotesize
\caption{Variant with elongation, $D=80$: Hodges--Lehmann estimate of the difference in $y_{\mathrm{any}}$ (first minus second) over the 24 functions, with 95\% bootstrap interval, uncorrected Wilcoxon $p$, and functions on which the first algorithm is better/worse}
\label{S-tab:d80oldb}
\begin{tabular}{@{}lcc@{}}
\toprule
Comparison & standard start & distant start \\
\midrule
MG-B vs. EMNA+AVS+AMS & $-$0.01 [$-$0.05, +0.03], $p=0.42$, 15/9 & +0.02 [$-$0.17, +0.11], $p=0.70$, 9/15 \\
MG-S vs. EMNA+AVS+AMS & 0.00 [$-$0.04, +0.06], $p=0.92$, 13/11 & $-$0.04 [$-$0.16, +0.02], $p=0.18$, 14/10 \\
MG-B vs. always-on & $-$0.14 [$-$0.26, $-$0.06], $p=6\times10^{-5}$, 19/5 & $-$0.64 [$-$1.18, $-$0.26], $p=1\times10^{-7}$, 24/0 \\
MG-S vs. MG-B & +0.02 [$-$0.02, +0.05], $p=0.29$, 9/15 & $-$0.07 [$-$0.15, +0.05], $p=0.19$, 16/8 \\
MG-B vs. AMaLGaM & +1.37 [+0.29, +1.87], $p=8\times10^{-6}$, 2/22 & +1.09 [+0.60, +2.13], $p=4\times10^{-6}$, 3/21 \\
EMNA+AVS+AMS vs. AMaLGaM & +1.39 [+0.33, +1.87], $p=8\times10^{-7}$, 1/23 & +1.22 [+0.67, +2.07], $p=6\times10^{-7}$, 2/22 \\
MG-B vs. CMA-ES & +3.03 [+0.43, +4.21], $p=0.0001$, 3/21 & +5.18 [+2.89, +6.86], $p=1\times10^{-7}$, 0/24 \\
AMaLGaM vs. CMA-ES & +1.18 [+0.06, +3.02], $p=0.04$, 7/17 & +3.54 [+1.86, +5.06], $p=1\times10^{-5}$, 1/23 \\
\bottomrule
\end{tabular}
\end{table}

\section{Results of the skewness-only predecessor}
\label{S-sec:pred}

The following tables and figures come from the first form of this work, in which the skewness was the only correction (the gated configuration SN, there called ASN+AVS), and are reproduced for reference. They were obtained with the implementation of that time (constrained search space with clipping, generations instead of evaluations as the budget, fixed thresholds $(2,6)$, decay 0.9) and with unpaired tests; they are not part of the evidence of the manuscript, but they show the same pattern: a clear gain when the optimum is distant, none when it is central, and no benefit from the gate for the skewness. All settings use the population size $N=\lfloor17+3D^{1.5}\rfloor$, truncation ratio 0.35, elitism, the same random seeds for all algorithms and 30 runs per setting. Algorithm names: UMDA$_{G}$ is the Gaussian EDA with diagonal covariance, EMNA the Gaussian EDA with full covariance, ASN the skewness-only algorithm without variance scaling, ASN+AVS with the improvement-triggered variance scaling (decay 0.9, multiplier in [1, 10]), ASN-fixed+AVS the fixed maximal skewness ($\iota_{t}=1$ from the second generation on) and ASN-moment+AVS a moment-matched magnitude. Comparisons are two-sided Mann--Whitney tests on the final error or on the generations to success (failures counted as 801) with Holm correction within the family given in each table note, at $\alpha=0.05$; in parentheses, ``+'' significantly better, ``$-$'' significantly worse, ``$\approx$'' no significant difference.

\subsection{Distant optimum}

Table \ref{S-tab:offc} gives the distant-optimum problems on which the skewness-only algorithm was first studied (initial distribution $\mathcal{N}(\mathbf{0},I)$, optimum at $s\mathbf{1}$, 30 runs, 800 generations; these problems were used in the design of the algorithm, so they are not an independent test). On the 30-dimensional sphere and rotated ellipsoid, EMNA+AVS solved no run because elitism keeps a best initial solution that the shrinking distribution needs many generations to reach, ASN+AVS solved all runs, and EMNA+AVS+AMS solved all runs faster; on the Rosenbrock function the mean shift is what makes the difference (7/30 runs for ASN+AVS against 30/30 for EMNA+AVS+AMS). The ablations ASN-fixed (skewness fixed at $\rho_{\max}$, no gate) and ASN-moment (skewness estimated from the third-moment vector $\mathbf{g}$ of Section \ref{sec:skew} of the manuscript, capped at $d_{\max}$) are included; the latter illustrates the noise argument of Section \ref{sec:skew} of the manuscript.
\begin{table*}[!htbp]
\caption{Distant optimum: successes and median generations of successful runs (30 runs)}
\label{S-tab:offc}
\centering \scriptsize
\begin{tabular}{@{}lccccccc@{}}
\toprule
$D$ & Function & EMNA+AVS & ASN+AVS & ASN-fixed+AVS & ASN-moment+AVS & EMNA+AVS+AMS & ASN+AVS+AMS \\
\midrule
10 & Sphere & 30/30, 290 & 30/30, 205 (+) & 30/30, 198 (+) & 30/30, 287.5 ($\approx$) & 30/30, 179 (+) & 30/30, 176.5 (+) \\
10 & Rotated ellipsoid & 30/30, 335.5 & 30/30, 247.5 (+) & 30/30, 249.5 (+) & 30/30, 327 ($\approx$) & 30/30, 218.5 (+) & 30/30, 210.5 (+) \\
10 & Rosenbrock & 22/30, 527 & 26/30, 410 (+) & 27/30, 457 (+) & 24/30, 450.5 ($\approx$) & 27/30, 317 (+) & 25/30, 316 (+) \\
10 & Rotated Rastrigin & 0/30 [2.98] & 1/30, 235 ($\approx$) & 3/30, 233 ($\approx$) & 3/30, 306 ($\approx$) & 0/30 [3.98] ($\approx$) & 0/30 [3.98] ($\approx$) \\
30 & Sphere & 0/30 [$2.59\times10^{4}$] & 30/30, 361.5 (+) & 30/30, 362.5 (+) & 16/30, 720.5 (+) & 30/30, 309.5 (+) & 30/30, 308 (+) \\
30 & Rotated ellipsoid & 0/30 [2,647] & 30/30, 494.5 (+) & 30/30, 500.5 (+) & 0/30 [$6.61\times10^{4}$] ($\approx$) & 30/30, 394.5 (+) & 30/30, 405 (+) \\
30 & Rosenbrock & 0/30 [6,610] & 7/30, 686 (+) & 3/30, 705 ($\approx$) & 0/30 [37.4] ($\approx$) & 30/30, 703.5 (+) & 30/30, 721.5 (+) \\
30 & Rotated Rastrigin & 0/30 [3.98] & 0/30 [3.98] ($\approx$) & 0/30 [3.98] ($\approx$) & 0/30 [4.97] ($\approx$) & 0/30 [5.97] ($\approx$) & 0/30 [4.97] ($\approx$) \\
\bottomrule
\end{tabular}
\par\vspace{2pt}\begin{minipage}{\textwidth}\scriptsize Initial distribution $\mathcal{N}(\mathbf{0},I)$, optimum at $s\mathbf{1}$ ($s=100$ for $D=10$, $s=30$ for $D=30$), 800 generations, success if $f<10^{-8}$; if no run succeeded, the median final error is given in brackets. In parentheses: comparison of the generations to success with EMNA+AVS (failures counted as 801; Mann--Whitney, Holm over the eight problems). The last two columns add the anticipated mean shift (AMS) of AMaLGaM \cite{bosman2013}; their parentheses compare EMNA+AVS+AMS with EMNA+AVS and ASN+AVS+AMS with ASN+AVS.\end{minipage}
\end{table*}

\subsection{Centered benchmark functions}

\begin{figure*}[!htbp]
\centering
\includegraphics[width=0.9\textwidth]{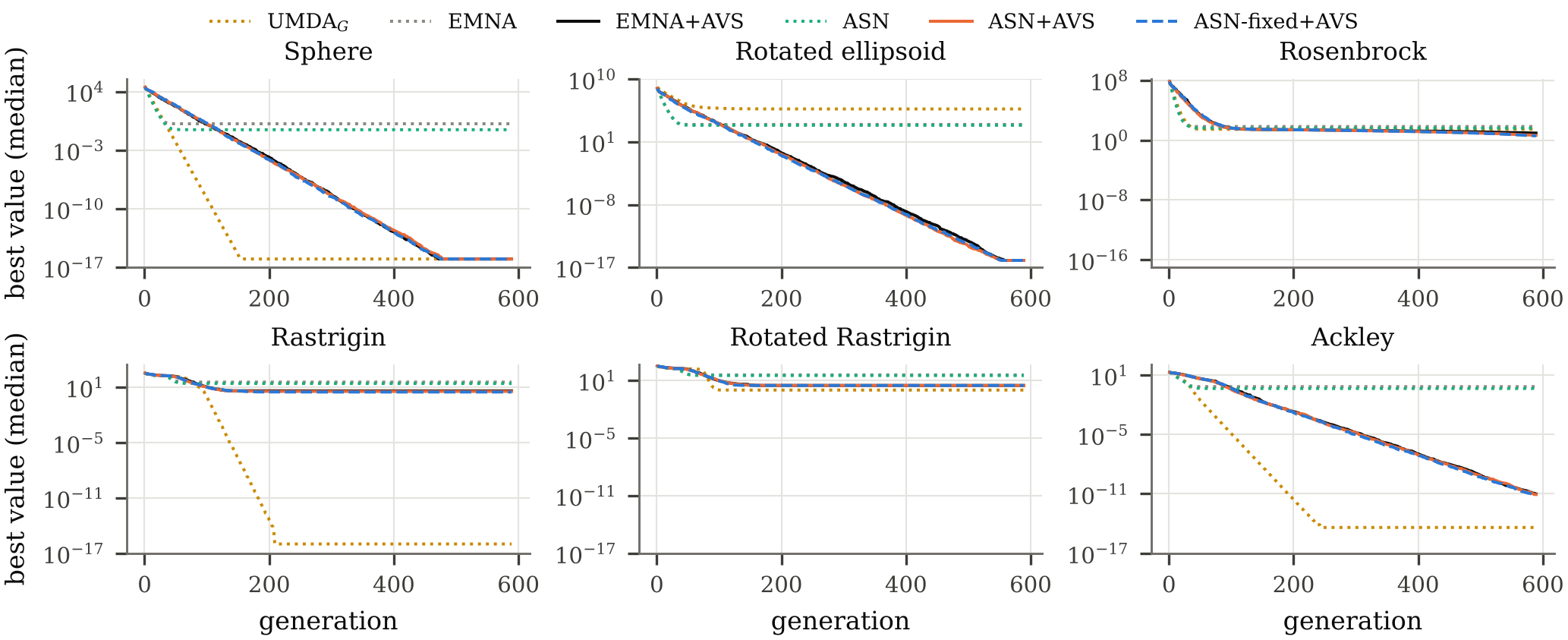}
\caption{Median best value on the centered functions ($D=30$, 30 runs). Values below $10^{-16}$ are drawn at $10^{-16}$.}
\label{S-fig:bench}
\end{figure*}

The six functions of Table \ref{S-tab:bench} were used with $D=30$, $N=509$, $M=178$ and 589 generations (300,310 evaluations); the initial population is uniform in the box and the optimum lies near its center (at $\mathbf{1}$ for Rosenbrock). Fig. \ref{S-fig:bench} shows the median convergence curves.
\begin{table*}[!htbp]
\caption{Final error on centered functions ($D=30$; mean $\pm$ standard deviation of 30 runs)}
\label{S-tab:bench}
\centering \footnotesize
\resizebox{\textwidth}{!}{\begin{tabular}{@{}lcccc@{}}
\toprule
Function & Range & UMDA$_{G}$ & EMNA & ASN \\
\midrule
Sphere & [$-$100, 100] & $6.10\times10^{-75}$ $\pm$ $2.29\times10^{-75}$ & 2.84 $\pm$ 3.27 & 1.47 $\pm$ 3.22 (+) \\
Rotated ellipsoid & [$-$100, 100] & $5.62\times10^{5}$ $\pm$ $2.29\times10^{5}$ & 4,433 $\pm$ 3,443 & 2,729 $\pm$ 1,977 ($\approx$) \\
Rosenbrock & [$-$30, 30] & 27.9 $\pm$ 0.0538 & 84.3 $\pm$ 49.8 & 49.8 $\pm$ 23.1 (+) \\
Rastrigin & [$-$5.12, 5.12] & 0.0995 $\pm$ 0.304 & 36.7 $\pm$ 10.7 & 28.0 $\pm$ 10.9 (+) \\
Rotated Rastrigin & [$-$5.12, 5.12] & 0.796 $\pm$ 0.843 & 36.4 $\pm$ 8.42 & 31.1 $\pm$ 11.6 ($\approx$) \\
Ackley & [$-$32, 32] & $4.00\times10^{-15}$ $\pm$ 0 & 0.746 $\pm$ 0.478 & 0.493 $\pm$ 0.402 ($\approx$) \\
\midrule
 &  & \textbf{EMNA+AVS} & \textbf{ASN+AVS} & \textbf{ASN-fixed+AVS} \\
Sphere & [$-$100, 100] & $5.57\times10^{-20}$ $\pm$ $2.35\times10^{-19}$ & $1.17\times10^{-19}$ $\pm$ $5.70\times10^{-19}$ ($\approx$) & $5.70\times10^{-21}$ $\pm$ $8.36\times10^{-21}$ ($\approx$) \\
Rotated ellipsoid & [$-$100, 100] & $8.43\times10^{-17}$ $\pm$ $2.36\times10^{-16}$ & $3.37\times10^{-17}$ $\pm$ $1.11\times10^{-16}$ ($\approx$) & $2.77\times10^{-16}$ $\pm$ $1.26\times10^{-15}$ ($\approx$) \\
Rosenbrock & [$-$30, 30] & 9.28 $\pm$ 2.80 & 4.43 $\pm$ 1.38 (+) & 3.63 $\pm$ 0.960 (+) \\
Rastrigin & [$-$5.12, 5.12] & 3.68 $\pm$ 1.83 & 3.62 $\pm$ 2.03 ($\approx$) & 3.32 $\pm$ 1.74 ($\approx$) \\
Rotated Rastrigin & [$-$5.12, 5.12] & 3.32 $\pm$ 1.44 & 3.52 $\pm$ 2.00 ($\approx$) & 2.59 $\pm$ 1.56 ($\approx$) \\
Ackley & [$-$32, 32] & $1.57\times10^{-11}$ $\pm$ $1.70\times10^{-11}$ & $2.68\times10^{-11}$ $\pm$ $4.29\times10^{-11}$ ($\approx$) & $1.40\times10^{-11}$ $\pm$ $2.12\times10^{-11}$ ($\approx$) \\
\bottomrule
\end{tabular}}
\par\vspace{2pt}\begin{minipage}{\textwidth}\scriptsize Upper block: no variance scaling; lower block: with AVS. In parentheses: comparison with EMNA (upper) or EMNA+AVS (lower) by the two-sided Mann--Whitney test with Holm correction over the six functions; ``+'' significantly better, ``$-$'' significantly worse, ``$\approx$'' no significant difference ($\alpha=0.05$).\end{minipage}
\end{table*}

\subsection{Distance to the optimum and variance control}

\begin{figure*}[!htbp]
\centering
\includegraphics[width=0.9\textwidth]{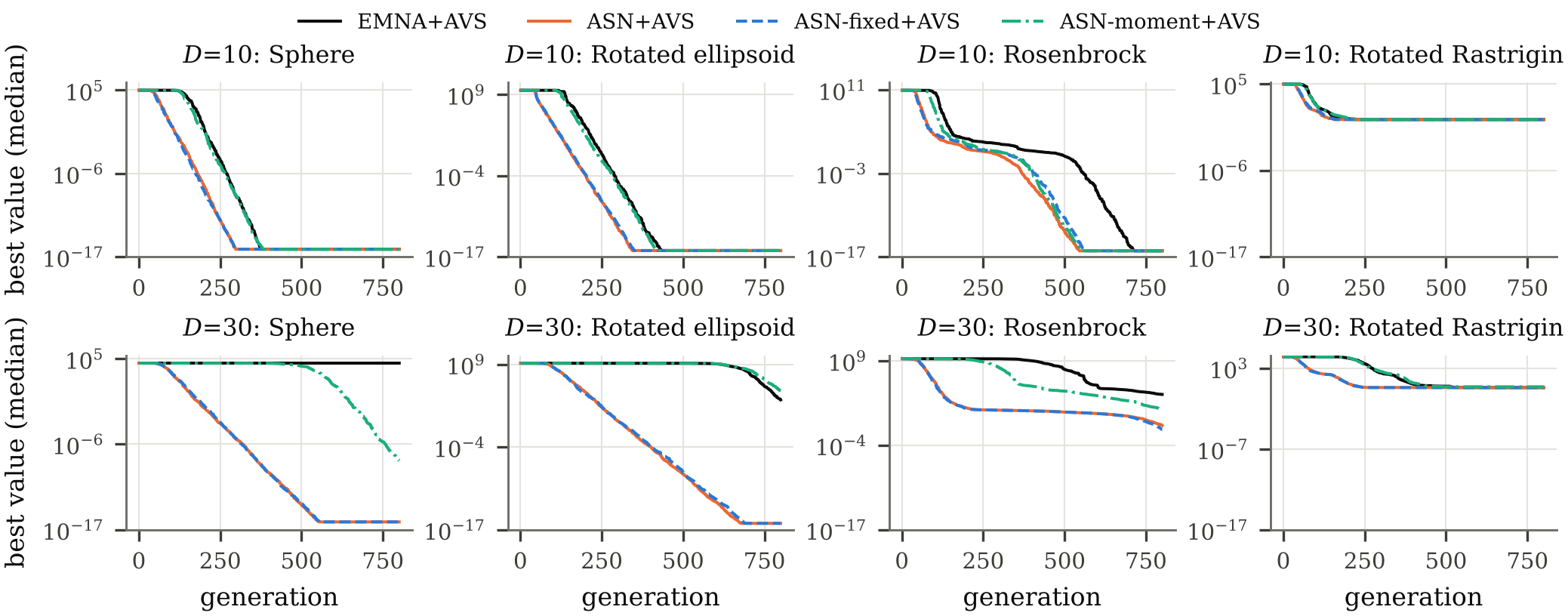}
\caption{Median best value when the optimum is far from the initial distribution (30 runs).}
\label{S-fig:off}
\end{figure*}

Table \ref{S-tab:sweep} repeats the shifted sphere of Table \ref{S-tab:offc} for optimum distances $s$ from 0 (optimum at the initial mean) to 100 with the same seeds and settings (initial distribution $\mathcal{N}(\mathbf{0},I)$, box $\pm10^{4}$, 800 generations). Table \ref{S-tab:ctl} varies the variance control of the Gaussian baseline on the eight distant problems of Table \ref{S-tab:offc}: no elitism, constant multipliers $\eta$, and a maturity-driven variance scaling (MVS). Fig. \ref{S-fig:off} shows the median curves on the distant problems.
\begin{table}[htbp]
\caption{Distance to the optimum: successes and median generations of successful runs (shifted sphere, 30 runs)}
\label{S-tab:sweep}
\centering \footnotesize
\begin{tabular}{@{}lccc@{}}
\toprule
$D$ & $s$ & EMNA+AVS & ASN+AVS \\
\midrule
10 & 0 & 30/30, 96.5 & 30/30, 89.5 ($\approx$) \\
10 & 1 & 30/30, 118.5 & 30/30, 105 (+) \\
10 & 3 & 30/30, 160.5 & 30/30, 142.5 (+) \\
10 & 10 & 30/30, 214 & 30/30, 163 (+) \\
10 & 30 & 30/30, 251 & 30/30, 184 (+) \\
10 & 100 & 30/30, 290 & 30/30, 205 (+) \\
30 & 0 & 30/30, 208.5 & 30/30, 207 ($\approx$) \\
30 & 1 & 30/30, 228 & 30/30, 221 ($\approx$) \\
30 & 3 & 30/30, 334 & 30/30, 264.5 (+) \\
30 & 10 & 21/30, 525 & 30/30, 308 (+) \\
30 & 30 & 0/30 & 30/30, 361.5 (+) \\
30 & 100 & 0/30 & 30/30, 427.5 (+) \\
\bottomrule
\end{tabular}
\par\vspace{2pt}\begin{minipage}{\columnwidth}\scriptsize Optimum at $s\mathbf{1}$, i.e., $s\sqrt{D}$ initial standard deviations from the initial mean along the gradient; other settings and the comparison in parentheses as in Table \ref{S-tab:offc} (Holm over the six distances of each dimension).\end{minipage}
\end{table}

\begin{table*}[!htbp]
\caption{Variance control of the Gaussian baseline on the distant problems of Table \ref{S-tab:offc}: successes and median generations of successful runs (30 runs)}
\label{S-tab:ctl}
\centering \scriptsize
\resizebox{\textwidth}{!}{\begin{tabular}{@{}clccccc@{}}
\toprule
$D$ & Function & EMNA+AVS, no elitism & EMNA, $\eta=2$ & EMNA, $\eta=4$ & EMNA+MVS & ASN+MVS \\
\midrule
10 & Sphere & 0/30, $f_{\mathrm{med}}=9.83\times10^{4}$ & 0/30, $f_{\mathrm{med}}=7.46\times10^{-6}$ & 0/30, $f_{\mathrm{med}}=9.43\times10^{4}$ & 27/30, 311 & 28/30, 326.5 ($\approx$) \\
10 & Rotated ellipsoid & 0/30, $f_{\mathrm{med}}=4.88\times10^{9}$ & 0/30, $f_{\mathrm{med}}=0.55$ & 0/30, $f_{\mathrm{med}}=2.46\times10^{9}$ & 29/30, 405 & 30/30, 458.5 ($\approx$) \\
10 & Rosenbrock & 1/30, 582 & 0/30, $f_{\mathrm{med}}=1.87\times10^{7}$ & 0/30, $f_{\mathrm{med}}=7.74\times10^{10}$ & 0/30, $f_{\mathrm{med}}=7.4$ & 0/30, $f_{\mathrm{med}}=9.5$ ($\approx$) \\
10 & Rotated Rastrigin & 0/30, $f_{\mathrm{med}}=9.84\times10^{4}$ & 0/30, $f_{\mathrm{med}}=65$ & 0/30, $f_{\mathrm{med}}=9.48\times10^{4}$ & 1/30, 604 & 2/30, 491 ($\approx$) \\
30 & Sphere & 0/30, $f_{\mathrm{med}}=2.61\times10^{4}$ & 0/30, $f_{\mathrm{med}}=2.58\times10^{4}$ & 0/30, $f_{\mathrm{med}}=2.55\times10^{4}$ & 30/30, 417 & 30/30, 400.5 ($\approx$) \\
30 & Rotated ellipsoid & 0/30, $f_{\mathrm{med}}=1.50\times10^{9}$ & 0/30, $f_{\mathrm{med}}=1.28\times10^{9}$ & 0/30, $f_{\mathrm{med}}=1.15\times10^{9}$ & 29/30, 460 & 30/30, 463 ($\approx$) \\
30 & Rosenbrock & 0/30, $f_{\mathrm{med}}=2.04\times10^{9}$ & 0/30, $f_{\mathrm{med}}=2.03\times10^{9}$ & 0/30, $f_{\mathrm{med}}=2.01\times10^{9}$ & 0/30, $f_{\mathrm{med}}=53$ & 0/30, $f_{\mathrm{med}}=48$ ($\approx$) \\
30 & Rotated Rastrigin & 1/30, 550 & 0/30, $f_{\mathrm{med}}=2.61\times10^{4}$ & 0/30, $f_{\mathrm{med}}=2.58\times10^{4}$ & 0/30, $f_{\mathrm{med}}=7.1$ & 0/30, $f_{\mathrm{med}}=8$ ($\approx$) \\
\bottomrule
\end{tabular}}
\par\vspace{2pt}\begin{minipage}{\textwidth}\scriptsize Settings and seeds as in Table \ref{S-tab:offc}; $f_{\mathrm{med}}$ is the median final error when no run succeeded. MVS: the variance multiplier is driven by the maturity statistic instead of by improvements, $s\leftarrow\min(s\,e^{0.3(\sqrt{m_{t}}-1)},10)$ if $m_{t}>1$ and $s\leftarrow\max(0.9s,1)$ otherwise (untuned). In parentheses: ASN+MVS against EMNA+MVS (generations, failures counted as 801; Mann--Whitney, Holm over the eight problems).\end{minipage}
\end{table*}

\subsection{Mature phase and noisy problems}

Table \ref{S-tab:local} starts the distribution at the optimum so that the run is mature throughout and compares the generations to $f<10^{-10}$; Table \ref{S-tab:noise} adds Gaussian noise to the function values (the pre-registered noise experiment of the earlier version).
\begin{table*}[!htbp]
\caption{Mature phase: successes and median generations to $f<10^{-10}$}
\label{S-tab:local}
\centering \footnotesize
\begin{tabular}{@{}lccccc@{}}
\toprule
$D$ & Function & EMNA
with AVS & ASN
with AVS & ASN-fixed
with AVS & Mean $\iota$ of ASN+AVS \\
\midrule
10 & Sphere & 30/30, 118 & 30/30, 113 ($\approx$) & 30/30, 119 ($\approx$) & 0.013 \\
10 & Rotated ellipsoid & 30/30, 159 & 30/30, 157 ($\approx$) & 30/30, 162 ($\approx$) & 0.009 \\
30 & Sphere & 30/30, 252 & 30/30, 255.5 ($\approx$) & 30/30, 253.5 ($\approx$) & 0.000 \\
30 & Rotated ellipsoid & 30/30, 337.5 & 30/30, 343 ($\approx$) & 30/30, 340.5 ($\approx$) & 0.000 \\
\bottomrule
\end{tabular}
\par\vspace{2pt}\begin{minipage}{\textwidth}\scriptsize Initial distribution $\mathcal{N}(\mathbf{0},I)$ centered at the optimum; 30 runs. In parentheses: comparison of generations with EMNA+AVS (Mann--Whitney, Holm over the eight comparisons).\end{minipage}
\end{table*}

\begin{table}[htbp]
\caption{Noisy problems: median noise-free error of the final mean (30 runs)}
\label{S-tab:noise}
\centering \footnotesize
\begin{tabular}{@{}lccc@{}}
\toprule
Setting & EMNA+AVS & ASN+AVS & ASN-fixed+AVS \\
\midrule
C, 10, S, $10^{-6}$ & $7.2\times10^{-8}$ & $8.0\times10^{-8}$ & $6.9\times10^{-8}$ \\
C, 10, S, $10^{-2}$ & $6.6\times10^{-4}$ & $7.4\times10^{-4}$ & $7.3\times10^{-4}$ \\
C, 10, E, $10^{-6}$ & $7.5\times10^{-8}$ & $7.8\times10^{-8}$ & $8.0\times10^{-8}$ \\
C, 10, E, $10^{-2}$ & $7.0\times10^{-4}$ & $8.1\times10^{-4}$ & $7.1\times10^{-4}$ \\
C, 30, S, $10^{-6}$ & $5.3\times10^{-8}$ & $4.8\times10^{-8}$ & $4.4\times10^{-8}$ \\
C, 30, S, $10^{-2}$ & $4.8\times10^{-4}$ & $5.2\times10^{-4}$ & $4.5\times10^{-4}$ \\
C, 30, E, $10^{-6}$ & $5.6\times10^{-8}$ & $4.7\times10^{-8}$ & $4.8\times10^{-8}$ \\
C, 30, E, $10^{-2}$ & $5.2\times10^{-4}$ & $4.7\times10^{-4}$ & $4.6\times10^{-4}$ \\
D, 10, S, $10^{-6}$ & $8.7\times10^{-8}$ & $8.9\times10^{-8}$ & $6.8\times10^{-8}$ \\
D, 10, S, $10^{-2}$ & $9.4\times10^{-4}$ & $7.9\times10^{-4}$ & $7.3\times10^{-4}$ \\
D, 10, E, $10^{-6}$ & $7.8\times10^{-8}$ & $6.3\times10^{-8}$ & $6.9\times10^{-8}$ \\
D, 10, E, $10^{-2}$ & $6.9\times10^{-4}$ & $7.1\times10^{-4}$ & $6.4\times10^{-4}$ \\
D, 30, S, $10^{-6}$ & $2.6\times10^{4}$ & $4.7\times10^{-8}$$^{\dagger}$ & $5.0\times10^{-8}$ \\
D, 30, S, $10^{-2}$ & $2.6\times10^{4}$ & $5.0\times10^{-4}$$^{\dagger}$ & $4.8\times10^{-4}$ \\
D, 30, E, $10^{-6}$ & 0.0252 & $4.3\times10^{-8}$$^{\dagger}$ & $4.7\times10^{-8}$ \\
D, 30, E, $10^{-2}$ & 0.0225 & $4.6\times10^{-4}$$^{\dagger}$ & $4.3\times10^{-4}$ \\
\bottomrule
\end{tabular}
\par\vspace{2pt}\begin{minipage}{\columnwidth}\scriptsize Setting: start (C centered as in Table \ref{S-tab:local}, 500 generations; D distant as in Table \ref{S-tab:offc}, 800 generations), $D$, function (S sphere, E rotated ellipsoid), $\sigma_{\varepsilon}$ of the additive noise $\sigma_{\varepsilon}\xi$, $\xi\sim\mathcal{N}(0,1)$, the same for the three algorithms. Entries: median of $f(\boldsymbol{\mu}_{T})-f^{\ast}$ for the mean $\boldsymbol{\mu}_{T}$ of the last generation. ASN+AVS and ASN-fixed+AVS do not differ significantly in any setting; $^{\dagger}$ ASN+AVS significantly better than EMNA+AVS (Mann--Whitney, Holm over the 16 settings).\end{minipage}
\end{table}

\subsection{Cost per generation}

Table \ref{S-tab:cost} lists the work and the parallel depth of each step of one generation. The mean shift of Section \ref{S-sec:gate} adds a rank-one term of the same $O(ND)$ order.
\begin{table}[htbp]
\caption{Cost per generation of the skewness-only algorithm (one run)}
\label{S-tab:cost}
\centering \footnotesize
\begin{tabular}{@{}llcc@{}}
\toprule
Step & Matrix operation & Work & Depth \\
\midrule
Evaluation & $\mathbf{f}=F(X)$ & $NC_{F}$ & --- \\
Selection & argsort, row gather & $O(N\log N+MD)$ & $O(\log N)$ \\
Mean, covariance & $X_{s}^{T}\mathbf{1}/M$, $X_{c}^{T}X_{c}/M$ & $O(MD^{2})$ & $O(\log M)$ \\
Cholesky & $L=\mathrm{chol}(\Sigma)$ & $D^{3}/3$ & $O(D)$ \\
Maturity, skewness* & solve $L[\mathbf{z},\mathbf{u}]=[\Delta\boldsymbol{\mu},\mathbf{P}]$ & $O(D^{2})$ & $O(D)$ \\
Sampling & $ZL^{T}$; $Z\hat{\mathbf{d}}$, $L\hat{\mathbf{d}}$, rank-one update* & $O(ND^{2})+O(ND)$ & $O(\log D)$ \\
Clip, elitism, AVS & element-wise & $O(ND)$ & $O(1)$ \\
\bottomrule
\end{tabular}
\par\vspace{2pt}\begin{minipage}{\columnwidth}\scriptsize * New relative to EMNA. Depth is the length of the critical path, that is, the running time with an unbounded number of processors.\end{minipage}
\end{table}

\subsection{Optimum close to a bound of a constrained search space}

\begin{table}[htbp]\centering\small
\caption{Optimum close to the bound: successes and median generations of successful runs (30 runs)}
\label{S-tab:s1}
\begin{tabular}{@{}cllccc@{}}
\toprule
$D$ & Function & $\delta$ & EMNA+AVS & ASN+AVS & ASN-fixed+AVS \\
\midrule
10 & Sphere & 0.01 & 30/30, 325.5 & 30/30, 239.5 ($\approx$, $\approx$) & 30/30, 203 \\
10 & Sphere & 1 & 30/30, 181 & 30/30, 157 ($\approx$, $\approx$) & 30/30, 151 \\
10 & Rotated ellipsoid & 0.01 & 30/30, 461 & 30/30, 303 ($\approx$, $\approx$) & 30/30, 293.5 \\
10 & Rotated ellipsoid & 1 & 30/30, 210 & 30/30, 192 ($\approx$, $\approx$) & 30/30, 189.5 \\
30 & Sphere & 0.01 & 0/30 [0.753] & 16/30, 774.5 ($-$, $+$) & 28/30, 732 \\
30 & Sphere & 1 & 30/30, 455.5 & 30/30, 329 ($\approx$, $+$) & 30/30, 309 \\
30 & Rotated ellipsoid & 0.01 & 0/30 [$4.50\times10^{4}$] & 0/30 [13.4] ($-$, $+$) & 0/30 [$3.15\times10^{-3}$] \\
30 & Rotated ellipsoid & 1 & 17/30, 650 & 30/30, 424.5 ($\approx$, $+$) & 30/30, 399.5 \\
\bottomrule
\end{tabular}
\par\vspace{3pt}\begin{minipage}{0.95\textwidth}\footnotesize Optimum at $(5-\delta)\mathbf{1}$ in the box $[-5,5]^{D}$ (samples clipped to the box), initial distribution $\mathcal{N}(\mathbf{0},I)$, 800 generations, seed 3, success if $f<10^{-8}$; if no run succeeded, the median final error is given in brackets. In parentheses: ASN+AVS against ASN-fixed+AVS and against EMNA+AVS (two-sided Mann--Whitney test on the final error, Holm correction over the 8 settings; $+$ better, $-$ worse, $\approx$ no significant difference at $\alpha=0.05$). Verdict of the pre-specified rule for ASN+AVS against ASN-fixed+AVS: opposite. Not specified in advance: on the generations to $10^{-8}$ (failures counted as 801; Mann--Whitney, Holm over the 8 settings) ASN-fixed+AVS was significantly faster in $D=10$, sphere, $\delta=0.01$; $D=30$, sphere, $\delta=0.01$; $D=30$, sphere, $\delta=1$; $D=30$, rotated ellipsoid, $\delta=1$, and ASN+AVS in none; the success counts differ significantly only for $D=30$, sphere, $\delta=0.01$ (Fisher, corrected $p=0.0073$).\end{minipage}
\end{table}
\begin{table}[htbp]\centering\small
\caption{Diagnostic of the two significant settings of Table \ref{S-tab:s1} ($D=30$, $\delta=0.01$): medians over the 30 runs (same seeds as Table \ref{S-tab:s1}; not pre-registered)}
\label{S-tab:s2}
\resizebox{\textwidth}{!}{\begin{tabular}{@{}llccccccc@{}}
\toprule
Function, algorithm & Quantity & $t=50$ & $t=100$ & $t=200$ & $t=300$ & $t=400$ & $t=600$ & $t=800$ \\
\midrule
Sphere, ASN+AVS & immaturity $\iota_t$ & 0.64 & 0.45 & 0.99 & 0.98 & 0.98 & 0.73 & 0.00 \\
 & distance of the mean & 4.57 & 3.04 & 1.43 & 0.697 & 0.258 & 0.0415 & $7.29\times10^{-6}$ \\
 & best value & 322 & 145 & 28.1 & 5.7 & 0.988 & 0.019 & $5.64\times10^{-9}$ \\[2pt]
Sphere, ASN-fixed+AVS & immaturity $\iota_t$ & 1.00 & 1.00 & 1.00 & 1.00 & 1.00 & 1.00 & 1.00 \\
 & distance of the mean & 4.67 & 2.95 & 1.3 & 0.517 & 0.217 & 0.0169 & $3.81\times10^{-7}$ \\
 & best value & 296 & 115 & 23.1 & 3.44 & 0.621 & $3.22\times10^{-3}$ & $1.65\times10^{-11}$ \\[2pt]
Rotated ellipsoid, ASN+AVS & immaturity $\iota_t$ & 1.00 & 0.94 & 0.98 & 1.00 & 1.00 & 0.48 & 0.51 \\
 & distance of the mean & 7.21 & 7.04 & 6.31 & 5.16 & 3.33 & 0.898 & 0.166 \\
 & best value & $2.52\times10^{6}$ & $5.93\times10^{5}$ & $1.07\times10^{5}$ & $3.16\times10^{4}$ & $9.92\times10^{3}$ & 755 & 13.4 \\[2pt]
Rotated ellipsoid, ASN-fixed+AVS & immaturity $\iota_t$ & 1.00 & 1.00 & 1.00 & 1.00 & 1.00 & 1.00 & 1.00 \\
 & distance of the mean & 7.22 & 6.97 & 6.24 & 3.64 & 1.88 & 0.405 & $3.24\times10^{-3}$ \\
 & best value & $2.21\times10^{6}$ & $4.77\times10^{5}$ & $6.78\times10^{4}$ & $1.53\times10^{4}$ & $3.18\times10^{3}$ & 114 & $3.15\times10^{-3}$ \\[2pt]
\bottomrule
\end{tabular}}
\par\vspace{3pt}\begin{minipage}{0.95\textwidth}\footnotesize Distance of the mean: largest coordinate distance between the mean of the selected solutions and the optimum. The final values of these runs are identical to those of Table \ref{S-tab:s1}.\end{minipage}
\end{table}

Tables \ref{S-tab:s1} and \ref{S-tab:s2} report the pre-registered near-boundary experiment of the skewness-only predecessor and its post hoc diagnostic: the optimum at $(5-\delta)\mathbf{1}$ in the box $[-5,5]^{D}$, where the clipping of the samples makes the selected set asymmetric. It is the only experiment of this work with an optimum near a bound of a constrained search space; the experiments of the manuscript are unconstrained. The fixed maximal skewness (no gate) was at least as good as the gated one in every setting and significantly faster in four, which agrees with family G4 of the manuscript. ASN+AVS is the skewness-only configuration (immaturity-gated skewness $\rho_{\max}=0.95$, AVS decay 0.9, thresholds $(\kappa_{0},\kappa_{1})=(2,6)$), ASN-fixed+AVS the same with the gate replaced by the fixed maximal skewness, and EMNA+AVS the Gaussian baseline.

\renewcommand{\refname}{References of the Supplementary Material}

\end{document}